\documentclass{aa}       

\usepackage{hyperref}
\usepackage{natbib}
\usepackage{graphicx}
\usepackage{txfonts}
\usepackage[normalem]{ulem}
\usepackage{xcolor}

\makeatletter
\renewcommand*\aa@pageof{, page \thepage{} of \pageref*{LastPage}}
\makeatother

\newcommand{\eP}{$e$ - $P$}
\newcommand{\beforeReferee}[1]{}
\newcommand{\afterReferee}[1]{{#1}}

\newcommand{\ignoreThis}[1]{ }
\def\quoting#1{`#1'}

\providecommand{\dt}[1]{{\tt #1}} 

\makeatletter
\DeclareRobustCommand*{\fieldName}[1]{%
  \begingroup\@fieldName\scantokens{\texttt{\small {#1}}\noexpand}\endgroup}
\begingroup\lccode`\~=`\_\relax
   \lowercase{\endgroup\def\@fieldName{\catcode`\_=\active \let~\_}}
\makeatother

\newcommand\gaia{\textit{Gaia}\xspace}
\newcommand\gdrone{\gaia~DR1\xspace}
\newcommand\gdrtwo{\gaia~DR2\xspace}
\newcommand\gdrearlythree{\gaia~EDR3\xspace}
\newcommand\gdrthree{\gaia~DR3\xspace}

\newcommand{\orcit}[1]{\protect\href{https://orcid.org/#1}{\protect\includegraphics[width=8pt]{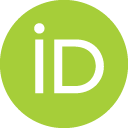}}}

\newcommand{\sourceId}[1]{\gaia DR3 #1}

\def\nss{non-single stars\xspace}\def\NSS{\nss}
\def\nssa{non-single star\xspace}\def\NSSa{\nssa}

\newcommand\GDRdocURL{\footnote{\url{https://gea.esac.esa.int/archive/documentation/GDR3/index.html}}}
\newcommand\TBOTable{\fieldName{nss_two_body_orbit}\xspace}
\newcommand\ACCTable{\fieldName{nss_acceleration_astro}\xspace}
\newcommand\NLSTable{\fieldName{nss_non_linear_spectro}\xspace}
\newcommand\VIMTable{\fieldName{nss_vim_fl}\xspace}
\newcommand\LPVTable{\fieldName{vari_long_period_variable}\xspace}
\newcommand\EBTable{\fieldName{vari_eclipsing_binary}\xspace}
\newcommand\APTable{\fieldName{astrophysical_parameters}\xspace}
\newcommand\APTableSupp{\fieldName{astrophysical_parameters_supp}\xspace}
\newcommand\MASSTable{\fieldName{binary_masses}\xspace}

\def\comments{true}

\ifx\comments\undefined
\def\NM#1{}
\def\AJ#1{}
\def\PP#1{}
\def\CB#1{}
\def\TM#1{}
\def\JSA#1{}
\def\FA#1{}
\def\EG#1{}
\else
\def\NM#1{{\color{blue} \textsl{\small $^{[NM]}[$#1]}}}
\def\AJ#1{{\color{red} \textsl{\small $^{[AJ]}[$#1]}}}
\def\PP#1{{\color{green} \textsl{\small $^{[PP]}[$#1]}}}
\def\CB#1{{\color{orange} \textsl{\small $^{[CB]}[$#1]}}}
\def\TM#1{{\color{purple} \textsl{\small $^{[TM]}[$#1]}}}

\def\JSA#1{{\color{gray} \textsl{\small $^{[JSA]}[$#1]}}}
\def\FA#1{{\color{pink} \textsl{\small $^{[FA]}[$#1]}}}
\def\EG#1{{\color{olive} \textsl{\small $^{[EG]}[$#1]}}}
\fi

\makeatletter
\renewcommand*\maketitle{%
  \thispagestyle{firstpage}
\begingroup
    \if@wideboxfn
    \setlength\bibindent{1.4\parindent}
    \else
    \setlength\bibindent{\parindent}
    \fi
    \renewcommand*\thefootnote{\@fnsymbol\c@footnote}%
    \renewcommand\@makefntext[1]{%
    \ifaa@longfn\hsize\textwidth\fi
    \noindent
    \hb@xt@\bibindent{\hss\@makefnmark\enspace}##1}
  \ifaa@twocolumn
  \begingroup
    \begin{aa@strip}
          \aa@maketitle
    \end{aa@strip}
    \@thanks	  	
  \endgroup
  \else
    \begingroup
      \let\thanks\footnote
      \aa@maketitle
    \endgroup
  \fi
\endgroup
  \setcounter{footnote}{0}%
}
\makeatother

\newcommand\hip{\textsc{Hipparcos}\xspace}
\newcommand\tyc{\textit{Tycho}}
\newcommand\tyctwo{\textit{Tycho}-2}

\newcommand\secref[1]{Sect.~\ref{#1}}

\newcommand\figref[1]{Fig.~\ref{#1}}

\newcommand\equref[1]{Eq.~\eqref{#1}}

\newcommand\tabref[1]{Table~\ref{#1}}

\bibpunct{(}{)}{;}{a}{}{,} 

\def\MK{$M_{\rm K}$}

\providecommand{\secref}[1]{Sect.~\ref{#1}}
\providecommand{\tabref}[1]{Table~\ref{#1}}
\providecommand{\figref}[1]{Fig.~\ref{#1}}
\providecommand{\equref}[1]{Eq.~\ref{#1}}

\def\deg{\degr}

\def\arcsec{\,$''$}
\def\ltsim{\ifmmode\stackrel{<}{_{\sim}}\else$\stackrel{<}{_{\sim}}$\fi}

\providecommand{\kms}{\,km\,s$^{-1}$}

\providecommand{\Msun}{\ensuremath{\,{\cal M}_{\odot}}\xspace}
\providecommand{\Mjup}{\ensuremath{\,{\cal M}_{\rm Jup}}\xspace}
\providecommand{\Mass}{\ensuremath{\mathcal{M}}}
\providecommand{\Rsun}{\ensuremath{\,\mathrm{R}_{\odot}}\xspace}

\def\a0{$A_{\rm 0}$}

\def\gmag{$G$\xspace}
\def\gbp{$G_{\rm BP}$}
\def\grp{$G_{\rm RP}$}

\def\apjl{ApJ}%
\def\apjs{ApJS}%
\def\ao{Appl.~Opt.}%
\def\aap{A\&A}%
\begin{document} 

   \title{Gaia Data Release 3: Stellar multiplicity, a teaser for the hidden treasure}



\author{
Authors: Gaia Collaboration
,         F.                         Arenou
,         C.                      Babusiaux
,       M.A.                        Barstow
,         S.                        Faigler
,         A.                       Jorissen
,         P.                       Kervella
,         T.                          Mazeh
,         N.                        Mowlavi
,         P.                        Panuzzo
,         J.                       Sahlmann
,         S.                         Shahaf
,         A.                       Sozzetti
,         N.                        Bauchet
,         Y.                       Damerdji
,         P.                         Gavras
,         P.                       Giacobbe
,         E.                         Gosset
,      J.-L.                      Halbwachs
,         B.                           Holl
,       M.G.                       Lattanzi
,         N.                        Leclerc
,         T.                          Morel
,         D.             Pourbaix$^\dagger$
,         P.                   Re Fiorentin
,         G.                       Sadowski
,         D.                  S\'{e}gransan
,         C.                         Siopis
,         D.                       Teyssier
,         T.                        Zwitter
,         L.                      Planquart
,     A.G.A.                          Brown
,         A.                      Vallenari
,         T.                         Prusti
,     J.H.J.                     de Bruijne
,         M.                       Biermann
,       O.L.                        Creevey
,         C.                      Ducourant
,       D.W.                          Evans
,         L.                           Eyer
,         R.                         Guerra
,         A.                         Hutton
,         C.                          Jordi
,       S.A.                        Klioner
,       U.L.                        Lammers
,         L.                      Lindegren
,         X.                           Luri
,         F.                        Mignard
,         C.                          Panem
,         S.                        Randich
,         P.                     Sartoretti
,         C.                       Soubiran
,         P.                          Tanga
,       N.A.                         Walton
,     C.A.L.                   Bailer-Jones
,         U.                        Bastian
,         R.                        Drimmel
,         F.                         Jansen
,         D.                           Katz
,         F.                    van Leeuwen
,         J.                         Bakker
,         C.                       Cacciari
,         J.                  Casta\~{n}eda
,         F.                      De Angeli
,         C.                      Fabricius
,         M.                      Fouesneau
,         Y.                     Fr\'{e}mat
,         L.                      Galluccio
,         A.                       Guerrier
,         U.                         Heiter
,         E.                         Masana
,         R.                       Messineo
,         C.                        Nicolas
,         K.                   Nienartowicz
,         F.                        Pailler
,         F.                         Riclet
,         W.                           Roux
,       G.M.                       Seabroke
,         R.                          Sordo
,         F.                   Th\'{e}venin
,         G.                   Gracia-Abril
,         J.                        Portell
,         M.                        Altmann
,         R.                         Andrae
,         M.                         Audard
,         I.                 Bellas-Velidis
,         K.                         Benson
,         J.                       Berthier
,         R.                         Blomme
,       P.W.                        Burgess
,         D.                       Busonero
,         G.                          Busso
,         H.                    C\'{a}novas
,         B.                          Carry
,         A.                        Cellino
,         N.                          Cheek
,         G.                     Clementini
,         M.                       Davidson
,         P.                     de Teodoro
,         M.               Nu\~{n}ez Campos
,         L.                     Delchambre
,         A.                       Dell'Oro
,         P.                         Esquej
,         J.    Fern\'{a}ndez-Hern\'{a}ndez
,         E.                         Fraile
,         D.                       Garabato
,         P.               Garc\'{i}a-Lario
,         R.                        Haigron
,       N.C.                         Hambly
,       D.L.                       Harrison
,         J.                  Hern\'{a}ndez
,         D.                     Hestroffer
,       S.T.                        Hodgkin
,         K.                     Jan{\ss}en
,         G.           Jevardat de Fombelle
,         S.                         Jordan
,         A.                  Krone-Martins
,       A.C.                      Lanzafame
,         W.                   L\"{ o}ffler
,         O.                        Marchal
,       P.M.                        Marrese
,         A.                       Moitinho
,         K.                       Muinonen
,         P.                        Osborne
,         E.                        Pancino
,         T.                        Pauwels
,         A.                   Recio-Blanco
,         C.                      Reyl\'{e}
,         M.                         Riello
,         L.                      Rimoldini
,         T.                       Roegiers
,         J.                        Rybizki
,       L.M.                          Sarro
,         M.                          Smith
,         E.                        Utrilla
,         M.                    van Leeuwen
,         U.                          Abbas
,         P.                \'{A}brah\'{a}m
,         A.                 Abreu Aramburu
,         C.                          Aerts
,       J.J.                         Aguado
,         M.                           Ajaj
,         F.                  Aldea-Montero
,         G.                      Altavilla
,       M.A.                    \'{A}lvarez
,         J.                          Alves
,         F.                         Anders
,       R.I.                       Anderson
,         E.                 Anglada Varela
,         T.                         Antoja
,         D.                         Baines
,       S.G.                          Baker
,         L.         Balaguer-N\'{u}\~{n}ez
,         E.                       Balbinot
,         Z.                          Balog
,         C.                        Barache
,         D.                        Barbato
,         M.                         Barros
,         S.                  Bartolom\'{e}
,      J.-L.                      Bassilana
,         U.                       Becciani
,         M.                     Bellazzini
,         A.                      Berihuete
,         M.                         Bernet
,         S.                        Bertone
,         L.                        Bianchi
,         A.                     Binnenfeld
,         S.                Blanco-Cuaresma
,         A.                        Blazere
,         T.                           Boch
,         A.                        Bombrun
,         D.                        Bossini
,         S.                     Bouquillon
,         A.                      Bragaglia
,         L.                       Bramante
,         E.                         Breedt
,         A.                        Bressan
,         N.                      Brouillet
,         E.                     Brugaletta
,         B.                    Bucciarelli
,         A.                        Burlacu
,       A.G.                      Butkevich
,         R.                          Buzzi
,         E.                         Caffau
,         R.                    Cancelliere
,         T.                  Cantat-Gaudin
,         R.                       Carballo
,         T.                       Carlucci
,       M.I.                      Carnerero
,       J.M.                       Carrasco
,         L.                    Casamiquela
,         M.                     Castellani
,         A.                  Castro-Ginard
,         L.                         Chaoul
,         P.                        Charlot
,         L.                         Chemin
,         V.                     Chiaramida
,         A.                      Chiavassa
,         N.                        Chornay
,         G.                      Comoretto
,         G.                       Contursi
,       W.J.                         Cooper
,         T.                         Cornez
,         S.                         Cowell
,         F.                          Crifo
,         M.                        Cropper
,         M.                         Crosta
,         C.                        Crowley
,         C.                        Dafonte
,         A.                     Dapergolas
,         P.                          David
,         P.                     de Laverny
,         F.                       De Luise
,         R.                       De March
,         J.                      De Ridder
,         R.                       de Souza
,         A.                      de Torres
,       E.F.                     del Peloso
,         E.                       del Pozo
,         M.                          Delbo
,         A.                        Delgado
,      J.-B.                        Delisle
,         C.                       Demouchy
,       T.E.                  Dharmawardena
,         S.                        Diakite
,         C.                         Diener
,         E.                      Distefano
,         C.                        Dolding
,         H.                           Enke
,         C.                          Fabre
,         M.                       Fabrizio
,         G.                       Fedorets
,         P.                       Fernique
,         F.                       Figueras
,         Y.                       Fournier
,         C.                         Fouron
,         F.                      Fragkoudi
,         M.                            Gai
,         A.               Garcia-Gutierrez
,         M.               Garcia-Reinaldos
,         M.              Garc\'{i}a-Torres
,         A.                       Garofalo
,         A.                          Gavel
,         E.                        Gerlach
,         R.                          Geyer
,         G.                        Gilmore
,         S.                         Girona
,         G.                      Giuffrida
,         R.                          Gomel
,         A.                          Gomez
,         J.     Gonz\'{a}lez-N\'{u}\~{n}ez
,         I.    Gonz\'{a}lez-Santamar\'{i}a
,       J.J.             Gonz\'{a}lez-Vidal
,         M.                        Granvik
,         P.                       Guillout
,         J.                        Guiraud
,         R.      Guti\'{e}rrez-S\'{a}nchez
,       L.P.                            Guy
,         D.                 Hatzidimitriou
,         M.                         Hauser
,         M.                        Haywood
,         A.                         Helmer
,         A.                          Helmi
,       M.H.                      Sarmiento
,       S.L.                        Hidalgo
,         N.                    H\l{}adczuk
,         D.                          Hobbs
,         G.                        Holland
,       H.E.                         Huckle
,         K.                        Jardine
,         G.                     Jasniewicz
,         A.           Jean-Antoine Piccolo
,     \'{O}.             Jim\'{e}nez-Arranz
,         J.              Juaristi Campillo
,         F.                          Julbe
,         L.                      Karbevska
,         S.                         Khanna
,         G.                     Kordopatis
,       A.J.                           Korn
,      \'{A}                 K\'{o}sp\'{a}l
,         Z.            Kostrzewa-Rutkowska
,         K.                 Kruszy\'{n}ska
,         M.                            Kun
,         P.                        Laizeau
,         S.                        Lambert
,       A.F.                          Lanza
,         Y.                          Lasne
,      J.-F.                     Le Campion
,         Y.                       Lebreton
,         T.                      Lebzelter
,         S.                         Leccia
,         I.                  Lecoeur-Taibi
,         S.                           Liao
,       E.L.                         Licata
,     H.E.P.                   Lindstr{\o}m
,       T.A.                         Lister
,         E.                        Livanou
,         A.                          Lobel
,         A.                          Lorca
,         C.                           Loup
,         P.                  Madrero Pardo
,         A.                Magdaleno Romeo
,         S.                        Managau
,       R.G.                           Mann
,         M.                       Manteiga
,       J.M.                       Marchant
,         M.                        Marconi
,         J.                         Marcos
,     M.M.S.                  Marcos Santos
,         D.                 Mar\'{i}n Pina
,         S.                       Marinoni
,         F.                        Marocco
,       D.J.                       Marshall
,         L.                    Martin Polo
,       J.M.             Mart\'{i}n-Fleitas
,         G.                         Marton
,         N.                           Mary
,         A.                          Masip
,         D.                        Massari
,         A.           Mastrobuono-Battisti
,       P.J.                       McMillan
,         S.                        Messina
,         D.                       Michalik
,       N.R.                         Millar
,         A.                          Mints
,         D.                         Molina
,         R.                       Molinaro
,         L.                     Moln\'{a}r
,         G.                         Monari
,         M.                    Mongui\'{o}
,         P.                    Montegriffo
,         A.                        Montero
,         R.                            Mor
,         A.                           Mora
,         R.                     Morbidelli
,         D.                         Morris
,         T.                       Muraveva
,       C.P.                         Murphy
,         I.                        Musella
,         Z.                           Nagy
,         L.                          Noval
,         F.                      Oca\~{n}a
,         A.                          Ogden
,         C.                      Ordenovic
,       J.O.                         Osinde
,         C.                         Pagani
,         I.                         Pagano
,         L.                      Palaversa
,       P.A.                        Palicio
,         L.                Pallas-Quintela
,         A.                         Panahi
,         S.                Payne-Wardenaar
,         X.          Pe\~{n}alosa Esteller
,         A.                  Penttil\"{ a}
,         B.                         Pichon
,       A.M.                     Piersimoni
,      F.-X.                         Pineau
,         E.                         Plachy
,         G.                           Plum
,         E.                         Poggio
,         A.                       Pr\v{s}a
,         L.                         Pulone
,         E.                         Racero
,         S.                        Ragaini
,         M.                         Rainer
,       C.M.                        Raiteri
,         P.                          Ramos
,         M.                   Ramos-Lerate
,         S.                         Regibo
,       P.J.                       Richards
,         C.                      Rios Diaz
,         V.                         Ripepi
,         A.                           Riva
,      H.-W.                            Rix
,         G.                          Rixon
,         N.                       Robichon
,       A.C.                          Robin
,         C.                          Robin
,         M.                        Roelens
,     H.R.O.                         Rogues
,         L.                     Rohrbasser
,         M.               Romero-G\'{o}mez
,         N.                         Rowell
,         F.                          Royer
,         D.                     Ruz Mieres
,       K.A.                        Rybicki
,         A.         S\'{a}ez N\'{u}\~{n}ez
,         A.        Sagrist\`{a} Sell\'{e}s
,         E.                       Salguero
,         N.                        Samaras
,         V.                Sanchez Gimenez
,         N.                          Sanna
,         R.                  Santove\~{n}a
,         M.                        Sarasso
,       M.S.                     Schultheis
,         E.                        Sciacca
,         M.                          Segol
,       J.C.                        Segovia
,         D.                         Semeux
,       H.I.                       Siddiqui
,         A.                        Siebert
,         L.                        Siltala
,         A.                        Silvelo
,         E.                         Slezak
,         I.                         Slezak
,       R.L.                          Smart
,       O.N.                         Snaith
,         E.                         Solano
,         F.                        Solitro
,         D.                         Souami
,         J.                        Souchay
,         A.                         Spagna
,         L.                          Spina
,         F.                          Spoto
,       I.A.                         Steele
,         H.             Steidelm\"{ u}ller
,       C.A.                     Stephenson
,         M.                   S\"{ u}veges
,         J.                         Surdej
,         L.                       Szabados
,         E.                   Szegedi-Elek
,         F.                          Taris
,       M.B.                         Taylor
,         R.                       Teixeira
,         L.                        Tolomei
,         N.                        Tonello
,         F.                          Torra
,         J.                Torra$^\dagger$
,         G.                 Torralba Elipe
,         M.                      Trabucchi
,       A.T.                        Tsounis
,         C.                          Turon
,         A.                           Ulla
,         N.                          Unger
,       M.V.                       Vaillant
,         E.                     van Dillen
,         W.                     van Reeven
,         O.                          Vanel
,         A.                      Vecchiato
,         Y.                          Viala
,         D.                        Vicente
,         S.                      Voutsinas
,         M.                         Weiler
,         T.                         Wevers
,      \L{}.                    Wyrzykowski
,         A.                         Yoldas
,         P.                          Yvard
,         H.                           Zhao
,         J.                          Zorec
,         S.                         Zucker
}


   \date{ }

\abstract
{
The \gdrthree Catalogue contains for the first time about eight hundred thousand solutions with either orbital elements or trend parameters for astrometric, spectroscopic and eclipsing binaries, and combinations of them. 
}
{
This paper aims to illustrate the huge potential of this large non-single star catalogue.
}
{
Using the orbital solutions together with models of the binaries, a catalogue of tens of thousands of stellar masses, \afterReferee{or lower limits}, partly together with consistent flux ratios, has been built. Properties concerning the completeness of the binary catalogues are discussed, statistical features of the orbital elements are explained and a comparison with other catalogues is performed.
}
{
Illustrative applications are proposed for binaries across the H-R diagram. The binarity is studied in the RGB/AGB and a search for genuine SB1 among long-period variables is performed. The discovery of new EL CVn systems illustrates the potential of combining variability and binarity catalogues. Potential compact object companions are presented, mainly white dwarf companions or double degenerates,
but one candidate neutron star is also presented.  
\beforeReferee{Towards the bottom of the main sequence, new binary ultracool dwarf candidates are discovered with their masses estimated.}
\afterReferee{Towards the bottom of the main sequence, the orbits of previously-suspected binary ultracool dwarfs are determined and new candidate binaries are discovered.}
The long awaited contribution of Gaia to the analysis of the substellar regime shows the brown dwarf desert around solar-type stars using true, rather than minimum, masses, and provides new important constraints on the occurrence rates of substellar companions to M dwarfs. 
\beforeReferee{Two new exoplanets are found and several dozens of candidates are identified, including one super-Jupiter orbiting a white dwarf.} 
\afterReferee{Several dozen new exoplanets are proposed, including two with validated orbital solutions and one super-Jupiter orbiting a white dwarf, all being candidates requiring confirmation.} 
Beside binarity,  higher order multiple systems are also found.
}
{
By increasing by more than one order of magnitude the number of known binary orbits, \gdrthree will provide, beside a rich reservoir of dynamical masses, an important contribution to the analysis of stellar multiplicity.
}

   
   \titlerunning{{\gdrthree} -- Stellar multiplicity} 
   \authorrunning{Gaia Collaboration et al.}

   \maketitle
 
 \ignoreThis{\color{blue}\tiny {\bf Note to the referee} : ``{\it Performance verification papers 
 by the full \gaia Collaboration give an overview of the science potential of the data release in a particular
 science domain. Without going into in-depth analysis, these papers provide a short introduction to the selected
 science topic and hence merely demonstrate the scientific quality of the data through some examples. 
 Performance verification papers are neither exhaustive nor definitive treatments of the science topics they
 address.}'' (extract from the Gaia documentation). As agreed with A\&A, we are submitting this paper based 
 on nearly final version of the Gaia catalogue to ensure the papers can appear on time together with the release. The expected changes would be minor, e.g. the correct references to the other A\&A DR3 papers or the full list of names of all DPAC authors.}

%
\section{Introduction}
%
The success of \gaia \citep{DR1-DPACP-18}, with parallaxes for around 1.5 billion sources, could overshadow how difficult the path to measuring the first stellar distances was. The two millennia during which this research has been unsuccessfully carried out have been littered with unrelated but equally fundamental discoveries. In particular, Herschel, following the suggestion by Ramponi to Galileo in 1611 \citep{2005JHA....36..251S}, observed pairs of stars in order to measure their differential parallaxes, but did not succeed. Instead, what he demonstrated for the first time, in 1802, was the existence of orbits for these stars, changing their nature from unrelated double stars to binaries, proving that the law of gravitation was universal.

After Bessel had obtained the first convincing parallax measurement in 1838, he also deduced in 1844 from the nonlinear proper motion of Sirius and Procyon that there could exist not only visual binaries but also invisible, now called astrometric, binaries. Astrometry and binarity have then been intimately linked from the start. Indeed, it was not until much later, by observing the periodic Doppler shift of Algol's lines, that Vogel correctly deduced in 1889 that this was due to its orbital motion, making Algol the first spectroscopic binary. In fact, Algol had been suspected by John Goodricke in 1782 to be periodically eclipsed, making this star also the first eclipsing binary \citep{1995hatp.book.....L}.

Since then, it has been realised that binary stars were not only important to derive their physical properties but also due to their fundamental role in stellar evolution; understanding the statistical properties of binary and multiple stars is then an important ingredient for the knowledge of our Galaxy. The properties of companions down to the substellar regime is another key to understanding stellar formation. Unfortunately, until now, too small samples, selection effects, and also the absence of the required astrometric precision, have always complicated the analysis of the various existing ground-based data. 

As a large survey, \gaia should be in an ideal place to bring a newer, much broader perspective to these fundamental topics.
What makes \gaia so unique is its ability to find, and above all to parameterise, most types of binaries at once, whether visual, astrometric, spectroscopic or eclipsing and even through stellar parametrisation, with a remarkable homogeneity of epoch, level of calibration accuracy, data reduction and process organisation. 

The \gaia precursor, \hip, had already discovered and measured double stars \citep{1997ESASP.402...13L}, mostly resolved ones but also several categories of unresolved astrometric binaries, which allowed to determine stellar masses \citep{1997ESASP.402..251S, 1997A&AS..122..571M} though for a small number of sources only. 

With the successive \gdrone \citep{DR1-DPACP-8}, DR2 \citep{DR2-DPACP-36} then EDR3 \citep{EDR3-DPACP-130}, the multiple stars had not yet been handled, the difficulty to analyse the single stars being already a challenge, these successive releases representing the improvement of the calibrations and source analysis. This does not mean that non-single stars were absent. Whether double or binaries, they are indeed present and processing them as single stars seriously degrades their results, although, fortunately, several flags in the \gaia Catalogue inform about the potential duplicity. 
Beside, the combination of these first \gaia releases with \hip data already allowed to detect long period binaries \citep{Kervella2019,2022A&A...657A...7K,Brandt2021}.

The advent of \gdrthree \citep{DR3-DPACP-185} now presents impressive new data products among which, to quote a few only, variability \citep{DR3-DPACP-162}, radial velocities \citep{DR3-DPACP-159}, astrophysical parameters \citep{DR3-DPACP-157} determined using either high-resolution (RVS) and low-resolution data \citep[BP-RP photometers,][]{EDR3-DPACP-118}, for a very large fraction of the Catalogue. \gdrthree also contains the first analysis of the unresolved binary star contents covering the typical binary classes (astrometric, spectroscopic, photometric) presented into several tables: two-body orbits, astrometric \afterReferee{or} spectroscopic accelerations, variable binaries. These tables contain the orbital or trend parameters of the found binaries. Above all this offers the prospect of deriving the physical properties of the individual components. Marginally, this should also improve the measurements of these systems in the main catalogue, with better astrometric parameters or systemic radial velocity.

Although the maturity of the analysis of \gaia data now makes it possible to obtain for the first time a multi-type catalogue of binaries much larger than has been compiled with difficulty over the previous centuries, it must be stressed that only a small fraction of the binary content of the main catalogue has been analysed for DR3. This data analysis is described in the documentation \citep{NSS-DR3-documentation}\GDRdocURL and the articles accompanying this data release, namely  \cite{DR3-DPACP-163, DR3-DPACP-176, DR3-DPACP-178, DR3-DPACP-170, DR3-DPACP-179}. 

The purposes of this publication are manyfold. It is first intended to describe the possible use cases of the catalogue, illustrating in particular the potential complementarity of the different data processing chains. This is essentially an appetiser that shows the quality of the data, highlighting the basic results that can be readily obtained, in particular estimating masses, these ones not being part of the non-single star tables. Beside, this performance verification paper acts as a final validation step before releasing the data. It is outside of the scope to explore the data in detail, nor to compare it to models, nor to confirm candidates of various kind, as this will be the goal of the scientific exploitation by the community, but we wish to facilitate this exploitation by describing what has been learned from the analysis so far.

We start by describing the data content, then useful statistical properties are clarified together with what is known about the selection function. We then focus on orbits, not acceleration solutions (for astrometry), nor trend solutions (for spectroscopic binaries) and propose for these \afterReferee{orbital solutions} a catalogue of masses. From this, we present an impressionistic panorama of the potential that lies inside this catalogue, concerning basic statistical properties and candidate sources of various types e.g. EL CVn, compact companions, white dwarfs or high mass ones, then ultracool dwarfs and substellar companions. Multiple systems are finally discussed.

%
\section{Data description}\label{ssec:data}
%
%
\subsection{Table contents}\label{sec:data_content}
%
The non-single star (NSS) tables are presented more by type of solution or period range rather than by kind of binaries. The first of the four tables,  \TBOTable, contains the orbital parameters for all three categories, astrometric, spectroscopic or eclipsing binaries, all being unresolved. 
The table \ACCTable contains accelerations or derivatives of it for sources that have an astrometric motion better described using a quadratic or cubic rather than a linear proper motion. Similarly, the \NLSTable are trend (long period) solutions of spectroscopic binaries.
The solutions in the \VIMTable table are different in that the photocentre displacement due to the photometric variability of one component of fixed binaries required the correction of the astrometric parameters (variable-induced movers fixed, \fieldName{VIMF}). A summary of the solutions 
is given \tabref{NSS-Table}.

\begin{table*}
 \caption{Content of the four non-single star tables.\label{NSS-Table}}
 \centering
  \begin{tabular}{llrl}
  \hline
  \hline
   Table & \fieldName{nss_solution_type} & Solutions & \multicolumn{1}{c}{Description } \\ 
  \hline 
\ACCTable &          \fieldName{Acceleration7} & 246\,947 & Second derivatives of position (acceleration)\\
          &          \fieldName{Acceleration9} &  91\,268 & Third derivatives of position (jerk)\\
\TBOTable &                \fieldName{Orbital} & 134\,598 & Orbital astrometric solutions\\
          & \fieldName{OrbitalAlternative*}    &      629 & Orbital astrometric, alternative solutions\\
          & \fieldName{OrbitalTargetedSearch*} &      533 & Orbital astrometric, supplementary external input list \\
          &        \fieldName{AstroSpectroSB1} &  33\,467 & Combined orbital astrometric + spectroscopic solutions \\
          & \fieldName{SB1} or \fieldName{SB2} & 186\,905 & Orbital spectroscopic solutions \\
          &       \fieldName{EclipsingSpectro} &      155 & Combined orbital spectroscopic + eclipsing solutions \\
          &        \fieldName{EclipsingBinary} &  86\,918 & \afterReferee{Orbits of} eclipsing binaries \\
\NLSTable &    \fieldName{FirstDegreeTrendSB1} &  24\,083 & First order derivatives of the radial velocity\\
          &   \fieldName{SecondDegreeTrendSB1} &  32\,725 & Second order derivatives of the radial velocity\\
\VIMTable &                   \fieldName{VIMF} &      870 & Variable-induced movers fixed\\
  \hline 
\end{tabular}
 \tablefoot{
 The number of solutions is larger than the number of sources. The type of solution \fieldName{OrbitalAlternative*} indicates solutions which are either \fieldName{OrbitalAlternative} or \fieldName{OrbitalAlternativeValidated}.}
\end{table*}

The astrometric orbits in the \TBOTable table have a \fieldName{nss_solution_type} 
name starting with \fieldName{Orbital*} and the orbital parameters are described at Appendix \ref{ssec:model_parameters_astrometry}. The spectroscopic binaries with either one component being
parametrised (\fieldName{SB1}) or both (\fieldName{SB2}) have their parameters described at Appendix  \ref{ssec:model_parameters_spectroscopy}
and short periods may have a circular solution (\fieldName{nss_solution_type} = \fieldName{SB1C}).
As a source may simultaneously be e.g. astrometric binary and spectroscopic binary, combined
solutions have been computed in some cases (\fieldName{nss_solution_type} = \fieldName{AstroSpectroSB1}). 
For the same reason, the \fieldName{EclipsingSpectro} solutions
are combinations of eclipsing and spectroscopic solutions.
However, when no combination has 
been done, then two solutions for the same source may be present in 
the \TBOTable table, i.e. a query by \fieldName{source_id} may return several solutions.
These multiple solutions may indicate either \beforeReferee{ternary}\afterReferee{triple} systems, or some inconsistency that users
may wish to sort out, and then possibly combine \afterReferee{these solutions offline}.

For the same reason, some sources may also have solutions in several tables simultaneously. 
To take an example, there are 160 eclipsing
binaries having also a \fieldName{VIMF} solution. As the \fieldName{VIMF} model should have improved their astrometric solution,
and distance of eclipsing binaries is of interest, this solution should in principle be preferred 
over the one given in the \fieldName{gaia_source} table.

This potential multiplicity of solutions for a given source
explains the total number of unique NSS sources being 813\,687 while the total number of NSS
solutions is larger, 839\,098. 

\begin{figure}[htb]\begin{center}
\includegraphics[width=1\columnwidth]{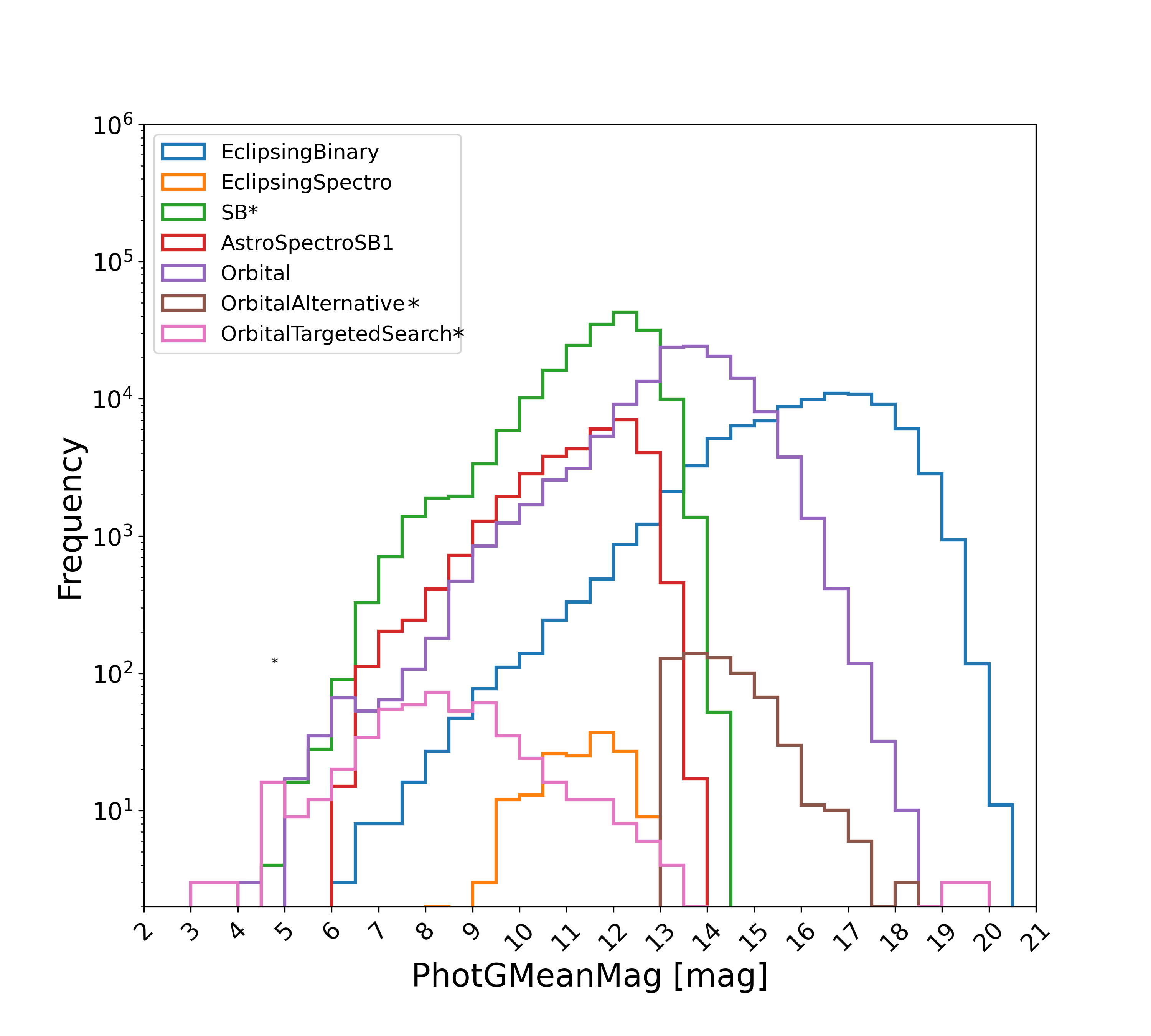}
\caption{Magnitude distribution for each solution type in the \TBOTable table.}\label{fig:dist-TBO-mag}
\end{center}\end{figure}

The distributions of the various orbital solutions with magnitude are shown \figref{fig:dist-TBO-mag}. As expected, 
the brightest are the \afterReferee{SB1 and SB2}, and consequently also their intersection with astrometric binaries, \fieldName{AstroSpectroSB1}, 
and with eclipsing binaries, \fieldName{EclipsingSpectro}. The orbital astrometric binaries, brighter than $G<19$, peak at $G \approx 14$ 
while the \fieldName{OrbitalTargetedSearch} span the entire magnitude range as the sources were given as input list.
The eclipsing binaries are the faintest. 
Note that the NSS eclipsing binaries are a small subset of the ones detected by photometry \citep{DR3-DPACP-170}, 
for which an orbital solution has been computed \citep{DR3-DPACP-179}, 
cf. the much more complete \fieldName{vari_eclipsing_binary} table.

\begin{figure}[htb]\begin{center}
\includegraphics[width=1\columnwidth]{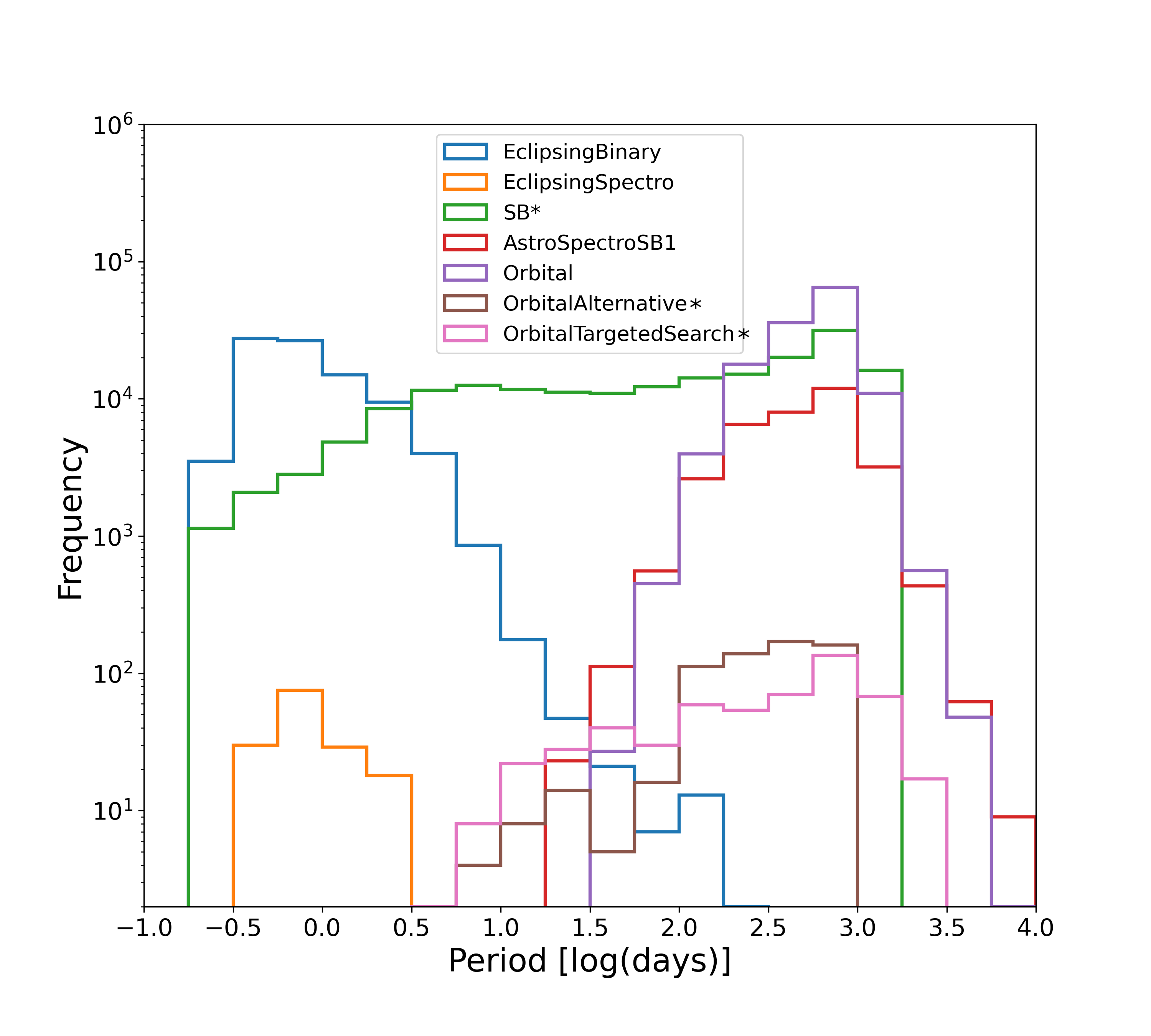}
\caption{Number of solutions for each solution type in the \TBOTable table as a function of period.}\label{fig:dist-TBO-period}
\end{center}\end{figure}

\begin{figure}[htb]
\includegraphics[width=1.05\columnwidth]{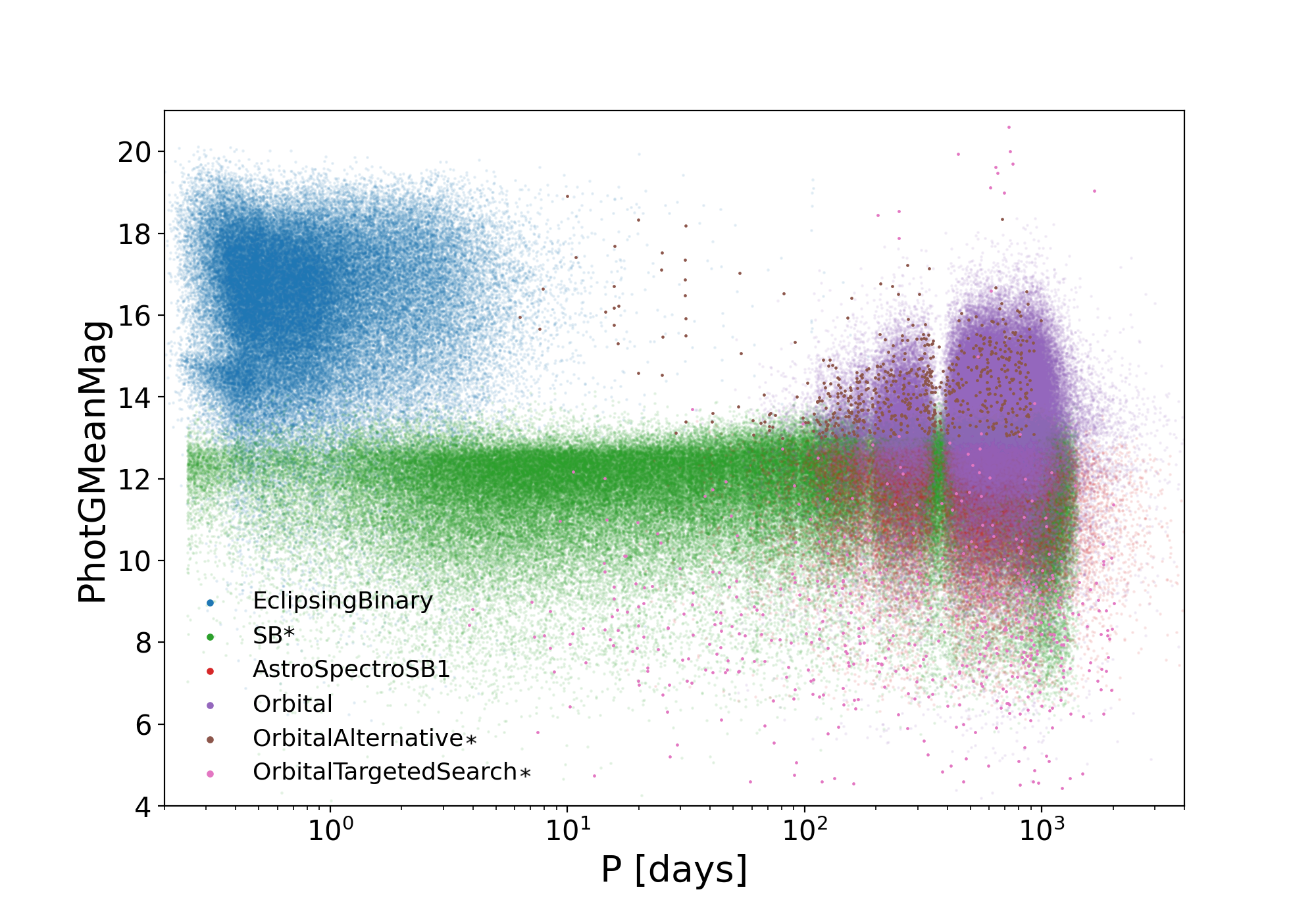}
\caption{\afterReferee{$G$ apparent magnitude vs period in the \TBOTable table.}}\label{fig:GlogP_TBO}
\end{figure}

The distribution of periods, by construction restricted to the \TBOTable table, is depicted in \figref{fig:dist-TBO-period}.
The short period eclipsing and long period astrometric binaries are nicely bridged by the SBs. In clear, within a few years \gaia
has covered the impressive $0.28-1500$ day period range (99\% CI) for thousands
of sources, that should prove very valuable for the statistics of binary properties.
\afterReferee{The coverage in the joint distribution of period and magnitude is qualitatively illustrated  \figref{fig:GlogP_TBO}}.

\begin{figure*}\begin{center}
\includegraphics[width=0.33\textwidth]{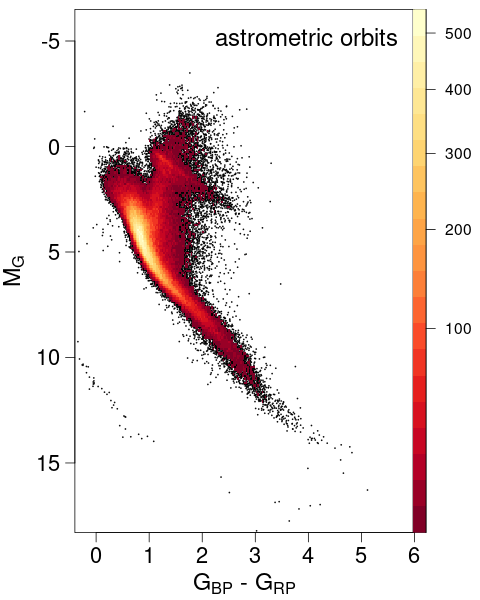}
\includegraphics[width=0.33\textwidth]{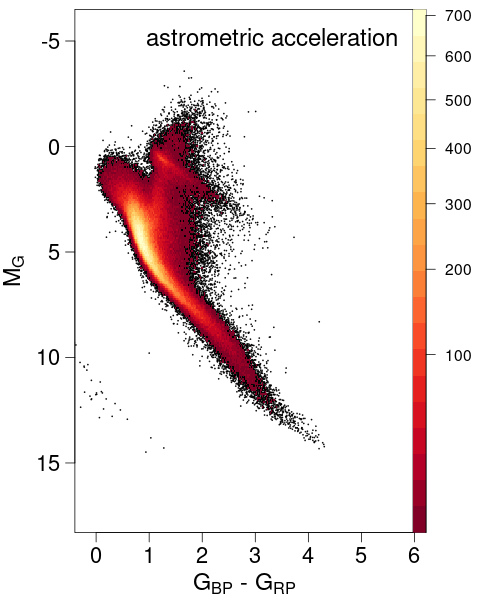}
\includegraphics[width=0.33\textwidth]{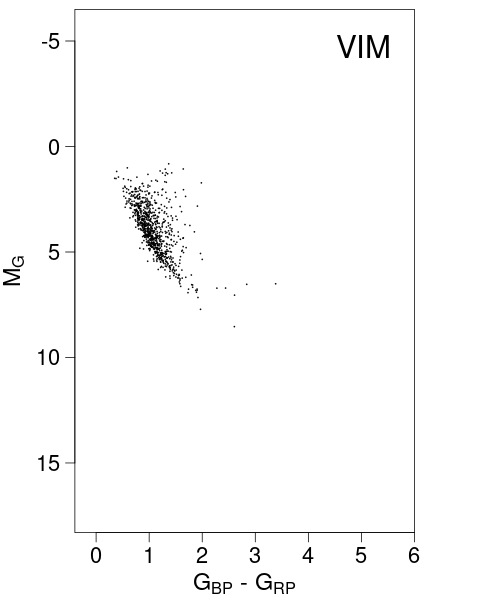}
\includegraphics[width=0.33\textwidth]{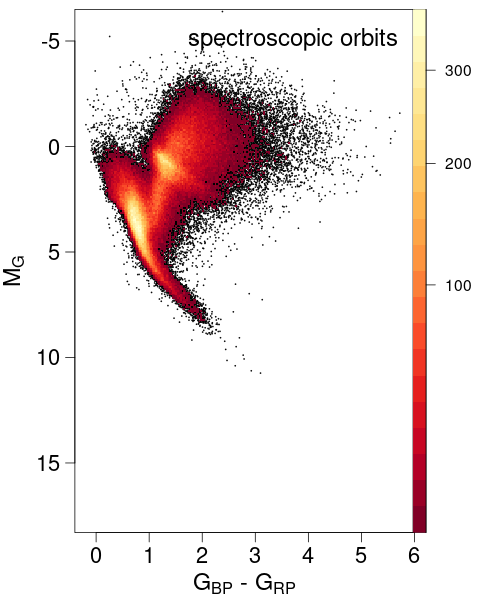}
\includegraphics[width=0.33\textwidth]{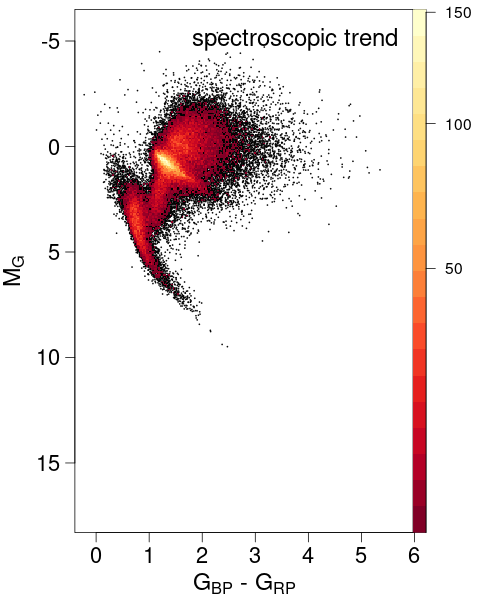}
\includegraphics[width=0.33\textwidth]{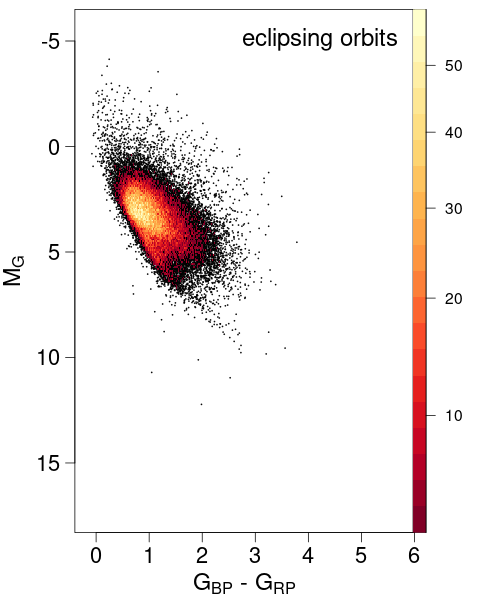}
\caption{\gaia\ H-R diagram, uncorrected for extinction, for all NSS solutions with a relative parallax error better than 20\%. No selection is done on the photometric quality. The colour scale represents the square root of the relative density of stars. 
{\it Top}: Astrometric binaries, (a): \afterReferee{\fieldName{all \fieldName{Orbital*} solutions plus \fieldName{AstroSpectroSB1}}}, (b): \fieldName{Acceleration} solutions, (c): \fieldName{VIMF}; 
{\it Bottom}: Spectroscopic binaries with (d): \fieldName{SB*} orbits and (e): \fieldName{NonLinearSpectro}, (f): eclipsing binaries. } 
\label{fig:hrds}
\end{center}\end{figure*}

The Hertzsprung-Russell diagrams (HRD) for the various categories are represented  \figref{fig:hrds}, for sources with a parallax S/N $>5$. Note that the used parallax
is the one from the NSS solution for what concerns the putative astrometric binaries,
while it is the one from the main catalogue for spectroscopic and eclipsing binaries.

%
\subsection{Table construction}\label{sssec:cu4nss_intro_inputfiltering}
%
Although we refer to the online documentation and the articles accompanying this data release
for a detailed understanding of the data processing, it is of interest to
describe how the \NSSa data have been obtained, starting with their input data selection,
as this is one first key to understand the \NSSa selection function.

The basic NSS processing selected its input sources from those that had a
bad goodness of fit (GoF) in the upstream results, either in the astrometric or in the spectroscopic processing, or that were detected as eclipsing variables; the
only exception is the \fieldName{OrbitalTargetedSearch}, \secref{sec:alternativeOrb},
where a predefined source list was given as input to the astrometric orbital fit, irrespective of their actual GoF in the single star solution.

\subsubsection{Astrometric binaries, main processing}\label{sssec:cu4nss_intro_inputfiltering_astro}

As \gdrthree is the first publication of NSS solutions, it was decided to
limit the content to the most significant ones, this criterion being relaxed for further releases, and the motivation for this will appear later.
The definition of the input source list started after \gdrtwo, where it was assumed that
the sources with \fieldName{ruwe} $>1.4$ represented a reasonable threshold for sources
with problematic astrometric solutions\footnote{cf. \href{https://gea.esac.esa.int/archive/documentation/GDR2/Gaia_archive/chap_datamodel/sec_dm_main_tables/ssec_dm_ruwe.html}{Gaia DR2 documentation}}. More recent analysis \citep[e.g.][]{2021arXiv211110380P} may now suggest that a
lower threshold could have been chosen, but this value also had the advantage to 
decrease the processing requirements.

To this \fieldName{ruwe} $> 1.4$ criterion, $G<19$ was added in order to keep the best signal-to-noise.
Obviously, the sample defined like this contains many contaminants, partially resolved
rather than unresolved sources.
In particular, for a double star with a projected separation between components
between $\approx 9$ mas and $\approx 0.27$\arcsec \citep{LL:LL-136}, depending on the magnitude difference,
the epoch position is not exactly on the photocentre\footnote{We coined the word \quoting{Gaiacenter} 
in \cite{2022A&A...657A...7K} by analogy with the \quoting{Hippacenter} defining the actual pointing 
of the epoch \hip observations of double stars \citep{1997A&AS..122..571M}}, so that
the astrometry of such sources is perturbed and the source is likely to have been selected.

Consequently the criteria \fieldName{ipd_frac_multi_peak} $\leq 2$ was added to avoid double stars
with a large separation and \fieldName{ipd_gof_harmonic_amplitude} $<0.1$ to reject pairs 
with smaller separations. The \fieldName{visibility_periods_used} $> 11$ criteria was also added
($>12$ for orbital solutions), in order to avoid spurious solutions. 
\footnote{Although the DoF is still large enough as there are on the average 
about 18 astrometric observations per visibility period, one may still be cautious
with solutions having a low number of visibility periods.} 

It was found however that the sample was still polluted so another criterion was
added, this time based on photometry, trying to avoid sources with light being contaminated
by a neighbour. For this purpose, it was made use of
the corrected BP and RP flux excess factor $C^*$ associated to its uncertainty $\sigma_{C^*}(G)$ 
as defined by \citet[][Eqs. 6 and 18]{EDR3-DPACP-117}, drastically keeping
sources with $|C^*|< 1.645 \sigma_{C^*}$ only.

\subsubsection{Astrometric binaries, alternative processing}\label{sec:alternativeOrb}
As described in \cite{DR3-DPACP-176}, alternative orbit determination algorithms have been run on two different input lists.
The first one is based on astrometric binaries that could not be successfully modelled by any model 
in the main processing pipeline, for which a more complex handling was attempted, \fieldName{nss_solution_type} = \fieldName{OrbitalAlternative*}. These sources originate from the same list as described in \secref{sssec:cu4nss_intro_inputfiltering_astro}.
The second one is constituted by a sample of sources with detected companions published in the literature,
\fieldName{nss_solution_type} = \fieldName{OrbitalTargetedSearch*}, where all sources have been kept for processing.

\subsubsection{Spectroscopic binaries}\label{sssec:cu4nss_intro_inputfiltering_spectro}
The selection of the sources that had to be treated by the spectroscopic binary pipeline was based on sources with enough measurements, a dispersion of these measurements large enough, rejecting stars of stellar type outside of range, viz. 
\fieldName{rv_renormalised_gof} $> 4$ and \fieldName{rv_nb_transits} $\geq 10$ 
and 3875 K $< $\fieldName{rv_template_teff} $< 8125$ K, or detected as SB2.

One may notice that there are more than six thousands sources
with a SB solution that have no average radial velocity in the
\fieldName{gaia_source} main catalogue. In that case (as in the other cases
where a SB solution is given), the \fieldName{center_of\_mass_velocity}
gives the systemic velocity. The absence of a mean RV for what concerns SB2s is normal, 
as the main spectroscopic processing did not compute this value. For SB1s,
it may be useful to note that the computation had not been performed 
for the sources that were considered either peculiar, or potentially SB2, 
too hot or with emission lines.
Consequently, when some SB results appear doubtful, it may then be useful to check
whether \fieldName{gaia_source.radial_velocity is NULL} for these sources.
More details are given in \citet{DR3-DPACP-178}.

\subsubsection{Eclipsing binaries}\label{sssec:cu4nss_intro_inputfiltering_eclipsing}
The input list for candidate eclipsing binaries contained about two millions 
sources that can be found in the \EBTable \gdrthree table. 
\afterReferee{Their selection is} described in \citet{DR3-DPACP-170}; see also the online documentation.
The selection of the subset therein for which an orbital solution has been computed is described in \citet{DR3-DPACP-179}.

%
\subsection{Output filtering}\label{sssec:cu4nss_intro_outputfiltering}
%
Once the first processing results were analysed, it appeared that the cleaning of the input
list had still left a very large fraction of spurious solutions.
This is why it was decided to keep the most significant solutions for \gdrthree:
a general filter was applied to keep 
those with goodness of fit smaller than 50 and significance $>5$
($>2$ for \fieldName{OrbitalTargetedSearch*}).  The \fieldName{significance} is 
computed as the S/N ratio of the semi-major axis for astrometric orbits, ($a_0/\sigma_{a_0}$),
on the S/N ratio of the acceleration module for acceleration solutions,
and of the semi-amplitude for spectroscopic binaries, ($K_1/\sigma_{K_1}$).
Supplementary filtering was applied during the processing or at post-processing
level as described for the various models below.

\paragraph{Astrometric binaries, acceleration solutions: }

One could naively hope that the estimated accelerations would allow to detect
binaries of intermediate period and provide some useful information about the binary,
e.g. the minimum mass producing the given acceleration on the primary. 
The situation appears actually more complex. The acceleration values themselves
are not discussed, and it can be seen that these solutions improve the baseline
solution, e.g. looking at a slightly thinner giant branch for an HRD produced using the
parallaxes from the acceleration solution, compared to those from the  main catalogue.
What is at stake is the interpretation of the acceleration term.

What happens is the combination of two effects. The first one 
originates from the organisation of
the NSS processing: acceleration solutions were attempted before any other solutions,
and kept if significant enough with a reasonable GoF. The (unwanted) effect is that some
solutions which could have received a full orbit parametrisation were not attempted and appear in the
NSS catalogue with an acceleration solution instead. The second effect is that an acceleration
term can be significant even for short periods or very long periods. This is
demonstrated by the analysis in \cite{LL:LL-136}. 

The following filtering has been applied \citep[see documentation and][for details]{DR3-DPACP-163}:
the sources which have been kept were those with significance $s>20$ and
$\varpi/\sigma_\varpi \; > \; 1.2 \; s^{1.05}$ and GoF $< 22$
for \fieldName{Acceleration7} and 
$\varpi/\sigma_\varpi \; > \; 2.1 \; s^{1.05}$and GoF $< 25$
for \fieldName{Acceleration9}.

Nevertheless, it is known that a large fraction of the acceleration solutions 
are not intermediate period binaries as one would expect, rather short or long periods instead.

\paragraph{Astrometric binaries, \fieldName{Orbital} solutions: }

The processing of orbital solutions starts by a period search.
Unfortunately, this may lead to the detection of periods related to the scan law, 
rather than due to some true periodic motion: partially resolved objects with fixed position may
give a signal depending on the scanning angle with respect to the orientation of the pair.
This problem is fully analysed in \cite{DR3-DPACP-164}. Consequently, most detected periods
below $\approx 100$ days were wrong, leading to solutions with huge and wrong mass functions.

\afterReferee{To circumvent this,} the following filtering was adopted \citep{DR3-DPACP-163}: parallax S/N $> 20000/\fieldName{period}$, 
\fieldName{significance} $s = a_0/\sigma_{a_0} > 5$ and $s > 158/\sqrt{\fieldName{period}}$,
\fieldName{eccentricity_error} $< 0.182*\log_{10}(\fieldName{period}) - 0.244$.

\paragraph{Astrometric binaries, alternative processing: }

Aggressive post-processing filtering approaches for both samples produced subsets of solutions that were assigned \fieldName{OrbitalAlternative*} and \fieldName{OrbitalTargetedSearch*} solution types, respectively, in the \gdrthree archive. For both cases, subsamples of sources that passed a variety of validation procedures were further assigned \fieldName{OrbitalAlternativeValidated} and \fieldName{OrbitalTargetedSearchValidated} solution types \citep[See][for details]{DR3-DPACP-176}. 

Inspection of the \fieldName{OrbitalAlternative} solutions reveals that the caveat of unrealistically large inferred companion masses at short orbital periods is not entirely removed. A few percent of spurious solutions still likely contaminates this sample.

\paragraph{Spectroscopic binaries: }

Only the sources with
GoF $<3$, $|\fieldName{center_of_mass_velocity}| < 1000$\kms, $K_1<250$\kms and \fieldName{efficiency} > 0.1
were kept, \afterReferee{where \fieldName{efficiency} is a measure of the correlation between parameters}..

One of the most important problems found after processing was 
the presence of many spurious SB detections 
with short periods. For this reason, the lower confidence threshold on the period 
was adapted depending on the period itself: it was set to 0.995 for $P<1$ d, 0.95 for $P>10$ d,
and $-0.045\log P + 0.995$ in between. For details on this and other
filtering during the spectroscopic processing, see \cite{DR3-DPACP-178}.

Despite all this, the comparison of NSS results with catalogues of known binaries shows that 
for a few percent of the SB1 solutions the period may still be wrong, mainly due to the sparse time sampling. When these sources have both an \fieldName{SB1} and \fieldName{Orbital} solution, such cases may be spotted 
by comparing the respective semi-amplitudes. Short periods with large \fieldName{ruwe} (e.g. $>1.4$) are frequently suspect; some may be the inner system of a \beforeReferee{ternary}\afterReferee{triple} system, but most may instead be some kind of 
aliases of a longer orbital solution.

\begin{figure}\begin{center}
\includegraphics[width=0.9\columnwidth]{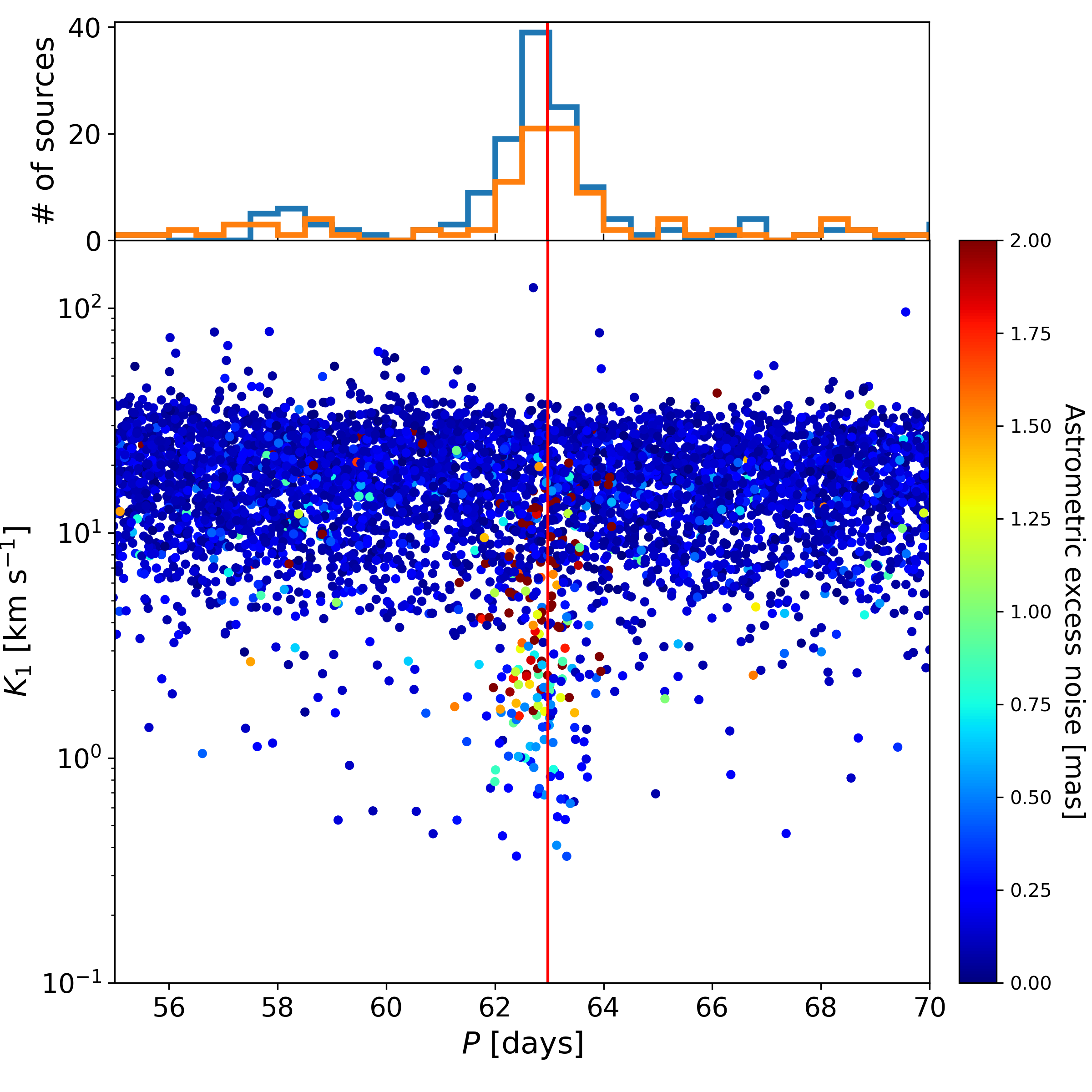}
\end{center}
\caption{The $K_1$ semi-amplitude versus period diagram of \fieldName{SB1} solutions, colour-coded according to their \fieldName{astrometric_excess_noise}. The diagram shows the presence of an overdensity of solutions at periods near the precession period (marked with a vertical line) with large astrometric excess noise. The histogram of the top shows the density of solutions with astrometric excess noise larger than 1 mas (blue line) and of those with \fieldName{ipd_frac_multi_peak} $>20$ (orange line).}
\label{fig:spurious_periods_sb1}
\end{figure}

Inspecting the SB1 solutions, it can be noted an overdensity of solutions with 
periods around the precession period (62.97 days), in particular by selecting sources with large astrometric excess noise (see \figref{fig:spurious_periods_sb1}).
These solutions are spurious, due to an offset in the astrometric coordinates, which generates a spurious variation of the computed epoch radial velocities, which depends on the scanning angle and thus with a periodicity linked to the precession of the satellite.
The inaccuracy of their astrometric coordinates is most probably due to the fact that they are partially resolved binaries/double stars, which is confirmed by the fact that we see this overdensity also selecting sources with \fieldName{ipd_frac_multi_peak} $>20$. \citet{DR3-DPACP-164} describes the effect of the scanning law on the NSS solutions in more detail.

Spurious SB1 solutions can be also generated by pulsation of the source, like in RR Lyrae and Cepheids. In many cases, the SB1 solution will have the same period of the pulsation, but in other cases, due to the sparse sampling, the pipeline can find a keplerian fitting solution at a different, typically shorter, period. During the release validation, SB1 solutions of sources identified by \gaia as RR Lyrae or Cepheids were removed from the release.

Another cause generating spurious SB1 solutions is the contamination from nearby brighter star. As explained in \citet{2021A&A...653A.160S}, and noted in \citet{2019MNRAS.486.2618B}, the RVS spectrum of 
sources extracted at a given transit can be contaminated by a nearby source, producing spurious values of the radial velocity. In \gdrthree, the RVS pipeline includes a deblending algorithm, which is however limited to spectra with overlapping windows \citep[see][for details]{2021A&A...653A.160S}. 

\paragraph{Eclipsing binaries: }

At post-processing, only the sources with
$0.2<$ \fieldName{efficiency} $\leq 1$ and \fieldName{g_rank} $\geq 0.6$
were kept, where the rank is a measure of the quality of the fit. 
See the online documentation \citep[][Sect.~7.6]{NSS-DR3-documentation} for details.

%
\section{Completeness}\label{sec:Completeness}
%
The resulting \NSSa dataset is the result of a selection process in three successive steps:
\begin{enumerate}
\item The selection of the input list, discussed \secref{sssec:cu4nss_intro_inputfiltering};
\item The sources for which the orbital motion can be preferentially detected by the processing;
\item The filtering done at post processing, indicated \secref{sssec:cu4nss_intro_outputfiltering} .
\end{enumerate}

In this section, we give some indications concerning the second step. One main
reason for the expected non-uniformity of orbit detections is the number of observations 
and their temporal distribution. As this is governed by the scanning law of the \gaia 
satellite, see e.g. \figref{fig:healpixObsRatio}, this should appear clearly on a sky plot, 
and this is discussed \secref{sssec:dky_distrib}.


However, even with a given set of observations, all orbits are not created equal. First, the period
distribution of astrometric orbits shows a prominent lack of solutions around one year,
which was expected for long due to the difficulty to decouple the orbital from
the parallactic effect.
There are other more subtle biases depending on the orbit itself which are discussed \secref{ssec:astrometric_orbit_sensitivity}.

The distribution of solutions is finally discussed within the 100 pc horizon at \secref{sssec:gcns} and the completeness is also studied for \hip stars \secref{compPMa}.

%
\subsection{Sky distribution}\label{sssec:dky_distrib}
%
Over the sky, the distribution of the various solution types shows the expected 
higher density along the galactic plane together with a larger excess at high ecliptic latitudes
around $l\pm 100\deg$.
The latter is due to a larger number
of observations, and thus to a larger probability of detecting periodically 
variable motions.

This tells however little about whether the (in)completeness is uniform over the sky.
Although we may have e.g. more eclipsing binaries among young stars, let
us assume for a moment that $F$, the true (unknown) fraction of binaries,
is uniform over the sky, and that our NSS samples are roughly complete up 
to some given magnitude $G_\mathrm{max}$. 

Dividing up the sky in healpix \citep{2005ApJ...622..759G} level 4 equal-area pixels, we note $N_j$ the number 
of sources up to $G<G_\mathrm{max}$ in the full \gaia catalogue in a given healpix cell $j$,
and $n_j$ the corresponding number of NSS of a given type up to $G<G_\mathrm{max}$. 
With $f={\rm med}(n_j/N_j)$ the empirical median of the ratio over the sky as estimate 
of $F$, we call \quoting{sky density factor} $d_j={\frac{n_j}{f \, N_j}}$. 
This factor gives the up or down factor of the average NSS fraction and 
should be a noisy value around 1 if $F$ is approximately constant over the sky.

\begin{figure*}[htb]
\centerline{
\includegraphics[scale=0.2]{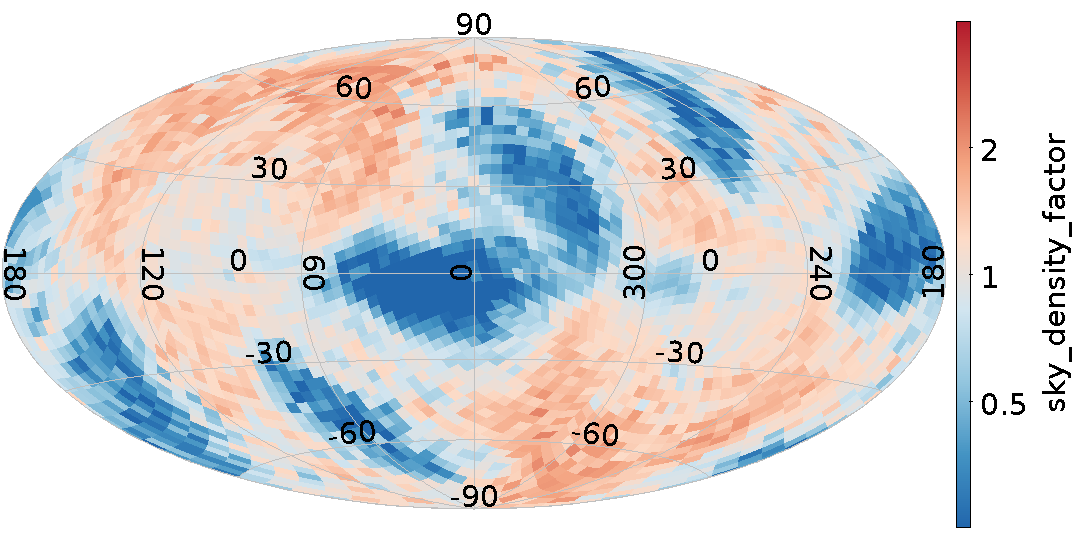}
\includegraphics[scale=0.2]{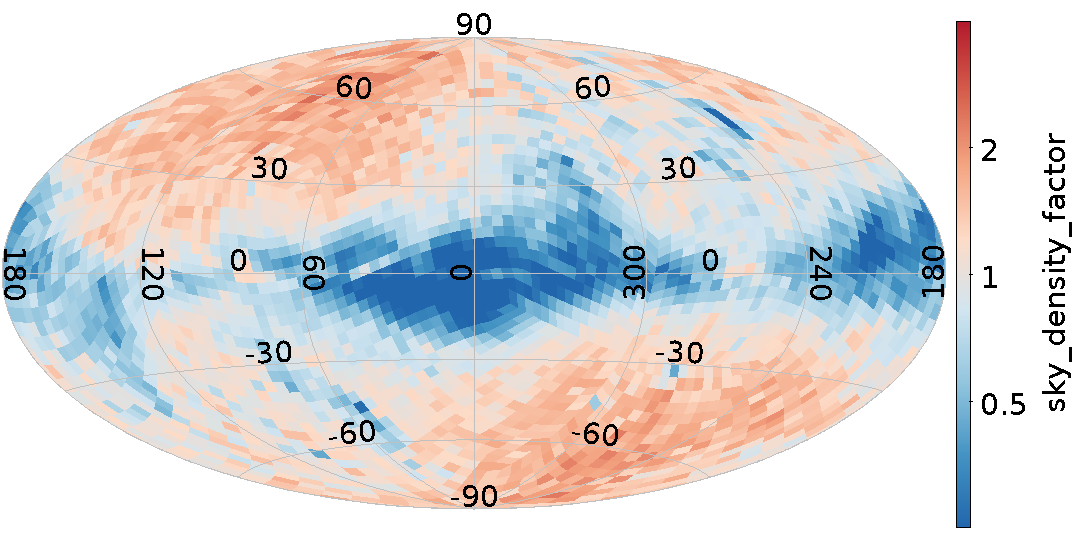}
}
\centerline{
\includegraphics[scale=0.2]{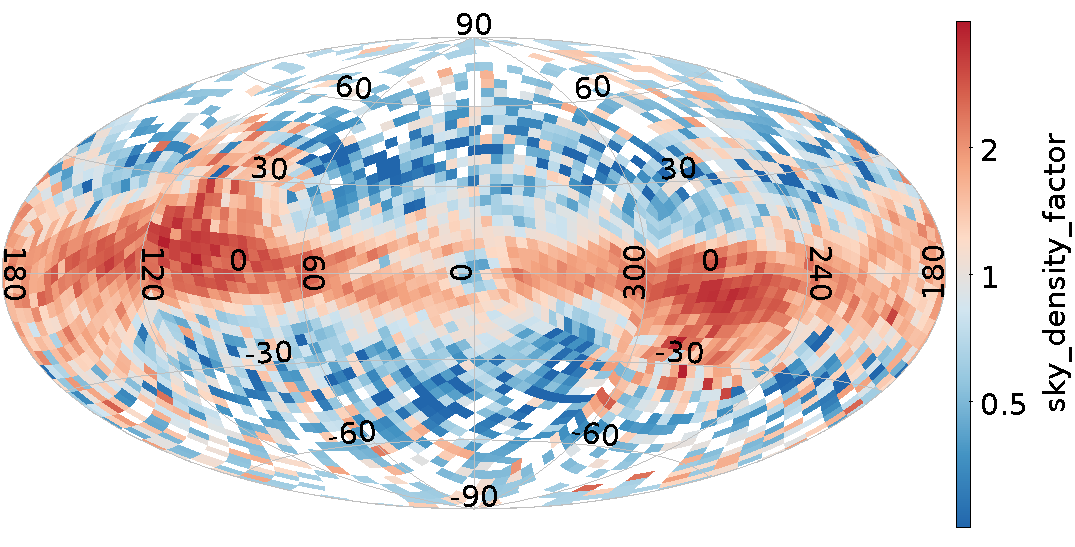}
\includegraphics[scale=0.2]{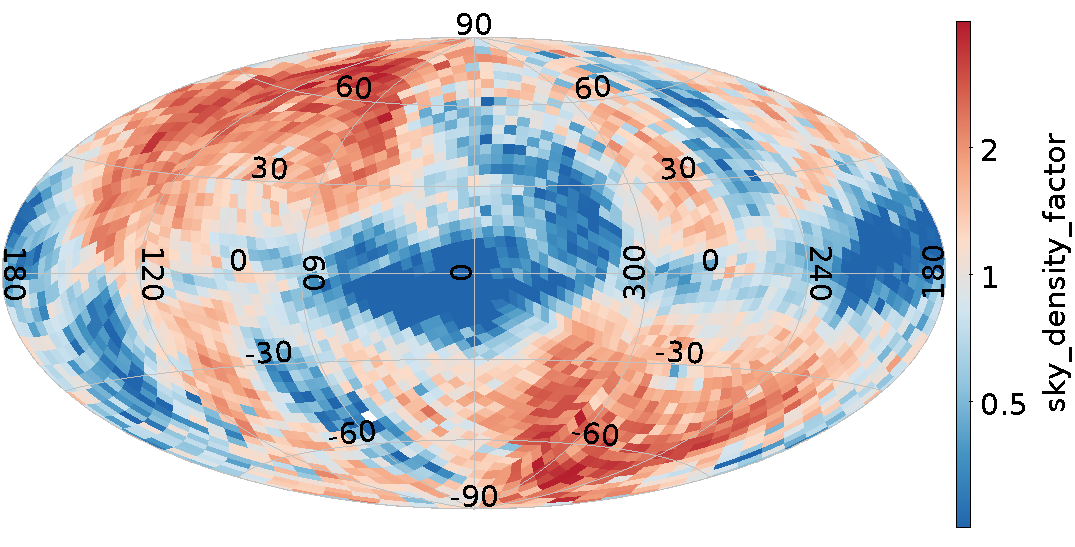}
}\caption[Density excess of solution types over the sky]
        {Sky density factor (galactic coordinates, healpix level 4, log scale, see text) 
        illustrating the excess or deficit factors of NSS sources compared to their sky average value.
        {\it Panel a}: \fieldName{SB*}, {\it Panel b}: \fieldName{Acceleration},
        {\it Panel c}: \fieldName{EclipsingBinary}, {\it Panel d}: \fieldName{Orbital*}.
        }
\label{fig:healpixSolTypes}
\end{figure*}

\begin{figure*}[htb]
\centerline{
\includegraphics[scale=0.2]{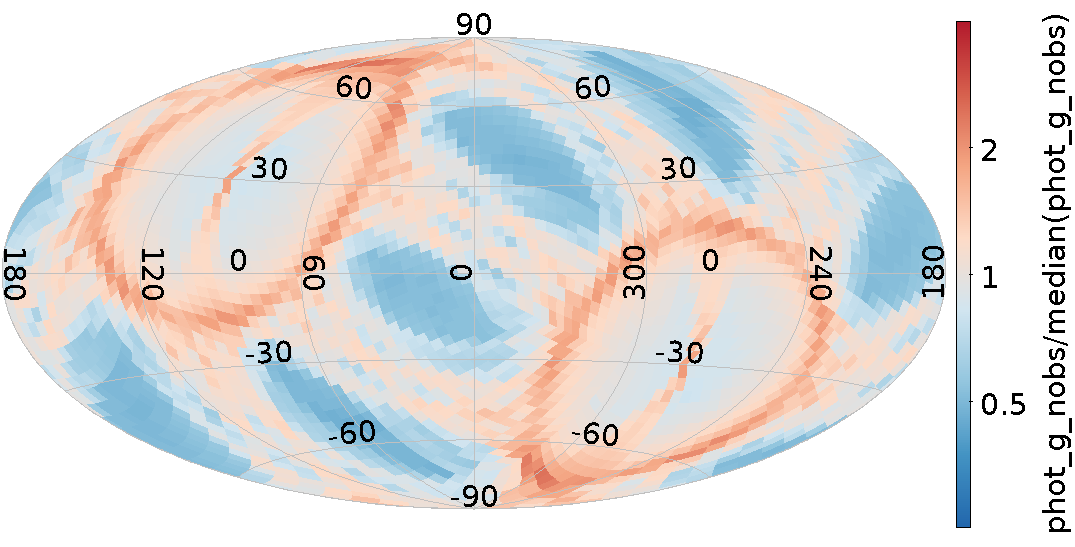}
\includegraphics[scale=0.2]{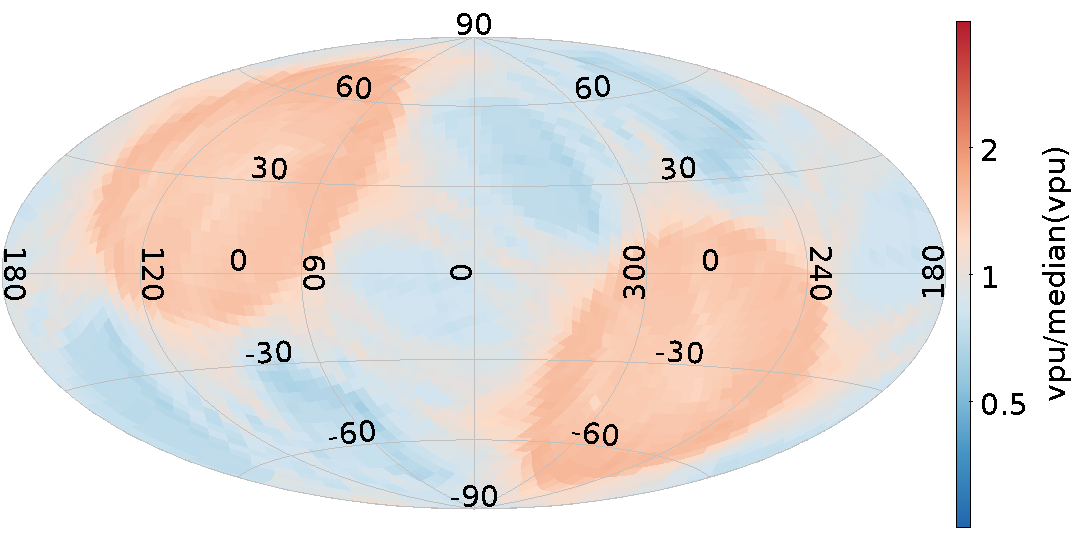}
}\caption[Ratio of number of observations]
        {Ratio of the number of photometric observations over their median values for $G<18$ ({\it left})
        and ratio of the number of visibility periods used in astrometry over their
        median values for $G<15$ ({\it right}), same colour scale (from $\frac{1}{4}$ to $4$) as \figref{fig:healpixSolTypes}.
        }
\label{fig:healpixObsRatio}
\end{figure*}

\figref{fig:healpixSolTypes} shows the sky density factor for several solution
types truncated up to a reasonable $G_\mathrm{max}$ value in healpix level 4 pixels.
As this density factor may be attributed to the number of observations
available, \figref{fig:healpixObsRatio} presents with the same scale 
for comparison purposes the ratio 
of useful observations over the sky for photometry and astrometry.

For all type of binaries, the expected deficit of sources near the
galactic center can be seen, due to both high density and poor coverage. 
The distribution for spectroscopic binaries 
($G_\mathrm{RVS-max}=12$, \figref{fig:healpixSolTypes}a) is also as expected with a 
smooth pattern of regions with higher numbers of Field of View transits.
The non uniformity is less expected for eclipsing binaries 
($G_\mathrm{max}=18$, \figref{fig:healpixSolTypes}c) with a slight excess
at the anticentre and an excess, larger than expected
from the number of transits, around $l\pm 100\deg$ towards high ecliptic latitudes.
For acceleration solutions ($G_\mathrm{max}=15$, \figref{fig:healpixSolTypes}b) 
there is a deficit in the galactic plane and an excess at high ecliptic latitudes.
This is worse for \fieldName{Orbital} solutions
($G_\mathrm{max}=15$, \figref{fig:healpixSolTypes}d), which may be due to the fact that 
orbital solutions require a number of observations larger than acceleration ones
as the number of parameters to determine is larger.
Again, the sky density factor is relative to the average over the sky, so that an 
excess in some regions may also, or rather, indicate a deficit in the rest of the sky.
Part of the explanation of the above features of the astrometric solutions
likely originates from the input source selection 
where sources suspected to be resolved doubles were excluded, which 
is more frequent in the galactic plane.

%
\subsection{Astrometric orbit detection sensitivity as a function of orbital inclination }\label{ssec:astrometric_orbit_sensitivity}
%
\gaia is observing sources with a cadence and scan angle $\psi$ determined by its scanning law. Depending on whether a binary system is seen face-on (inclination $i=0\degr$ or $i=180\degr$) or edge-on ($i=90\degr$), the detection probability of the astrometric orbit varies. An edge-on orbit that is oriented North-South and is being observed only with 1D astrometry along the East-West axis is undetectable. That extreme example does not occur for \gaia, but it illustrates that we can expect a continuous variation as a function of inclination angle, with edge-on orbits having lowest detection probability.

To obtain an empirical estimate of the expected dependency, we simulated 50000 circular orbits ($e=0$, $\omega=0$) with the following fixed parameters: distance 20 pc; period of 500 days; primary mass 1\Msun; companion mass 1\Mjup, and hence a semi-major axis of the host's orbit of $a_0=0.059$ mas. The ascending node $\Omega$ was uniformly distributed. We simulated inclinations such that $\cos{i}$ is uniformly distributed, as expected for isotropic orbit orientation in space.

\begin{figure}[htb]\begin{center}
\includegraphics[width = \columnwidth,trim=0 0cm 0 0cm, clip]{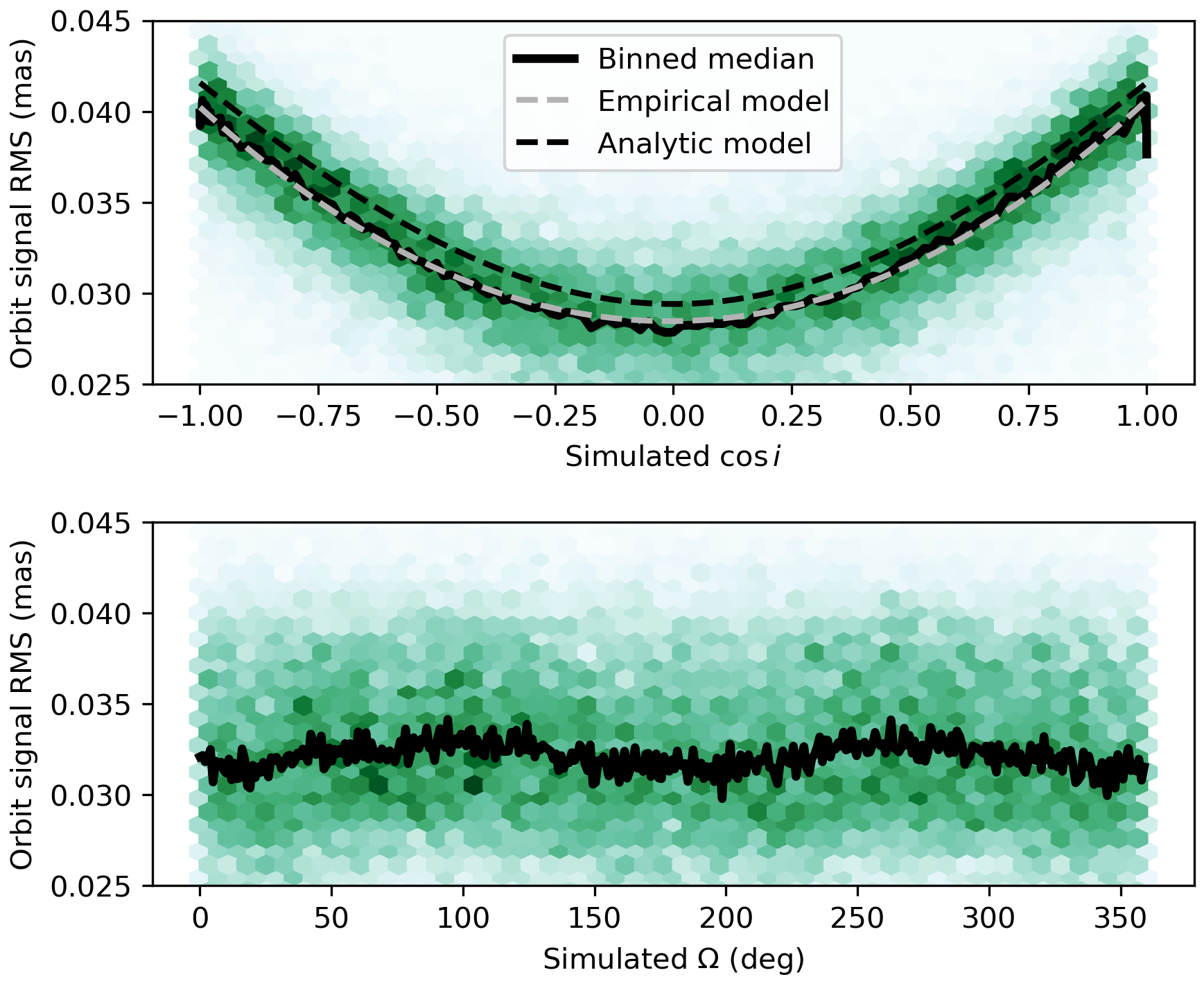}
\caption{Density histograms of simulated orbit signal dispersion as a function of $\cos{i}$ ({\it top}) and $\Omega$ ({\it bottom}). The black solid curve shows the running median value. In the top panel the empirical and analytic model is shown as dashed grey and black line, respectively. Edge-on orbits have $\cos{i}=0$ and face-on configurations have $|\cos{i}\,|=1$.}
\label{fig:nss_inclination_sensitivity_simulation}\end{center}\end{figure}

To each orbit we assigned a realistic \gdrthree time sampling with associated scan angles, randomly retrieved from $\sim$1\,000 real sources distributed over the entire sky with the aim of averaging scan-law dependent effects. We then computed the RMS dispersion of the AL signal $w_\mathrm{k1}$ (\equref{eq:k1_model}) caused by the astrometric orbit only, i.e.\ neglecting proper and parallactic motion. That dispersion shows a clear dependence on inclination angle, see \figref{fig:nss_inclination_sensitivity_simulation}, with the expected minimum for edge-on orbits. The vertical scatter is caused by the variation in the number of assigned \gaia observations and their scan-angle distribution for a particular time series realisation. The dependence on ascending node (\figref{fig:nss_inclination_sensitivity_simulation}, bottom) is much weaker but noticeable. Because we limited our simulation to circular orbits, there is no dependence on the argument of periastron.

In Appendix~\ref{sec:appendix_analytical_signal_dispersion}  we analytically derive the following expression (Eq.~\ref{eq:rmsInclDep}) for the RMS of $w_\mathrm{k1}$ as a function of $\cos{i}$ which is valid for one-dimensional along-scan observations as used for DR3 \citep{EDR3-DPACP-128} under the assumption of circular orbits and random scan angles:
\begin{equation}
\text{RMS}(w_\mathrm{k1})(x)  = \frac{a_0}{2} \, \sqrt{ 1 +  x^2 } \quad \text{with} \quad x=\cos{i}.
\end{equation}
This dependency is shown as \quoting{Analytic model} in \figref{fig:nss_inclination_sensitivity_simulation}. A fit with a quadratic polynomial is also shown as \quoting{Empirical model}. The analytic model reproduces the data very well, both in absolute amplitude and in shape, except for a small amplitude offset which probably reflects that the \gaia scan angles are not random and sometimes are restricted in range.


Because the amplitude of the orbit signal is the principal factor in deciding whether an orbit can be detected\footnote{This holds when neglecting complications with e.g.\ periods around one year due to crosstalk with parallax or scan-angle dependent effects \citep{DR3-DPACP-164}.} there is no need to simulate the complete processing chain \citep{DR3-DPACP-163, DR3-DPACP-176}. Our simulations demonstrate that the signal of a face-on orbit is $\sqrt{2}$ larger than that of an edge-on orbit, which means that the former is more likely to be detected.


%
\subsection{The \gaia catalogue of nearby stars}\label{sssec:gcns}
%
Together with {\gdrearlythree} was published a clean catalogue of 331\,312 
sources within 100 pc of the Sun \citep[][GCNS]{EDR3-DPACP-81}. This
catalogue would represent a useful subset for the completeness analysis.

As the NSS parallaxes of \fieldName{Orbital} or acceleration solutions may supersede the 
EDR3 ones, it is of interest to analyse first what would their impact be on 
the GCNS content.
One finds 116 orbital sources that would now enter GCNS using the following query:

{\small\begin{verbatim}  
SELECT NSS.source_id, GS.phot_g_mean_mag, NSS.parallax, 
NSS.parallax_error, GS.parallax as gs_parallax, 
GS.parallax_error as gs_parallax_error 
FROM user_dr3int6.nss_two_body_orbit NSS, 
user_dr3int6.gaia_source GS
LEFT JOIN external.gaiaedr3_gcns_main_1 GCNS ON 
NSS.source_id = GCNS.source_id
WHERE GCNS.source_id IS NULL 
AND NSS.source_id = GS.source_id 
AND NSS.parallax > 10
\end{verbatim}}\noindent

Using a similar query, 89 sources with an acceleration solution would enter GCNS,
i.e. a total of 205 sources. These numbers would change by 13\% only if we had taken
a $1\sigma$ margin, so the random errors have a weak influence on this.

Conversely, one may count sources which should no more belong to GCNS
according to their new parallax:

{\small\begin{verbatim}  
SELECT NSS.source_id, GS.phot_g_mean_mag, NSS.parallax, 
NSS.parallax_error, GS.parallax as gs_parallax, 
GS.parallax_error as gs_parallax_error
FROM user_dr3int6.nss_two_body_orbit NSS, 
user_dr3int6.gaia_source gs,
external.gaiaedr3_gcns_main_1 GCNS
WHERE NSS.source_id = GS.source_id AND
NSS.source_id = GCNS.source_id AND NSS.parallax < 10
\end{verbatim}}\noindent
amounting to 415 sources for orbital solution plus 413 sources for
acceleration solutions, a total of 828 sources.

Although these numbers are not significant compared to the total size
of the GCNS, they represent 9\% of the orbital + acceleration solutions which 
may no more be in the GCNS (4723+4523 = 9246 astrometric NSS sources are in the GCNS) 
which is not negligible,
while 2\% may now enter. This means that any study of the NSS completeness 
within the GCNS should use the NSS parallax rather than the one from the main catalogue.

One may also note that the balance between the number of NSS that would
be rejected from GCNS compared to the numbers that would enter
illustrates one adverse effect of the random errors that is known for long
\citep{1913MNRAS..73..359E, 1953stas.book.....T}. 
The parallaxes of NSS sources managed as single star in DR3 have a significant error, now much  
reduced in the NSS tables; this, added to the asymmetric distribution 
of the parallaxes made that binary sources 
preferentially entered GCNS that should not have belong to it.
Noting also that the DR3 NSS catalogue contains only a small fraction
of the actual unresolved astrometric binaries, then,
if the GCNS is used to compute a binarity fraction, the above remark 
indicates that the GCNS has a small positive bias. 

\begin{figure*}[htb]
\centerline{
\includegraphics[scale=0.3]{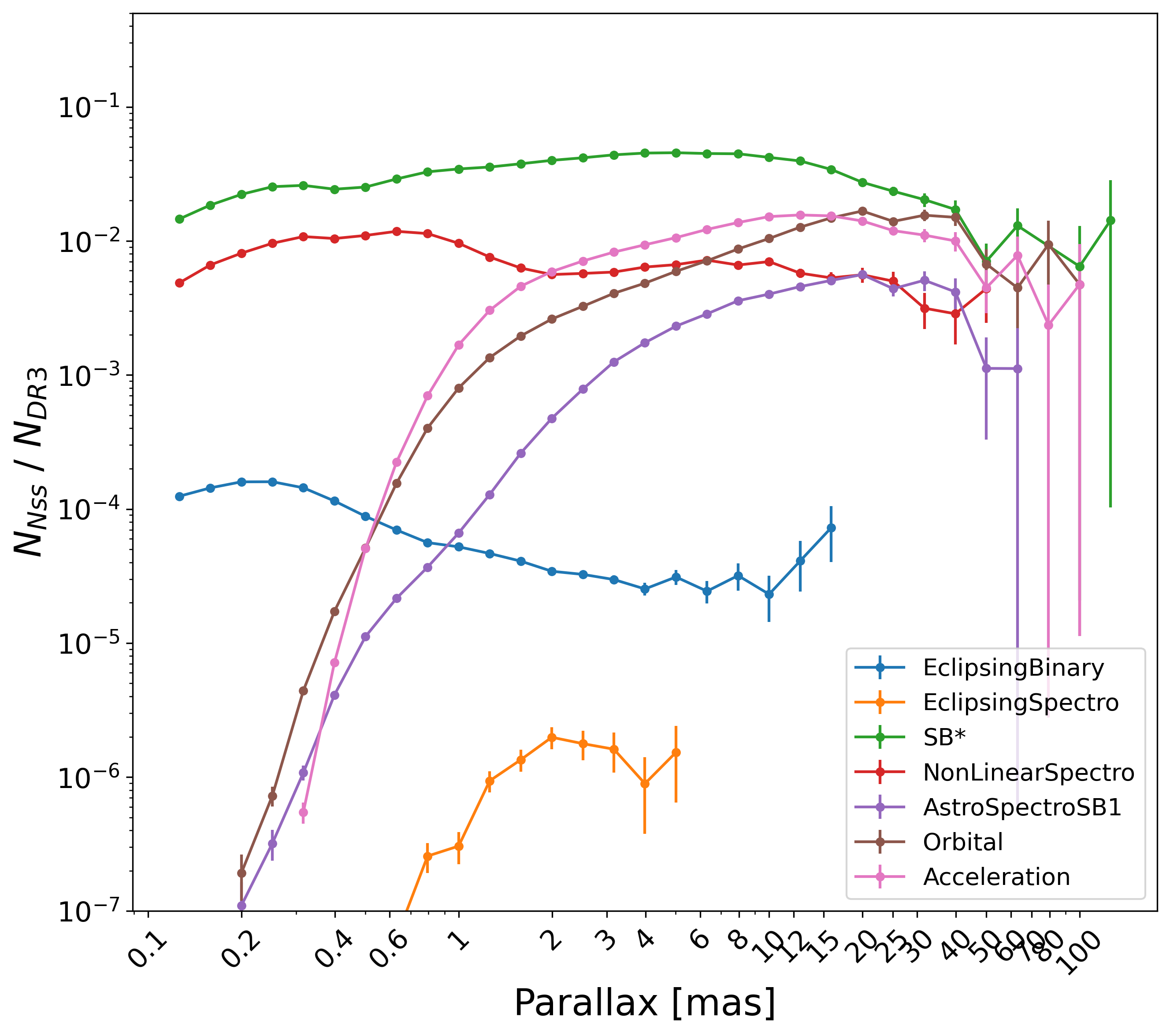}
\includegraphics[scale=0.3]{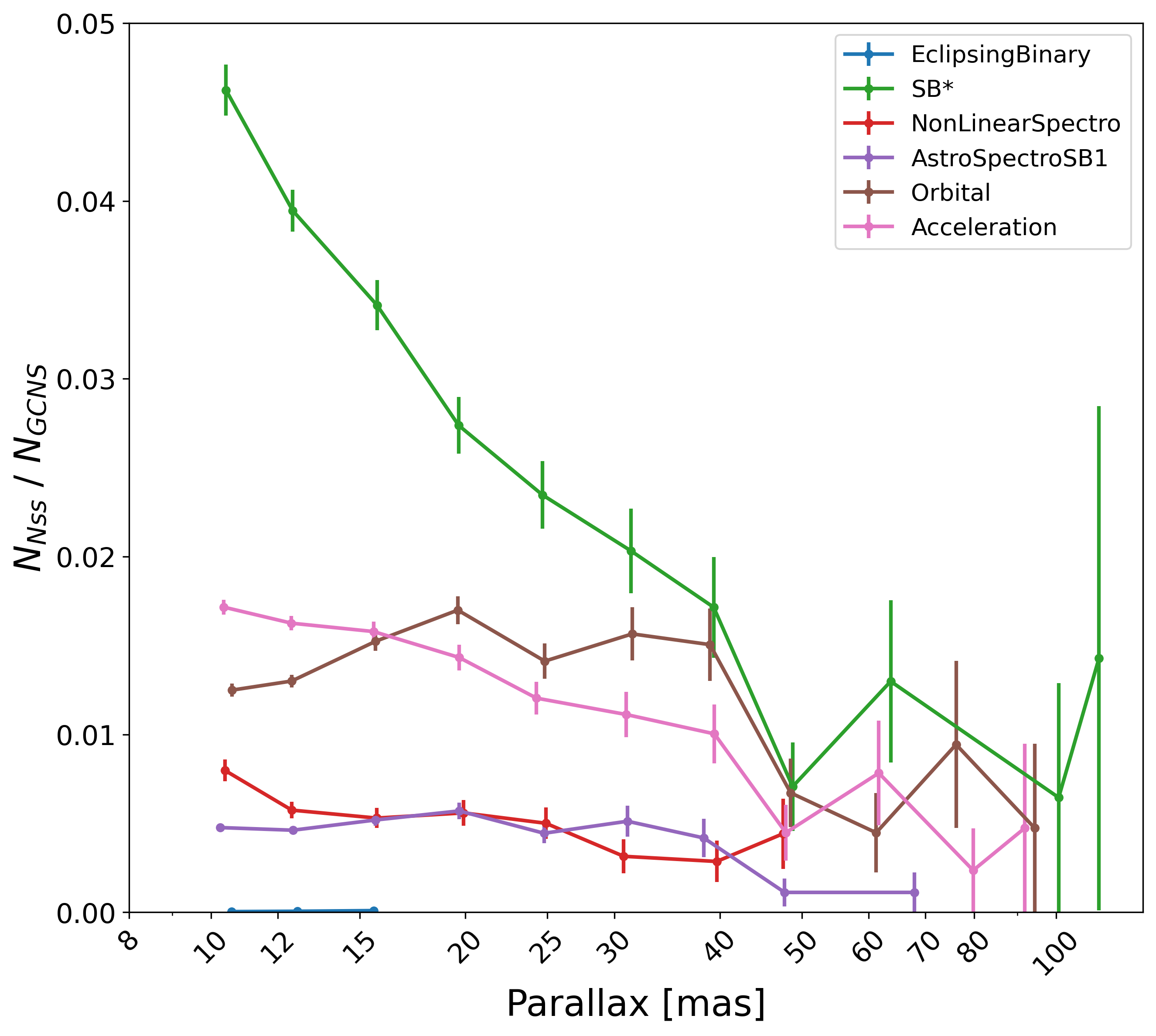}
}\caption[NSS fraction vs parallax]
        {Fraction of NSS solutions among EDR3 sources vs parallax ({\it left})
        and fraction of NSS sources in GCNS ({\it right}).
        \afterReferee{In both figures, we added \fieldName{AstroSpectroSB1} counts to \fieldName{Orbital} counts
        and to \fieldName{SB*=SB1+SB2} counts beside counting them individually, and we restrict the ratios to $G_\mathrm{RVS}<12$ sources only for \fieldName{SB*} and \fieldName{NonLinearSpectro}, to $G<19$ for \fieldName{Orbital} and \fieldName{Acceleration} solutions and to $G<20$ for eclipsing binaries.}
        }
\label{fig:ratioNSSplx}
\end{figure*}

As a clarification of the GCNS content using the NSS parallax is outside
the scope of this article, we keep the GCNS as reference in what follows. 
We show the fraction of NSS sources among $G<19$ GCNS sources as a function of parallax, for all solution types, 
\figref{fig:ratioNSSplx} right. In these figures and the following
we add the \fieldName{AstroSpectroSB1} counts both to orbital solution counts and SB counts, beside counting
them independently \afterReferee{and, for the comparison to be fair, 
we restricted the ratios to the typical magnitude ranges used 
respectively for astrometric, spectroscopic or eclipsing binaries.} 

What first appears is the \afterReferee{conspicuous} increase of the fraction of SB up to 100\,pc. \afterReferee{One reason for this may be the transition from the $G_\mathrm{RVS}<12$ population of dwarfs to giants, as can be seen \figref{fig:hrds}d, with the latter having a better intrinsic RV precision at a given apparent magnitude \citep{DR3-DPACP-159}, and thus a larger binary detection probability; a difference in the binary fraction between dwarfs and giants cannot be excluded however.}
Second, contrarily to what could have been expected, the total of 
orbital and acceleration solutions, about 3\%, appears roughly constant with distance
in the GCNS, despite all the complex filtering that had been applied. 
For comparison, the fraction of NSS among DR3 sources,
\figref{fig:ratioNSSplx} left, shows logically the drop of astrometric solutions 
with distance beyond 100~pc
while the spectroscopic binaries (SB+\NLSTable) do not vary so sharply.
We keep from this that even if the absolute value of the astrometric binary fraction
is difficult to extrapolate after all the filtering done, the fact that it looks
roughly uniform with distance in a first approximation in the GCNS sample could
make this sample useful to study the properties of the astrometric binaries.

\begin{figure*}[htb]
\centerline{
\includegraphics[scale=0.3]{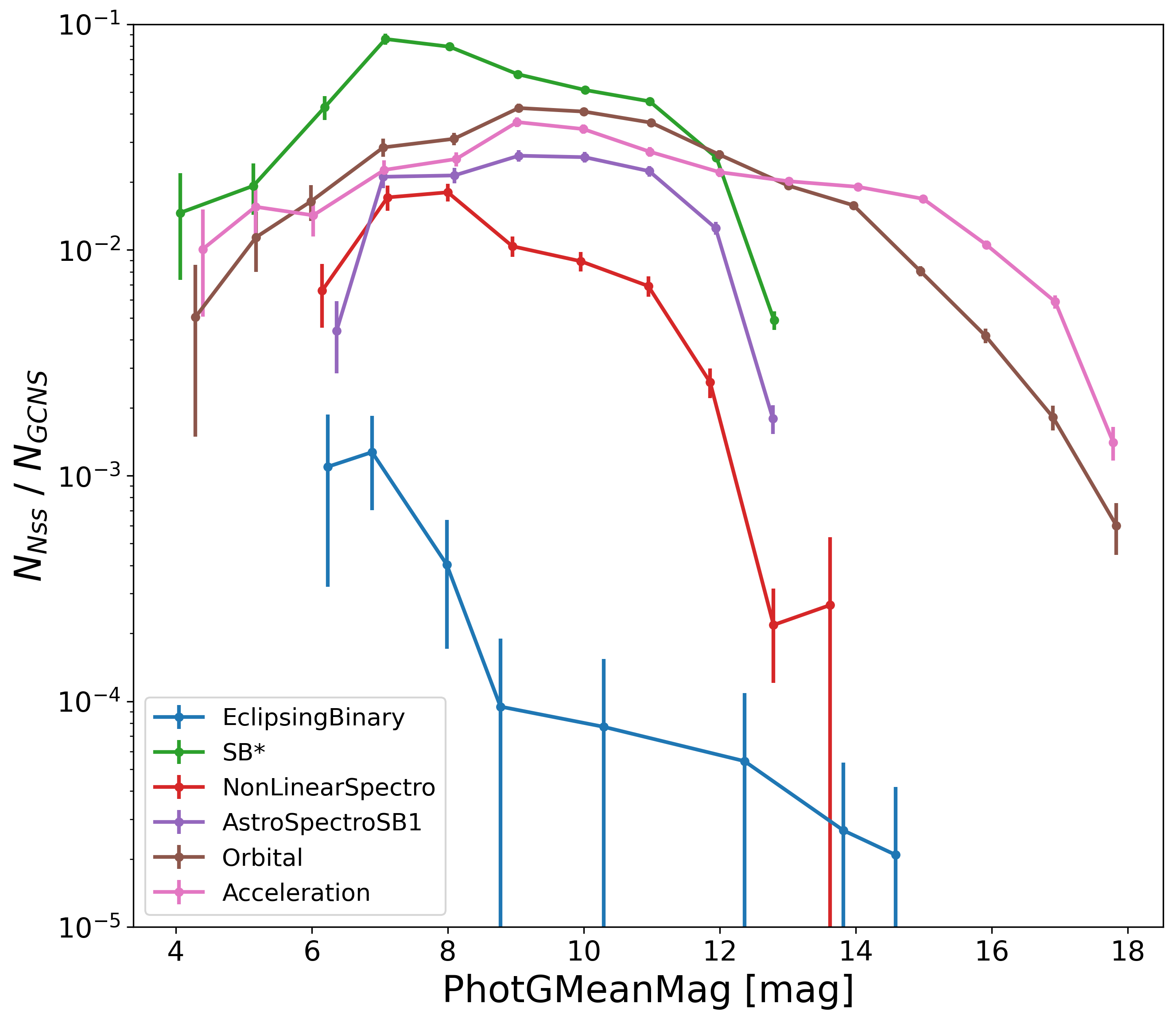}
\includegraphics[scale=0.3]{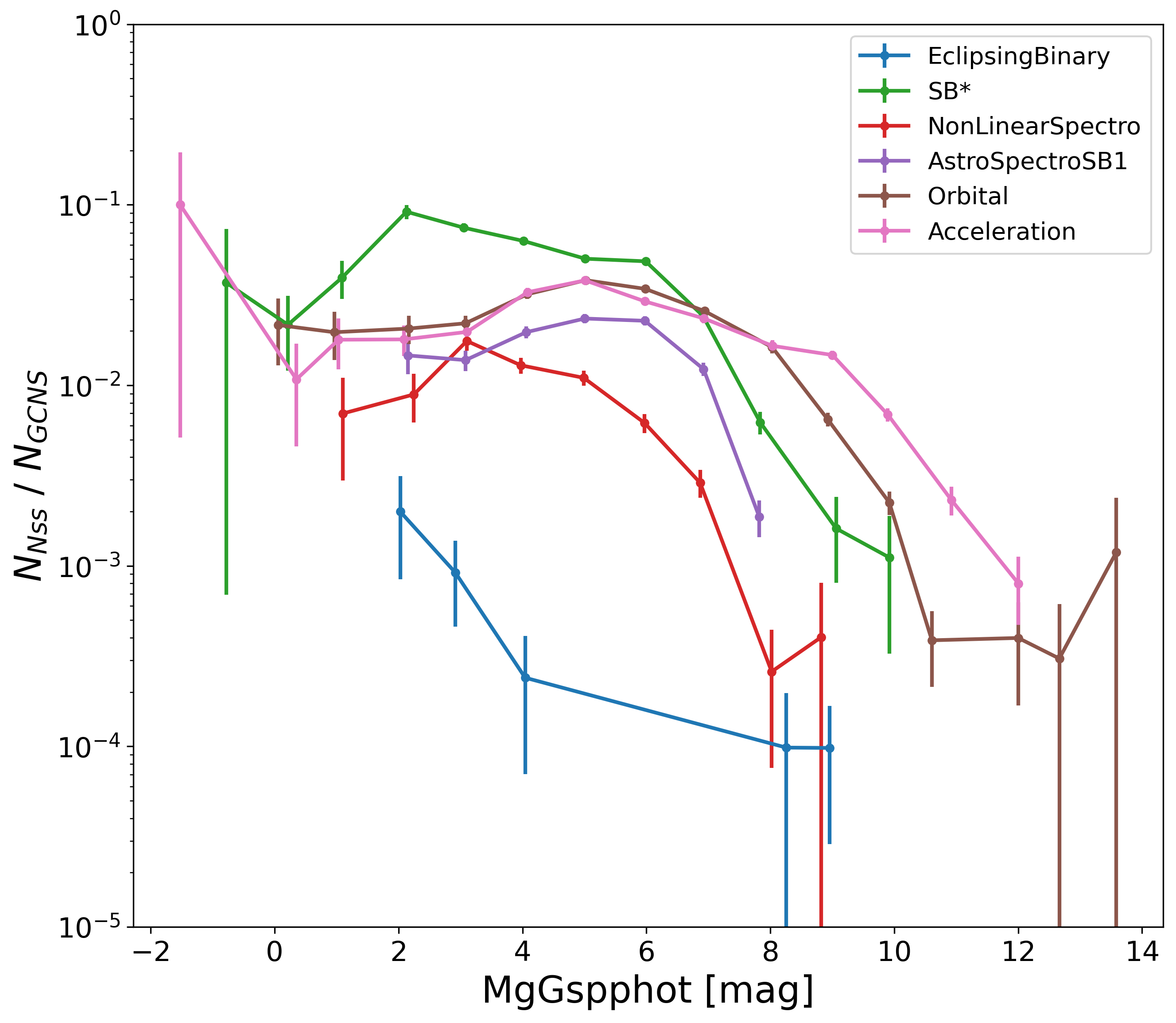}
}\caption[NSS fraction vs magnitudes]
        {Fraction of NSS solutions among GCNS sources vs $G$ apparent magnitude ({\it left})
        and vs $G$ GSP-Phot absolute magnitude ({\it right}). \afterReferee{The same constraints as mentioned in \figref{fig:ratioNSSplx} have been applied.}
        }
\label{fig:ratioNSSmag}
\end{figure*}

Consequently, the fraction of NSS among GCNS may give some useful insight,
and \figref{fig:ratioNSSmag} represents this ratio vs $G$ apparent and absolute magnitude
of the pair respectively\footnote{The absolute magnitude \fieldName{mg_gspphot} originates
from the General Stellar Parametrizer from Photometry (GSP-Phot) 
which computed the astrophysical parameters of stars from the low-resolution 
BP/RP spectra and is available in the \APTable table.}.

%
\section{Caveats}\label{sec:validation}
%

Many validations have been performed and described in the catalogue documentation
\citep{NSS-DR3-documentation}, 
and accompanying papers, \cite{DR3-DPACP-163, DR3-DPACP-176, DR3-DPACP-178, DR3-DPACP-179},
and the independent validation of all catalogues, \cite{DR3-DPACP-127}. 
Elsewhere in this article we also check the distribution tails of some parameters
which allowed us to discover undesired aspects and we indicate ways to circumvent them.
We describe here two supplementary tests that draw attention to some properties of
the catalogue, the first analysing the
distribution of orbital parameters, the second comparing the results to 
binaries detected externally.

%
\subsection{Distributions and biases of astrometric orbit parameters}\label{ssec:astrometric_orbit_distributions}
%
Under the assumption that the orbits of binary systems are randomly oriented, we can infer the expected distributions in the geometric elements of the corresponding astrometric orbits, i.e.\ the inclination $i$, argument of periastron $\omega$, and longitude of the ascending node $\Omega$\footnote{\afterReferee{These Campbell elements were computed from the Thiele-Innes coefficients ($A,B,F,G$), which are part of the archive table, using standard formulae \citep[e.g.][]{DR3-DPACP-163}, software tools being available at \url{https://www.cosmos.esa.int/web/gaia/dr3-software-tools}}}. In an ideal experiment, we expect to recover uniform distributions in $\cos{i}$, $\Omega$, and $\omega$. Here we inspect the observed distributions of these parameters in DR3.

\subsubsection{Observed distributions of geometric elements in DR3 solutions}
\begin{figure*}[htb]\begin{center}
\includegraphics[width = \linewidth]{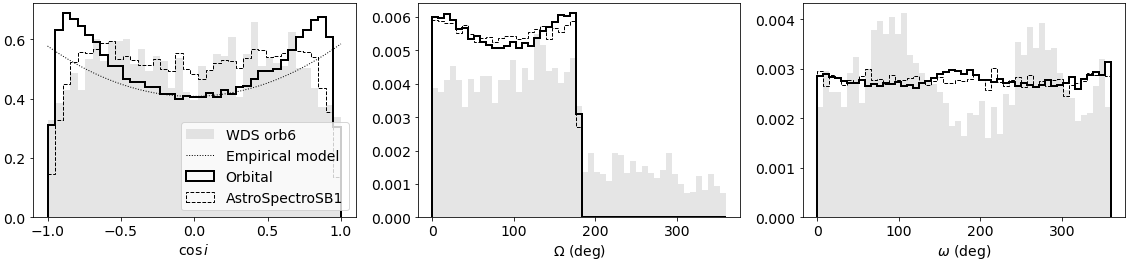}
\caption{Normalised distributions of $\cos{i}$ ({\it left}), $\Omega$ ({\it middle}), and $\omega$ ({\it right}) parameters. \fieldName{Orbital} (solid lines, 122\,989 entries) and \fieldName{AstroSpectroSB1} (dashed lines, 29\,770 entries) solutions with $P<1\,000$\,d are shown. The orb6 solutions from the literature (3\,405 entries, without filter on period) are shown in grey. In the {\it left panel}, the dotted line shows the empirical model defined in \secref{ssec:astrometric_orbit_sensitivity}, which was re-scaled on the 5 central histogram bins. In the right panel, we have suppressed the circular solutions with $\omega=0$.}
\label{fig:orbit_param_distributions_wds_comparison}\end{center}\end{figure*}

Figure \ref{fig:orbit_param_distributions_wds_comparison} shows the distributions of $\cos{i}$, $\Omega$, and $\omega$ for the solution types \fieldName{Orbital} and \fieldName{AstroSpectroSB1}.  
To mitigate effects related to incomplete orbit coverage, we selected solutions with orbital periods shorter than 1\,000 days, which roughly corresponds to the DR3 timespan.

For \fieldName{Orbital} solutions there is a strong modulation in $\cos{i}$. Although the expected suppression of edge-on orbits  is present, the observed distribution deviates significantly from the empirical model defined in \secref{ssec:astrometric_orbit_sensitivity}. For progressively face-on configurations with increasing $|\cos{i} \, |$ there is an excess of solutions compared to the model. Beyond the modes $|\cos{i} \, |\gtrsim0.85$ the number of detected almost-face-on orbits drops sharply and much below the expected level. We also observe a smooth modulation of the $\Omega$ distribution\footnote{The ascending node extracted from the Thiele-Innes coefficients of astrometric orbits is constrained to $\pm180\degr$. By convention, the value between 0 and $180\degr$ is chosen.} with a single minimum at $\Omega=90\degr$ and a bimodal modulation of the $\omega$ distribution with minima at $\omega=90\degr$ and $270\degr$.

For \fieldName{AstroSpectroSB1} solutions, resulting from the combined analysis of \gaia astrometry and RVs, the $\cos{i}$ distribution shows a good agreement with the empirical model for edge-on and intermediate configurations without region of excess detections. However, there is also a clear lack of face-on orbits compared to the empirical expectation. This is influenced by the decreasing orbital RV signature towards face-on orbits. Since \fieldName{AstroSpectroSB1} solutions require independent detections in both astrometry and RV, the lack of face-on orbits can be expected. The modulation in $\Omega$ is similar to \fieldName{Orbital} solutions but weaker\footnote{The $\Omega$ parameter is only constrained by the astrometry data. When deriving it from the \fieldName{AstroSpectroSB1} Thiele-Innes coefficients we have neglected the additional RV information that would have allowed us to compute it unambiguously.} and there is no apparent modulation in the $\omega$ distribution. 

\begin{figure}[htb]\begin{center}
\includegraphics[width = 0.7\columnwidth]{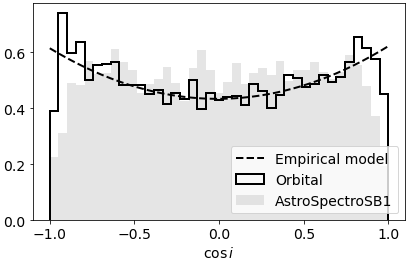}
\caption{Normalised distributions of $\cos{i}$ within 200 pc for \fieldName{Orbital} (solid line, 9106 entries) and \fieldName{AstroSpectroSB1} (grey-filled, 5735 entries) solutions with $P<1\,000$ d and $\varpi>5$ mas. The dashed line shows the empirical model defined in \secref{ssec:astrometric_orbit_sensitivity}.}
\label{fig:orbit_param_distributions_200pc}\end{center}\end{figure}

In \figref{fig:orbit_param_distributions_200pc} we show the $\cos{i}$ distributions for systems within 200 pc, where the signal-to-noise 
is on average higher and the astrometric-orbit detection can be expected to be more complete. This is confirmed by the \fieldName{Orbital} solutions that follow the empirical model nicely across most of the inclination range. This validates our model for the inclination-dependent detection efficiency of astrometric surveys (\secref{ssec:astrometric_orbit_sensitivity}). The $\Omega$ and $\omega$ distributions for this subset of solutions are approximately uniform. 

We have inspected but do not discuss other astrometric solution types here because those have fewer ($<1\,000$) entries and are therefore less suitable for distribution analyses.

\subsubsection{Origins of the geometric element biases}
Concentrating on the \fieldName{Orbital} solutions, we identify three main deviations from the expected uniform distributions in the low signal-to-noise regime, comprising most solutions and therefore dominating the overall distributions in \figref{fig:orbit_param_distributions_wds_comparison}: (a) a pronounced suppression of face-on orbits; (b) a smooth modulation of the $\Omega$ distribution with a single minimum; (c) a bimodal modulation of the $\omega$ distribution.

In Appendix \ref{ssec:thiele_innes_biases} we identify the origin of features (a) and (b) in the linear fit of the Thiele-Innes coefficients to noisy data and reproduce these biases qualitatively in simulations. The noise bias in the recovered inclination shifts solutions away from face-on configurations leading to the observed excess at intermediate inclinations\footnote{Since the survey is not volume-limited, \gaia's sensitivity variation in principle leads to an expected excess of face-on orbit detections. We believe that such effects are secondary in the context of DR3.}. A modulation akin to feature (c) can also be caused by noise biases, however, with a 90\degr\ phase shift. In Appendix \ref{ssec:omega_selection_effect} we show that feature (c) is instead explained by the application of a semi-major axis significance threshold when selecting the solutions to be published.

\subsubsection{Geometric elements from Monte-Carlo resampled Thiele-Innes coefficients}
\begin{figure}[htb]\begin{center}
\includegraphics[width = 0.7\columnwidth]{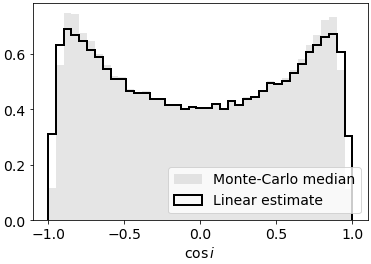}
\caption{Normalised distributions of $\cos{i}$ for non-circular \fieldName{Orbital} solutions with $P<1\,000$ d (121\,207 entries). The linearised and Monte-Carlo estimates are shown as solid line and filled grey, respectively.}
\label{fig:cosi_distribution_change_mc}\end{center}\end{figure}

Instead of using the linearised formulae \citep[e.g.][]{DR3-DPACP-163} for converting $A,B,F,G$ values and uncertainties to $a_0$, $i$, $\Omega$, $\omega$, one can use Monte-Carlo resampling which accounts more accurately for the parameter correlations (Appendix \ref{ssec:thiele_innes_uncertainties}). As an example of potential effects this may have, we computed an alternative estimate of the orbital inclination for individual solutions as the median of the resampled Monte-Carlo distribution. The difference between linearised and Monte-Carlo estimates on the inclination distribution is shown in \figref{fig:cosi_distribution_change_mc}, where we see that the apparent depletion of face-on orbits is more pronounced when applying the resampling. We note that the resampled distributions of $a_0$, $i$, $\Omega$, $\omega$ are seldom Gaussian and the median value is not always a good representation. Whether it is advisable to use the linearised estimate or Monte-Carlo resampling depends on the particular problem and individual orbital solution.

\subsubsection{Comparison with known astrometric orbits}
In \figref{fig:orbit_param_distributions_wds_comparison} we also show the distributions of geometric elements compiled in the "Sixth Catalog of Orbits of Visual Binary Stars" \citep[orb6, ][]{Hartkopf:2001wi}\footnote{We retrieved \url{http://www.astro.gsu.edu/wds/orb6/orb6orbits.sql} on 2022-02-11 and did not remove orbits with two independent solutions or apply any other filters.}.
The orb6 inclination distribution is bimodal with modes at $|\cos{i} \, | \simeq 0.5$, which could be interpreted as the same signature of lacking face-on orbits as for \gaia \fieldName{Orbital} but setting in  \emph{earlier}. The comparison with the simulated inclination biases in  \figref{fig:simulation_cosi_biases} then would lead to the interpretation that the average signal-to-noise is higher for the \gaia orbits than for the orb6 solutions. However, we caution that the orb6 dataset is of heterogeneous nature and such comparisons have to be done more carefully by accounting for differences in period range, significance, and other factors.\\
The orb6 $\Omega$ distribution does not seem to exhibit the minimum at $\Omega=90\degr$ seen for \gaia \fieldName{Orbital}. In contrast, the orb6 $\omega$ distribution shows clear modes at
$\omega=90\degr$ and $270\degr$, 
i.e.\ shifted by $90\degr$ relative to \gaia \fieldName{Orbital}. Our simulations in \figref{fig:simulation_omega_bias} reproduce the peak location for orb6 orbits but not for \fieldName{Orbital} solutions.

It is clear that the factor $>$40 increase in astrometric orbit solutions delivered by \gdrthree\ compared to orb6 will make a multitude of population-level studies possible and push forward in our understanding of stellar binary systems.

\subsubsection{Recommendations}
The observed features in the distributions of $i$, $\omega$, and $\Omega$ are the result of variations in the survey's detection sensitivity, of selection effects, and of biases that are introduced in the astrometric non-single star processing. Their presence is not specific to \gaia and astrometric orbits in the literature show similar features. The geometric elements of DR3 orbits are encoded in the Thiele-Innes coefficients and different conversion methods can be applied, depending on the use-case and individual solution. Both the distribution features and the conversion aspects have to be considered in scientific analyses of \gdrthree\ orbital parameters and their distributions.

%
\subsection{Proper motion anomaly of \hip stars of the NSS sample}
%
\subsubsection{Comparison sample}
\afterReferee{In this Section we} compare the properties of the \hip stars based on the proper motion anomaly (hereafter PMa) approach (\citealt{2022A&A...657A...7K}; see also \citealt{Brandt2021}) and the NSS analysis.
The PMa approach is described in details in \citet{Kervella2019}. The general principle is to look for a difference in proper motion (PM) between the long-term PM computed from the \hip (epoch 1991.25; \citealt{book:newhip}, see also \citealt{1997ESASP1200.....E}) and \gdrthree \citep[2016.0;][]{EDR3-DPACP-130} astrometric $(\alpha, \delta)$ positions on the one hand and the individual short-term PM vector from the \gdrthree catalogue on the other hand.
For a single star, the long-term PM  is identical to the short-term PM measured by \gaia, as its space velocity is constant with time.
For a binary star, the short-term PM includes in addition the tangential component of the orbital velocity of its photocentre.
As the latter is changing with time over the orbital period of the system, a deviation appears between the short-term and long-term PMs of the star, due to the curvature of its sky trajectory.
The proper motion anomaly (PMa), namely, the difference between the short-term and long-term PM, is therefore an efficient and sensitive indicator to detect non-single stars. 

In order to compare the NSS catalogue with the PMa approach, we cross-matched the NSS catalogue with the PMa catalogue\footnote{Available through the CDS/VizieR service as catalogue \fieldName{J/A+A/657/A7/tablea1}} of \citet{2022A&A...657A...7K}, which covers 116\,343 Hipparcos stars.
This resulted in a list of 2\,767 common targets with astrometric NSS \fieldName{Acceleration7} or \fieldName{Acceleration9} solutions and 5\,416 stars with \fieldName{Orbital}, \fieldName{AstroSpectroSB1} or \fieldName{OrbitalTargetedSearch*} orbital solutions.
In addition, 4\,385 \hip targets are listed in the NSS tables with \fieldName{EclipsingBinary} (photometric), \fieldName{SB1} or \fieldName{SB2} (radial velocity) solutions.
Overall, 12\,568 \hip/PMa stars have an entry in the NSS catalogue, that is, 10.8\% of the \hip/PMa catalogue.

\subsubsection{Completeness of the NSS sample for \hip stars}\label{compPMa}

The \gaia stars that are present in the NSS catalogue were selected based on criteria on parameters from their single-star solutions, tailored to identify the most probable binaries.
For the astrometric solutions based on astrometry, this includes the presence of a \fieldName{ruwe} higher than 1.4 in their single-star solution.
As pointed out by \citet{2020MNRAS.496.1922B} and \citet{2021ApJ...907L..33S}, this criterion is efficient to identify the stars that host partially resolved companions. Furthermore, based on the PMa analysis, the binary fraction was found to remain high for \fieldName{ruwe} values lower than 1.4 by \citet{2022A&A...657A...7K} with, e.g., 30\% of the stars with \fieldName{ruwe} $\approx1.2$ exhibiting a PMa S/N > 3 (their Fig.~11).
As a consequence, the degree of completeness of the star sample present in the NSS is likely relatively low, due to its selection threshold on the \fieldName{ruwe} value.
To estimate the completeness of the NSS for the \hip stars, we first applied to the PMa catalogue the same selection criteria as the NSS input sources  (Sect~\ref{sssec:cu4nss_intro_inputfiltering_astro}), except the condition \fieldName{ruwe} $>1.4$, resulting in a subsample of 92\,240 stars (79.3\%). Within this subsample, 28\,111 stars are high probability astrometric binaries as their PMa S/N > 3.
Restricting the count to the NSS stars that have an astrometric solution (\fieldName{Acceleration7}, \fieldName{Acceleration9}, \fieldName{Orbital}, \fieldName{AstroSpectroSB1} or \fieldName{OrbitalTargetedSearch*}), we therefore obtain a completeness level of the NSS catalogue relative to the PMa catalogue of $8\,183 / 28\,111 = 29.1\%$.

However, this high-level estimate based on global target numbers does not directly reflect the actual efficiency of the NSS reduction to detect that a star is a binary or not, compared to the PMa technique.
To estimate this efficiency, we consider the same initial sample, following the NSS selection criteria including \fieldName{ruwe} $>1.4$, and we derive the fraction of stars with an NSS solution within this common sample.
The results are listed in \tabref{PMa-NSS-Tab}. Overall, the astrometric solutions provided in the NSS catalogue represent 41\% of the potential binaries present in the NSS reference sample, compared to 92\% for the PMa catalogue.

In summary, due to the stringent selection of the solutions for the NSS, the catalogue comprises approximately 40\% of the binaries from the \hip-\gaia PMa catalogue that were potentially detectable from \gaia astrometry alone.

\begin{table}
 \caption{Comparison of the PMa and NSS astrometric detection rate on the common \hip star sample.
 \label{PMa-NSS-Tab}}
 \centering
  \begin{tabular}{lrr}
  \hline
  \hline
    & Number & Fraction  \\ 
  \hline  \noalign{\smallskip}
Objects eligible to NSS \& PMa & 14\,748 & 100.0 \% \\
PMa S/N<3 and absent from NSS   &  2\,254 & 15.3 \% \\
PMa S/N>3 and absent from NSS      &  7\,320 & 49.6 \% \\
PMa S/N<3 and present in NSS      &  950 & 6.4 \% \\
PMa S/N>3 and present in NSS &  4\,224 & 28.6 \% \\
  \hline  \noalign{\smallskip}
Total non-single stars (PMa or NSS) & 12\,494 & 100.0 \% \\
Non-single stars detected from PMa & 11\,544 & 92.4 \%\\
Non-single stars present in NSS & 5\,174 & 41.4 \%\\
  \hline 
\end{tabular}
\end{table}

\subsubsection{Statistics of proper motion anomaly of NSS targets}
\label{Sect:PMA}

The PMa is an efficient tracer of the presence of a massive orbiting companion, but its sensitivity is limited by two aspects.
Firstly, the time baseline between \hip and \gaia (24.75\,years), although long, significantly reduces the PMa signature of companions with orbital periods longer than approximately three times the \hip-\gaia time, that is, 75\,years.
Secondly, the fact that the \gdrthree proper motions are the result of an averaging over a time window of 34\,months strongly smears out the signature of companions with orbital periods shorter than approximately 4\,years.
In summary, the PMa technique is most sensitive for companions with orbital periods between $\approx 4$ and 75\,years.
On the other hand, the capacity to determine orbital solutions directly from \gaia astrometry (or radial velocity) time series is significantly higher for binaries with periods shorter than the \gdrthree measurement window. The longer periods remain detectable, mostly up to about twice the measurement window. However, the astrometric displacement of \afterReferee{long-period binaries} is generally detected only as an acceleration, hence without a period determination.

\begin{figure*}[ht]
\centering
\includegraphics[width=16cm]{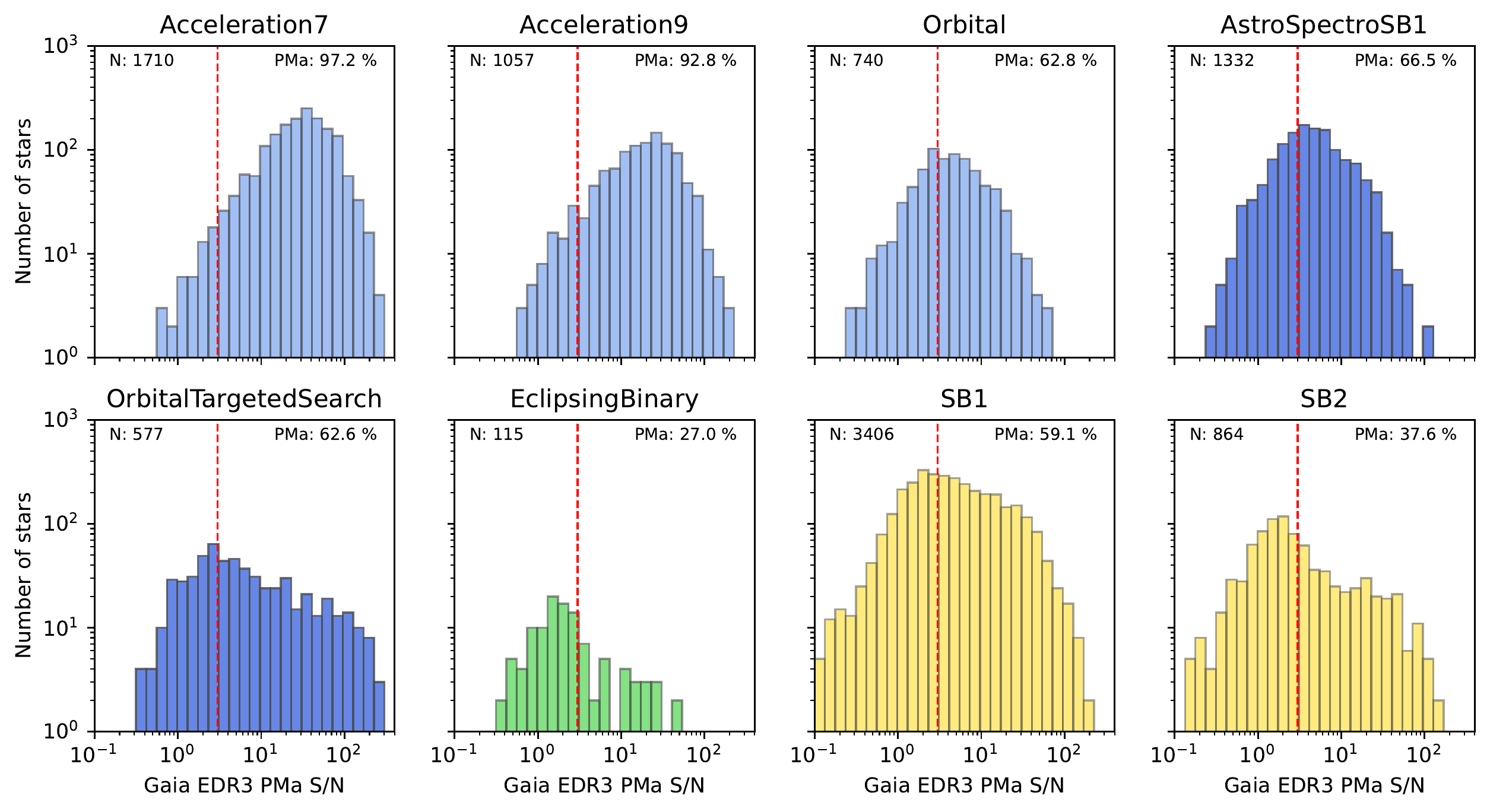}
\caption{Histogram of the number of NSS stars with different solution types that are present in the \hip catalogue, as a function of the S/N of their \gdrthree proper motion anomaly from \citet{2022A&A...657A...7K}. The total number of targets $N$ and the fraction of stars with a PMa S/N larger than 3 is displayed in each panel. \label{NSS-Histogram-PMa}}
\end{figure*}

Figure~\ref{NSS-Histogram-PMa} shows the histograms of the number of NSS stars with different kinds of solutions, as a function of their PMa signal-to-noise ratio.
The five histograms that are colour-coded in blue correspond to NSS solutions that include the \gdrthree astrometry either exclusively (\fieldName{Acceleration7}, \fieldName{Acceleration9}, \fieldName{Orbital}) or in conjunction with spectroscopic radial velocities (\fieldName{AstroSpectroSB1}) or previously known substellar orbital parameters (\fieldName{OrbitalTargetedSearch*}).
The eclipsing binary stars (\fieldName{EclipsingBinary}; green colour) are characterized from the \gaia photometric data, and the spectroscopic binaries (\fieldName{SB1}, \fieldName{SB2}; yellow colour) rely on the spectroscopic radial velocities measured by the \gaia RVS \citep{DR2-DPACP-46,DR2-DPACP-54}.

\subsubsection{Orbital periods and sensitivity}

Almost all the \hip targets with an \fieldName{Acceleration7} or \fieldName{Acceleration9} solution show a significant PMa signal. This behavior is expected for two reasons:
(1) The NSS astrometric solutions have been selected among the \gaia sources with a \fieldName{ruwe} larger than 1.4. This favors partially resolved binary stars, that often have orbital periods within the sensitivity range of the PMa technique.
(2) For longer orbital periods than the \gaia measurement window, the PMa and the acceleration are physically similar quantities, both related to the curvature of the sky trajectory of the star.

\begin{figure}
\includegraphics[width=\hsize]{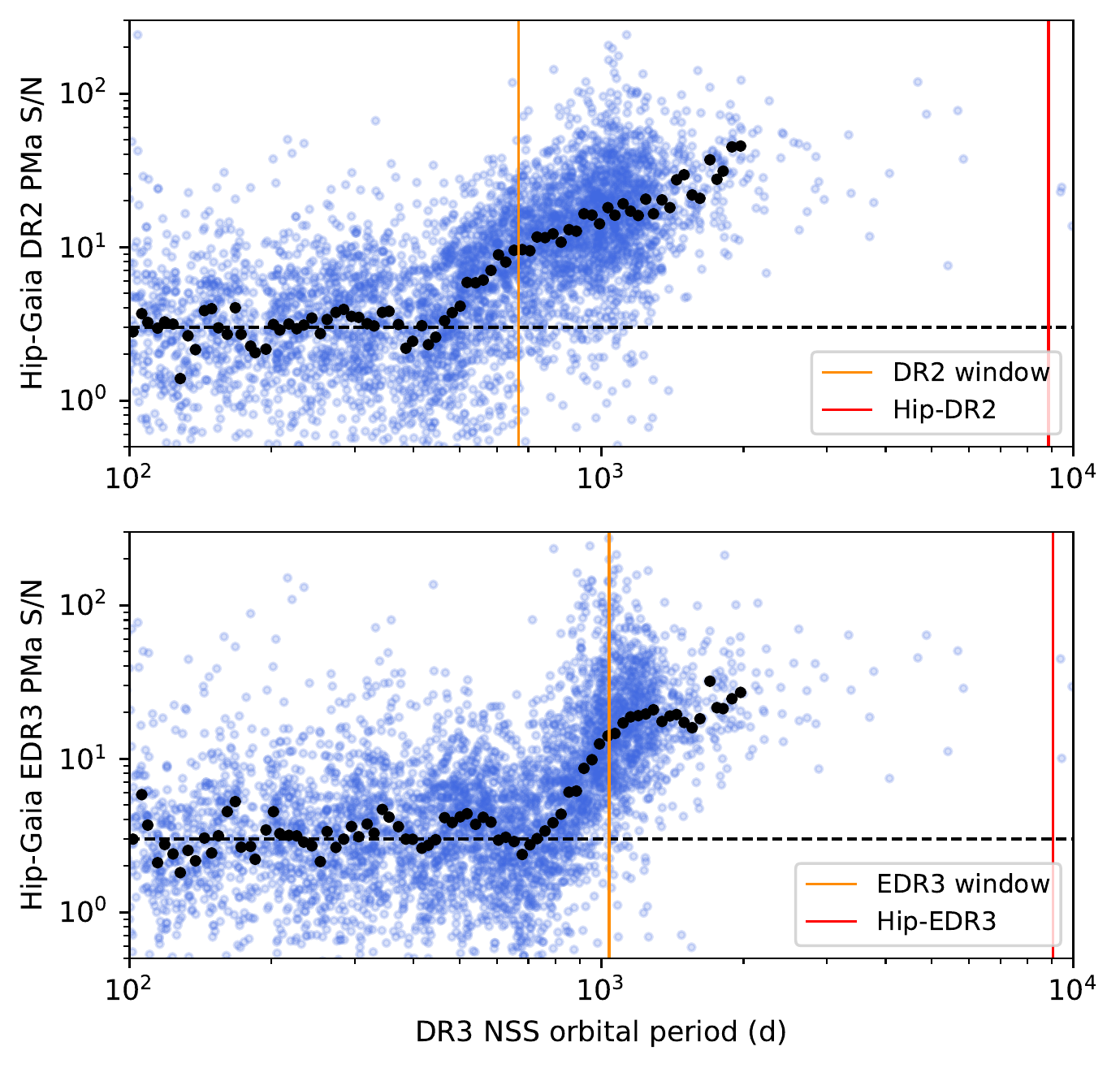}
\caption{Proper motion anomaly S/N as a function of the NSS catalogue orbital period for the DR2 PMa ({\it top panel}); from \cite{Kervella2019}) and the EDR3 PMa ({\it bottom panel}); from \cite{2022A&A...657A...7K}). The horizontal dashed line indicates the PMa S/N=3 significance limit. \label{PMa-TBOperiod-DR2-DR3-SNR-short}}
\end{figure}

The NSS catalogue stars with \fieldName{Orbital} or \fieldName{AstroSpectroSB1} solutions generally have shorter orbital periods than the \gdrthree time window.
Due to the time smearing of the \gdrearlythree proper motions, this usually prevents the production of a clear signature in PMa.
Nevertheless, approximately two thirds of the stars of these NSS classes exhibit a significant PMa signal with S/N > 3 (\figref{NSS-Histogram-PMa}).
As shown in \figref{PMa-TBOperiod-DR2-DR3-SNR-short}, the longer \gdrearlythree time window compared to the DR2 results in a decrease of the PMa S/N for the binaries whose orbital period is shorter than $\approx 1\,000$\,days.
This is caused by the stronger time smearing of the astrometric signal by the integration window in the \gdrthree compared to that of the DR2, that is not compensated by the increase in measurement accuracy in the \gdrthree.
The systems with shorter orbital periods than the \gaia integration window exhibit a median PMa S/N $\approx 3$. This indicates that despite the smearing, statistically the mean \gaia PM vector still contains a significant signature of the binarity. The vast majority of \gaia NSS targets with orbital periods longer than the \gaia time window (both for the DR2 and EDR3) exhibit a significant PMa S/N $> 3$.

\subsubsection{Long-term \hip-\gaia proper motion}

\begin{figure*}
\includegraphics[width=9cm]{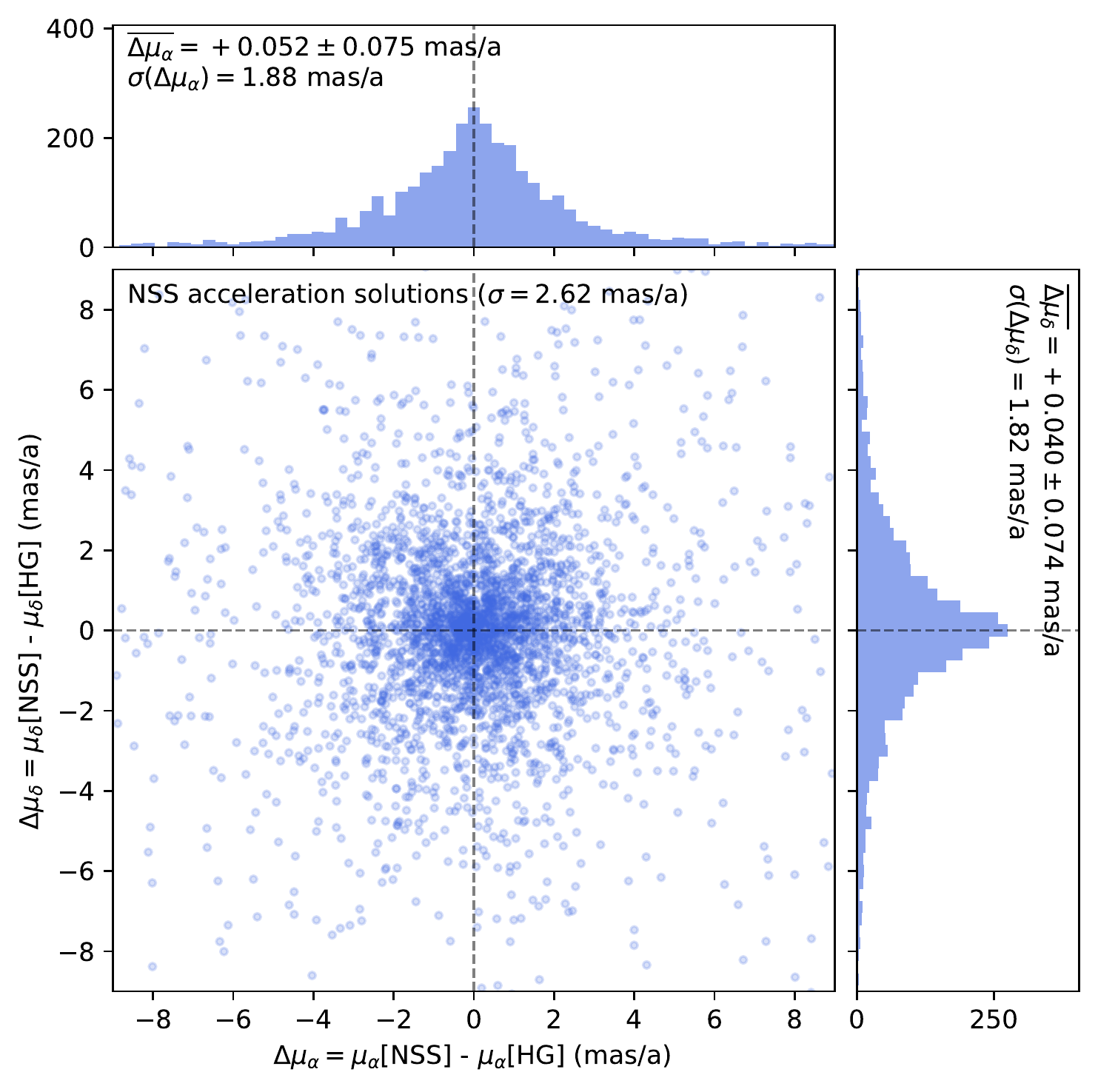}
\includegraphics[width=9cm]{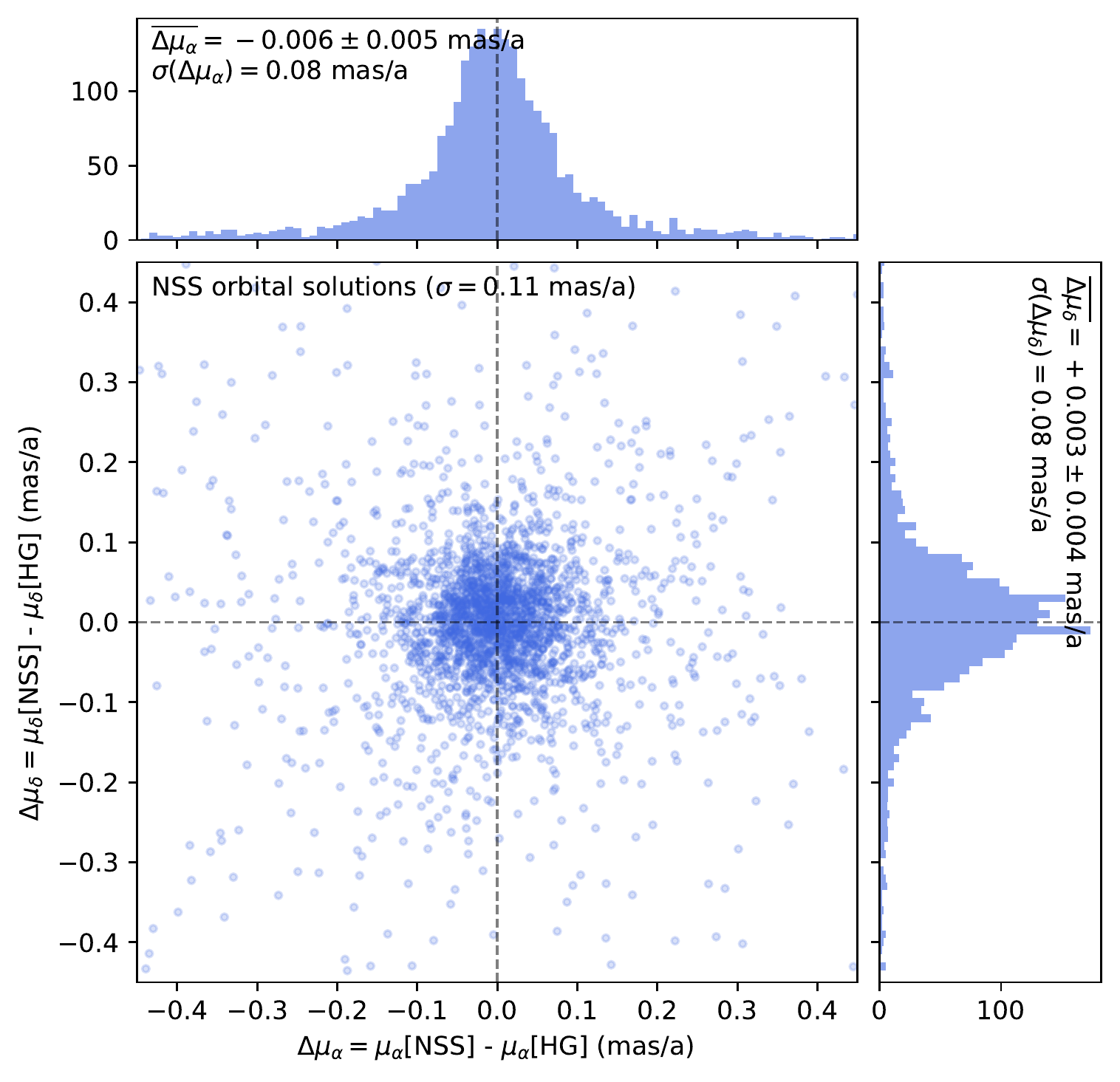}
\caption{Comparison of the long-term proper motions determined from the \hip and \gdrthree positions $\mu_\mathrm{HG}$ with the \gdrthree proper motions $\mu_\mathrm{NSS}$ for NSS stars with acceleration solutions ({\it left panel}) and orbital solutions ({\it right panel}). Note the different scales. \label{HG-NSS-comparison}}
\end{figure*}

We here compare the long-term proper motion deduced from the difference in position between the \hip (1991.25) and \gdrthree (2016.0) epochs by \citet{2022A&A...657A...7K} (hereafter $\mu_\mathrm{HG}$) with the short-term proper motion as determined in the NSS catalogue ($\mu_\mathrm{NSS}$).
Figure~\ref{HG-NSS-comparison} shows the observed differences $\Delta \mu = \mu_\mathrm{NSS} - \mu_\mathrm{HG}$ between these two quantities for the \hip catalogue stars with either accelerations (\fieldName{Acceleration7}, \fieldName{Acceleration9}) or orbital (\fieldName{Orbital}, \fieldName{AstroSpectroSB1}) solutions in the NSS.
There is a significantly larger divergence of the long-term proper motions between the stars with NSS accelerations only for which $\sigma(\Delta \mu) \approx 2.6$\,mas\,a$^{-1}$, than for the stars with an orbital solution for which $\sigma(\Delta \mu) \approx 0.1$\,mas\,a$^{-1}$.
The relatively poor agreement for the NSS acceleration stars may be explained by the fact that the measurement of the curvature of the sky trajectory is significantly easier with the longer \hip-\gaia temporal baseline.
For the full NSS orbital solutions, the agreement between the \hip-\gaia PM and the NSS PM is remarkably good, demonstrating that the orbital fit procedure does not introduce systematic biases on the estimation of the mean PM value.

%
\section{Catalogue of masses}
%
As the \fieldName{nss_two_body_orbit} table gives access to the orbital parameters
only, it was found desirable to provide an estimate of the masses, of the flux ratio,
or to lower and upper limits of these, when this was possible. We describe in this Section
the construction and content of the table \MASSTable which has been made 
available in the \gaia Archive.
%
\subsection{Computation of the masses}\label{Sect:masscat}
%
The astrometric binaries give access to an astrometric mass function which depends on the flux ratio ($F_2/F_1$) of the components: 
\begin{equation}
(\Mass_1+\Mass_2)\left(\frac{\Mass_2}{\Mass_1+\Mass_2}-\frac{F_2/F_1}{1+F_2/F_1}\right)^3=\frac{(a_0/\varpi)^3}{(P/365.25)^2} \,\,\, .
\label{eq:astroMF}
\end{equation}

While the spectroscopic binaries provide a spectroscopic mass function which also depends on the inclination:
\begin{equation}
f(\Mass)=\frac{\Mass_2^3 \sin^3{i}}{(\Mass_1+\Mass_2)^2}=1.0385\times10^{-7} K_1^3 (1-e^2)^{3/2} P
\,\,\, ,
\label{eq:spectroMF}
\end{equation}
\noindent with $P$ the period in days and $K_1$ the semi-amplitude of the primary in \kms.
For \fieldName{AstroSpectroSB1}, we have access to the Thiele Innes coefficients instead of $K_1$ which leads to the equivalent formula:
\begin{equation}
\frac{\mathcal{M}_2^3 \sin^3{i}}{(\mathcal{M}_1+\mathcal{M}_2)^2}= \frac{(C^2+H^2)^{3/2}}{(P/365.25)^2}.
\label{eq:spectroMFTI}
\end{equation}
The inclination can be provided by an astrometric solution or an eclipsing one. Without the inclination, Eq.~\ref{eq:spectroMF} only leads to a minimum mass function information. When a \fieldName{SB2} solution is provided, we have access to the mass ratio $\Mass_2/\Mass_1=K_1/K_2$. 
When a system has a \fieldName{SB2} solution and either an astrometric solution or an eclipsing one, the primary mass can be derived directly from the binary orbital parameters. 

Two estimates of $\Mass_1$ are provided in the \gdrthree\ by the \fieldName{FLAME} module \citep{DR3-DPACP-157}: \fieldName{mass_flame}, based on GSP-Phot parameters and available in the \APTable table and \fieldName{mass_flame_spec}, based on GSP-Spec parameters and available in the \APTableSupp table. 
However the \fieldName{FLAME} masses have three main limitations for our NSS sample: they are based on the parallax from the main catalogue while we now have a more accurate estimate for all astrometric solutions; they also assume a null flux ratio, which we know is not adapted for a significant fraction of the NSS solutions, in particular the SB2 ones; they are not available for stars with masses smaller than 0.5\Msun. We therefore implemented a specific code to derive the mass of the primary that allows to play with the luminosity ratio and is described in detail in Appendix~\ref{sec:homemademass}. These masses are only derived for stars on the main-sequence as estimations for evolved stars are degenerate \citep[e.g.][]{DR3-DPACP-157}.
They are called hereafter \quoting{IsocLum}. For white dwarfs, we simply assumed a fixed mass of 0.65$\pm0.16$\Msun \citep{2012ApJS..199...29G}.

The \afterReferee{uncertainties} on the masses and flux ratios obtained are derived using a Monte Carlo simulation of 1\,000 points. We take into account the covariance matrix of the orbital parameters. For $a_0$ as well as for the \fieldName{AstroSpectroSB1} spectroscopic part $a_1=\sqrt{C^2+H^2}/\sin{i}$, we use a Gaussian distribution with a local linearisation error estimation as Monte Carlo techniques are not adapted to the Thiele Innes coefficients
\citep[see ][]{DR3-DPACP-127}. Only sources with a \dt{significance} $>5$ are \afterReferee{present in NSS solutions} so that a Gaussian distribution of the semi-major axis \afterReferee{errors} is a reasonable assumption. The \afterReferee{uncertainties} for the SB2 and eclipsing solutions have been re-scaled according to their goodness-of-fit. For the IsocLum masses we use the full mass distribution function as we have it available. We then compute the 16th and 84th quantiles (corresponding to $\pm 1\sigma$) of the derived parameter distributions to estimate the lower/upper values respectively. The direct values are provided whenever applicable for \dt{m1}, \dt{m2}, \dt{fluxratio}. 

When combining two NSS solutions, we only use those with periods and eccentricities compatible within 5$\sigma$, assuming an uncertainty of 0.1 on the eccentricity for sources with a fixed eccentricity. A weighted mean of the periods and eccentricities of both solutions is then used in Eqs.~\ref{eq:astroMF} and \ref{eq:spectroMF}.
For the combination of astrometric and spectroscopic solutions, the primary mass is tested for different flux ratios by steps of 0.01, then for each of these, the secondary mass is derived from Eq.~\ref{eq:spectroMF} and the flux ratio from Eq.~\ref{eq:astroMF}; only solutions which are consistent with the tested flux ratio are kept and when no solution is consistent, the closest one is used.  

For \fieldName{Orbital} solutions, only upper and lower values can be derived as the flux ratio is not known. Different flux ratios are tested by steps of 0.01. The lower (resp. upper) secondary mass value is computed from the mass distribution obtained with the lower flux ratio (resp. upper).
\beforeReferee{flux ratio tested $>0$, the secondary mass derived is tested to be consistent with a secondary star on the main sequence to be accepted. }\afterReferee{The solution with \fieldName{fluxratio} $=0$ is always kept, as soon as the primary star magnitude is compatible with the isochrones. For the other flux ratios tested, the secondary mass derived is accepted if consistent with a secondary star on the main sequence. In practice, the Monte Carlo masses of the secondary lead to a range of possible absolute magnitudes from the isochrones, which, for the flux ratio tested, are converted into an absolute magnitude of the system which is accepted when it is at less than 3 sigma from the absolute magnitude of the system measured by Gaia.} In some cases, this leaves no flux ratio \beforeReferee{accepted}\afterReferee{kept}. These can be either pre-main sequence stars, in which case our masses are invalid, or \beforeReferee{ternary}\afterReferee{triple} systems with a primary which needs a flux ratio but a secondary mass not consistent with it. To handle this second option, the minimum flux ratio \beforeReferee{used to accept this star}\afterReferee{compatible with a primary on the main sequence} is used to derive the primary mass but the secondary mass is derived with $F_2/F_1=0$. These cases can be isolated with a `\fieldName{fluxratio_upper is NULL}' query. 
No limit is tested on the maximum flux ratio for white dwarfs. 
For SB1 solutions, the lowest valid flux ratio is used to derive the primary mass and the lower secondary mass value is derived on the distribution assuming $\sin i=1$. For eclipsing binaries, the flux ratio is fixed by \fieldName{g_luminosity_ratio}.

The catalogue of masses we derive is available in the \gaia Archive table \MASSTable. A summary of its content is presented in \tabref{tab:nssmass}. 
We selected only sources with a signal-to-noise ratio higher than 5 on the astrometric semi-major axis and on the spectroscopic primary semi-amplitude as well as a signal-to-noise ratio higher than 2 for the eclipsing binary and astrometric $\sin{i}$ and the spectroscopic secondary semi-amplitude.  For \fieldName{AstroSpectroSB1} solutions, both the signal-to-noise for the spectroscopic part, computed as the one of $\sqrt{C^2+H^2}$, and of $a_1$ are checked to be higher than 5. If not, the \fieldName{AstroSpectroSB1} is treated as an \fieldName{Orbital} solution only. \fieldName{OrbitalAlternative} solutions with $\log_{10}(\fieldName{parallax}/\fieldName{parallax_error})< 3.7-1.1 \log_{10}(\fieldName{period})$ 
have been excluded. 
76 sources are duplicated, having both an astrometric solution and either an eclipsing binary (6) or a \fieldName{SB2} solution (70) one, with the astrometric solution period being larger than the other one by more than 10 sigma. For sources with both an \fieldName{SB1} and an \fieldName{Orbital} solution, only the \fieldName{Orbital} solution has been kept.

\begin{table}
\caption{Content of the Catalogue of masses.}
{
\begin{tabular}{lcccc} 
\hline\hline
Combination Method & Number & $\Mass_1$ & $\Mass_2$ & $F_2/F_1$\\
\hline
Orbital+SB2               & 23 & \checkmark & \checkmark & \checkmark\\
EclipsingSpectro(SB2)     & 3 & \checkmark & \checkmark &\\
Eclipsing+SB2             & 53 & \checkmark & \checkmark &\\
AstroSpectroSB1+M1        & 17578 & & \checkmark & \checkmark\\
Orbital+SB1+M1            & 1513 & & \checkmark & \checkmark \\
EclipsingSpectro+M1       & 71 & & \checkmark & \\
Eclipsing+SB1+M1          & 155 & & \checkmark &\\
SB2+M1                    & 3856 & & \checkmark & \\
Orbital+M1                & 111792 & \multicolumn{3}{c}{lower/upper} \\
SB1+M1                    & 60271 & & lower &\\
\hline
\end{tabular}
}
\tablefoot{The full table is available in the \gaia table \MASSTable. 
}
\label{tab:nssmass}
\end{table}

\begin{figure}\begin{center}
\includegraphics[width=0.9\columnwidth]{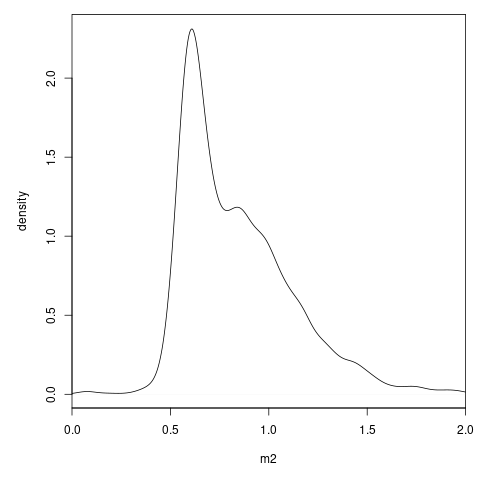}
\caption{Distribution of the secondary mass of astrometric solutions with \fieldName{fluxratio_upper} $=0$ in \tabref{tab:nssmass}.}
\label{fig:fr0m2dens}
\end{center}\end{figure}

A particularly interesting subset of this table are the astrometric solutions with \fieldName{fluxratio_upper} $=0$. There is only one star (\object{\sourceId{4288765058313593856}}) for which the secondary mass is sufficiently small ($0.57$\Msun) compared to the primary (1.26\Msun) so that \fieldName{fluxratio_upper} $=0$ is compatible with the secondary star being on the main sequence. 
The others are systems for which the secondary mass solutions for flux ratio $>0$ did not have the mass compatible with any of the flux ratio tested. The secondary mass distribution of these is presented in \figref{fig:fr0m2dens}. There are 12 stars with a secondary mass smaller than the minimum mass handled by the isochrones of 0.1$\Msun$ and a low mass for the primary too. Three other stars with low mass secondaries could be either \beforeReferee{ternary}\afterReferee{triple} or pre-main-sequence stars for which the primary mass is not correct. The most predominant peak is the one of the white dwarfs at $\Mass_2=0.61$\Msun which has a standard deviation of 0.07\Msun. Note that some white dwarf companions should actually have a flux ratio $>0$, such as \object{\sourceId{6416572288572864512}} which is an \fieldName{AstroSpectroSB1} with a significant flux ratio; the primary mass has been estimated as the one of a metal-poor star due to its location on the left of the main sequence; if it had been solved as an astrometric solution only it would have had \fieldName{fluxratio_upper} $=0$ and a secondary mass under-estimated. A long tail of high mass secondaries is also present. They can be compact objects but also \beforeReferee{ternary}\afterReferee{triple} stars for which the single primary star hypothesis was not valid (See \secref{triage}) or primary stars who started to evolve or metal-poor giants for which the primary mass is not correct.

%
\subsection{Masses using external data}
%
In this section we illustrate how other mass estimates can be obtained thanks to 
various kinds of combinations with external data.

\subsubsection{External SB2}

Combining astrometric orbits with spectroscopic ones from large surveys
is not recent, and it was e.g. done with \hip \citep{2000IAUS..200P.135A}.
Once the inclination known from the astrometric orbit, the semi-amplitudes from the 
spectroscopic orbit allows to determine simultaneously the masses and luminosities
of the two components.
Recently, APOGEE DR17 data has been used to detect 8\,105
SB2 or higher order systems \citep{kounkel21}. Once the needed number of epochs
will be available, individual masses and magnitude differences will be obtained
for the sources having an NSS \fieldName{Orbital*} solution. Here, we just take
the example of \object{\sourceId{702393458327135360}}, with 
$K_1=19.53\pm$0.95 \kms\ and $K_2=21\pm$1.1 \kms.
The masses are
found to be $\Mass_1 = 1.14 \pm 0.38\Msun$ and $\Mass_2 = 1.06 \pm 0.35\Msun$
with a $0.567\pm0.071$ flux ratio.

\subsubsection{Occultations}

Occultations represent a neglected way to constrain the sum of masses of binaries, thanks to the measurement of their separation at a given epoch. 

We illustrate this with \object{\sourceId{3162827836766605440}} (HIP 36189) which is a $V\approx 6.5^m$ K giant that has been discovered as double thanks to an occultation by (704) Interamnia on 23 March 2003. Its acceleration had been detected in \citet{Kervella2019}, \citet{Brandt2021}, and \citet{2022A&A...657A...7K}. 
\cite{2014IJAA....4...91S} indicate a  $\rho= 12 \pm 3$ mas separation while a more precise indication is given by \cite{2020MNRAS.499.4570H}, $\rho= 13.0 \pm 0.7$ mas with position angle $\theta = 231.9 \pm 4.0 \deg$.
\cite{2014IJAA....4...91S} evaluate the magnitude difference between components to about 1.5, to which we attribute a 0.2 mag uncertainty, accounting in particular for the observation in a band different from the \gmag band.

This source has received an \fieldName{Orbital} solution with a 2.6~yr period. From the combined information,
the masses of the components are found to be $\Mass_1=3.9\pm2.2\Msun$ and $\Mass_2=3.5\pm1.6\Msun$.

\subsubsection{One SB1 Cepheid}\label{Sect:Cepheids}

Although it is known that many Cepheids are in binary systems \citep[e.g. ][]{2019A&A...623A.116K}, not many orbits are present in the \NSSa Catalogue.
On the spectroscopic orbit side, the \gdrthree data processing did not include yet 
the simultaneous handling of orbit and Cepheid pulsations, so that is the latter only 
that were detected. 
Consequently, these solutions were filtered out from the catalogue to avoid any
confusion. On the astrometric orbit side, only one deserves attention which received an 
\fieldName{Orbital} solution.

\sourceId{470361114339849472} = \object{RX Cam} is known to be a G2Ib+A0V spectroscopic binary from \cite{1996A&AS..116..497I}. 
The \gaia solution has a period $999\pm 104$~d and an eccentricity $0.514\pm 0.049$,
consistent at $1\sigma$ with respectively $1113.8\pm 0.5$~d and $0.459\pm 0.007$ 
from \cite{2013A&A...550A..70G}. The inclination is $i=113\fdg 5\pm 1\fdg 7$.
We may safely assume that there is no flux contribution (in the $G$ band) from the companion, as confirmed by the difference between the semi-major
axis of the primary and that of the photocentre $a_1-a_0=0.04\pm 0.12$ au.
Using the semi-amplitude $K_1=14.27\pm 0.11$ \kms\ from \cite{2013A&A...550A..70G}
and the estimation 
of the primary mass from \cite{2019A&A...623A.116K}, $\Mass_1=5.40\pm 0.81$\Msun,
we obtain $\Mass_2=2.87\pm 0.34$\Msun.

%
\section{Special binaries in the HRD}
\begin{figure}[htb]
\includegraphics[width=1.05\columnwidth]{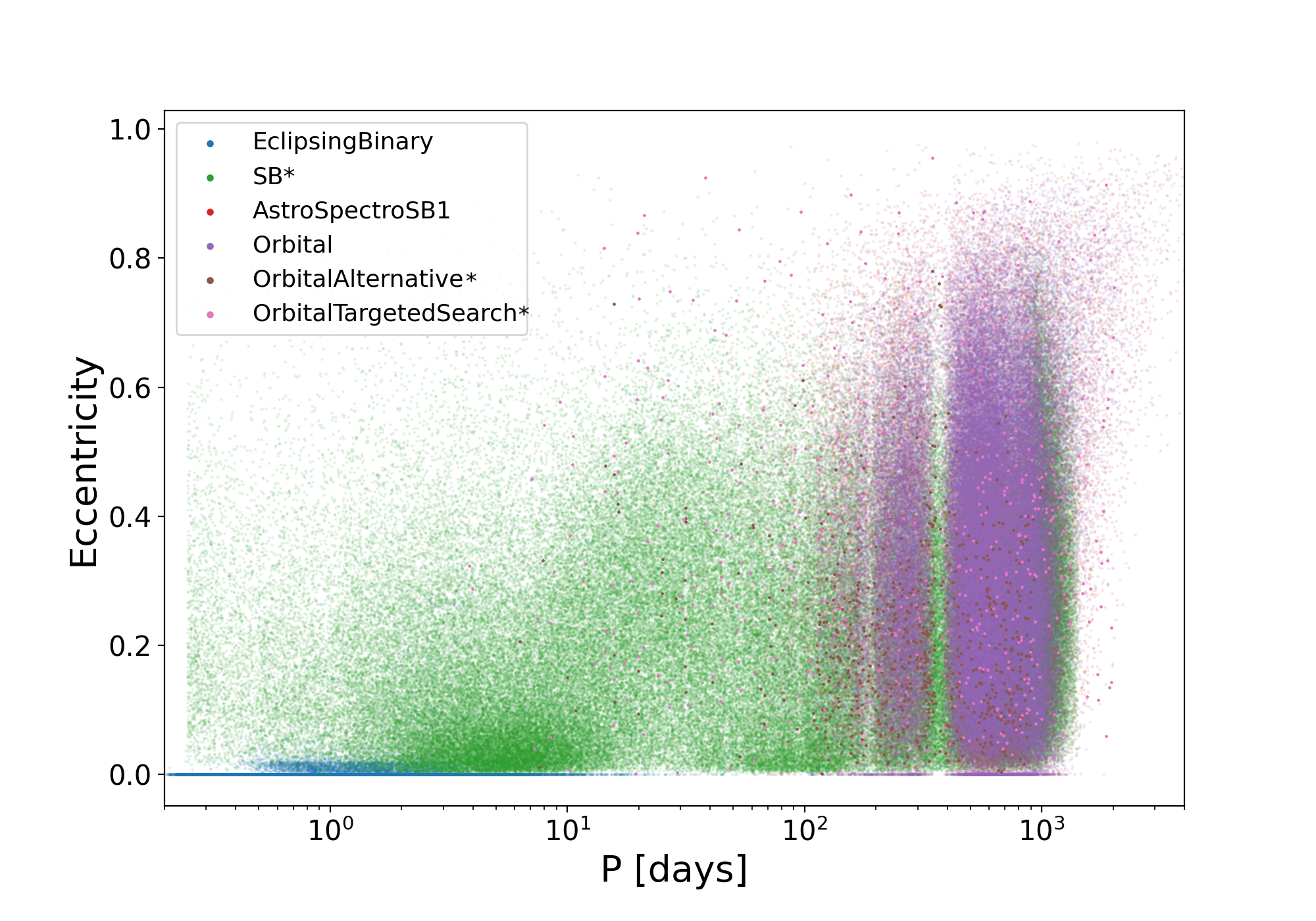}
\caption{Eccentricity vs period for most solutions with orbits.}\label{fig:elogP_TBO}
\end{figure}

%
In this section, we select several illustrative cases where the NSS catalogue
can provide a 
useful insight into populations of the Hertzsprung-Russell diagram (HRD).
\afterReferee{For further reference we present in \figref{fig:elogP_TBO} the period-eccentricity
diagram for the NSS solutions with orbits.}

%
\subsection{Spectroscopic binaries along the main sequence}
\label{Sect:SB_MS}

This section presents and briefly comments the eccentricity-period (hereafter \eP) diagrams 
of SB1s along the main sequence, defined as  $-7.5+10\;(G_{\rm BP,0}-G_{\rm RP,0}) < M_{G,0}$ (\figref{Fig:HRD_SB1_MS}), with the photometry being de-reddened in the same way as for the mass derivation (see Appendix \ref{sec:homemademass}). 
Stars along the main sequence are divided according to $(G_{\rm BP,0}-G_{\rm RP,0})$ bins, as indicated in \figref{Fig:HRD_SB1_MS}.

\begin{figure}[htb]
\includegraphics[scale=0.6, angle=0]{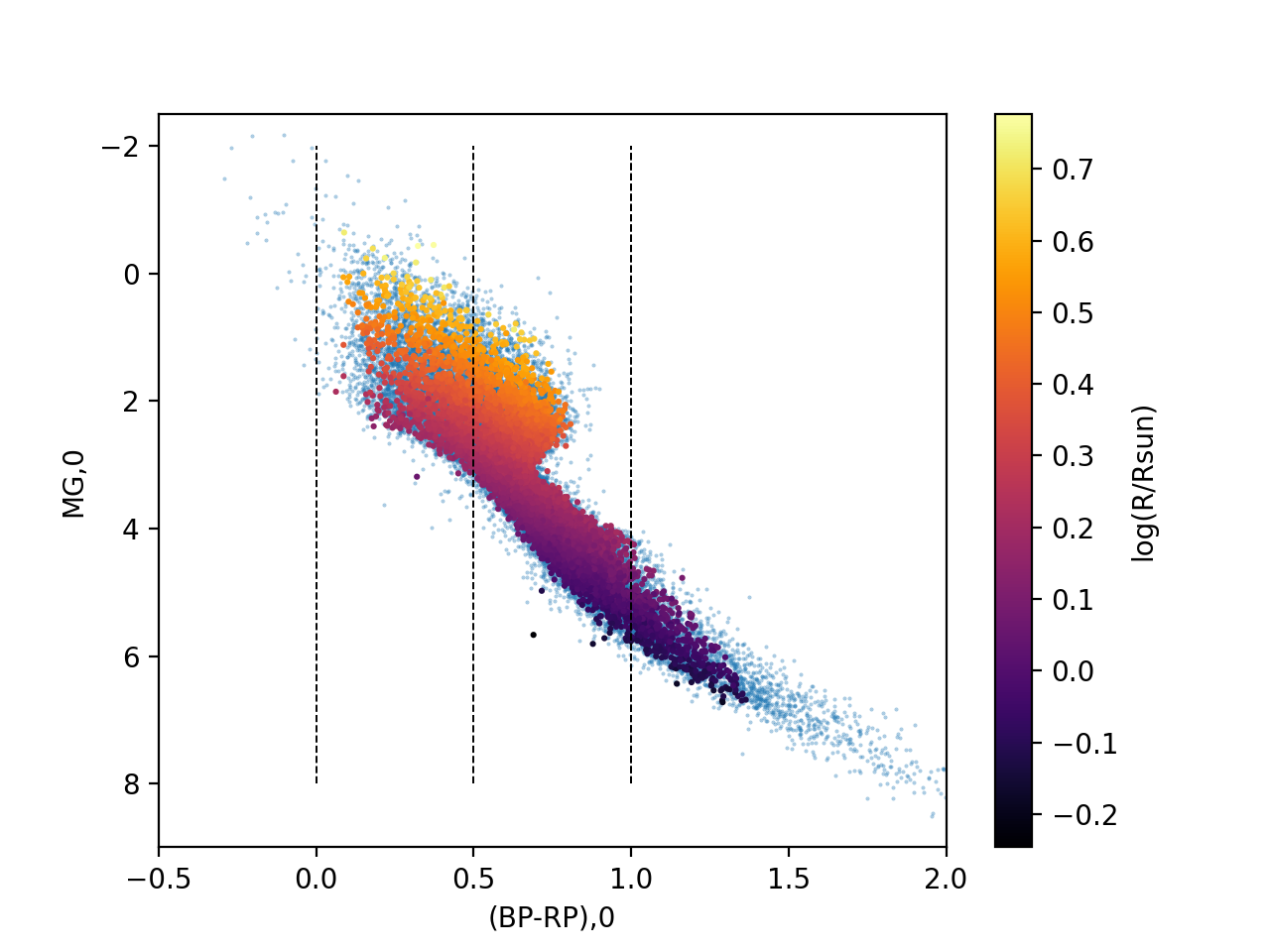}
\caption[]{Location in the dereddened HRD of the $(G_{BP,0}-G_{RP,0})$ bins used in \figref{Fig:e-P_MS}. Small blue dots correspond to the SB1 not selected by our selection criteria. The radius is the \fieldName{FLAME} estimate. 
\label{Fig:HRD_SB1_MS}
}
\end{figure}

\begin{figure*}[htb]
\includegraphics[scale=0.494, angle=0]{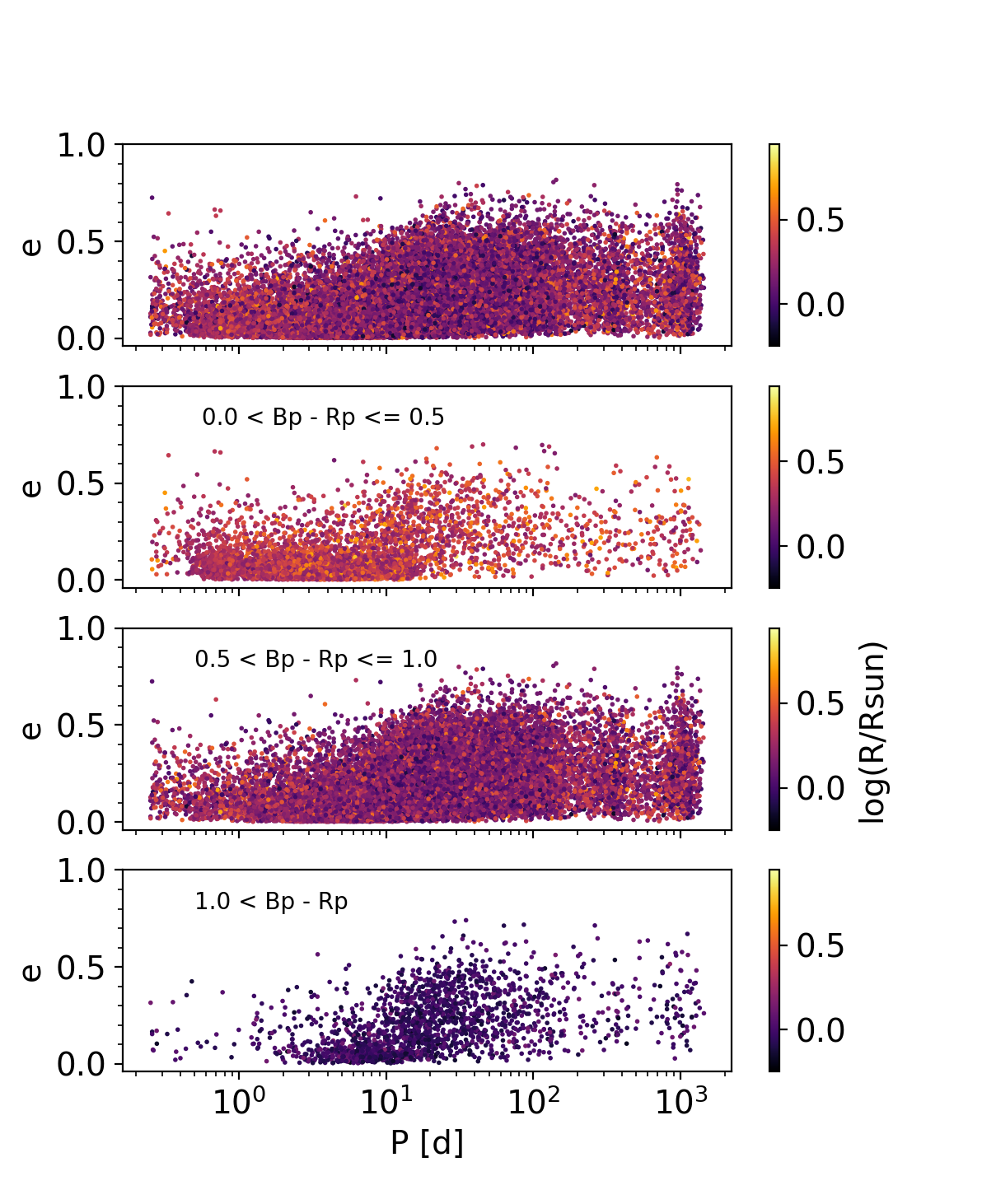}
\hspace*{-1.6cm}
\includegraphics[scale=0.475, angle=0]{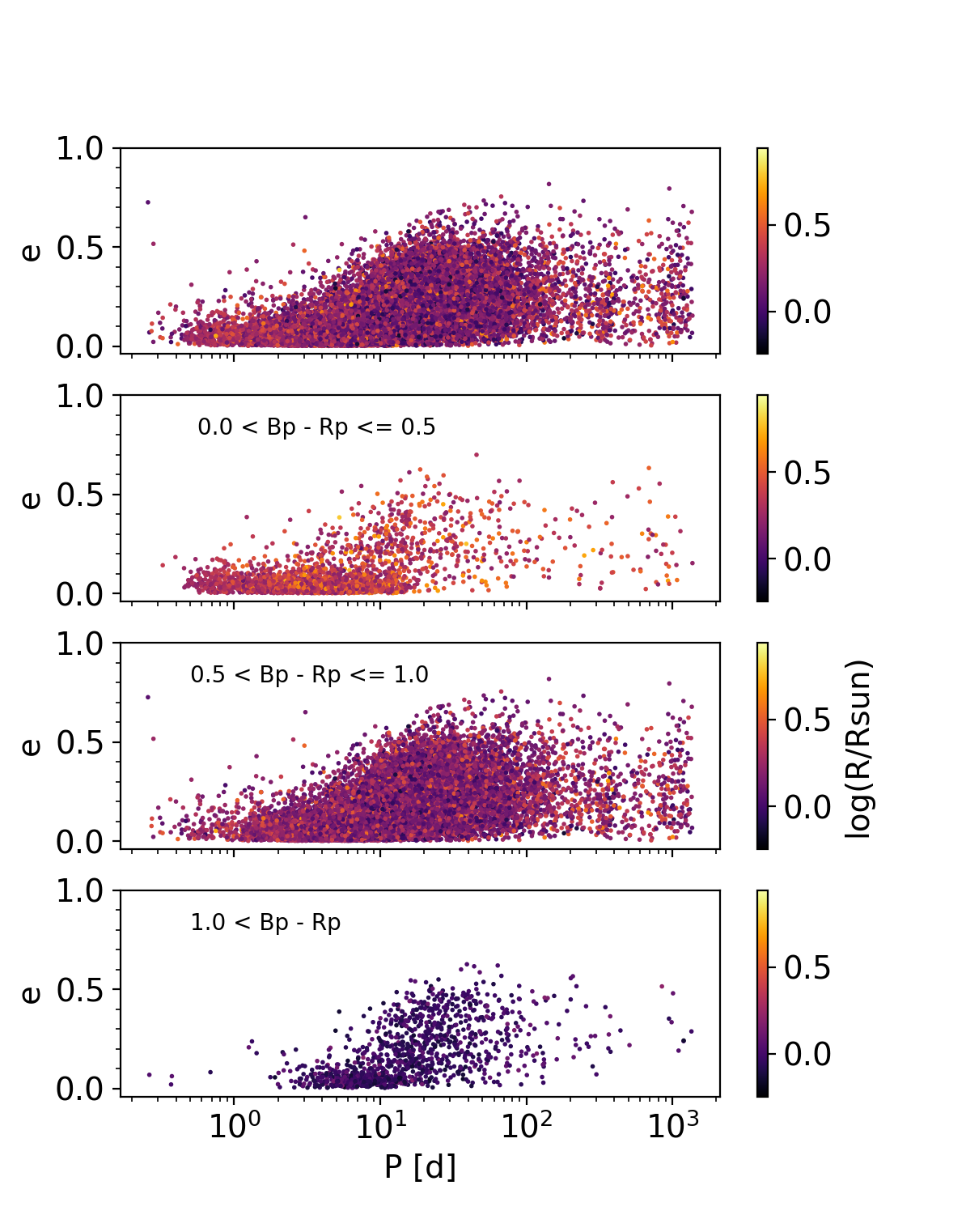}
\hspace*{-1.5cm}
\includegraphics[scale=0.46, angle=0]{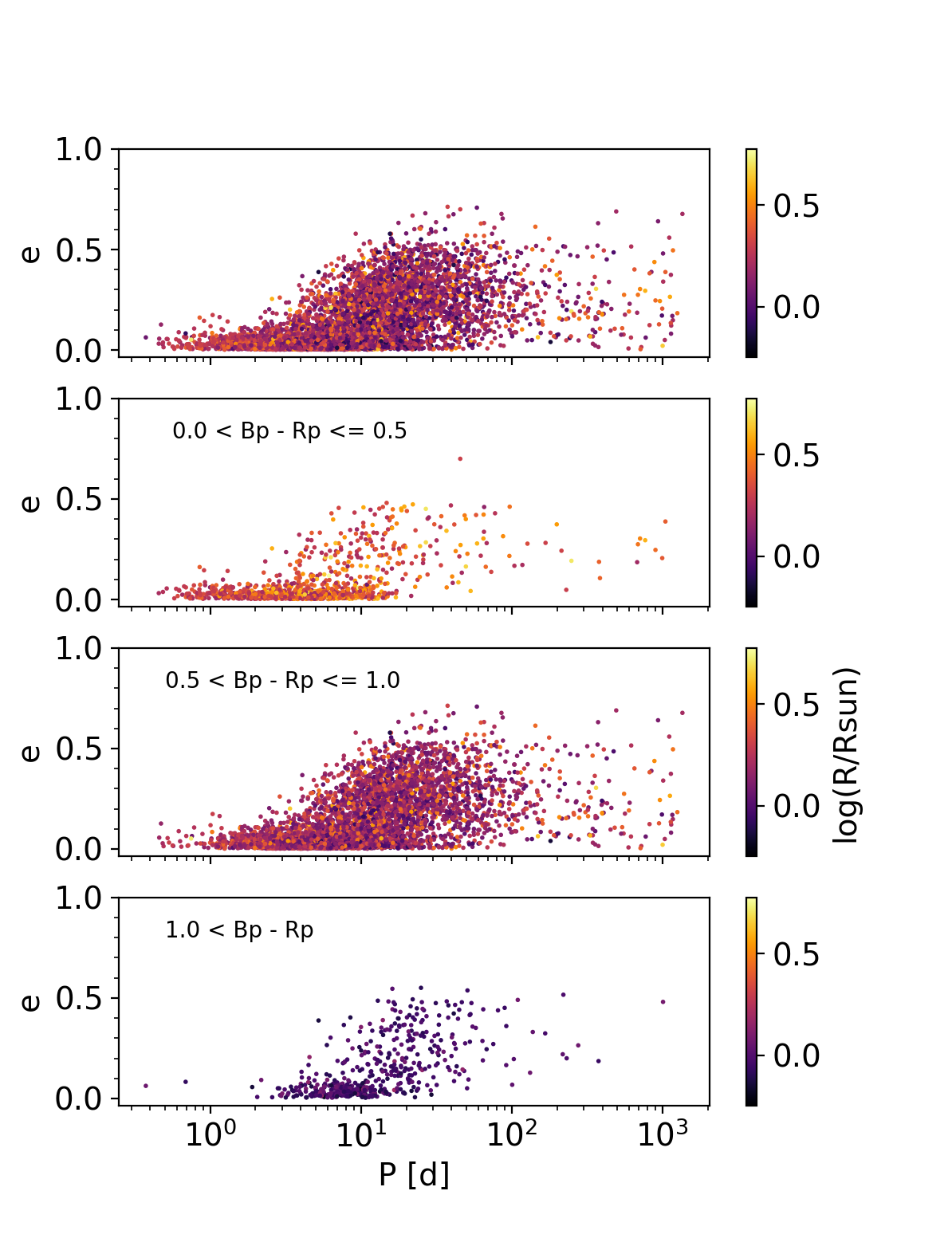}
\caption[]{The \eP\ diagram for SB1s along the main sequence, 
filtered according to  significance factors larger than 10, 20 or 40 (from {\it left} to {\it right}), and for different 
$(G_{\rm BP,0}-G_{\rm RP,0})$
spans ({\it top} to {\it bottom}). Note how the long-period tail gets more populated when the significance is allowed to be smaller ({\it left panels}), since long-period orbits have smaller $K_1$ on average, and hence smaller significances  $K_1/\sigma_{K_1}$. But at the same time the populated region becomes almost rectangular ({\it top left panel}), which appears quite unusual. There is a drop in the number of systems at $P = 0.5$~y due to insufficient sampling at this specific period. The color codes for the \fieldName{FLAME} radius estimate.
\label{Fig:e-P_MS}
}
\end{figure*}

The \eP\ diagrams along the main sequence are displayed in \figref{Fig:e-P_MS}.
Because of the aliasing problems faced by the SB1 processing (see \secref{sssec:cu4nss_intro_outputfiltering}), these diagrams are shown for different filtering based on the significance of the RV semi-amplitude, namely $K_1/\sigma_{K_1}$ larger than 10, 20 or 40 (from left to right). This filtering removes both high-eccentricity short-period orbits and long-period orbits. The former filtering is on purpose to remove possibly spurious solutions, while the disappearance of the long-period solutions is a side effect due to the fact that long periods have on average smaller $K_1$ and thus smaller significances as well.
Nevertheless, this filtering has the consequence of revealing the shape expected for any \eP\ diagram, namely short-period orbits being almost exclusively circular below a given \quoting{transition} period \citep[see e.g.,][for a detailed discussion]{Mazeh2008}.

The \eP\ diagrams along the main sequence, when ordered according to bins of increasing $G_{\rm BP,0}-G_{\rm RP,0}$ and properly filtered on a significance larger than 40 (right panels of \figref{Fig:e-P_MS}), reveal that this transition period does not vary strongly between the various $G_{\rm BP,0}-G_{\rm RP,0}$ bins, contrarily to the situation prevailing along the giant branch, as discussed below (\secref{Sect:HRDgiants}). \citet{Mazeh2008} has reviewed the processes shaping \eP\ diagrams, with the conclusion that the constancy of the transition period along the main sequence would naturally result if the circularisation occurred during the pre-main-sequence phase, when the stars were large, following a suggestion by \citet{ZahnBouchet1989} for F, G and K stars from 0.25 to 1.25~\Msun.  Their transition period does stay constant along the main sequence at about 8~d. The transition period observed on \figref{Fig:e-P_MS} seems a bit shorter though.  
x
\citet{Mazeh2008} also argues that short-period binaries (i.e., below the transition period) with non-circular orbits could result from a third distant companion pumping eccentricity into the binary orbit. However, at the time being, due to the confusion caused by possible period aliasing among short-period SB1 systems, this possibility cannot be tested with confidence.

%
\subsection{Binaries along the RGB/AGB}
\label{Sect:HRDgiants}
%

The goal of this section is to show that the transition period between circular and non-circular orbits is increasing with radius and luminosity along the red giant branch (RGB) and asymptotic giant branch (AGB). To select stars on these branches, it is more efficient to use the 2MASS colour - magnitude diagram ($J-K,$ \MK) rather than the usual Gaia colour - magnitude diagram. We used the 2MASS cross-match with EDR3 available within the data archive\footnote{\url{https://gea.esac.esa.int/archive/documentation/GEDR3/Gaia_archive/chap_datamodel/sec\_dm\_crossmatches/ssec\_dm\_tmass\_psc\_xsc\_best\_neighbour.html}} and used the following criteria to select giants: 
\begin{equation}\label{Crit:RGB_AGB_criterion}
J-K > 0 \mathrm{~and~} M_K < 0,
\end{equation}
as illustrated below.

\subsubsection{Period - radius diagram}

\begin{figure*}[htb]\centering
\includegraphics[scale=0.36]{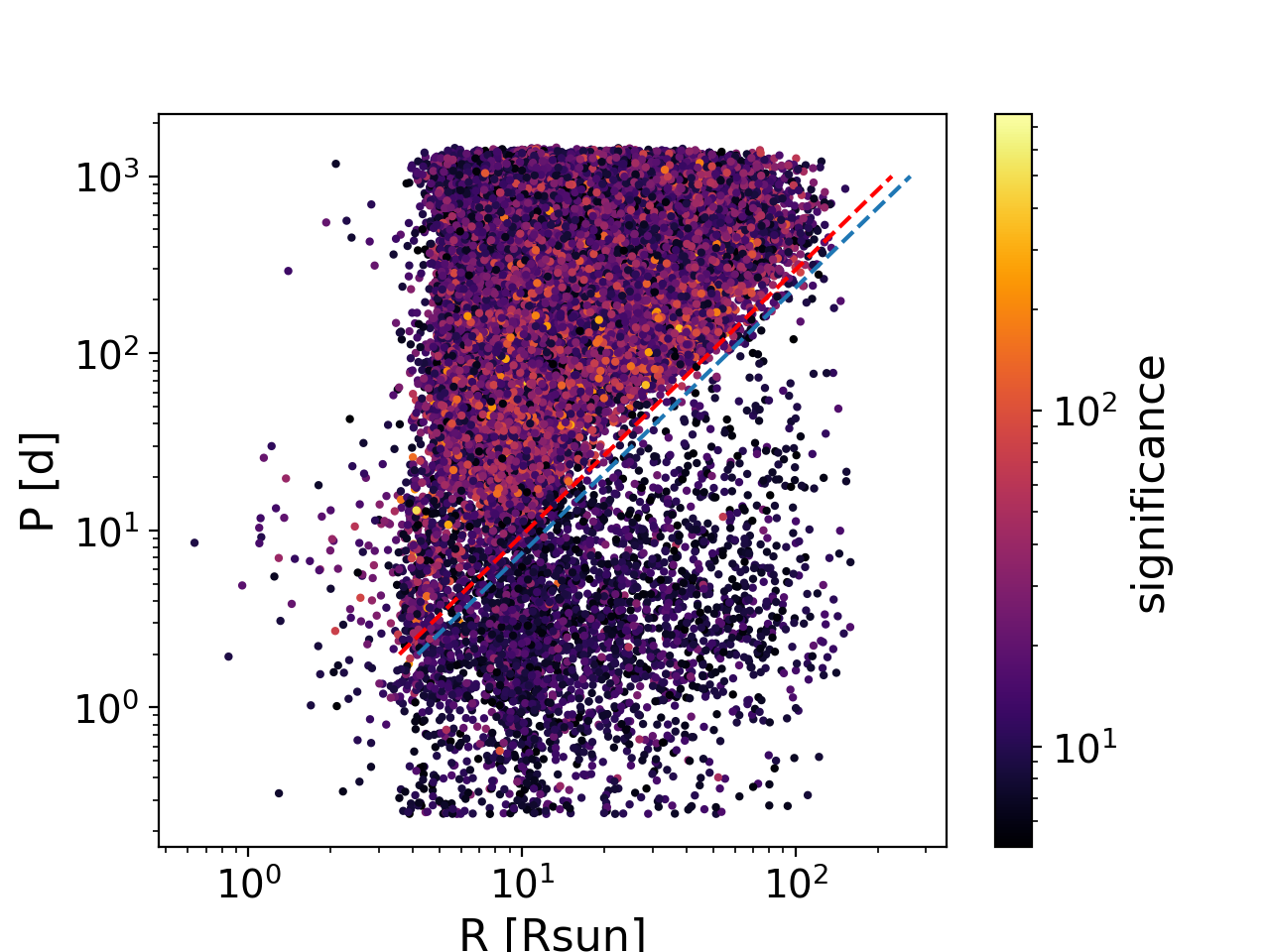}
\includegraphics[scale=0.36]{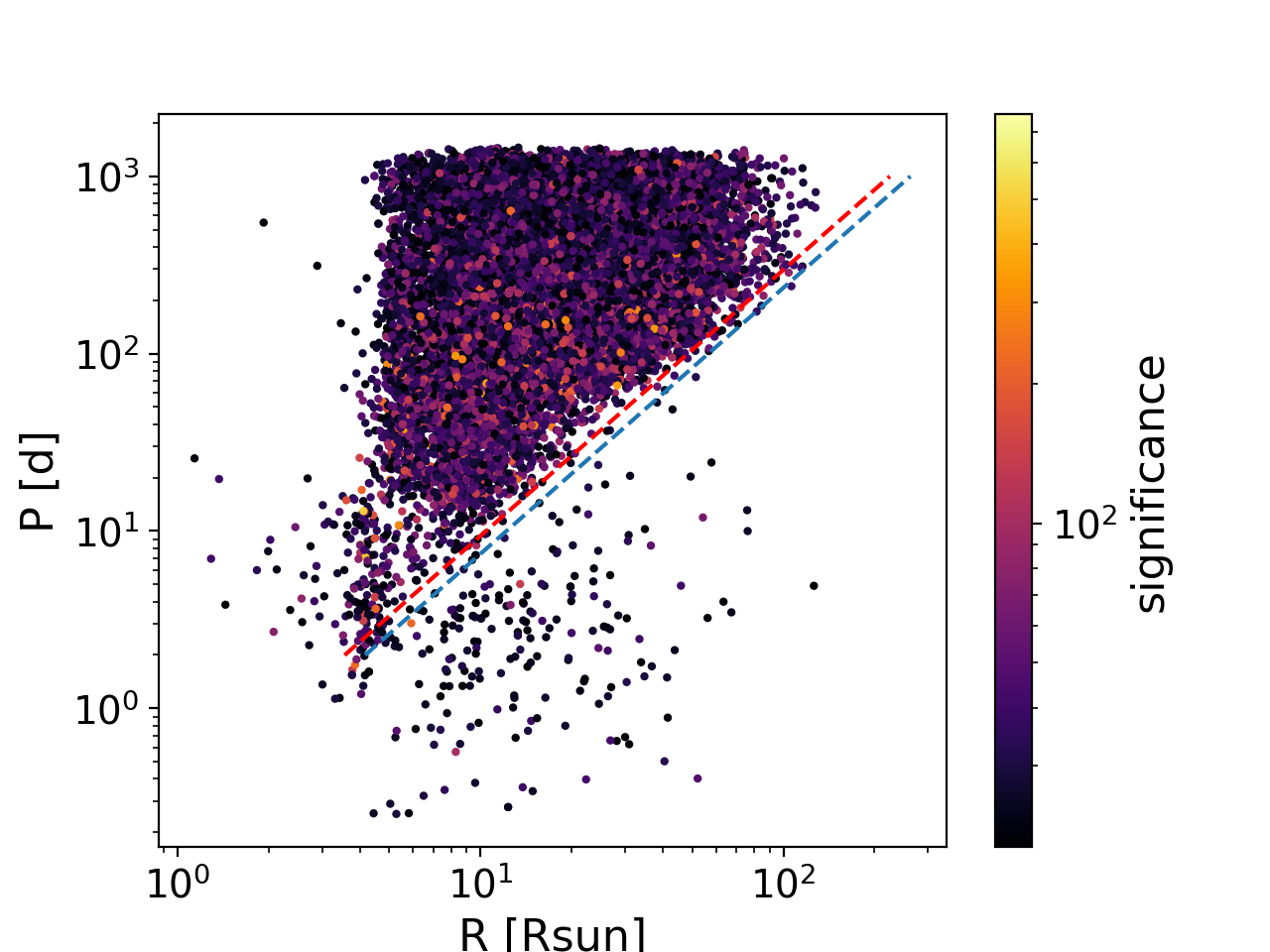}
\includegraphics[scale=0.36]{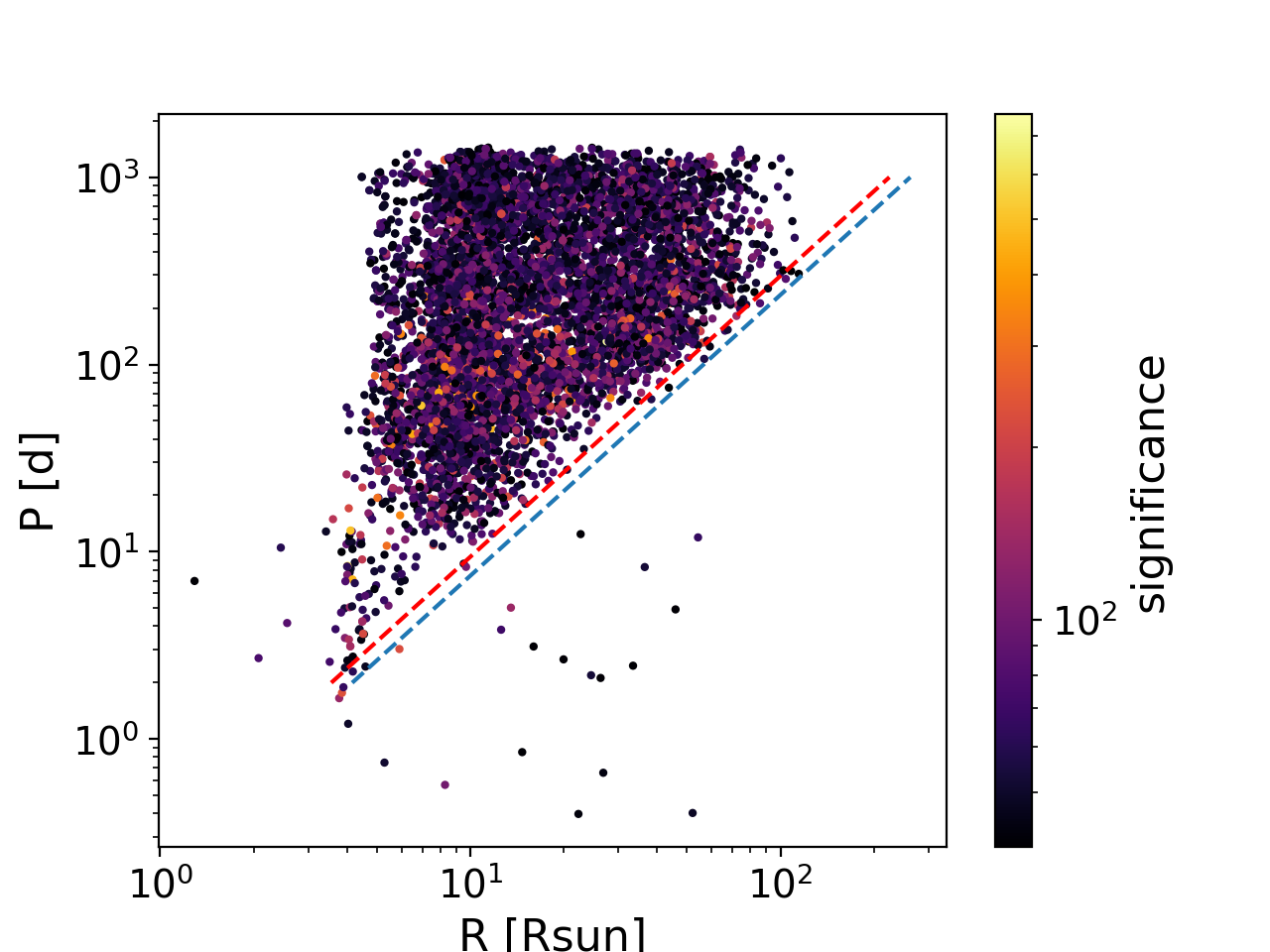}
\caption[]{The period - radius
diagram for all \fieldName{SB1} 
solutions falling along the RGB/AGB, according to the criterion \eqref{Crit:RGB_AGB_criterion}, and with a radius available from \fieldName{radius_flame}. The dashed lines correspond to the threshold periods expressed by Eq.~\ref{Eq:Pthres} for $\Mass_1 = 1.3\Msun$ and $\Mass_2 = 1.0\Msun$ (red dashed line) and $\Mass_1 = 1.3\Msun$ and $\Mass_2 = 0.2\Msun$ (cyan dashed line). {\it Left (a)}: unfiltered, 44\,706 \fieldName{SB1} solutions (among which 3\,056 unphysical, i.e., below the cyan dashed line); {\it middle (b)}: filtered by significance $K_1/\sigma_{K_1}>20$, 
 27\,404 solutions are rejected and 17\,302 are kept (among which 214 unphysical); {\it right (c)}: filtered by significance $>40$, 37\,850 solutions are rejected and 6\,856 are kept (among which 21 unphysical).
}
\label{Fig:P-R}
\end{figure*}

The existence of a circularisation threshold period in the \eP\ diagram was demonstrated by \citet{Pourbaix2004} in their Fig.~4 \citep[see also][]{ZahnBouchet1989,Verbunt1995,Mazeh2008,Jorissen2009,Escorza2019C}. Its analytic form in a period - radius diagram may be easily obtained from the simple expression of the Roche radius $R_R$ around the star of mass $\Mass_1$ with a companion of mass $\Mass_2$ \citep{Paczynski1971}: 
\begin{equation}
\label{Eq:Pac}
R_{R,1} = a\; \left(0.38 +0.2\log \frac{\Mass_1}{\Mass_2} \right).
\end{equation}
Substituting Kepler's third law, and assuming that the period  threshold (expressed in days) corresponds to the situation where the star radius is equal to the Roche radius, one finds after some algebra:
\begin{eqnarray}
\label{Eq:Pthres}
\log (P^{\rm thresh}_d / 365.25)  &=& \phantom{-}(3/2)\; \log (R_1\; /\; 216 \;\Rsun) \nonumber \\
& & - (1/2) \;\log (\Mass_1+\Mass_2) \nonumber \\
& &- (3/2) \; \log \left(0.38 + 0.2 \; \log \frac{\Mass_1}{\Mass_2}\right)\\
&
\label{Eq:c1}
\equiv&\phantom{-}(3/2)\; \log (R_1\; /\; 216 \;\Rsun)  \\
 & &- (3/2)\; c_1,\nonumber
\end{eqnarray}
\noindent where $c_1$ only depends on the masses.
These thresholds are represented on \figref{Fig:P-R} as dashed lines (corresponding to different choices for the pair $\Mass_1, \Mass_2$).
That figure presents all SB1 solutions 
falling along the RGB/AGB as defined above based on the ($J-K,$ \MK) colour - magnitude diagram. 
Figure~\ref{Fig:P-R}  reveals however that there are many SB1 solutions involving giant stars that do not fulfill the condition $P \ge P^{\rm thresh}$ expressed by Eqs.~\ref{Eq:Pthres} and \ref{Eq:c1}, especially when the significance $K_1/\sigma_{K_1}$ is smaller than 40.

\subsubsection{\texorpdfstring{\eP, $P$ - $f(\Mass)$}{e-P, P-f(\Mass)} diagrams}
 
\begin{figure*}[htb]
\includegraphics[scale=0.5]{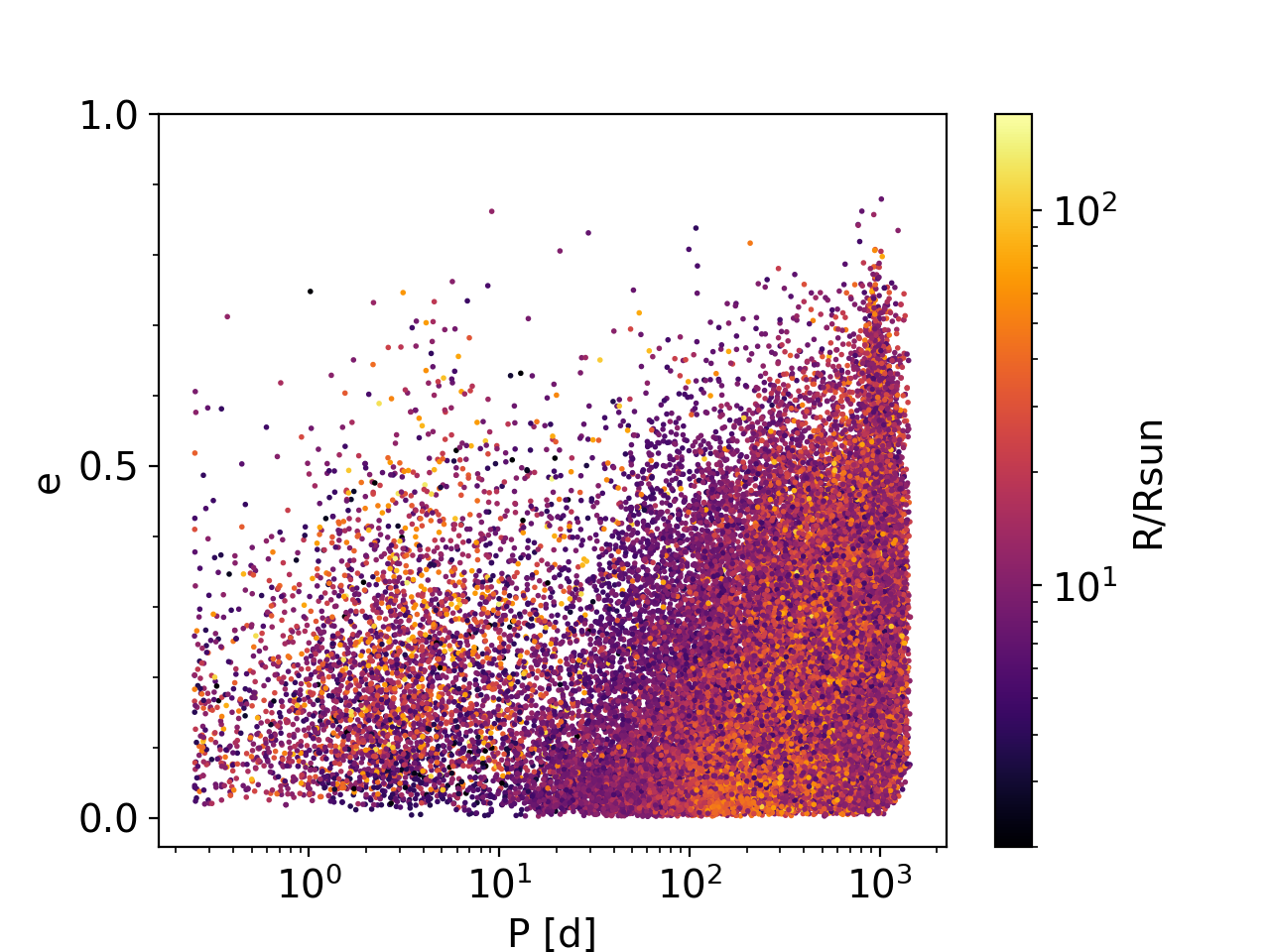}
\includegraphics[scale=0.5]{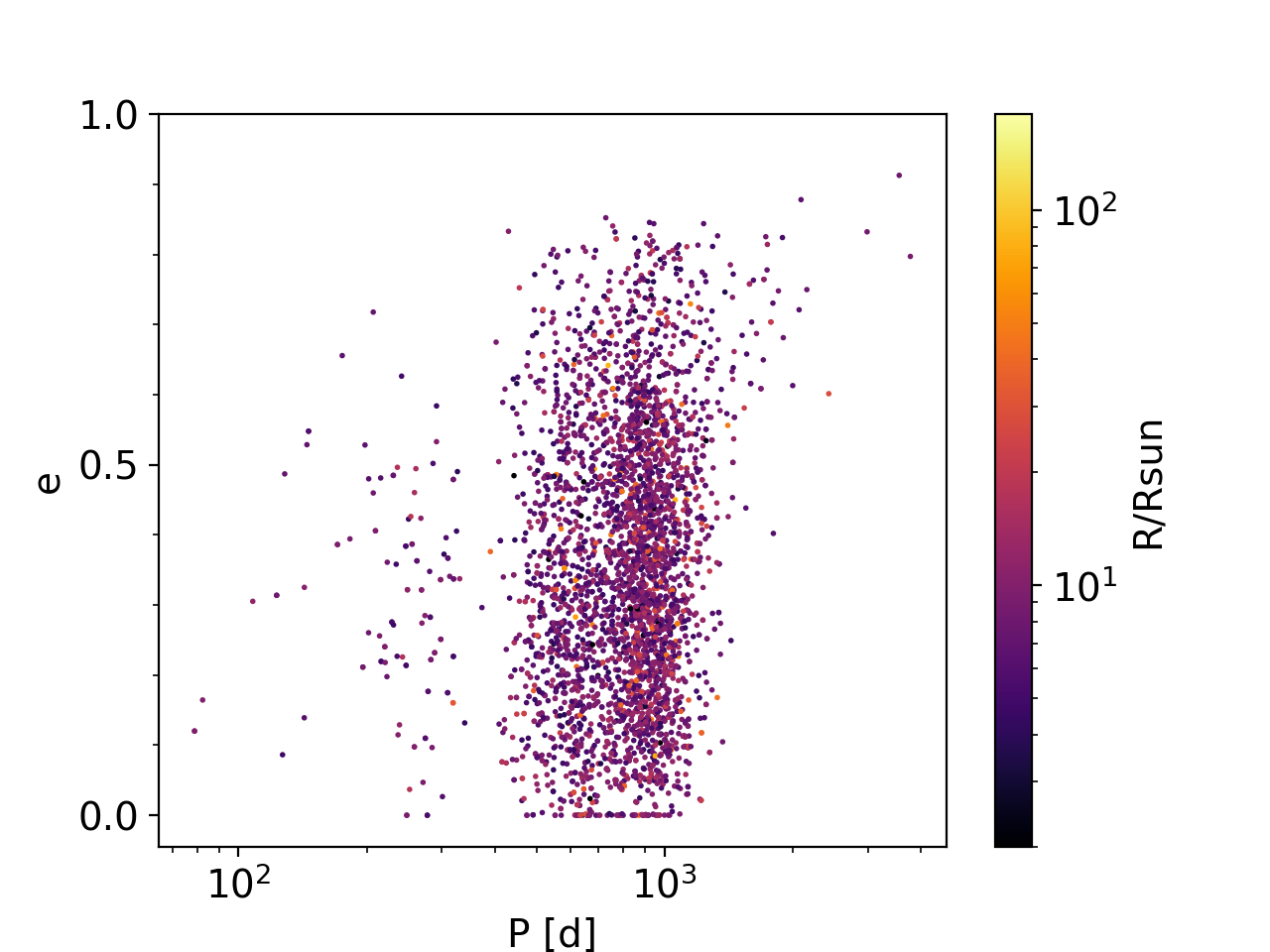}
\caption[]{The \eP\
diagrams for all \fieldName{SB1} ({\it left panel}: unfiltered; adequately filtered \eP\
diagrams for \fieldName{SB1} with an RGB/AGB primary is presented in \figref{Fig:e-P-clean}) and astrometric ({\it right panel}) solutions falling along the RGB/AGB, according to the criterion \eqref{Crit:RGB_AGB_criterion}, and with a radius available from \fieldName{radius_flame}. Note the restricted period scale of the astrometric binaries as compared to the \fieldName{SB1}, and the lack of binaries with periods close to 1~yr among astrometric binaries.
}
\label{Fig:SB1-e-P}
\end{figure*}

 Figure~\ref{Fig:SB1-e-P} presents the \eP\ diagram for the same set of SB1 solutions 
 (left panel) as shown on \figref{Fig:P-R}a, as compared to astrometric binaries along 
 the RGB/AGB (right panel). The difference between the period range covered by \fieldName{SB1} and astrometric solutions is striking. Since most astrometric orbits have periods longer than about 200~d, they clearly satisfy the criterion expressed by the dashed line on \figref{Fig:P-R} and do not overfill their Roche lobe, contrarily to the short-period \fieldName{SB1} solutions.

\begin{figure*}[htb]
\includegraphics[scale=0.37]{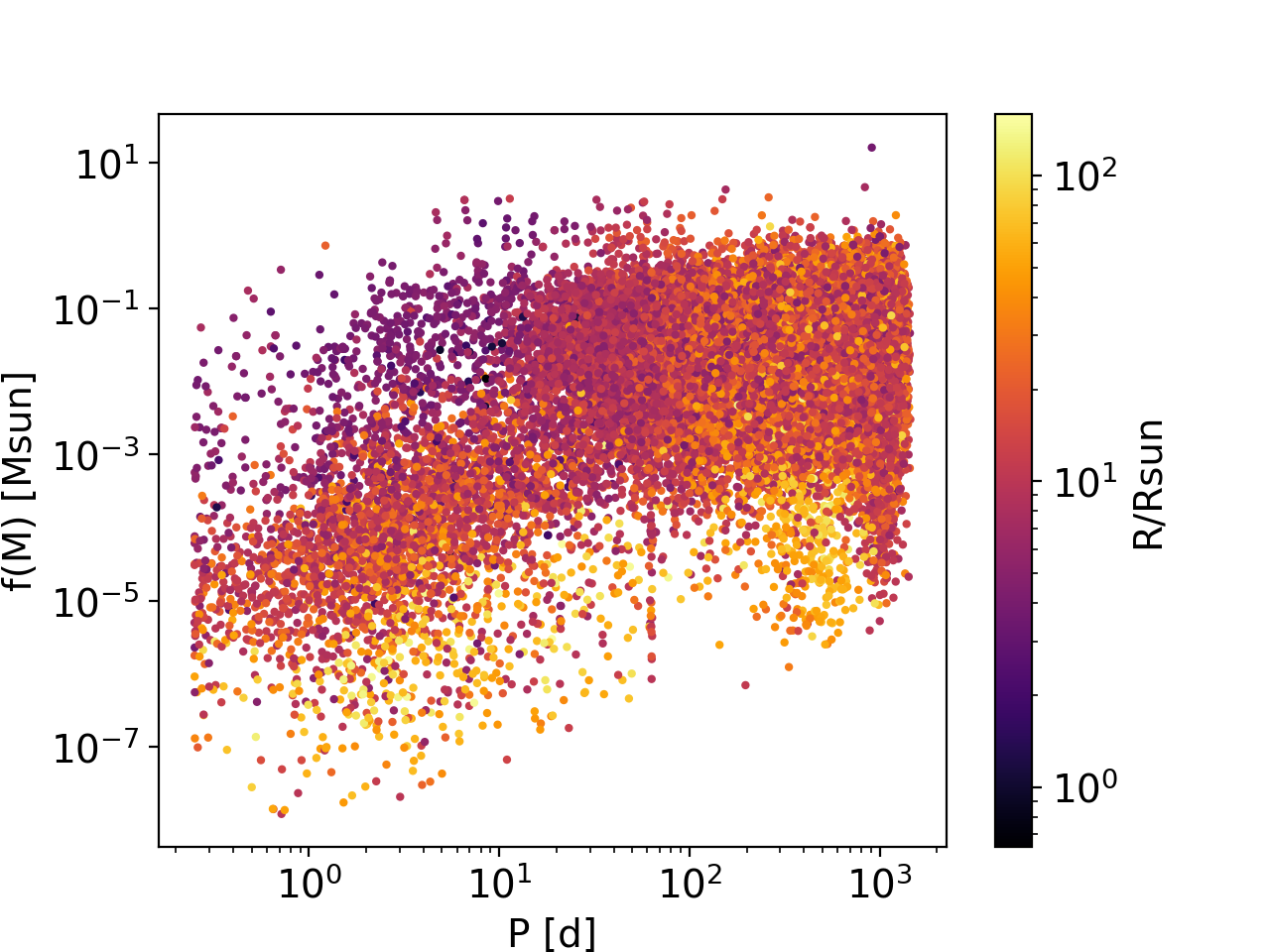}
\includegraphics[scale=0.37]{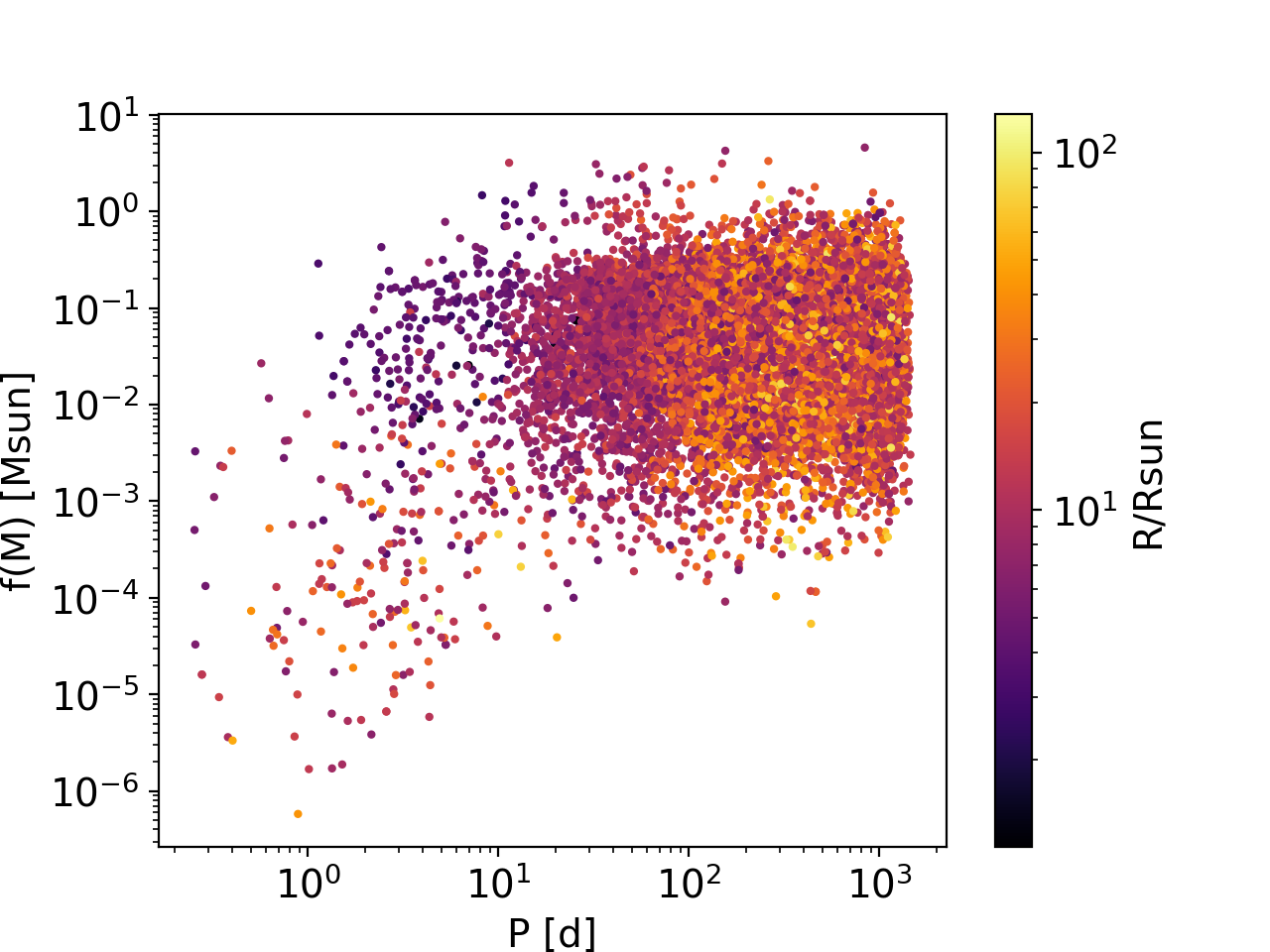}
\includegraphics[scale=0.37]{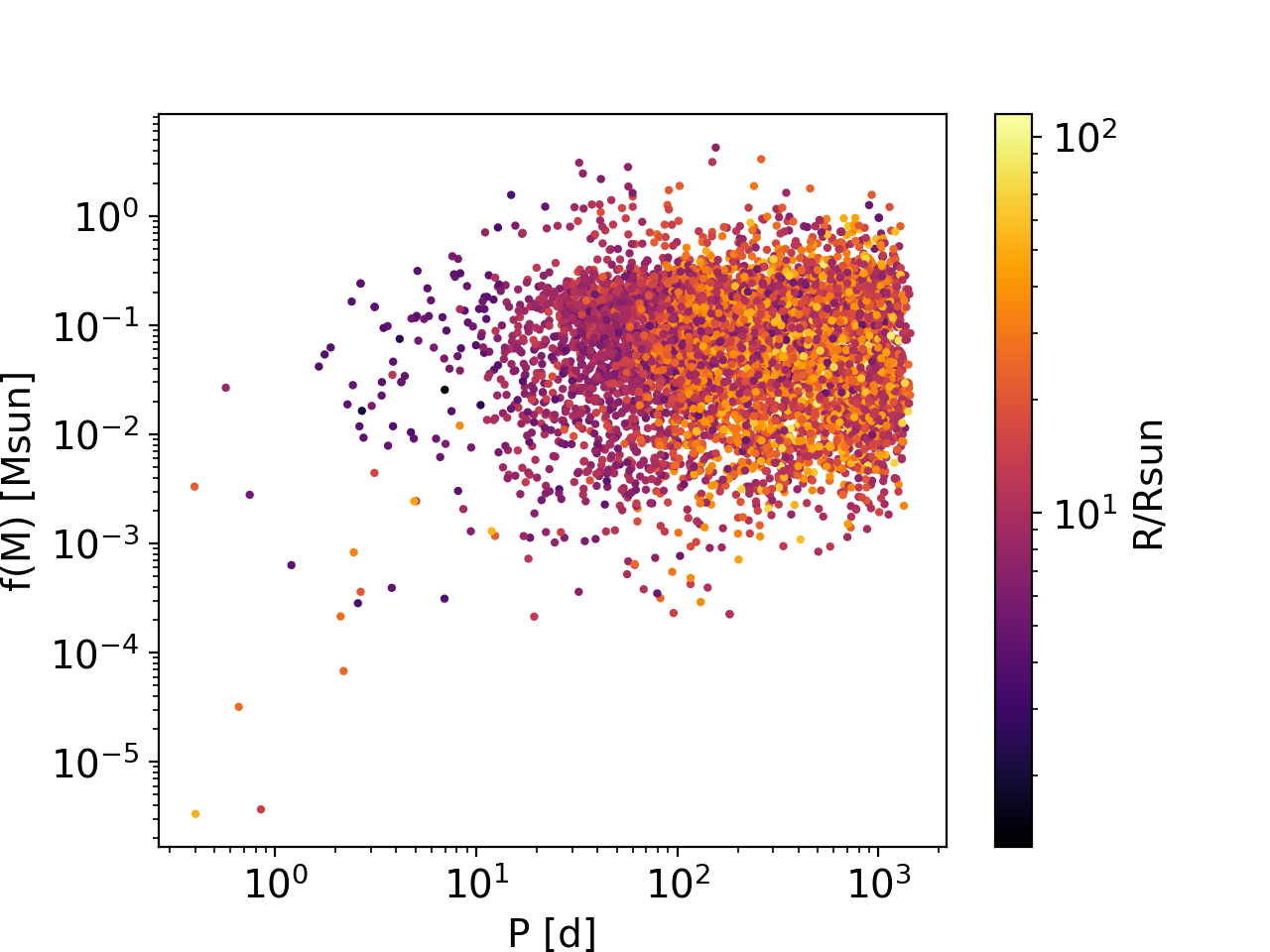}
\caption[]{
$P$ - $f(\Mass)$ diagrams for \fieldName{SB1} along the RGB/AGB.  
{\it Left (a)}: unfiltered. The yellow tail extending down to $f(\Mass) \sim 10^{-5}$\Msun at periods in the range 300 -- 800~d corresponds to pseudo-orbits associated to long-period pulsators (see \secref{Sect:LPVs}). {\it Middle (b)}: filtered by significance $>20$ and $>40$ ({\it right (c)}).
}
\label{Fig:SB1-P-fM}
\end{figure*}

 Figure~\ref{Fig:SB1-P-fM}a is similar to \figref{Fig:SB1-e-P} but replacing eccentricities by mass functions, and revealing again two populations of \fieldName{SB1} solutions, the short-period ones being characterized by very small mass functions $f(\Mass)$. The origin of this population of short-period SB1 solutions among RGB/AGB stars clearly needs clarification. In the following, we show that they are associated with poorly defined solutions. It appears indeed that almost all unphysical \fieldName{SB1} solutions may be eliminated by using the same purely observational criterion as used in \secref{Sect:SB_MS}, and 
based on the value of the significance of the \fieldName{SB1} solution
(available in Table \TBOTable from the \gaia archive), i.e. $K_1/\sigma_{K_1}$, $K_1$ being the semi-amplitude of the first component.

Almost no outlier remains in the  $P - R$ (\figref{Fig:P-R}c) and $P$ - $f(\Mass)$ (\figref{Fig:SB1-P-fM}c) diagrams when that significance exceeds 40, 
a few outliers remain when it exceeds 20 but many more solutions are kept, as may be seen from \tabref{Tab:cleaning_SB1_AGB}.
That table shows that the gradual disappearance of unphysical \fieldName{SB1} solutions as the significance increases corresponds to a real filtering out of unphysical solutions since the fraction of remaining unphysical solutions truly diminishes as the significance increases (passing from 6.8\% in the absence of any filtering to 0.3\% when the significance threshold is set at 40; see \tabref{Tab:cleaning_SB1_AGB}). The drawback of a filtering on significance is however that it tends to filter out as well  solutions with long orbital periods as those have on average smaller $K_1$ values (this was very clear as well from \figref{Fig:e-P_MS}). Alternatively, if one does not want to loose the long-period orbits as a result of filtering on significance, filtering is also possible by using the physical condition $P \ge P^{\rm thresh}$ -- Eq.~\ref{Eq:Pthres} -- with appropriate mass values; however \figref{Fig:P-R} reveals that for systems with periods above 10~d, the boundary between physical and unphysical systems does not depend sensitively upon the choice of $\Mass_1, \Mass_2$.


\begin{table}
\caption[]{
Sizes of the \fieldName{SB1} sample involving RGB/AGB primaries for different filtering on the significance threshold. 
The numbers of rejected and kept sources are given as a function of the significance $K_1/\sigma_{K_1}$. The column labelled `unphys.'  lists the number of sources which would have a Roche filling factor  larger than unity  (or $P < P^{\rm thresh}$ in \equref{Eq:Pthres} thus falling below the cyan dashed line on \figref{Fig:P-R}). The column labelled `fraction' gives the ratio `unphys.'/`accepted' (expressed in \%).} 
\label{Tab:cleaning_SB1_AGB}
\begin{tabular}{ccrrll}
\hline\hline
significance & rejected & accepted & unphys. & (\% )\\
\hline\\
all   & 0      & 44706  & 3056 &(6.8\%)\\
$>20$ & 27404  & 17302  &  214 &(1.2\%)\\
$>40$ & 37850  &  6856  &   21 &(0.3\%)\\
\hline\\
\end{tabular}
\end{table}

\begin{figure}[htb]
\includegraphics[scale=0.65, angle=0]{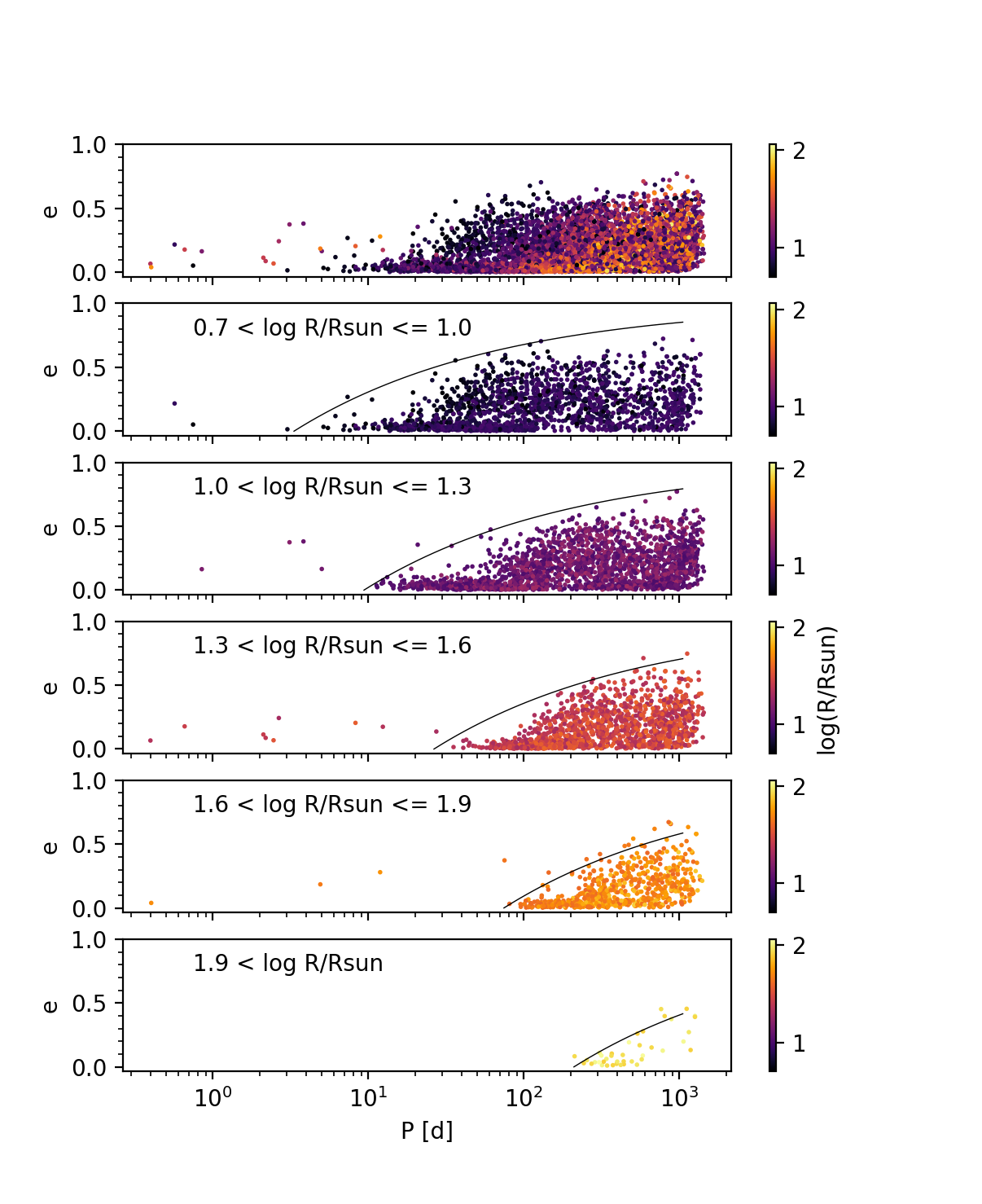}
\caption[]{The \eP\ diagram for \fieldName{SB1} systems along the giant branch, 
filtered according to significance factors larger than 40, for various radius spans. 
The top panel is the full sample. The solid black lines 
correspond to the 
loci such that $P (1-e)^3$ 
is constant 
(see text). 
The sample sizes are, from top to bottom, 1960, 2358, 1643, 737, and 40. The location in the HRD of the \fieldName{SB1} systems with $0.7 < \log(R/\Rsun) \le 1.0$ ({\it second panel from top}), $1.3 < \log(R/\Rsun) \le 1.6$ ({\it fourth panel from top}) and $1.9 < \log(R/\Rsun)$ ({\it bottom panel}) is shown in \figref{Fig:HRD_giants}.
\label{Fig:e-P-clean}}
\end{figure}

Now that the sample of RGB/AGB stars has been adequately cleaned of its unphysical orbits, it is possible to investigate the properties of the \eP\ diagram for giant stars. Figure~\ref{Fig:e-P-clean} presents those for bins of increasing radii (as taken from the corresponding \fieldName{radius_flame} field), as indicated on the figure labels. As expected from the dashed lines in \figref{Fig:P-R}, the minimum period increases with increasing radii. 
In the following discussion, we adopt $\Mass_1 = 1.3\Msun$ and $\Mass_2 = 1.0\Msun$ (corresponding to the red dashed line on \figref{Fig:P-R}, and $c_1 = -0.274$ in Eq.\,\ref{Eq:c1}), as these values match well the
observed trend. The above value for $c_1$ combined with the upper bound of the radius range adopted in each panel of \figref{Fig:e-P-clean}
defines the lower bound on the orbital period $P_{\rm min}$ for $e = 0$. It appears that 
the  upper envelope of the data cloud observed in each panel of \figref{Fig:e-P-clean} is well fitted by the empirical relation $P = P_{\rm min} (1-e)^3$, represented by the solid black lines in \figref{Fig:e-P-clean}, as already found by \citet{Pourbaix2004} in their analysis of the {\it Ninth Catalogue of Spectroscopic Binary Orbits} (their Fig.~5). Despite the fact that this curve matches rather well the uppermost data points in almost all panels, it must be stressed that there seems to be no physical justification for this specific analytical form.
A closer look at each of these panels reveals however an interesting sub-structure. At the shortest periods, each panel is dominated by a large amount of (nearly) circular orbits caused by circularisation operating in those systems where the giant stars with their convective envelope are close to filling their Roche lobe. As shown above,  $P_{\rm min}$ in each panel actually refers to systems where the giants with the shortest radius in the considered range fill their Roche lobe \citep[see e.g.][for details]{Verbunt1995,Mazeh2008}.  



\begin{figure}[htb]
\includegraphics[scale=0.6, angle=0]{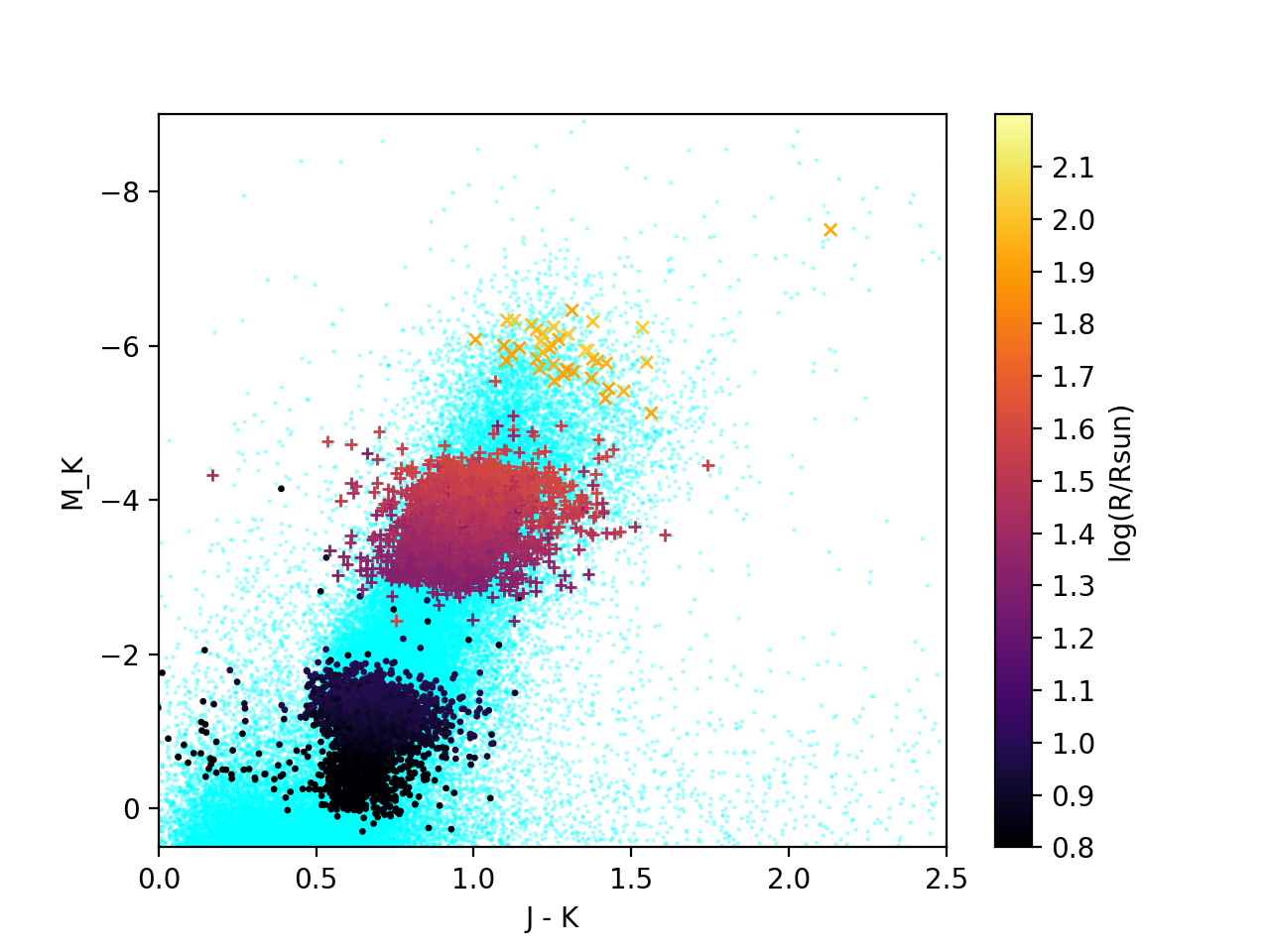}
\caption[]{
\label{Fig:HRD_giants}
Location in the HRD of three among the samples displayed in \figref{Fig:e-P-clean}, namely $0.7 < \log(R/\Rsun) \le 1.0$ (dots), $1.3 < \log(R/\Rsun) \le 1.6$ (plusses) and $1.9 < \log(R/\Rsun)$ (crosses). Small cyan dots correspond to the \fieldName{SB1} not
selected by our selection criteria. See \tabref{Tab:SB1giants} for a full discussion of the properties of the yellow crosses.}
\end{figure}

\subsubsection{A search for genuine SB1 among giants}
\label{Sect:LPVs}

\begin{figure}[htb]
\vspace*{-6cm}
\hspace*{-1.5cm}
\includegraphics[scale=0.65, angle=0]{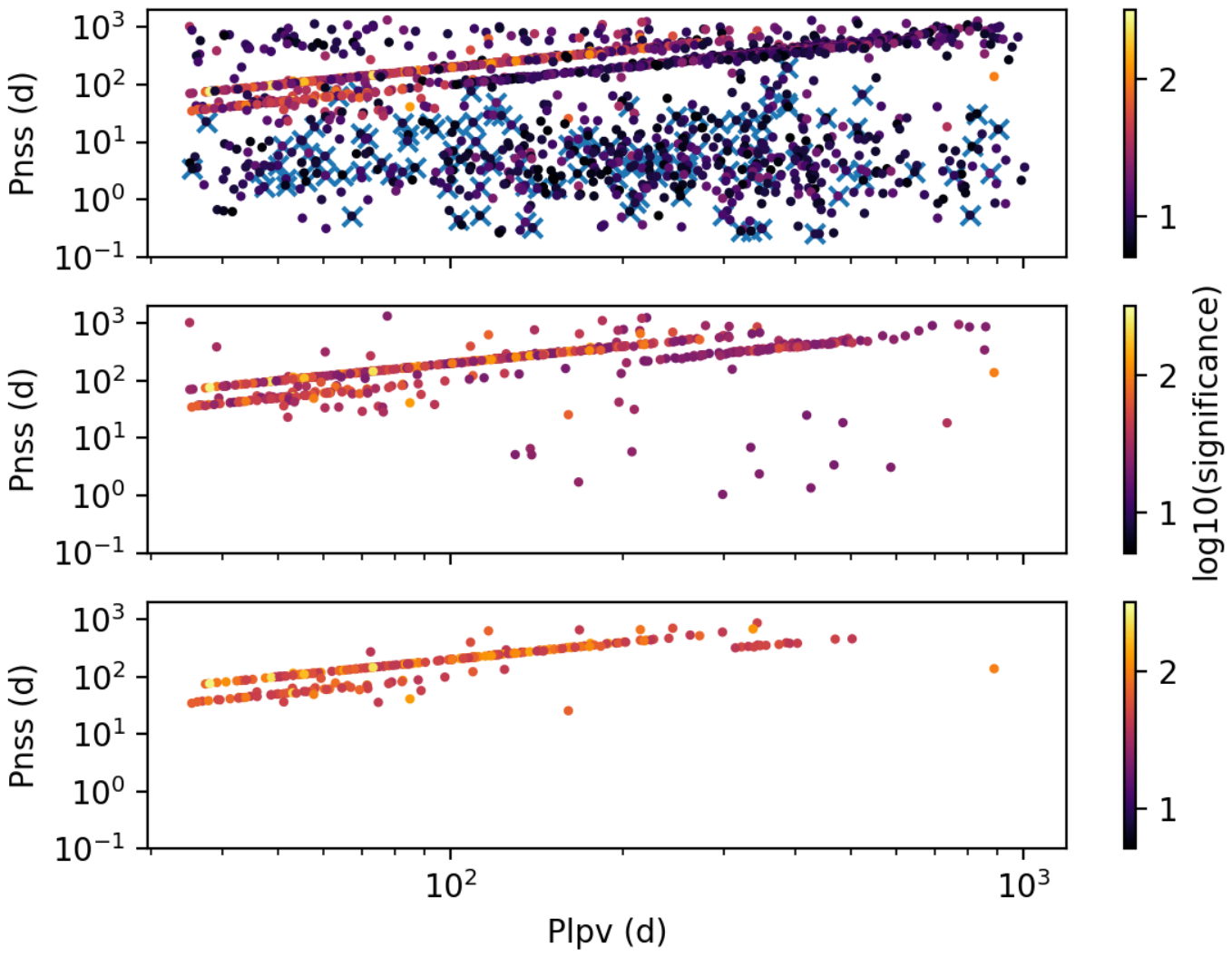}
\vspace*{-6cm}
\caption[]{Orbital period from the \TBOTable table versus photometric period from the \LPVTable table. The {\it top, middle} and {\it bottom panels} correspond to different filtering based on the \fieldName{SB1} significance factor (respectively, larger than 5 -- default in the NSS table --, larger than 20, and larger than 40). The two sequences observed in all panels correspond to $P_{\rm lpv}/P_{\rm nss} = 0.5$ (ellipsoidal variables; upper sequence),  $P_{\rm lpv}/P_{\rm nss} = 1$ (LPVs or rotational modulation in a synchronised system; lower sequence). 
In the {\it top panel}, the crosses denote NSS solutions for which the Roche-lobe filling factor is above unity, and are thus unphysical. The filtering with significance larger than 40 makes them disappear almost entirely.
}
\label{Fig:PNSS_PLPV}
\end{figure}

\begin{figure}[htb]
\includegraphics[scale = 0.55]{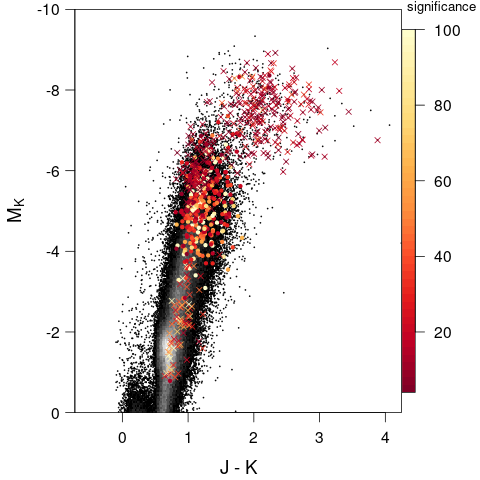}
\caption[]{Location in the infrared colour - magnitude diagram of the stars with $0.45 \le P_{\rm lpv}/P_{\rm nss} \le 0.55$ (ellipsoidal variables) from \tabref{Tab:LPV} (dots). Stars with $0.95 \le P_{\rm lpv}/P_{\rm nss} \le 1.05$  are represented by crosses. They appear in two different locations, among LPVs with low orbital significance on one hand, and among less luminous giants with much larger orbital significance perhaps suggesting starspot modulation or short-period pulsators. The dots ($P_{\rm lpv}/P_{\rm nss} \sim 0.5$) fall in between these two groups, as they are located just below the tip of the RGB. 
}
\label{Fig:HRD_lpv}
\end{figure}

The identification of SB1 among highly evolved giants and long-period variables (LPVs) is 
made difficult by the envelope pulsation \citep{Alvarez2001,Hinkle2002,Jorissen2004,JorissenWien2019}. Hence other methods have been used \citep{Jorissen2008,Sahai2008,Decin2020,Ortiz2021} to identify true binaries. \gaia, with its survey combining radial velocity and photometry data, offers however exquisite prospects to disentangle pulsational from orbital RV variations. In that respect, the bottom panel of \figref{Fig:e-P-clean} offers an interesting benchmark sample of giant stars with $R > 80$\Rsun with a SB1-like signature in their RVs. There are 40 such giants if the significance threshold is set at 40.\footnote{Interested readers may set the significance threshold at 20 instead to get more \fieldName{SB1}-like solutions (namely about 100), especially with long periods for the reasons explained in the text, or use instead the physical filtering $P \ge P^{\rm thresh}$. 
}
No \fieldName{Orbital} neither \fieldName{AstroSpectroSB1} solutions are present in \gdrthree among those giants with large radii (compare with \figref{Fig:SB1-e-P}). \tabref{Tab:SB1giants} lists their main  properties, while their location in the HRD is shown as the yellow crosses in \figref{Fig:HRD_giants}.  

As mentioned above, since there is the risk that some of these \fieldName{SB1} solutions may be caused by envelope pulsation mistaken as SB1s, a proxy for the amplitude of the photometric variation in the $G$ band has been listed as well, namely
\fieldName{phot_g_mean_mag_error} from the \fieldName{gaia_source} table.
A crossmatch has been performed as well with table \LPVTable and the photometric period, whenever available, has been listed in the column $P_{\rm lpv}$. It appears that only one LPV is present in this list (a carbon star also known as V688 Cas), as confirmed by its largest $\Delta G$ value in \tabref{Tab:SB1giants}. Since $P_{\rm lpv} = P_{\rm nss}$ for this star, the RV variations are not due to orbital motion but to envelope pulsations. Many other such cases will be discussed below (\tabref{Tab:LPV}). There are only 4 other stars appearing in the \LPVTable\ table in common with \tabref{Tab:SB1giants}, and these four have the unexpected property that $P_{\rm lpv} = 0.5\; P_{\rm nss}$, with a moderate $\Delta G$ value (on the order of 0.001 -- 0.003~mag). We argue below in relation with \tabref{Tab:LPV} that these are ellipsoidal variables, thus true binaries where the giant primary is close to filling its Roche lobe. Based on the fact that these ellipsoidal variables identified in table \LPVTable\ have small eccentricities ($e < 0.1$), we suspect that \tabref{Tab:SB1giants} contains many more such ellipsoidal variables, namely those with $e < 0.1$ and $\Delta G > 0.001$ ~mag, flagged as  \quoting{Ell. var.?} in the last column of \tabref{Tab:SB1giants}.

\begin{table*}[!htp]
\caption[]{
Source id and basic parameters for \fieldName{SB1} and acceleration  solutions with significance larger than 40 for giants with $R > 80$\Rsun (bottom panel of \figref{Fig:e-P-clean}). The  radius $R$ is from \fieldName{FLAME}. The column labeled $\Delta G$ gives a proxy of the variability in the $G$ band (see text). The column labeled $P_{\rm lpv}$ lists the period obtained by the photometric analysis (table \LPVTable), whereas $P_{\rm nss}$ corresponds to the period of the \fieldName{SB1} orbit. 
\label{Tab:SB1giants}
}\small
\begin{tabular}{rrclrrrrrl}
\hline
\multicolumn{1}{c}{source\_id} & \multicolumn{1}{c}{$R$} & \multicolumn{1}{c}{significance} & \multicolumn{1}{c}{$\Delta G$} & \multicolumn{1}{c}{$G$} &\multicolumn{1}{c}{$e$}& $P_{\rm nss}$ &  $P_{\rm lpv}$ & \multicolumn{1}{c}{Alt. id} & Rem.\\
& (\Rsun) & & (mag) & (mag) && \multicolumn{1}{c}{(d)} & \multicolumn{1}{c}{(d)} \\
\hline
\smallskip\\
\noalign{\centering{SB1}}\\
\hline\\
1825471125022885760 & 111 & 62 &0.0006&11.74&0.19& 475 &  \\
1963830094814564992 & 98 & 71  &0.0025&9.48 &0.06& 340 & 183 & &Ell. var.\\
1972501599433801856 & 104 & 82 &0.0026&11.21&0.04& 286 & 143 & &Ell. var.\\
1993611806061037824 & 91 & 45  &0.0011&11.66&0.05& 255 &   &&  Ell. var.?\\
1996190371286907904 & 81 & 51  &0.0005&11.65&0.10& 368 &   \\
2022016864326072832 & 81 & 53  &0.0002&12.25&0.13& 1182 &  \\
203396083342181248 & 79 & 55   &0.0004&11.48&0.28& 851 &    \\
2072346498024572672 & 83 & 49  &0.0006&11.67&0.40& 805 &   \\
2153213619706962304 & 85 & 97  &0.0008&8.92 &0.02& 395 &   & BD+56 2152\tablefootmark{a} & \\
2179330422489474304 & 100 & 63 &0.0007&11.27&0.40& 1261 &  \\
2189793031540178560 & 109 & 46 &0.0010&11.33&0.09& 582 &  &&  \\
2190661233409369728 & 81 & 179 &0.0008&9.75 &0.26& 538 &  \\
2198983058969830272 & 91 & 148 &0.0019&10.03&0.06& 573 &  &&  Ell. var.?\\
2203704946009240576 & 85 & 78  &0.0007&11.05&0.15& 663 &   \\
3023454391367052928 & 79 & 52  &0.0023&11.52&0.02& 513 &   &&  Ell. var.?\\
3385138711262550144 & 107 & 48 &0.0012&10.34&0.20& 1063 &  \\
3441375569926160768 & 82 & 72  &0.0004&8.58 &0.45& 1115 & & BD+26 935\tablefootmark{b}  \\
4309778580925549312 & 86 & 78  &0.0024&11.75&0.08& 212 &  106 & & Ell. var.\\
4538064682637397504 & 81 & 55  &0.0005&10.47&0.02& 440 & &&  \\
465787042893453696\tablefootmark{c} & 87 & 58   &0.0107&11.71&0.09& 366 & 375 & V688 Cas\tablefootmark{d} & Mira\\  
468328667095902720 & 82 & 109  &0.0008&11.31&0.01& 374 &  &&   \\
4479122750503280512 & 100 & 64 &0.0009&10.89&0.11& 307 &  \\
519141188227099776 & 89 & 46   &0.0004&11.04&0.02& 420 &   &&   \\
5340777165298298880 & 83 & 53  &0.0014&11.57&0.02& 273 &   &&  Ell. var.?\\
5354875859285271936 & 89 & 50  &0.0003&9.56 &0.45& 764 &   \\
5355550645876247808 & 95 & 49  &0.0014&12.44&0.04& 399 &  &&  Ell. var.? \\
5405499126983935872 & 83 & 40  &0.0018&13.07&0.03& 243 &  122 & &Ell. var.\\
5406434021101010176 & 102 & 40 &0.0010&10.28&0.01& 319 & &&  \\
5604143357268838400 & 80 & 53  &0.0007&9.97 &0.04& 327 &&&    \\
5697523299266655104 & 81 & 108 &0.0007&10.69&0.28& 585 &  \\
5697806664034217216 & 98 & 65  &0.0009&10.59&0.37& 889 &   \\
5796104824632537600 & 86 & 104 &0.0018&10.12&0.17& 556 &  \\
5806597567164955776 & 88 & 55  &0.0018&10.94&0.09& 433 &   &&  Ell. var.?\\
5835609040544745344 & 108 & 40 &0.0007&11.46&0.09& 315 &  &&  \\
5847195453486047616 & 105 & 56 &0.0009&9.74 &0.13& 782 &  \\
5854104780242997504 & 92 & 49  &0.0003&12.22&0.04& 495 &  &&  \\
5878621900997292800 & 92 & 114 &0.0004&12.20&0.02& 534 &  &&  \\
5888442292197648000 & 115 & 40 &0.0018&12.08&0.04& 305 & &&  Ell. var.? \\
6012575363926683648 & 84 & 44  &0.0014&12.43&0.04& 439 &   &&  Ell. var.?\\
6056302632126821760 & 84 & 43  &0.0005&12.05&0.39& 1261 &  \\
6057537697261452288 & 86 & 86  &0.0004&11.02&0.01& 343 &   &&  \\
992206959423861888 & 94 & 43   &0.0022&9.72 &0.27& 1152 &   \\
\medskip\\
\hline\\
\noalign{\centering{Acceleration solution}}\\
\hline\\
6665511449204071424 & 92 &  28 &  0.0013 &4.19 &   - & - & - & 	HD 190421\\
\smallskip\\ \hline
\end{tabular}
\tablefoottext{a}{Limb-darkened diameter of 0.393~mas from \citet{Cruzalebes2019}, or 96\Rsun with the \gdrthree parallax of $\varpi = 0.44$~mas}
\tablefoottext{b}{Limb-darkened diameter of 0.672~mas from \citet{Cruzalebes2019}, or 106\Rsun with the \gdrthree parallax of $\varpi = 0.68$~mas}
\tablefoottext{c}{Since $P_{\rm nss} = P_{\rm lpv}$, the RV variations are due to atmospheric pulsations rather than to orbital motion}
\tablefoottext{d}{also CGCS 396 in the General Catalog of Cool Galactic Carbon Stars, flagged as Mira variable.}
\end{table*}

The longest period found in \tabref{Tab:SB1giants} is 1261\,d, a value well in line with the \gdrthree time span, but short with respect to the periods expected among evolved giants (consider for instance the 17~yr period found by \citealt{JorissenWien2019} for the carbon Mira V Hya).
 Such long periods are not detectable at the current stage of the \gaia mission neither as SB1 nor as astrometric binaries. Nevertheless, one may look for acceleration solutions (there is only one solution from table \ACCTable matching the giant criteria defining \tabref{Tab:SB1giants}; about acceleration solutions, see \secref{Sect:PMA} and \citealt{Wielen1999,Makarov2005,Frankowski2007,Kervella2019,2022A&A...657A...7K,Brandt2021}) or for solutions with  a trend in the RVs  
 (121 solutions for giants found in \NLSTable, not listed here).

\subsubsection{Combining photometry and spectroscopy to diagnose RV variations in giants: pulsation, ellipsoidal variables, and rotational modulation}

Initially with the aim to further investigate how many LPVs may be mistaken as SB1, we searched for targets in common  between \fieldName{SB1} from the NSS \TBOTable table and LPVs as provided by the variability study in the \LPVTable table. But this cross-match has revealed some surprises.
The query 

\noindent\dt{SELECT  * from gaiadr3.\TBOTable TBO, gaiadr3.\LPVTable LPV where \fieldName{LPV.source_id} = \fieldName{TBO.source_id} and \fieldName{LPV.frequency} is not null}

\noindent yields 1869 entries, as shown on \figref{Fig:PNSS_PLPV}. The three panels differ in terms of the level of filtering applied on the \fieldName{SB1} significance parameter, as defined above: $>5$ (default, top panel), $>20$ (middle panel), and $>40$ (bottom panel). Striking are the two straight sequences observed in all three panels. The upper sequence corresponds to $P_{\rm lpv}/P_{\rm nss} = 0.5$ (as expected for ellipsoidal variables), whereas the lower sequence  corresponds to $P_{\rm lpv}/P_{\rm nss} = 1$ (as expected for pulsating stars or rotational
modulation in a synchronised system). The lower sequence is further made of two distinct clumps, one at short periods ($P_{\rm lpv} \la 100$~d; starspot modulation on a spin-orbit synchronised star?) and the other at long periods ($200 \la P_{\rm lpv} \la 1\,000$~d; LPVs). They will be discussed in turn in what follows. 
\medskip\\
\noindent {\it Ellipsoidal variables}
\medskip\\
\noindent
Besides the obvious property of their light-to-RV period ratio equal to 0.5, the ellipsoidal-variable sequence is further confirmed from its following properties: (i) small eccentricities ($e \la 0.1$), (ii) large filling factors  ($R_1/R_{R,1} \ga 0.65$ from Eq.~\ref{Eq:Pac}, adopting radii from \fieldName{FLAME} and the same typical masses as above -- $\Mass_1 = 1.3$~\Msun, $\Mass_2 = 1.0$~\Msun) whenever available, 
and (iii) small $G$ amplitudes ($0.01 \le \Delta G \le 0.1$~mag; see \tabref{Tab:LPV}). 
The  ellipsoidal nature of these stars has been confirmed from the comparison between the light and RV curves. 
As expected, the maximum light indeed occurs at the quadratures, when the RV is maximum or minimum.
In \figref{Fig:e-P-clean}, at any given radius range, these ellipsoidal variables are located in the nearly circular tail of each panel. 
The full list of ellipsoidal variables is not provided here as the reader may easily obtain it from the ADQL query mentioned at the beginning of this section and filtering on $P_{\rm lpv}/P_{\rm nss}$ around 0.5 (370 stars in the inclusive range 0.45 -- 0.55, most of them having significances in excess of 20). The first part of \tabref{Tab:LPV} nevertheless lists a few examples, randomly selected. Figure~\ref{Fig:HRD_lpv} shows the position in the 2MASS infrared colour-magnitude diagram ($M_{K}, J-K$)  of the 370 stars with $0.45 \le P_{\rm lpv}/P_{\rm nss} \le 0.55$  (dots). These ellipsoidal variables are located from the tip of the RGB\footnote{$M_{\rm K,RGB-tip} = -6.49$ as derived from \citet{Lebzelter2019} who find $K_{\rm RGB-tip} = 12$ for the LMC and considering its distance modulus $18.49 \pm 0.09$~mag \citep{deGrijs2017}.} to 3 magnitudes below. 
We note that some among these stars might be young, pre-main sequence stars. \object{\sourceId{2162167694508896128}} = \object{V1540 Cyg} listed in \tabref{Tab:LPV} is one such case (on \figref{Fig:HRD_lpv}, it is located at $M_K = -6.2$ and $J - K = 1.41$).

\begin{table*}
\renewcommand{\tabcolsep}{3pt}
\caption{\label{Tab:LPV}
A few illustrative examples of ellipsoidal variables ($P_{\rm lpv}/P_{\rm nss} = 0.5$) mistaken as LPVs in the \LPVTable table, LPVs with a pseudo \fieldName{SB1} orbit
   ($P_{\rm lpv}/P_{\rm nss} = 1$, $\Delta G > 0.1$~mag, $P_{\rm lpv} > 180$~d) in the \TBOTable table
and short-period ($P_{\rm lpv} < 100$~d)  \quoting{LPVs}
with $P_{\rm lpv}/P_{\rm nss} = 1$. Radii are the \fieldName{FLAME} DR3 estimates (see text for how the filling factors $R/R_R$ were estimated). The column labelled $\Delta G$ lists the field \fieldName{mad_mag_g_fov} (median absolute deviation) from table \fieldName{vari_summary}.
}
{\small 
\begin{tabular}{rrrcrrrrrlrrrrrrrrr}
\noalign{\mbox{}}\\
\hline
\noalign{\mbox{}}\\
\multicolumn{1}{c}{Gaia DR3 id} & \multicolumn{1}{c}{$P_{\rm lpv}$} &\multicolumn{1}{c}{$P_{\rm nss}$}   & \multicolumn{1}{c}{$P_{\rm lpv}/P_{\rm nss}$}& \multicolumn{1}{c}{signif.}  & \multicolumn{1}{c}{$K_1$} & \multicolumn{1}{c}{$e$} & \multicolumn{1}{c}{$\Delta G$} & \multicolumn{1}{c}{$f(\Mass)$} & \multicolumn{1}{c}{$R/R_R$} &  \multicolumn{1}{c}{$R$} & Alt. id. \\
          & \multicolumn{1}{c}{(d)} &\multicolumn{1}{c}{ (d)} &  && \multicolumn{1}{c}{(km/s)} & & \multicolumn{1}{c}{(mag)} & \multicolumn{1}{c}{(\Msun)} &&  \multicolumn{1}{c}{(\Rsun)}\\
\hline
\smallskip\\
\noalign{\hfill Ellipsoidal variables \hfill\mbox{} }\\
\object{2162167694508896128}\tablefootmark{a} & $143\pm16$ & $289\pm0.2$ & 0.49 & 249 & 34.6 & 0.02 & 0.02 & 1.25 & & & V1540 Cyg\\
\object{5871624883899265280} & $48 \pm3$   & $95.1\pm0.1 $& 0.50  & 101& 22.0 & 0.03 & 0.06 & 0.105 & 0.92& 43\\
\object{449088171382718848}  & $178\pm50$ & $356.7\pm1.5$& 0.50  & 73 & 12.8 & 0.01 & 0.09 & 0.078 &  & \\
\object{528840770565380352}  & $54 \pm4$ & $108.6\pm0.1$& 0.50  & 89 & 31.7 & 0.01 & 0.06 & 0.358 & 0.88 & 45\\
\object{1837292073273265920} & $58 \pm3$   & $116.1\pm0.1$& 0.50  & 86 & 34.1 & 0.04 & 0.06 & 0.477 &  & \\
\object{4305358093199399168} & $42 \pm2$   & $84.8\pm 0.1$& 0.50  & 89 & 27.3 & 0.03 & 0.05 & 0.178 & 0.74 & 32\\
\object{6653811713476525440} & $63 \pm4$   & $125.4\pm0.6$& 0.50  & 19 & 11.4 & 0.16 & 0.09 & 0.019 & 0.67 &37\\
\object{5933194270923372288} & $10 \pm17$  & $219.4\pm0.4$& 0.50  & 63 & 21.8 & 0.05 & 0.05 & 0.234 &  & \\
\object{5998937575770407936} & $69 \pm9$   & $137.1\pm0.2$& 0.50  &50 & 21.7 & 0.05 & 0.07 & 0.144 & 0.75 & 45\\
...\\
\hline\smallskip\\
\noalign{\hfill Large-amplitude ($\Delta G > 0.1$~mag) LPVs (Mira or SRa,b) with a pseudo SB1 orbit \hfill\mbox{}}\\
\object{3029929312263388416}& $ 330\pm39$ &$	329\pm	3	$ &    1.002&	27& 5.6&		0.27&		0.27&		0.0057&			&		&	  \\
\object{5861476288517412096}& $ 351\pm29$ &$	350\pm	4	$ &    1.003&	22& 7.4&		0.14&		0.25&		0.0144&			&		&	\\
\object{4498570706006456320}& $ 196\pm11$ &$	196\pm	 2	$ &    1.003&	 9& 3.5&		0.36&		0.21&		0.0008&			&		&	\\
\object{6635121600650977280}& $ 310\pm42$ &$	309\pm	2	$ &    1.003&	18& 2.9&		0.29&		0.13&		0.0007&		-&				75\\
\object{185224454669173120} & $ 491\pm114$&$490\pm	9	$ &    1.003&	 9& 2.6&		0.34&		0.29&		0.0009&			&		&	\\
\object{5522324157261027968}& $ 321\pm37$ &$	320\pm 3	$ &    1.003&	23& 5.3&		0.29&		0.26&		0.0046&			&		&	\\
\object{5428546471231540608}& $ 455\pm60$ &$	454\pm	6	$ &    1.003&	 7& 2.8&		0.33&		0.24&		0.0009&			&		&	\\
\object{463720476424410624} & $ 353\pm24$ &$	352\pm	3	$ &    1.004&	34& 7.5&		0.17&		0.26&		0.0152&			&		&	\\
\object{6358622017131465728}& $ 181\pm7$ & $	180\pm	2	$ &    1.004&	 10& 6.4&		0.17&		0.54&		0.0048&		-&			138\\
\object{2180493018598279296}& $ 178\pm25$ &$	177\pm	2	$ &    1.004&	11& 2.3&		0.09&		0.12&		0.0002&			&		&	\\
\object{5318375436185802368}& $ 262\pm32$ &$	260\pm	2	$ &    1.005&	12& 2.6&		0.33&		0.18&		0.0004&		-&			97\\
\object{5522970154700635392}& $ 412\pm37$ &$410\pm	8	$ &    1.005&	14& 5.6&		0.04&		0.49&		0.0076&		-&			84\\
\object{1989628623330891904} & $419\pm18$ & \, \, \, \, \, \, \, \, \, \, \, \, \, \, [$25\pm0.02$]\tablefootmark{b}& \,  [16.7]\tablefootmark{b} & 21 &  6.6 & 0.39 & 0.60& 0.0006 &  &  & \\
...\\
\hline
\smallskip\\
\noalign{\hfill Genuine binaries among LPVs ($P_{\rm lpv}/P_{\rm nss} \ne 0.5$ or 1; see text) \hfill\mbox{}}\\
\object{5341773936978279296} & $220\pm61$ & $1252\pm113$ & 0.18 & 21 & 7.7 & 0.25 & 0.05 & 0.053 & 0.28 &  73\\
\object{5597415372601747456 }& $168\pm54$ & $656\pm6$ & 0.26 & 43 & 11.1 & 0.36 & 0.03 & 0.076 & 0.33 & 56\\
\object{5414646307794529792} & $196\pm42$ & $753\pm17$ & 0.26 & 34 & 6.2 & 0.4 & 0.04 & 0.015 &  &  & \\
\object{5875470387113872768} & $214\pm28$ & $746\pm16$ & 0.29 & 29 & 15.3 & 0.35 & 0.09 & 0.229 &  &  &\\ 
\object{5347893273248921984} & $279\pm75$ & $913\pm40$ & 0.31 & 29 & 13.3 & 0.07 & 0.03 & 0.220 &  &  & \\
\object{5404683839108805248} & $215\pm28$ & $662\pm1$ & 0.32 & 89 & 11.5 & 0.33 & 0.11 & 0.088 &  &  & \\
\object{5796098502440628864} & $244\pm28$ & $701\pm4$ & 0.35 & 59 & 6.5 & 0.08 & 0.08 & 0.020 &  &  & \\
\object{1642955252784454144} & $374\pm93$ & $503\pm6$ & 0.74 & 33 & 6.0 & 0.28 & 0.04 & 0.010 &  &  & \\
\object{304717076269774336}  & $310\pm102$ & $158\pm1$ & 1.96 & 22 & 12.0 & 0.01 & 0.10 & 0.028 & 0.39 &  26\\
\object{6661657003818388480} & $197\pm92$ & $42\pm0.1$ & 4.66 & 32 &  26.3 & 0.11 & 0.04 & 0.078 & 0.25 &  7\\
\object{187075684355571200}  & $161\pm64$ & $26\pm0.01$ & 6.29 & 73 &  22.9 & 0.08 & 0.05 & 0.031 &  &  & \\
\object{5473442554645523712} & $209\pm57$ & $32\pm0.02  $ & 6.62 &  30 & 36.4 & 0.06 & 0.04 & 0.157 &  &  & 
\\
\hline
\smallskip\\
\noalign{\hfill Short-period ($P_{\rm lpv} < 100$~d) light and RV variations with  $P_{\rm lpv}/P_{\rm nss} = 1$: starspot modulation or short-period pulsators ? \hfill\mbox{}}\\
\object{5498026500770376576}& $ 46\pm1$ 	 &$45.7\pm 0.12 $&     1.003&	10& 4.9&		0.41&		0.14&		0.0005&			&		&	\\
\object{4498425604828703104}& $ 50\pm4$ 	 &$49.57\pm 0.05$&     1.003&	39&32.5&		0.03&		0.08&		0.18&		0.24&				7\\
\object{3047643956417931264}& $ 64\pm1 $  &$63.4\pm 0.1$  &     1.003&	13\tablefootmark{c}& 8.3&		0.21&		0.24&		0.0036&		1.44\tablefootmark{c}&				52\\
\object{5637220068643463424}& $ 57\pm4$ 	 &$57.17\pm 0.03$ &    1.004&	66&33.6&		0.04&		0.06&		0.2240&		0.27&			9\\
\object{5883587875302126592}& $ 78\pm4$ 	 &$78.1\pm  0.1 $ &    1.004&	16&12.2&		0.42&		0.08&		0.0128&			&		&	\\
\object{5701792904776417152}& $ 43\pm4$ 	 &$43.38\pm 0.01$ &    1.005&	85&25.2&		0.02&		0.06&		0.0718&			&		&	\\
\object{1869696952997313664}& $ 53\pm3$ 	 &$52.56\pm 0.05$ &    1.005&	31&22.5&		0.13&		0.08&		0.0610&		0.26&			8\\
\object{5235000057883364864}& $ 43\pm1$ 	 &$42.5\pm 0.03 $ &    1.005&	30&35.6&		0.10&		0.06&		0.1969&		0.40&				11\\
...\\
\hline
\end{tabular}
\tablefoot{
\tablefoottext{a}{This is likely a young star, of \quoting{Orion-variable} type (V1540 Cyg).}
\tablefoottext{b}{This star has been added to the Mira category despite its $P_{\rm lpv}/P_{\rm nss}$ ratio vastly different from 1, to illustrate some kind of aliasing problems (see text)}
\tablefoottext{c}{Doubtful case: significance is only 13 and filling factor is 1.4.}
} }
\end{table*}

\noindent{\it Long-period variables}
\medskip\\
\noindent
The transition between dots and crosses in \figref{Fig:HRD_lpv}   corresponds to the transition across the RGB tip. Above the RGB tip, 
most stars from the \LPVTable table belong to the sequence $P_{\rm lpv}/P_{\rm nss}$ around unity. They correspond to LPVs with a RV variation caused by the envelope pulsation. Although displayed in \figref{Fig:HRD_lpv}, the full source list is not given here as they are easily obtained in a way similar to that discussed above for ellipsoidal variables. \tabref{Tab:LPV} nevertheless lists a few examples, randomly selected. 

These LPVs are easily identified by their velocity semi-amplitudes smaller than 10~km~s$^{-1}$ (in that sense, they differ markedly from the ellipsoidal variables which generally have much larger $K_1$ values) and periods in excess of 180~d, as expected for Mira pulsations \citep[][]{Alvarez2001,Hinkle2002}. Hence, given these relatively small values of $K_1$, the significance of the SB1-like solution (namely $K_1/\sigma_{K_1}$) may in several cases be smaller than 20, but the identity of the NSS and LPV periods is {\it per se} an indication of the reliability of the RV model. We note that the filling factor has no meaning in this stellar category since there is no true orbit associated. Mira variables are recognized as well by their large amplitude in $G$ ($>0.1$~mag). 
The pseudo-eccentricities found by \citet{Hinkle2002} for Miras and semi-regular variables were clustered around 0.35, with a few cases below 0.1 as well. Here the pseudo eccentricities range all the way from 0.09 to 0.48 (\tabref{Tab:LPV}).
Furthermore, it has been checked that the maximum RV is reached at phase 0.8 while maximum light is reached at phase 1.0, a phase lag expected for Mira pulsators. 
Furthermore, since for these stars $K_1$ is relatively small and $P$ is long, the pseudo mass functions are consequently smaller than $10^{-3}\Msun$, with some values as small as $10^{-5}\Msun$, in agreement with the findings of \citet{Hinkle2002} (their Table~2) for Mira and semi-regular variables.  LPVs with low values of $f(\Mass)$ are most clearly seen on \figref{Fig:SB1-P-fM}a as the yellow tail extending down to $f(\Mass) \sim 10^{-5}\Msun$ for periods between 300~d and 1\,000~d. On that same figure, a lot more stars with large radii ($R > 100$\Rsun) are found at shorter periods, but those are spurious \quoting{SB1-like} solutions since their periods do not even match the LPV one, and when available, their filling factors are above unity, which is non-physical. Their mass functions are quite small either (down to $10^{-8}\Msun$). Therefore these targets cannot be genuine binaries.
\medskip\\
\noindent{\it Genuine binaries}
\medskip\\
\noindent
Genuine binaries among Miras are expected to have orbital periods much larger than currently detectable by \gdrthree (as is the case for instance for the carbon Mira \object{V~Hya} quoted above). 
Intriguingly, several genuine SB1 have nevertheless been found among stars with $P_{\rm lpv} > 150$~d, a property generally associated to LPVs. In the fourth part of \tabref{Tab:LPV} are listed 14 SB1s selected among the 1189 \fieldName{SB1} solutions in common with the \LPVTable table. These SB1s have a significance larger than 20, the 1$\sigma$ confidence range of $P_{\rm lpv}/P_{\rm nss}$ falling outside the ranges 0.45 -- 0.55 and 0.9 -- 1.1 (to avoid SB1-like variations caused by pulsations), $P_{\rm nss} > 20$~d and $P_{\rm lpv} > 150$~d.
Their RV curves  where checked visually and showed no peculiarity that would make the solution dubious. This visual inspection nevertheless revealed that some kind of aliasing problems remain with the NSS SB1 periods. The star \object{\sourceId{1989628623330891904}} was originally considered as a possible genuine binary among LPVs, since $P_{\rm lpv} = 419$~d as compared to $P_{\rm nss} = 25$~d. However, the visual inspection of the RV curve revealed that the LPV period of 419~d is clearly present in the RV curve, although it was not selected by the period-selection algorithm, which gave no warning about a possible problem with that solution (significance = 2, period confidence = 1.000, ruwe = 1.09) except for the goodness-of-fit of 2.5. Therefore, that star has been added to the \quoting{Large-amplitude LPVs} section.
\medskip\\
\noindent{\it Rotational modulation on a spin-orbit synchronised star}
\medskip\\
\noindent
The final category of interest in \tabref{Tab:LPV} contains targets with short periods (i.e., $P \la 100$ d) on the 
$P_{\rm lpv}/P_{\rm nss} = 1$ sequence. They are listed in the fourth part of \tabref{Tab:LPV} and identified in the HRD of \figref{Fig:HRD_lpv} as the crosses at the bottom of the giant branch. Contrarily to the situation prevailing for ellipsoidal variables and long-period variables discussed above, the phase lag between velocity and light curves now appears to be anything between 0 and $\pi$. For this reason, their light variation could be due to starspot modulation on a spin-orbit synchronised primary star  \citep[e.g.,][]{Mazeh2008}. Less likely, they could be small-amplitude pulsating stars.

%








%
\subsection{Identifying EL CVn systems in \gaia data}\label{sec:elcvn}
%
EL CVn systems are short-period eclipsing binaries (EBs) consisting of an A/F-type main-sequence (MS) primary and a low-mass pre-helium white-dwarf (pre-He-WD) secondary. These systems are a result of mass transfer from the evolved pre-He-WD progenitor to the currently observed primary star \citep[e.g.][]{kerkwijk10, maxted11, rappaport15}. EL CVn systems are at a rare stage of binary evolution in which the young pre-He-WD is bloated, with a radius of up to $\sim0.5\Rsun$, and hotter than the more luminous A/F-type primary. As a result, such systems, harboring a low-mass white-dwarf (WD) precursor, are detectable even in ground-based photometric surveys. EL CVn's with smaller and cooler He-WD secondaries can be detected in space photometry \citep{faigler15} . Consequently, 10, 18 and 36 such systems were discovered in the {\it Kepler} \citep{kerkwijk10,carter11,breton12,rappaport15,faigler15}, WASP \citep{maxted11,maxted14a}, and PTF \citep{roestel18} photometric surveys, respectively.

The detection of these systems in photometric surveys is based on identifying an eclipsing-binary folded light curve with a \quoting{boxy} deeper eclipse (steep ingress and egress and a flat bottom) and a shallower eclipse with a limb-darkening curved bottom. In an EL CVn, the deeper \quoting{boxy} eclipse is actually the secondary eclipse (total eclipse of the pre-He-WD secondary by the MS primary), while the shallower eclipse is the primary eclipse (pre-He-WD transit of the primary star). This is because the pre-He-WD secondary is hotter than the primary. Such photometric detections usually require confirmation through follow-up spectroscopic radial-velocity (RV) observations, that enable identifying the light-curve primary and secondary eclipses from the RV-curve phase.

\begin{figure}[htb]\begin{center}
\includegraphics[scale=0.38, trim={1.5cm 1.8cm 0 1.0cm}, clip]{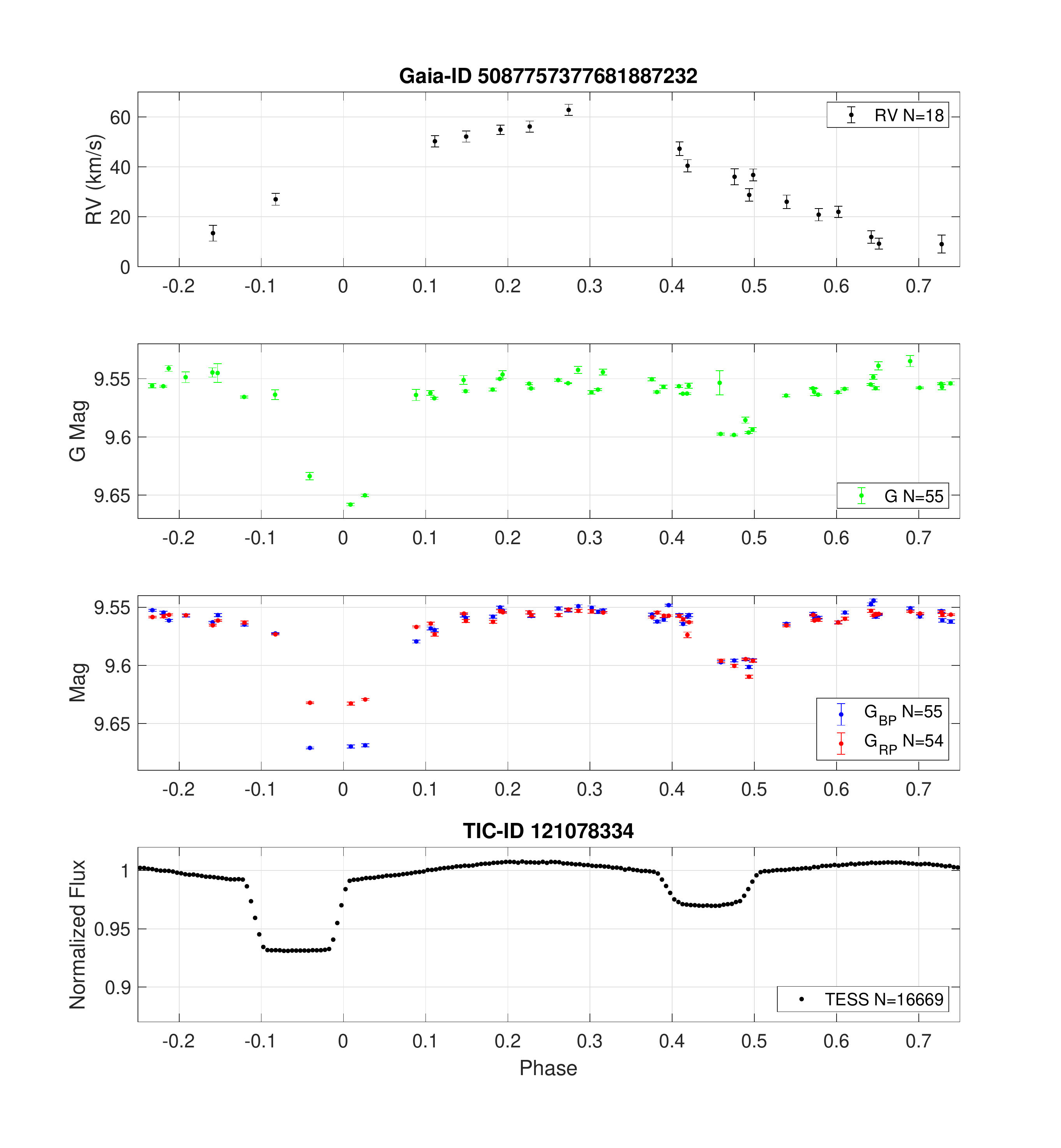}
\caption{
Folded \gaia \gmag, \gbp, and \grp\ photometry and RV data of HD 23692, together with TESS binned data. Top panel shows the \gaia RV data. 
Second panel presents the \gaia \gmag data. Third panel shows the \gaia \gbp\ and \grp\ data, with medians shifted to the \gaia \gmag median, for clarity. Bottom panel presents the TESS data binned to $200$ phase bins. All plots were folded using the \gaia EB-model period and deeper-eclipse epoch as the folding period and phase zero, respectively. Note that the primary eclipse is at phase {\it 0.5}, while the secondary eclipse is at phase {\it zero}. Observed TESS-eclipse phase drift is due to more than $1300$-day of delay from the last \gaia point to the first TESS point. For clarity, the three bottom panels use the same y-axis scale.
}
\label{fig:HD}
\end{center}\end{figure}

The \gaia data, however, enable direct detection of EL CVn systems by combining the \gaia photometry and RV data. Figure~\ref{fig:HD} shows the folded \gaia \gmag, \gbp, and \grp\ photometry and RV data, for a known EL CVn-type system \citep[HD 23692,][]{maxted14a}, together with the Transiting Exoplanet Survey Satellite \citep[TESS;][]{ricker14} binned data. Detrending of the TESS data was done using cosine detrending following \cite{faigler15}. 
The \gaia EB-model period and deeper-eclipse epoch were used as the folding period and phase zero, respectively. The RV plot enables identifying the phase-zero eclipse as the {\it secondary} eclipse, and the 0.5-phase eclipse as the {\it primary} eclipse. The figure shows that for this system, indeed the secondary eclipse is \quoting{boxy} shaped and deeper than the primary one, the main signatures of an EL CVn system. In addition, we see that the \gbp\ secondary eclipse is much deeper than the \grp\ one, an additional indication for the high temperature of the secondary.

\subsubsection{Sorting through the \gaia data}
\label{sec:sample}
To build the initial sample, from which we can identify EL CVn systems, we selected from the \gdrthree data systems with:
\begin{enumerate}
    \item An eclipsing-binary solution from \gaia photometry, and
    \item A spectroscopic-binary (SB) solution (\fieldName{SB1}, \fieldName{SB1C}, \fieldName{SB2} or \fieldName{SB2C}) derived from the \gaia RV data, and
    \item An orbital-frequency difference between the EB and SB solutions smaller than $\frac{1}{100} \, \rm{d}^{-1}$, and
    \item An orbital period shorter than 2~d.
\end{enumerate}
The maximum orbital-frequency difference was selected as significantly larger than the inverse of data time span ($\sim 1\,000 \, \rm{d}$), a rough estimate for the orbital-frequency uncertainty lower limit.
Limiting the orbital-frequency difference to $\frac{1}{500}
\, \rm{d}^{-1}$ yielded the same sample.
An orbital-period limit of 2 days was chosen since most discovered EL CVn systems are below it \citep[See Fig.~5 of][]{roestel18}.
These criteria resulted in an initial sample of 1174 systems.

\begin{figure}[htb]
\centering
{\includegraphics[scale=0.55, trim={1.0cm 0.3cm 0 0.5cm}, clip]{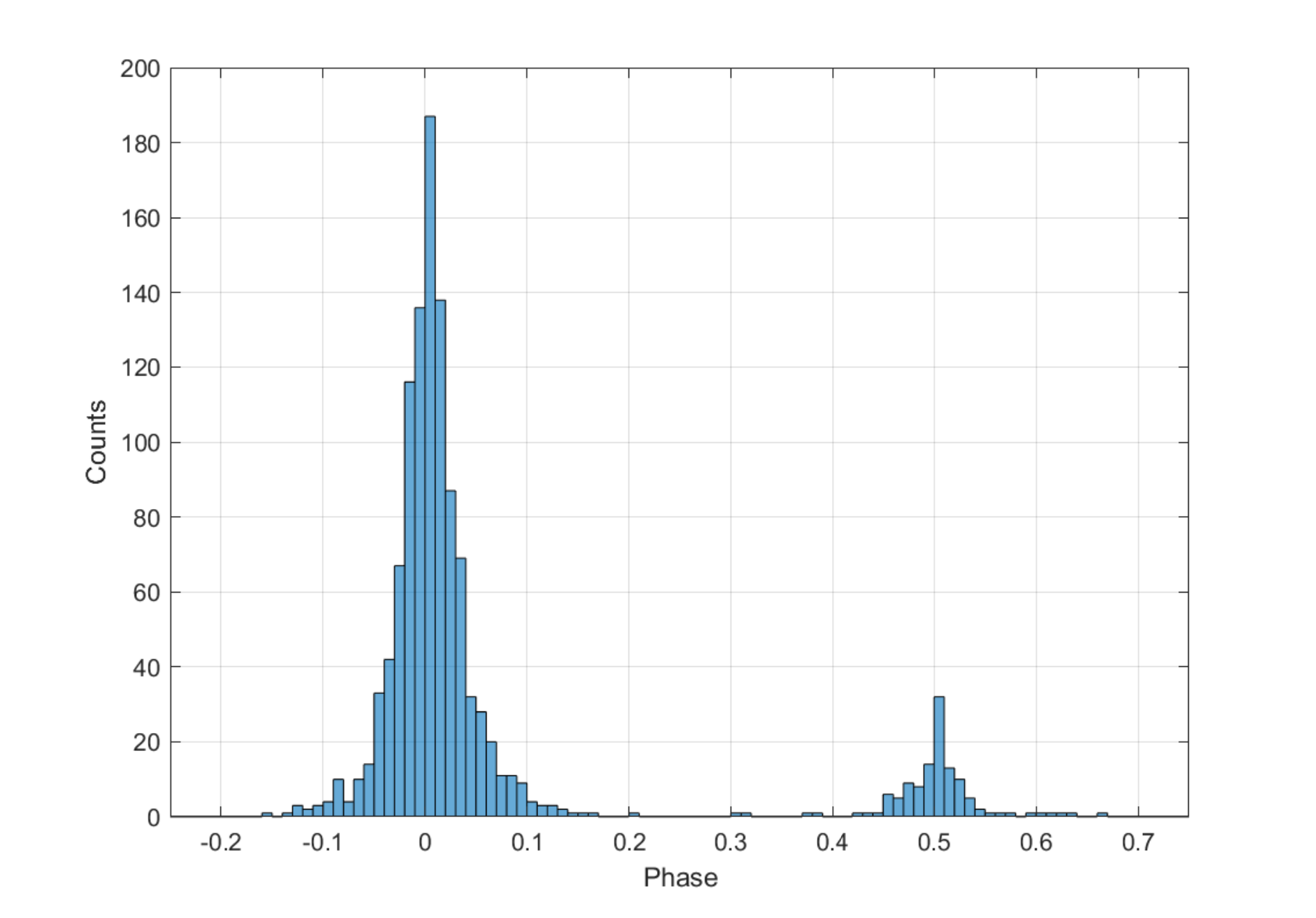}}
\caption{
Histogram of phase difference between SB-model primary eclipse and EB-model deeper eclipse, for 1174 stars in our sample.
}
\label{fig:phaseHist}
\end{figure}

Next, we calculated the phase difference between the SB-model primary eclipse and the EB-model deeper eclipse, for all stars in our sample. For a common binary, for example one consisting of two MS stars, we expect this phase difference to be zero. However, for an EL CVn, in which the secondary is hotter hence the secondary eclipse is deeper, we expect the phase difference to be $\sim0.5$, assuming a small eccentricity. Figure~\ref{fig:phaseHist} shows the phase-difference histogram of our initial sample, with a main peak at phase zero, and a much smaller peak at phase $0.5$, as expected.

Based on this, we selected an initial list of EL CVn candidates with:
\begin{enumerate}
    \item Eclipses phase difference in the $0.4-0.6$ range, and
    \item Eccentricity smaller than 0.3, and
    \item EB-model eclipse-depth difference with a signal-to-noise ratio (SNR) larger than 5.
\end{enumerate}
The eclipse-depth difference SNR was required since our method relies on reliably identifying a secondary eclipse that is significantly deeper than the primary one.
These criteria yielded 16 systems.

Finally, we visually inspected the \gaia photometry and RV data and models of the 16 systems in our initial list, and identified 5 systems as the most promising EL CVn candidates.


\subsubsection{Five EL CVn-type candidates}
\label{sec:cand}
\begin{figure*}[!hp]
\centering
\resizebox{18cm}{11.5cm}
{
\includegraphics[scale=0.4, trim={1.5cm 1.0cm 2.0cm 1.5cm}, clip]{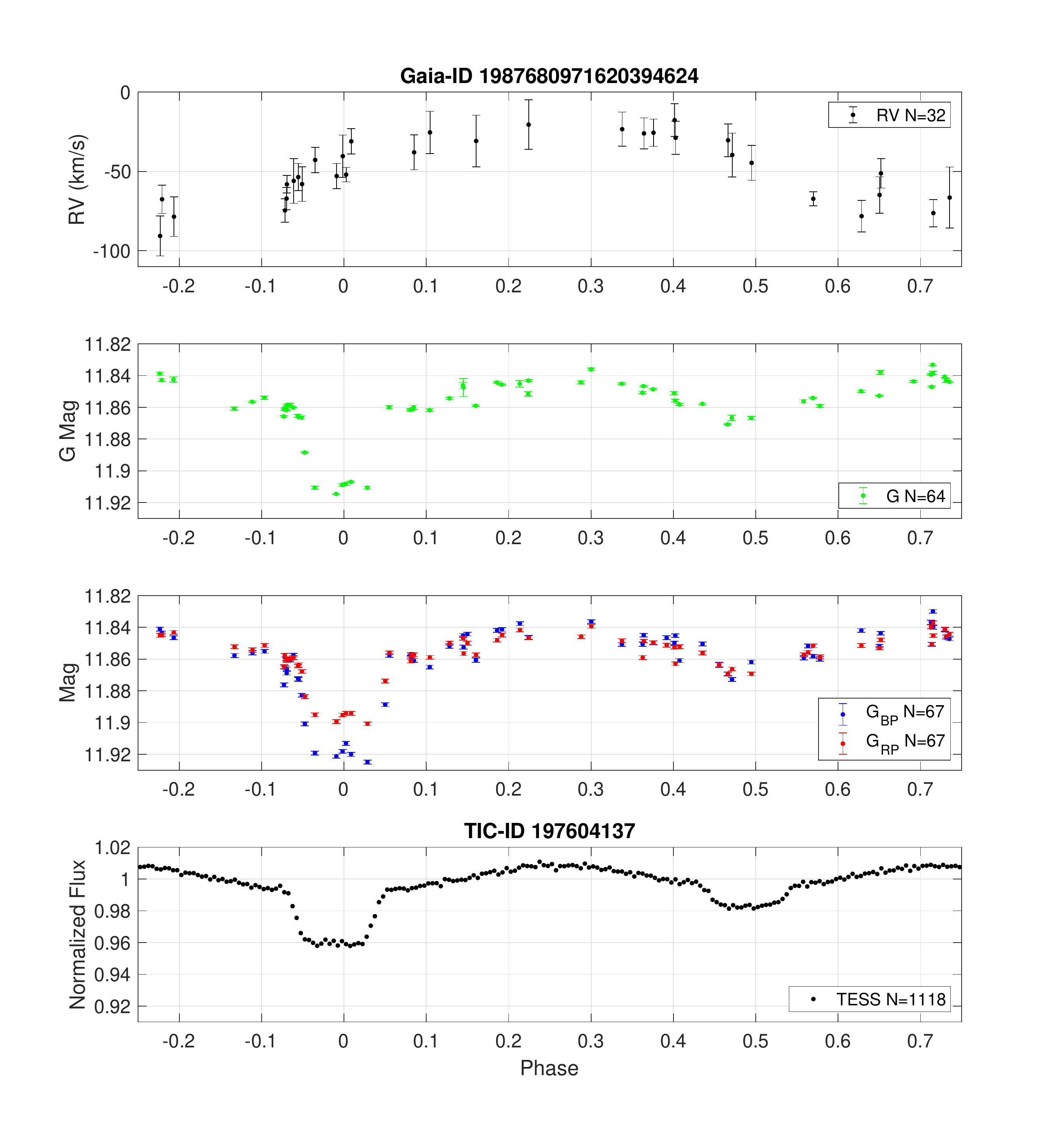}
\includegraphics[scale=0.4, trim={1.5cm 1.0cm 2.0cm 1.5cm}, clip]{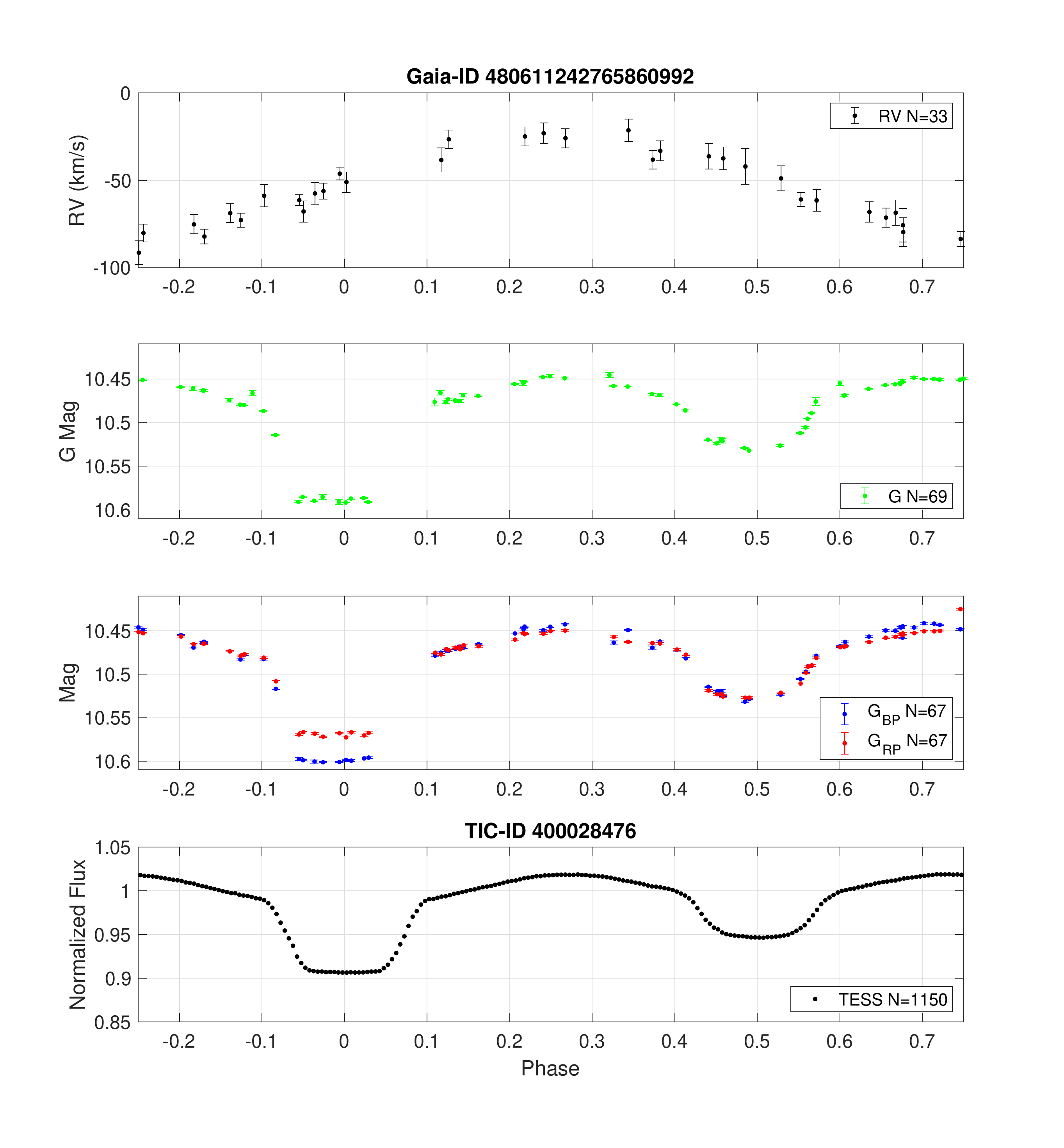}  
}

\centering
\resizebox{18cm}{11.5cm} 
{
\includegraphics[scale=0.4, trim={1.5cm 1.5cm 2.0cm 1.0cm}, clip]{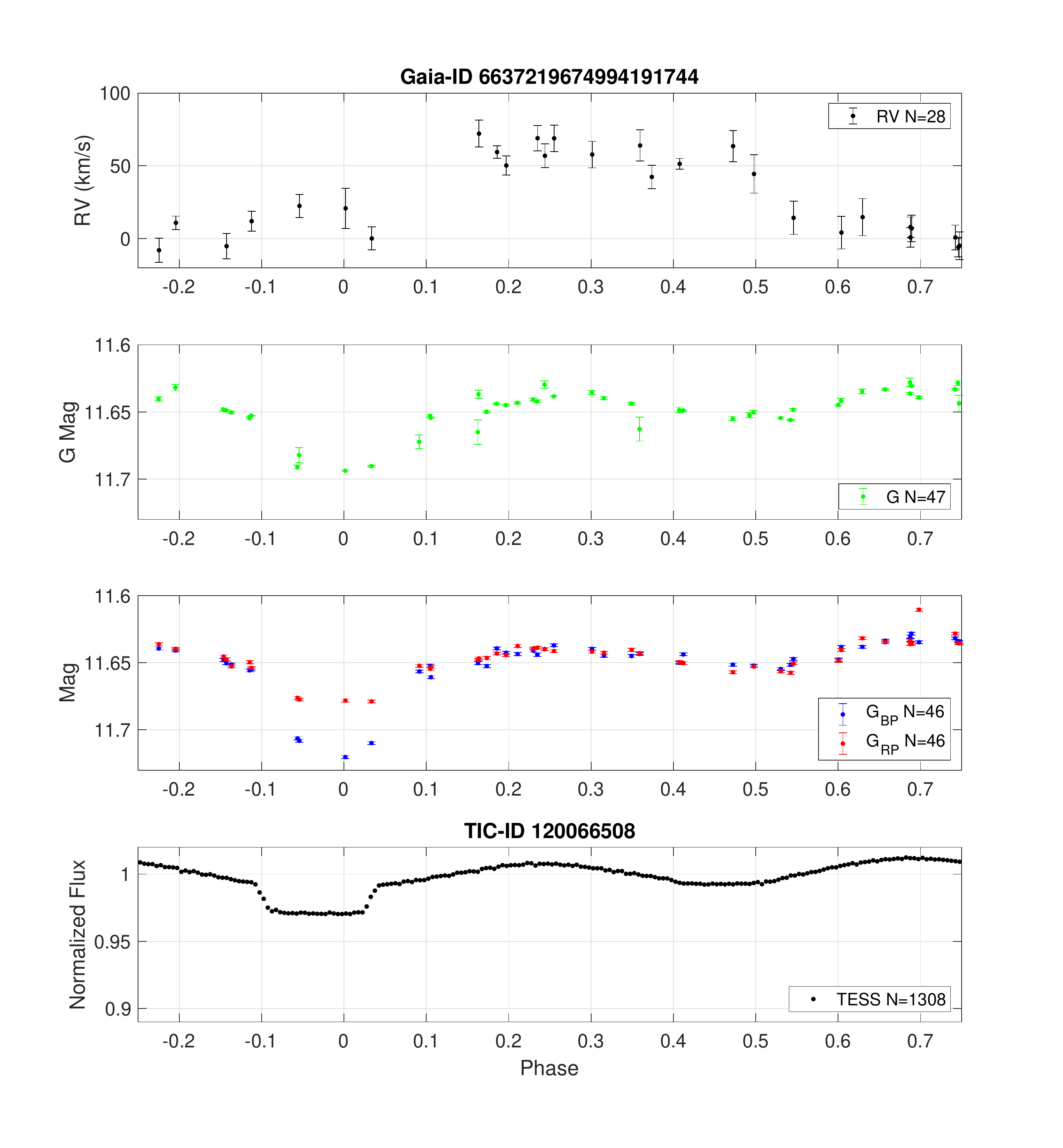}  
\includegraphics[scale=0.4, trim={1.5cm 1.5cm 2.0cm 1.0cm}, clip]{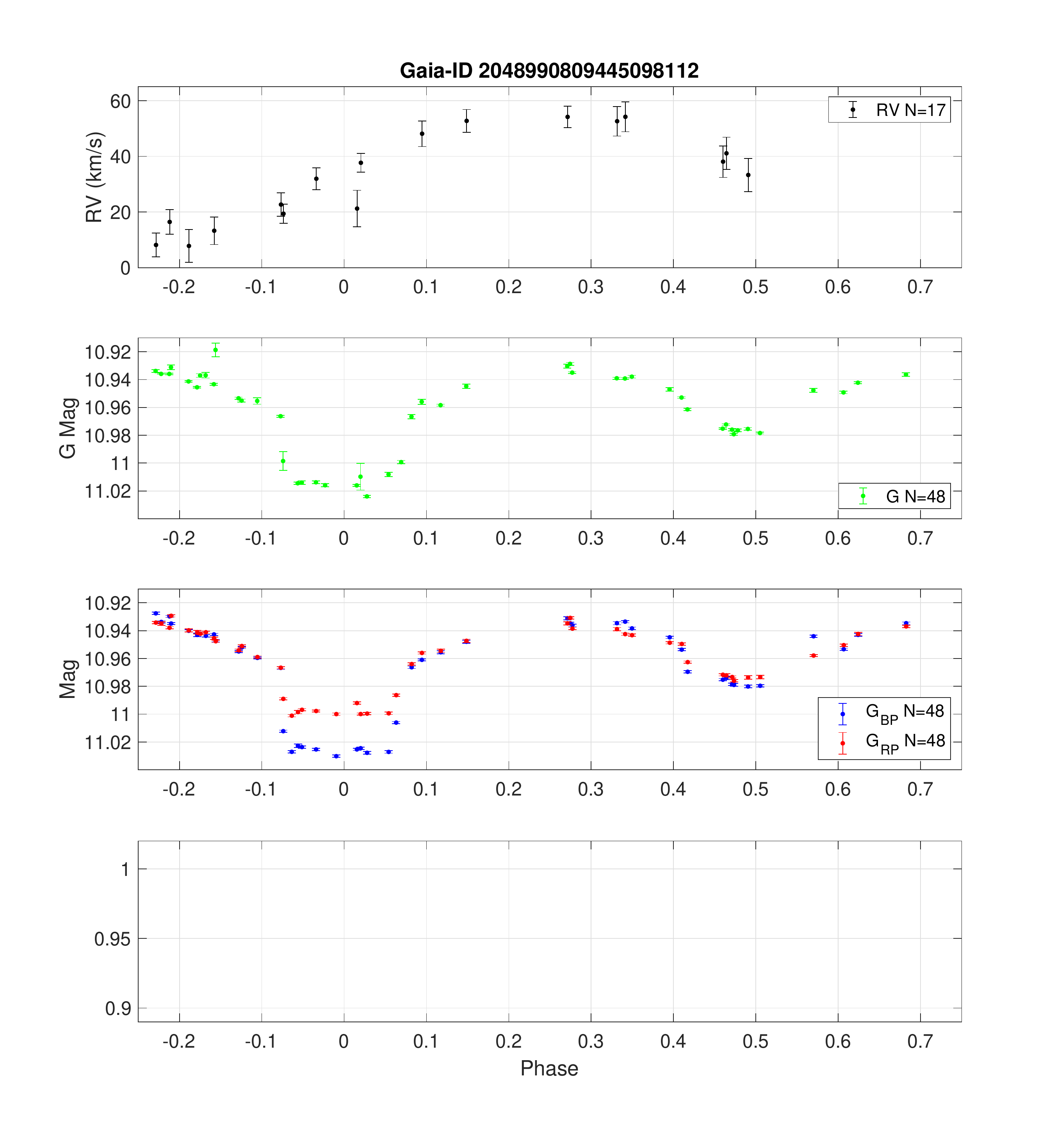}
}
\caption{
\gaia and TESS data of four new EL CVn candidates using \figref{fig:HD} presentation. We could not find TESS photometry for \object{\sourceId{2048990809445098112}}, so its TESS panel was left empty.
}
\label{fig:ELCVn4}
\end{figure*}

After identifying the $5$ EL CVn candidates we realized that all have a \gaia \fieldName{SB1} model, and one of them is actually a known EL CVn-type system \citep[\sourceId{5087757377681887232} (G5087); TIC-ID 121078334; \object{HD 23692}; ][]{maxted14a}, which is shown in \figref{fig:HD}. This system serves as an initial validation for our discovery method and the rest of the candidates.
Figure~\ref{fig:ELCVn4} shows the \gaia and TESS data (except for \object{\sourceId{2048990809445098112}} (G2048), for which we could not find TESS photometry) of the four new candidates using the \figref{fig:HD} presentation.
A selected set of \gaia parameters of the $5$ candidates are listed in \tabref{tab:ELCVn}.

In principle, one can use the rich \gaia photometry, RV, and derived properties, together with the TESS data, to fit an astrophysical model and estimate the pre-He-WD mass, radius, effective temperature and other properties. Such an analysis was performed by e.g. \cite{maxted14a}, \cite{faigler15} and \cite{roestel18} for multiple EL CVn systems they identified. However, this analysis is beyond the scope of this \quoting{teaser} paper.


\newcommand*{\thead}[1]{#1}
\begin{table*}[htb]
\caption{ Parameters of $5$ EL CVn candidate systems.
	}
    \resizebox{\textwidth}{!}{
    \small
	\begin{tabular}{rrrrrr}
        \hline 
		\hline
		 \thead{\gaia $\rm ID$} & \thead{\footnotesize 5087757377681887232} & \thead{\footnotesize 1987680971620394624} & \thead{\footnotesize 480611242765860992} & \thead{\footnotesize 6637219674994191744} & \thead{\footnotesize 2048990809445098112} \\
        \hline 
        \thead{TIC ID} & 121078334 & 197604137 & 400028476 & 120066508 & 378080617\\
        \thead{Other ID} & \object{HD 23692} & \object{TYC 3615-2289-1} & \object{TYC 4106-576-1} & \object{TYC 8757-1969-1} & \object{TYC 3135-13-1}\\
        \thead{ GSP-Phot $T_{\rm{eff}}$ (K)} & 7360 & 7504 & 7470 & NA & 7148\\
        \thead{ GSP-Phot $\log{g}$ (dex)} & $3.98_{0.03}^{0.05}$ & $4.01_{0.03}^{0.01}$ & $4.08_{0.004}^{0.005}$ & NA & $3.77_{0.01}^{0.03}$\\
        \thead{ Parallax (mas)} & $3.447\pm0.063$ & $1.169\pm0.012$ & $2.647\pm0.013$ & $1.197\pm0.020$ & $1.251\pm0.013$ \\
        \thead{ Radius \fieldName{FLAME} (\Rsun)} & $1.975_{0.050}^{0.053}$ & $2.132_{0.047}^{0.047}$ & $1.803_{0.038}^{0.037}$ & NA & $3.006_{0.067}^{0.067}$\\
         %
         %
         %
        \thead{ \gmag/\gbp/\grp\ (mag)} & 9.56 / 9.70 / 9.29 & 11.86 / 12.05 / 11.52 & 10.47 / 10.65 / 10.16 & 11.64 / 11.86 / 11.28 & 10.95 / 11.13 / 10.64 \\
        \thead{ $\rm RV$/\gmag/\gbp/\grp\ points} & 18 / 55 / 55 / 54 & 32 / 64 / 67 / 67 & 33 / 69 / 67 / 67 & 28 / 47 / 46 / 46 & 17 / 48 / 48 / 48\\
        \thead{EB period (day)} & $0.92859504\pm0.00001141$ & $1.16189659\pm0.00002522$ & $0.64410724\pm0.00000429$ & $1.27337301\pm0.00005718$ & $1.36819733\pm0.00005130$ \\
        \thead{EB $T_0$ (BJD)} & $2457389.055\pm0.004$ & $2457389.680\pm0.013$ & $2457388.877\pm0.003$ & $2457389.784\pm0.019$ & $2457388.853\pm0.010$ \\
        \thead{SB1 $K_1$ \kms)} & $24.28\pm0.94$ & $30.42\pm3.57$ & $28.72\pm1.47$ & $32.28\pm1.77$ & $22.28\pm1.67$ \\
        \hline   
	\end{tabular}
	}
	\\
	\vspace{3mm}
	\label{tab:ELCVn}
\end{table*}

%
\subsection{Ultracool dwarf binaries}\label{ssec:hrd_compact_ucd_binaries}
%
Multiplicity studies of ultracool dwarfs (UCDs, which comprise both very low-mass stars and brown dwarfs) at separations $\lesssim$1\,au have for long been hampered by the relative faintness of those objects and the associated observational limitations. The separation and mass (or magnitude)-ratio distribution of known UCD binaries therefore carries a significant observational bias, which is also affecting the estimated UCD binary fraction of 10-30\% \citep[e.g.][]{Burgasser:2007ix}. 
Surveys of small samples indicate that the occurrence of compact UCD binaries is significant \citep{Blake:2010lr,Bardalez-Gagliuffi:2014aa,Sahlmann:2014aa} and that many of those systems have photocentre orbit amplitudes in the range of mas. It has therefore been predicted that \gaia astrometry will eventually characterize hundreds of UCD binary orbits \citep{Sahlmann:2014ab} and significantly improve our knowledge on the occurrence and properties of compact UCD binaries.

\begin{figure}[htb]
\begin{center}
\includegraphics[width=\columnwidth]{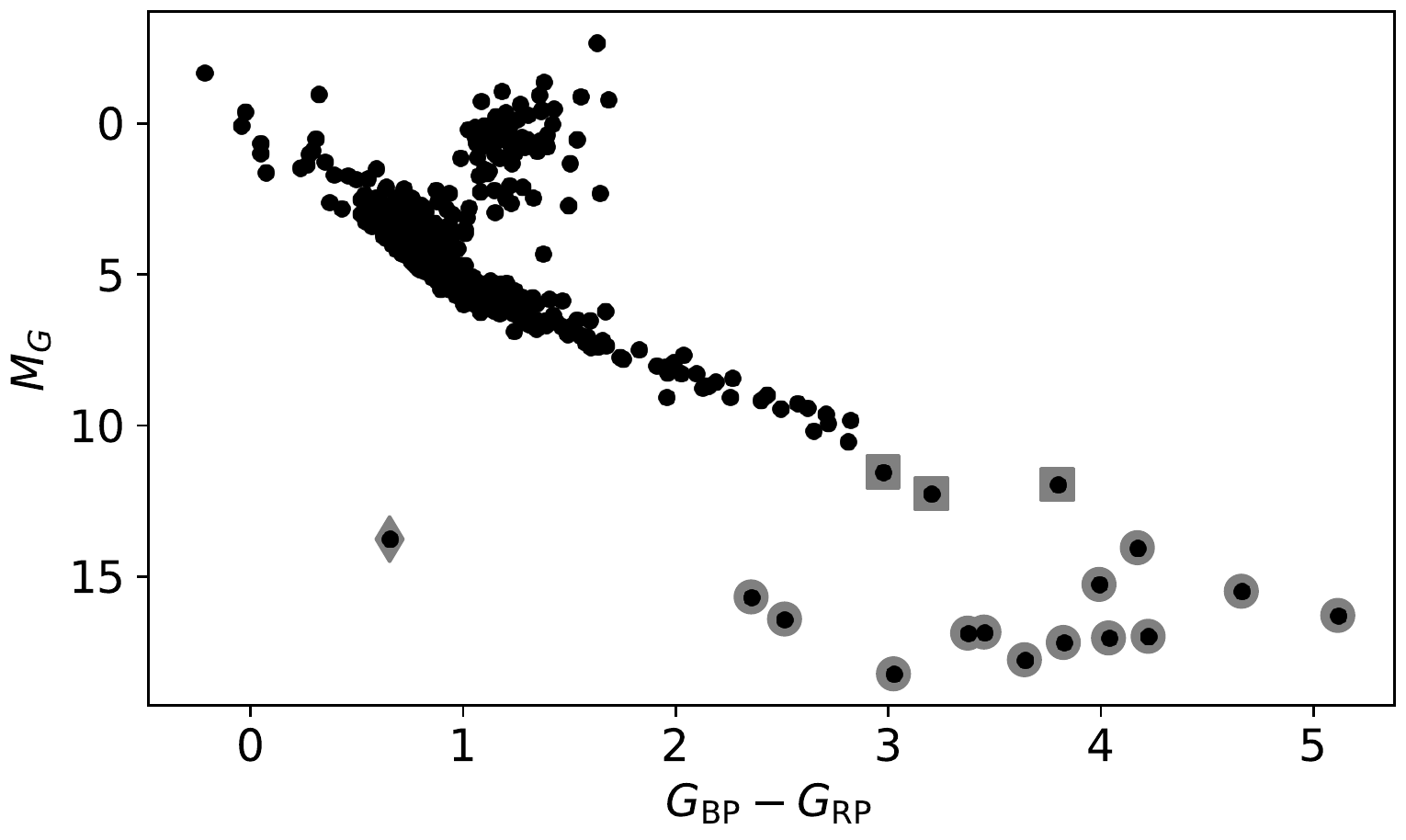}
\caption{Colour-magnitude diagram of 533 sources with \fieldName{OrbitalTargetedSearch*} solutions (black circles). The larger grey circles indicate sources that are listed in the \gaia Ultra-cool Dwarf Sample \citep{Smart:2019aa}, the grey squares indicate red sources discussed in the text, and the diamond indicates the white dwarf discussed in \secref{ssec:exoplanets_wd}.}\label{fig:hrd_compact_ucd_binaries_cmd}
\end{center}\end{figure}

Here we have a first look at the UCD orbits in \gdrthree. Figure \ref{fig:hrd_compact_ucd_binaries_cmd} shows the \gaia colour-magnitude diagram of all sources with \fieldName{nss_solution_type} = \fieldName{OrbitalTargetedSearch*}. 
Large grey circles indicate the known UCD sources from the \gaia Ultra-cool Dwarf Sample \citep[GUCD, ][]{Smart:2019aa}, where we used the \fieldName{gaiaedr3.dr2_neighbourhood} table to match a \fieldName{source_id} from \gdrtwo\ to \gdrthree.

\begin{figure}\begin{center}
\includegraphics[width=\columnwidth]{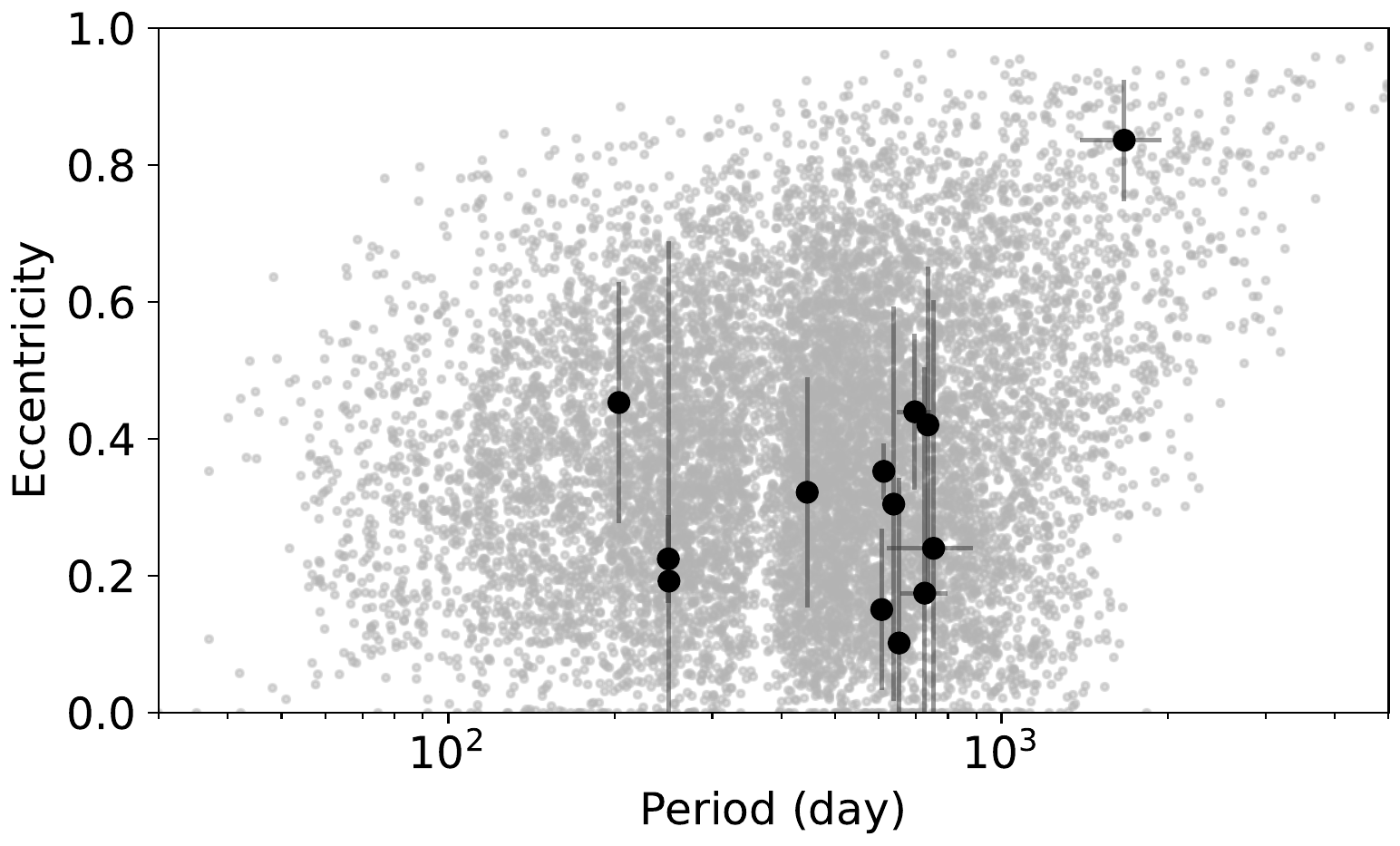}
\caption{Orbital eccentricity as a function of period for the UCD systems listed in Table \ref{tab:compact_ucd_binaries} (black symbols). The parameters of \fieldName{Orbital} solutions within 200 pc are shown in grey.}\label{fig:nss_compact_ucd_binaries_period_ecc}
\end{center}\end{figure}

The only non-single star solutions for GUCD sources are found in the \TBOTable table with solution type \fieldName{OrbitalTargetedSearch*}. No GUCD could be matched with an \fieldName{Orbital} solution or an acceleration solution. This is mainly because a large fraction ($\sim$75\%) of GUCDs are fainter than the $G<19$ cut-off for processing with the nominal NSS pipelines \citep{DR3-DPACP-163}. As described in \citet{DR3-DPACP-176}, the full GUCD sample was however included in the targeted search for orbital signals.

Thirteen sources with DR3 orbits are part of the GUCD catalogue, and only 5 of those are brighter than $G=19$. Table \ref{tab:compact_ucd_binaries} lists the GUCD sources with DR3 orbital solutions.

\begin{table*}[htb]
\caption{Identifiers and basic properties of 13 UCD binaries in DR3. The \quoting{Name} column corresponds to the SHORTNAME in the GUCD catalogue. The uncertainties of $a_0$ were computed using linear error propagation. The three additional sources at the bottom of the table are the reddest \fieldName{OrbitalTargetedSearch} sources at the bottom of the main sequence that are not in the GUCD catalogue.}
\small
\begin{tabular}{rrrlrl}
\hline
\hline
          Gaia DR3 &                Name & \fieldName{nss_solution_type} &         Period (d) &    $a_0$ (mas) & Notes \\
\hline
\object{4997505546262260096} &              J0031-3840 & OrbitalTargetedSearch & ${1665.4}\pm{281.5}$ & ${23.96}\pm{3.47}$ &  New binary candidate\\
\object{2576389458819793920} &              J0106+0557 & OrbitalTargetedSearch &   ${726.2}\pm{71.0}$ &  ${4.48}\pm{0.60}$ &    New binary candidate   \\ 
\object{5182151481717042944} &              J0320-0446 & OrbitalTargetedSearch &    ${250.6}\pm{1.7}$ &  ${7.75}\pm{2.76}$ &      Known RV binary\tablefootmark{d} \\
\object{3269943938874146688} &              J0344+0111 & OrbitalTargetedSearch &   ${652.9}\pm{23.4}$ &  ${7.55}\pm{0.42}$ &     Known spectral binary\tablefootmark{e}\\
 \object{144711230753602048} &              J0435+2115 & OrbitalTargetedSearch &   ${607.4}\pm{25.2}$ &  ${5.48}\pm{0.34}$ &      New binary candidate\\
\object{3361210791323909504} &              J0659+1717 & OrbitalTargetedSearch &  ${753.6}\pm{132.8}$ &  ${7.90}\pm{1.55}$ &   New binary candidate\\
 \object{933054951834436352} &              J0805+4812 & OrbitalTargetedSearchValidated &   ${735.9}\pm{23.0}$ & ${14.18}\pm{2.06}$ &    Known astrometric binary\tablefootmark{a}  \\
\object{5514929155583865216} &              J0823-4912 & OrbitalTargetedSearchValidated &    ${250.0}\pm{1.2}$ &  ${4.99}\pm{0.21}$ &      Known astrometric binary\tablefootmark{b}    \\
\object{1610979010812148224} &              J1429+5730 & OrbitalTargetedSearch &    ${445.4}\pm{8.5}$ &  ${4.64}\pm{0.46}$ &     New binary candidate\\
\object{4406489184157821952} &              J1610-0040 & OrbitalTargetedSearch &    ${612.5}\pm{7.7}$ &  ${9.41}\pm{0.25}$ &      Known astrometric binary\tablefootmark{c} \\
\object{6797628972554531840} &              J2026-2943 & OrbitalTargetedSearch &   ${638.3}\pm{14.5}$ &  ${5.69}\pm{1.05}$ &     Known spectral binary\tablefootmark{f}\\
\object{1754495583527340416} &              J2036+1051 & OrbitalTargetedSearch &   ${696.7}\pm{50.4}$ &  ${4.57}\pm{0.53}$ &    New binary candidate\ \\
\object{6616442994033876480} &       J2200-3038A\!\!\! & OrbitalTargetedSearch &    ${203.5}\pm{1.9}$ &  ${3.84}\pm{0.68}$ &     New binary candidate\tablefootmark{g} \\
\hline
\object{4963614887043956096} & J0219-3925 & OrbitalTargetedSearch &    ${538.0}\pm{3.0}$ &  ${1.41}\pm{0.05}$ &      New binary candidate\tablefootmark{h}    \\
  \object{43574131143039104} &                 LHS1610 & OrbitalTargetedSearch &  ${10.6}\pm{0.0}$ & ${1.39}\pm{0.04}$ &    Known RV binary\tablefootmark{i} \\
\object{5600272625752039296} &        L 601-78 A  & OrbitalTargetedSearch &  ${14.3}\pm{0.0}$ & ${1.87}\pm{0.07}$ &   New binary candidate\\
\hline
\end{tabular}
\label{tab:compact_ucd_binaries}
\tablefoot{References: 
\tablefoottext{a}{\cite{2020MNRAS.495.1136S}}
\tablefoottext{b}{\cite{Sahlmann:2013ab}}
\tablefoottext{c}{\cite{Dahn:2008fk}}
\tablefoottext{d}{\cite{Blake:2008fr}}
\tablefoottext{e}{\cite{Bardalez-Gagliuffi:2014aa}}
\tablefoottext{f}{\cite{Bardalez-Gagliuffi:2014aa, Gelino:2010aa}}
\tablefoottext{g}{This would be a new inner binary as component A in a known 1\arcsec-wide binary \citep{Burgasser:2006ww, Smart:2019aa}}
\tablefoottext{h}{This would be a new inner binary in a known 4\arcsec-wide binary \citep{Artigau:2015vp}}
\tablefoottext{i}{\cite{Winters:2018wm}}
}
\end{table*}

There are three binaries with previously published astrometric orbit solutions. These are J0805+4812 \citep{2020MNRAS.495.1136S}, J0823-4912 \citep{Sahlmann:2013ab, Sahlmann:2015ab}, and J1610-0040 \citep{Dahn:2008fk, Koren:2016ua}. Generally, the DR3 orbit parameters agree with these published solutions.

The J0320-0446 binary has a published RV orbit and both the \gaia period and eccentricity agree well. The \gaia solution indicates an almost edge-on configuration, in agreement with the expectations from the RV modelling by \citet{Blake:2008fr}. The Thiele-Innes coefficients for this low-eccentricity solution are highly correlated, which leads to correlated and skewed distributions when resampling the geometric parameters. This has to be accounted for when using the \gaia solution parameters for estimating the companion mass in particular.

Two sources in this list have only been identified as spectral binaries, i.e. typically L+T-dwarf systems in which the companion's spectrum is discernible in the combined-light near-infrared spectrum. In principle, the \gaia orbital solutions will make it possible to constrain the masses of the binary components. Since the orbital parameters refer to the system's photocentre, however, the applicable constraints depend on the brightness- and mass-ratio of the system which usually are not determined by \gaia. Therefore, external information and assumptions have to be incorporated. As an example, we explore the orbit of J2026-2943, which has previously been identified as a spectral binary with components of spectral types L1+T6 \citep{Bardalez-Gagliuffi:2014aa, Gelino:2010aa}. If we assume a field-age mass of $0.080\pm0.005\Msun$ for the L1 primary and that the light contribution of the T6 companion is negligible in the $G$ band, then the \gaia orbit solution implies a companion mass of ${0.041}^{+0.016}_{-0.009}\Msun$, where we accounted for all parameter covariances using Monte-Carlo resampling. This mass estimate is compatible with expectations for a T6 brown dwarf and adds a valuable entry in the list of low-mass systems with dynamically-determined masses \citep{Dupuy:2019aa, 2020MNRAS.495.1136S}. Examples of  combining astrometric orbit solutions with RVs, spectral-binary indicators, and spatially-resolved observations to investigate the physics of UCDs can be found in the literature \citep[e.g.\ ][]{Garcia:2017aa, 2021MNRAS.500.5453S, 2020MNRAS.495.1136S, Brandt:2020ve}.

For the remaining 7 sources we have not found previously-identified multiplicity indicators in the literature. These are therefore potentially new UCD binary discoveries made in \gdrthree. These could be confirmed by independent observations of their spectroscopic properties or with RV monitoring. Because of its long period and proximity, the J0031-3840 system has an estimated relative semi-major axis of $\sim$60 mas and could possibly be spatially resolved with specialised instruments, which can give access to model-independent mass determinations.

As expected, the UCD binaries discussed above are compact systems with estimated relative separations $\lesssim$1.5 au because a large fraction of their orbits is covered by the DR3 data. Their period-eccentricity distribution is essentially unaffected by the complications in terms of mass determination discussed above and can be used to investigate UCD formation mechanisms or dynamical histories \citep[e.g.\ ][]{Dupuy:2011uq}. Figure \ref{fig:nss_compact_ucd_binaries_period_ecc} shows these parameters in comparison with the \fieldName{Orbital} solutions. The high eccentricity of the long-period solution for J0031-3840 may be affected by the incomplete orbit-coverage with \gaia data, which tends to push eccentricity up as also shown by the \fieldName{Orbital} solutions.

Comparison with \citet[][Fig. 18, which includes a few sources in common]{Dupuy:2017aa} shows that these \beforeReferee{distributions of} generally intermediate eccentricities are in agreement. Importantly, \gaia is filling in the period-range of $\sim0.5-5$ years which so far is sparsely populated because of the resolution-limit of direct-imaging instruments. This will help to build statistically-robust samples of UCD binaries that can be used for comparison to stellar  binaries.

Finally, we inspected the orbits of the three reddest objects that are not in the GUCD catalogue, which are highlighted with squares in \figref{fig:hrd_compact_ucd_binaries_cmd}:
\begin{description}
  \item[J0219-3925] This is \object{2MASS J02192210-3925225}, which was characterized as a young late-M dwarf with a wide (4\arcsec) substellar companion by \citet{Artigau:2015vp}. The \gaia astrometric orbit corresponds to an inner companion to the M-dwarf, which (if dark) could have a mass as low as ${11.2}\pm{0.9}$\Mjup for the primary mass of ${113}\pm{12}$\Mjup \citep{Artigau:2015vp}. Another possible explanation is a more massive companion, whose light contribution dilutes the photocentre orbit. Auxiliary data or observations have to be considered to better characterise the inner companion.
  \item[LHS1610] This mid-M dwarf was identified as an RV binary by \cite{Winters:2018wm} and the \gaia astrometry independently confirms the eccentric 10.6 day orbit. The Gaia solution indicates a close-to edge-on configuration and a companion-mass estimate of ${0.052}^{+0.004}_{-0.004}\Msun$ when assuming a primary mass of ${0.17}\pm0.02\Msun$, i.e.\ it establishes the substellar nature of the companion.
  \item[L 601-78 A] This is the primary component of the wide binary \object{L 601-78}. It was identified as a lens candidate for  mass determination using astrometric microlensing by \gaia \citep{Kluter:2020ux}.
\end{description}


%
\section{Compact-object companions}\label{compacts}
%

In what follows we point out a few channels of identifying binary candidates in our tables that might harbour compact-object companions. The discussion is meant to suggest further study that is needed in order to confirm the nature of these binaries. We first underline the method used for astrometric orbits, then discuss white dwarf secondaries, then larger masses, including those found using SB1 orbits.

%
\subsection{Astrometric binaries with compact-object companions using the triage algorithm}\label{triage}
%
To identify unresolved astrometric binaries that are likely to host a compact object as their faint binary companion, we use the triage classification of \citet{shahaf19}. The algorithm divides the astrometric binaries into three classes:
\begin{itemize}
\item class-I binaries, where the companion is most likely a single MS star (but can also be a close binary or a compact object), 
\item class-II binaries, where the companion cannot be a single MS star; therefore, it is most likely a close MS binary (but can also be a compact object), and 
\item class-III binaries, where the companion cannot be a single MS star or a close MS binary; therefore, these systems most likely host a compact object as secondary. 
\end{itemize}

The distinction between the three classes depends on the value of the newly defined astrometric mass ratio function (AMRF),  $\mathcal{A}$, given by
\begin{equation}
    \label{eq:amrf}
    \mathcal{A} \equiv \frac{a_0}{\varpi} \bigg(\frac{\Mass_1}{\Msun}\bigg)^{-1/3} \bigg(\frac{P}{\textrm{yr}}\bigg)^{-2/3} \, ,
\end{equation}
%
where $a_0$ is the derived angular semi-major axis of the photocentric orbit, $\varpi$ is the parallax, $\Mass_1$ is the mass of the primary star, and $P$ is the orbital period of the binary.

AMRF, $\mathcal{A}$, can be written as a function of the mass ratio $q=\Mass_2/\Mass_1$  and the luminosity ratio of the two components of the astrometric binary $\mathcal{S}=F_2/F_1$ as
\begin{equation}
    \mathcal{A}= \frac{q}{(1+q)^{2/3}}\,\bigg(1 - \frac{\mathcal{S}(1+q)}{q(1+\mathcal{S})} \bigg)\, ,
\end{equation}
\citep[see details in ][]{shahaf19}. 
%
The luminosity ratio modelling is based on the \citet{mamajek13} main-sequence colour and effective temperature tables,
assuming no extinction. 
%
As emphasized by \cite{shahaf19}, there are two limiting AMRF values: the maximal value for a single MS companion (class I sources), and the maximal value assuming a close binary of two identical MS stars as the secondary companion (class II sources). 
%

\begin{figure*}[htb]\begin{center}
    \includegraphics[width=2\columnwidth]{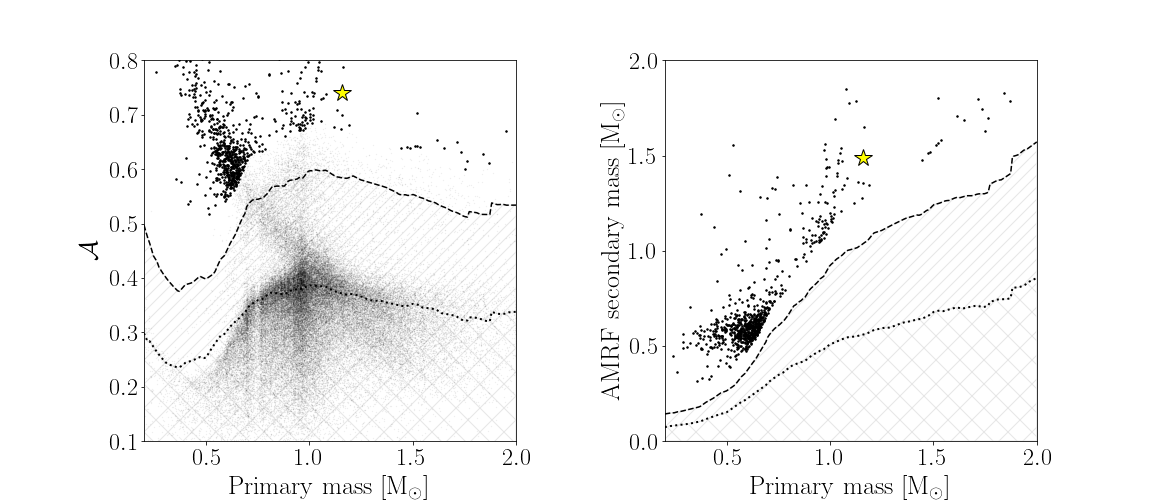}
\caption{Triage of the \gaia astrometric binaries, with classes I to III from bottom to top delineated by dotted and dashed lines
respectively. \textit{Left}: AMRF vs.~mass of the primary for the astrometric solutions of this paper. The systems with a probable compact-object companion (class III) are presented as bold black points (see text). For reference, the rest of the sample is plotted as thin black points.  \textit{Right}: Masses of the compact companions, derived from the AMRF, as a function of their primary mass. The rest of the systems do not appear in this panel.
Yellow star is \afterReferee{the neutron star candidate  \object{\sourceId{5136025521527939072}} discussed in \secref{neutron}}.}
  \label{fig:GaiaTriage}
\end{center}\end{figure*}

Figure~\ref{fig:GaiaTriage} 
presents the derived AMRF values for the 
{\it Gaia} unresolved astrometric binaries presented in this work, with primaries in the range of $0.2{-}2.0$\Msun \afterReferee{ and significance larger than $20$}, using the primary mass values reported in \tabref{tab:nssmass}. Systems with AMRF values that exceed the maximal possible value for \beforeReferee{ternary}\afterReferee{triple} stars by more than $4\sigma$, where the uncertainty $\sigma$ was calculated by propagating the uncertainties of the quantities in \equref{eq:amrf}.
They are presented in \afterReferee{Fig.}~\ref{fig:GaiaTriage} as bold black points and are likely to host a compact-object companion. 
%
%
Since the light contribution of the companions of these systems is negligible, namely $\mathcal{S}{\simeq}0$, we can derive the mass ratio and then the mass of the secondary of these systems. Their mass values are presented in the right panel of the figure. 
%

\begin{figure*}[htb]\begin{center}
    \includegraphics[width=2\columnwidth]{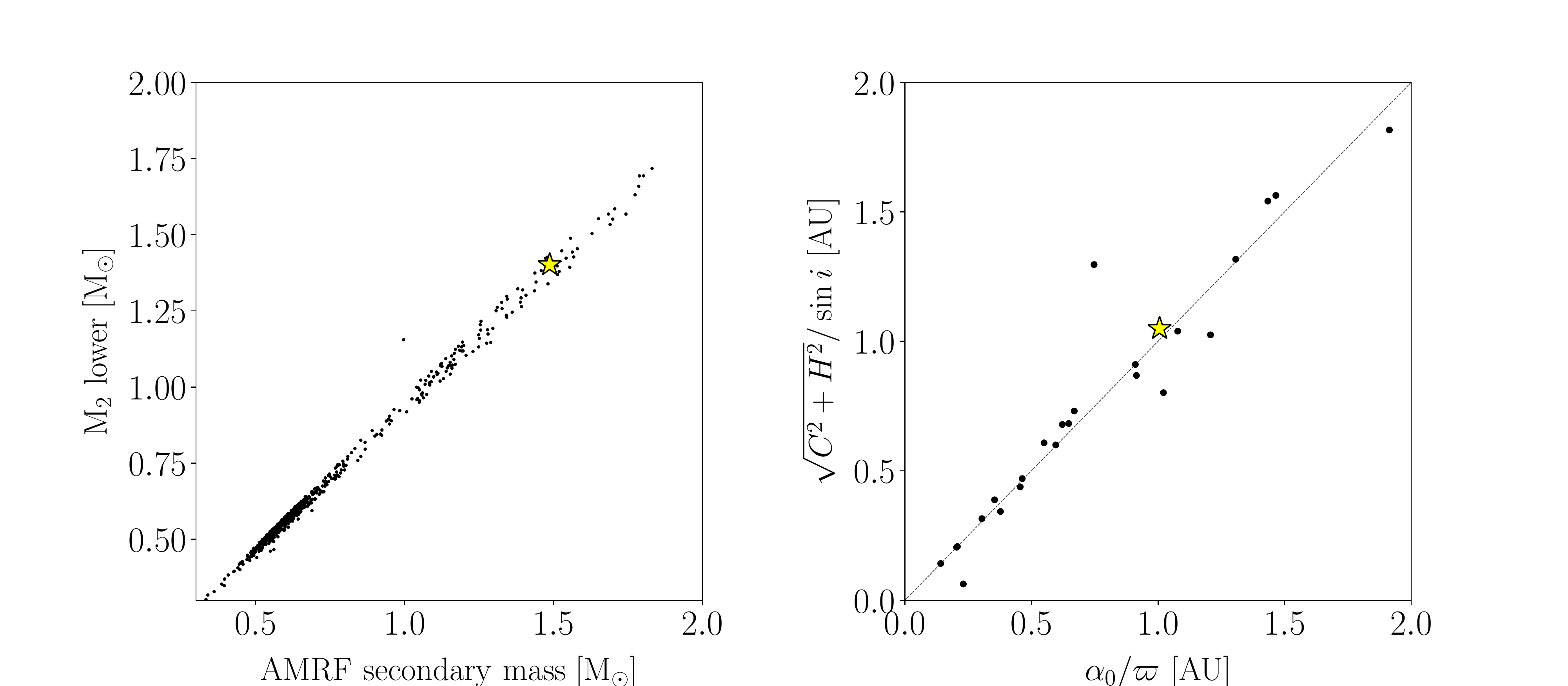}
\caption{\textit{Left}: Comparison between the minimum companion mass, derived in \secref{Sect:masscat}, and the AMRF-derived masses for the compact-object candidates. The sources where the AMRF mass is smaller than the minimum mass are \fieldName{AstroSpectroSB1} with actually a non-negligeable flux ratio.
\textit{Right}: Comparison between semi-major axes derived for the \fieldName{AstroSpectroSB1} systems using the astrometric parameters (horizontal axis) and the spectroscopic parameters (vertical axis). 
Yellow star is the \sourceId{5136025521527939072}.
  }
  \label{fig:QA}
\end{center}\end{figure*}

%
We compare the obtained masses with the conservative lower-limit estimate of the companion mass (\secref{Sect:masscat}), in the left panel of \figref{fig:QA}. Additionally, because for faint companions the photocentre coincides with the position of the primary star, we expect  the semi-major axes derived from the astrometry and the RV spectroscopy to be similar, as demonstrated in the right panel of \figref{fig:QA} for the \fieldName{AstroSpectroSB1} cases.

Figure~\ref{fig: AMRF mass dist} presents a mass histogram of the secondaries of the astrometric binaries assumed to have a compact companion, up to $2.1$\Msun. It seems as if most of the companions are white dwarfs, with a clear narrow peak at $\sim 0.6\Msun$, as is the case for field white dwarfs. Obviously, the secondary mass population is heavily biased by the way it was derived, so any astrophysical interpretation should be done carefully. 
Anyway, the circularisation at large periods is striking, \figref{fig:AMRF_Pe_diagram}.

\begin{figure}[htb]\begin{center}
    \includegraphics[width=\columnwidth]{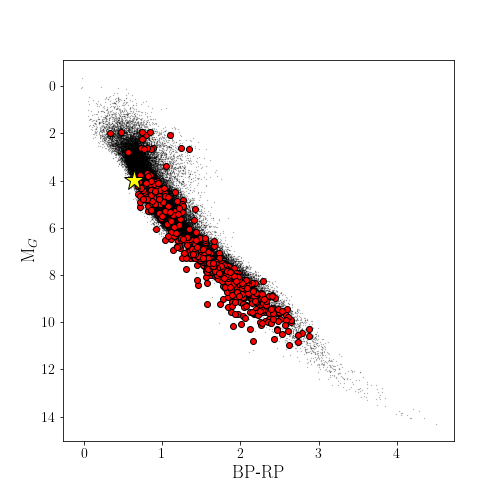}
\caption{Class-III CMD. Black points - Orbital and AstroSpectro, for reference. Red points are AMRF class-III. The yellow star is  \sourceId{5136025521527939072}.}
  \label{fig: AMRF cmd}
\end{center}\end{figure}

\begin{figure}[htb]\begin{center}
    \includegraphics[width=\columnwidth]{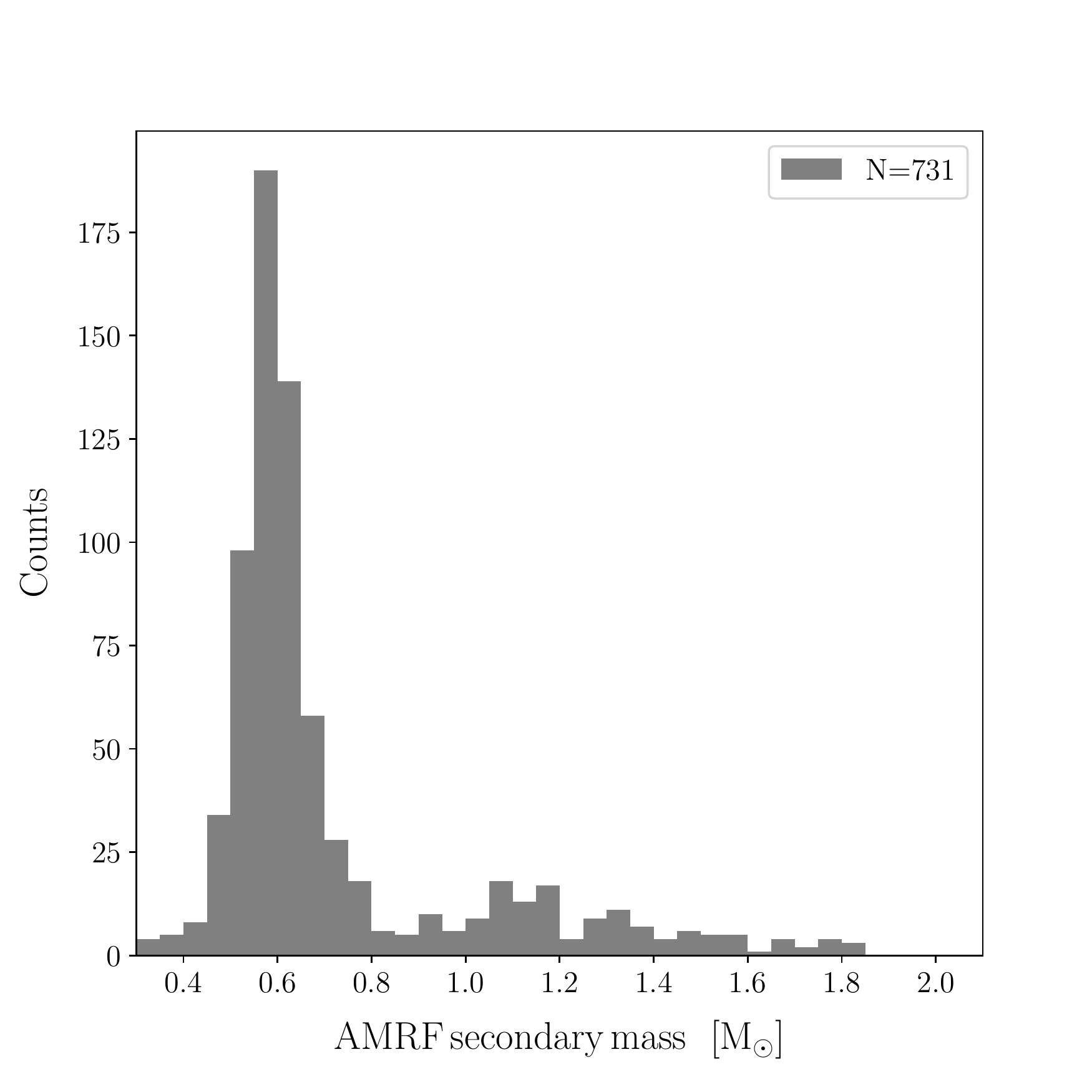}
\caption{ A histogram of the companion masses for compact object candidates, up to a mass of $2.1$\Msun.}
  \label{fig: AMRF mass dist}
\end{center}\end{figure}

\begin{figure}[htb]\begin{center}
    \includegraphics[width=\columnwidth]{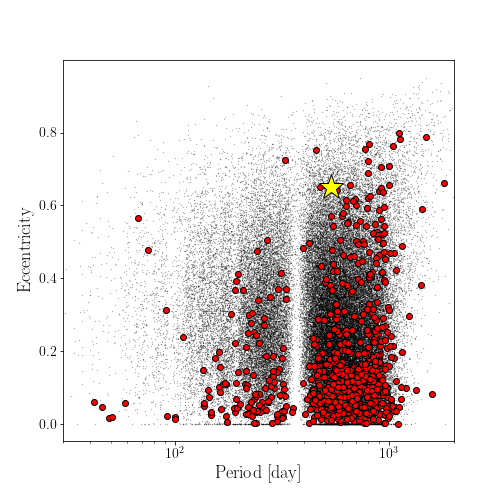}
\caption{ \eP\ diagram. Black points - \fieldName{Orbital} and \fieldName{AstroSpectro}, for reference. Red points are AMRF class-III. Yellow star is the \sourceId{5136025521527939072}.}
  \label{fig:AMRF_Pe_diagram}
\end{center}\end{figure}

The histogram of secondary masses, \figref{fig: AMRF mass dist} 
has been limited to 2\Msun. There are however \afterReferee{4} sources with a
larger mass (identified as \fieldName{AMRFClassIII} in the table),
but we have indications that they may be artefacts.

%
\subsection{A closer look at the astrometric binaries with white-dwarf companions}
%


White dwarfs are often present in binary systems. The co-eval context with their companion star makes such systems important benchmarks for understanding stellar evolution. WDs in wide systems can be detected relatively easily through searches for common proper motion objects  \citep[e.g.\ ][]{elbadryrix2018}.
However, in closer systems the WD can be very difficult to detect due to the overwhelming brightness of the companion, particularly for early spectral types. In that case, the WD may be hidden until revealed by astrometric motion, spectroscopic variability or a photometric excess. A classic example is the discovery of Sirius B \citep{bessel1844}, which is both an astrometric and spectroscopic binary \citep[e.g.][]{barstow2005, bond2017}. 
Sirius represents a class of binary systems, the Sirius-type binaries, consisting of a primary star of spectral type earlier than late-K and a WD. Examples of such systems have often been identified by flux excesses at wavelengths shorter than those spanned by the primary, in the UV, EUV or X-ray \citep[see e.g.][]{barstow1994}. A fraction have subsequently been resolved by space-based observations \citep[e.g.][]{barstow2001}.

The number of Sirius-type binaries known lies in the 10s of objects, with selection effects playing a strong role in identification of an individual system. For example, unless resolved, the flux of any cool white dwarf will always be buried in the light of the primary companion. Whether or not a system can be resolved depends on the separation and distance of the components. Binary systems comprising 2 WDs (double degenerates) or a WD with an M dwarf companion are easier to find as they will sit closer to the main WD cooling tracks in the H-R diagram. 
Even so, the nature of any double degenerates might not be apparent if the stars have similar or featureless spectra. The SPY survey sought to identify double degenerates from radial velocity variations finding 39 double degenerate systems and 46 WDs with cool companions from a sample of 643 stars \citep{napiwotzki2020}. 

The \gaia stellar catalogue is an enormous resource that will potentially increase the number of known binary systems with WD components by at least an order of magnitude. In \gdrtwo and \gdrearlythree the number of known WDs expanded by such a factor \citep{gf2019, gf2021}. While many WD+M systems can by identified by their location in the HRD, double degenerates will typically overlap strongly with the isolated WD cooling tracks and Sirius-like systems largely overlap with the main sequence.
The release of eclipse, astrometry and spectroscopy data related to the identification of NSS in \gdrthree is an important step forward.
In principle, these new resources can be used to search for the presence of WDs in a variety of binary systems.

%
\subsubsection{Selection of candidate binary systems with WD components}
%

The wavelength coverage of the \gaia Radial Velocity Spectrometer is not optimised for the study of WDs, which typically have broad absorption lines in their photospheres, when any are present.
Therefore, we would expect only the primary stars in Sirius-like binaries will be detected as radial velocity variables. Hence, they will most likely be found in the sample of \fieldName{SB1} systems.
If the secondary is a WD, it will not contribute significantly to the brightness of the system in the \gaia bands. However, if hot enough, there may be a measurable excess at shorter wavelengths. In addition, the secondary mass will lie between $\approx 0.4-1.4\Msun$, where the Chandrasekhar limit defines the upper bound.

Binaries with WD components might also be detected astrometrically or as eclipsing binaries. 
Therefore, we have assembled an initial list of potential candidate binary systems with WD components from the \NSS , selected for all these possibilities, using the following \fieldName{nss_solution_type} keywords: \fieldName{AstroSpectroSB1}, \fieldName{SB1C}, \fieldName{SB1}, \fieldName{Orbital}, \fieldName{OrbitalAlternative}, \fieldName{Eclipsing*}, \fieldName{OrbitalTargetedSearch}. Effectively, this is all \NSSa\ systems except for SB2 binaries.
We did not apply any quality selection criteria except to reject any objects with \gmag greater than 20.
This selection yielded a total of 355\,524 \NSS for further analysis.
The AMRF sample, based on \fieldName{AstroSpectroSB1} and \fieldName{Orbital} solutions alone, is a subset of these candidates for which we have good estimates of the component masses.

\begin{figure}[htb]
\centering
    \includegraphics[width=\columnwidth]{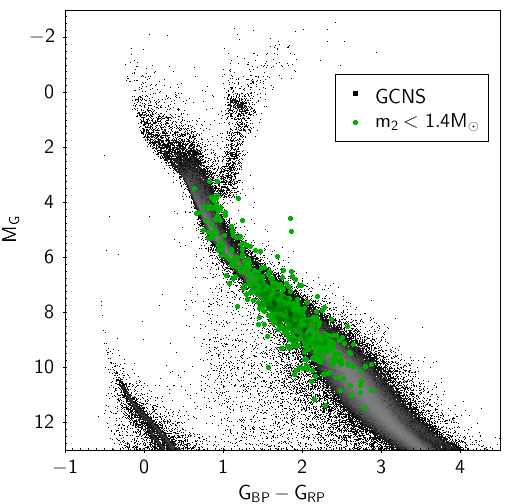}
\caption{\label{fig:binHRD}H-R diagram of the candidate \NSS with WD components with a secondary mass solution below the Chandrasekhar limit ($<1.4\Msun$, green data points). These are compared with the locus of stars in the GCNS (grey data points).}
\end{figure}

In the search for WD companions, we use the AMRF data to restrict the mass range of compact objects to lie below the 1.4\Msun Chandrasekhar limit. Applying this criterion yields 676 objects which are shown in green in \figref{fig:binHRD} and compared to the locus of the GCNS (grey data points). 
The secondary mass alone cannot be used to definitively determine that these are WD components, as F, G, K and the earliest M main sequence stars also occupy this mass range. However, several factors indicate that these objects are highly likely to be Sirius-type binaries. 
Any unresolved main sequence binaries should appear to be over luminous. Evolution on the main sequence and the range of possible luminosity combinations may prevent a clear separation of binaries and isolated stars, but unresolved main sequence binaries should appear as SB2 systems in the \NSSa\ \gaia\  catalogue.
Finally, as noted above, the mass distribution for the secondaries in these unresolved astrometric binaries has a strong peak at 0.55-0.6\Msun, corresponding to the known peak of the mass distribution of field WDs (\figref{fig: AMRF mass dist}). This is in contrast to the smooth \beforeReferee{IMF}\afterReferee{mass distribution} expected for main sequence stars.

\begin{figure}[htb]\begin{center}
    \includegraphics[width=\columnwidth]{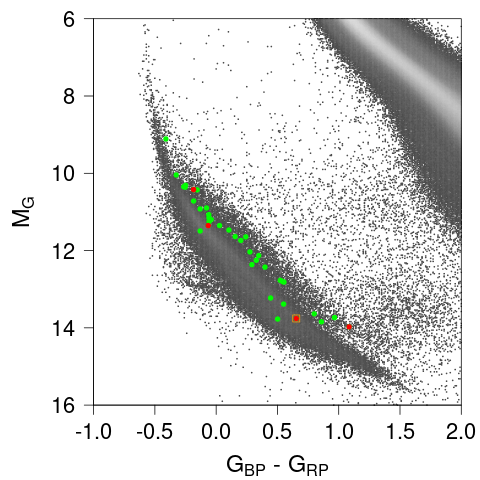}
\caption{WD with orbital solutions (green points, corrected from extinction) overplotted
on the \gdrthree low extinction HRD. Most of the points lie above the hydrogren sequence. The red dots and WD 0141-675 (orange square) are discussed \secref{ssec:exoplanets_wd}.}
  \label{fig:double_degenerates}
\end{center}\end{figure}

While the vast majority of selected objects lie on the main sequence, \afterReferee{as shown in \figref{fig:binHRD}, } there are 39 objects that match WD colours and magnitudes, indicating a WD primary, shown in  \figref{fig:double_degenerates}. The majority of these objects lie above the main concentration of the isolated white dwarfs. Therefore, these systems are highly likely to be double degenerates, where the brightness of two unresolved combines to yield an apparent excess in luminosity. 

\begin{figure}[htb]
\centering
{\includegraphics[width=\columnwidth]{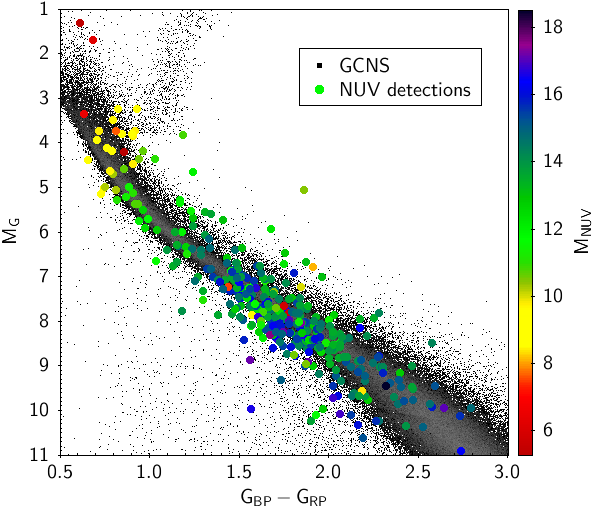}}
\caption{\label{fig:NUVHRD}\gmag vs \gbp\,--\,\grp\ H-R diagram showing the cross-match between the GALEX GR6+7 AIS catalogue with the GCNS (small symbols) and the candidate list of binaries with WD components with computed secondary star masses below the Chandrasekhar limit ($\rm {<1.4\Msun}$). The symbols are colour coded with the absolute NUV magnitude as indicated by the side-bar.}
\end{figure}

\begin{figure}[htb]
\centering
{\includegraphics[width=\columnwidth]{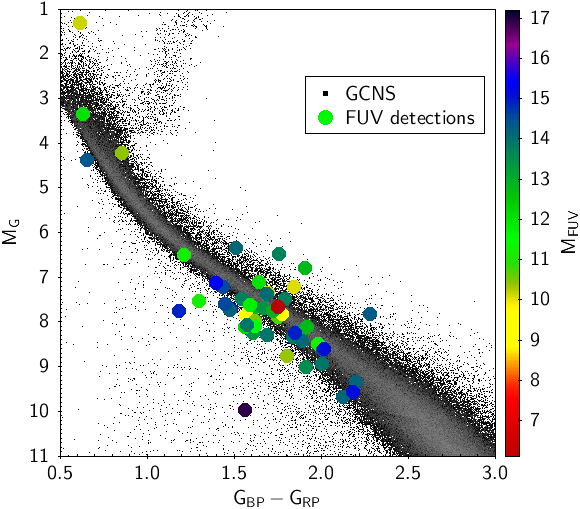}}
\caption{\label{fig:FUVHRD}\gmag \ vs  \gbp\,--\,\grp\ H-R diagram showing the cross-match between the GALEX GR6+7 AIS catalogue with the GCNS (small symbols) and the candidate list of binaries with WD components with computed secondary star masses below the Chandrasekhar limit ($\rm {<1.4\Msun}$). The symbols are colour coded with the absolute FUV magnitude as indicated by the side-bar.}
\end{figure}

As discussed above, when we know the secondary mass and the binary is not revealed as an SB2 system, we can be very confident that the secondary is a WD and, therefore, the binary a Sirius-type system. However, it is likely that there are many more Sirius-type systems in the non-\fieldName{SB2} sample.
An unresolved binary system can be revealed by a flux excess in a waveband where the contribution from the primary star is expected to be weak. A number of Sirius-type binary discoveries have been made by detecting the WD in the EUV or UV wavebands. The GALEX mission surveyed most of the sky in two broad FUV and NUV bands. Cross-correlating the GALEX data base with the \gdrthree non-\fieldName{SB2} binaries will potentially reveal the Sirius-type systems with a hot WD component.
This is illustrated by applying this to the astrometric binaries in the sample. \figref{fig:NUVHRD} and \figref{fig:FUVHRD} show the \gmag \ vs \gbp\,--\,\grp\ H-R diagram for the cross-match of the GALEX GR6+7 AIS catalogue with this sample for NUV and FUV bands respectively. The small grey data points are the GCNS stars in each figure while the coloured symbols are the NUV (\figref{fig:NUVHRD}) and FUV detections (\figref{fig:FUVHRD}), colour-coded by the absolute NUV or FUV magnitude, as indicated in the side bar.
\afterReferee{These magnitude ranges can be compared to the typical values for white dwarfs in the optical colour/absolute magnitude diagram, as indicated in \figref{fig:double_degenerates}.}
The absolute NUV magnitude correlates well with the \gbp\,--\,\grp\ colour, an indicator of the temperature of the primary star. However, a few systems show a strong NUV excess for the systems with cooler main sequence primaries, indicated by red to yellow colours compared to green to blue in that region of the diagram.
Hence, the integrated NUV flux is not generally a good indicator of the presence of the white dwarf.
In contrast, there is little, if any, correlation between the absolute FUV magnitude and \gbp\,--\,\grp\ colour, indicating that the FUV is a better discriminator than the NUV, the measured magnitudes potentially providing an estimate of the WD temperature. However, this can only be applied to the relatively few stars that are sufficiently hot to have a measurable FUV flux. 
We note that the FUV magnitude is a not a completely unique indicator of the presence of a WD, as coronally active main sequence stars can also generate an enhancement in the total FUV flux through the strong \ion{C}{iv} (154.8nm \& 155.0nm) and \ion{He}{ii} (164nm) emission lines.

\begin{figure}[htb]
\centering
{\includegraphics[width=\columnwidth]{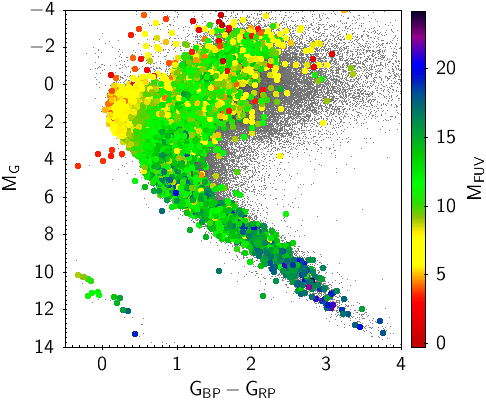}}
\caption{\label{fig:FUVFULLHRD}\gmag \ vs \gbp\,--\,\grp\ H-R diagram showing the cross-match between the GALEX GR6+7 AIS catalogue with the full list of non-\fieldName{SB2} binaries selected here (large symbols). The symbols are colour coded with the absolute FUV magnitude as indicated by the side-bar. The small grey data points are the full list of non-\fieldName{SB2} binaries.}
\end{figure}

We also cross-matched the full \gdrthree non-\fieldName{SB2} binaries with the GALEX GR6+7 AIS catalogue.
The results are shown in \figref{fig:FUVFULLHRD}, with 29\,000 stars of the list of 355\,000 objects having a GALEX FUV counterpart.
The FUV detections in the WD region of the H-R diagram provide an indication of the range of FUV absolute magnitudes, between $\approx 8 - 20$, that correspond to WDs. Many of the FUV counterparts in \figref{fig:FUVFULLHRD} have similar magnitudes, but without further information it is not possible to categorically identify these as WDs and rule out alternative explanations for the FUV flux.

A more detailed modelling of the predicted primary and WD fluxes across the expected range of temperatures should be able to refine the discriminatory power of this approach to selecting Sirius-type systems, but that is beyond the scope of this \gdrthree companion paper.

%
%




%
\subsection{A binary with a dormant neutron star with an 
\texorpdfstring{\fieldName{AstroSpectroSB1}}{AstroSpectroSB1} orbit?}\label{neutron}
%

Only a few tens of dynamically confirmed Galactic stellar black holes (BH) and neutron stars (NS) are known to reside in binary systems. They are discovered either by  
their X-ray emission, fueled by mass transfer from their non-compact stellar companions
\citep[e.g.,][]{fabian89, remillard06, orosz07, ziolkowski14}, or, in the case of active pulsars, by their radio pulsed emission.

Obviously, most BH in binaries were not detected yet,
because their optical counterparts are well within their Roche lobes, so mass is not transferred and X-rays are not generated, making these systems dormant
\citep[see discussion on the frequency of such systems and the prospect of their detection with \gaia astrometry by][]{breivik17, mashian17, shao19, wiktorowicz19}.
Similar arguments apply to dormant NS --- the pulsation phase lasts  for 10--100 million years only, when the pulsar is young \citep{kaspi06, bransgrove18} --- and to binaries with white-dwarf (WD) companions (see discussion above).
Such dormant binaries can be found by the orbital motion of the primary, detected either by astrometry or RVs. 
In all three classes of dormant companions, the challenge is to identify the binaries, estimate the mass of the unseen companion, and rule out a faint MS companion.

One of the systems identified by the triage algorithm as having a compact companion, \object{\sourceId{5136025521527939072}}, is in fact an \fieldName{AstroSpectroSB1} binary, with a primary mass of $1.2 \Msun$ and secondary mass of $1.5 \Msun$, consistent with a binary having a dormant neutron star with a period of $536$ days. Its location is marked in the pertinent figures of the triage analysis above.

\begin{figure}
\centering
{\includegraphics[width=0.9\columnwidth]{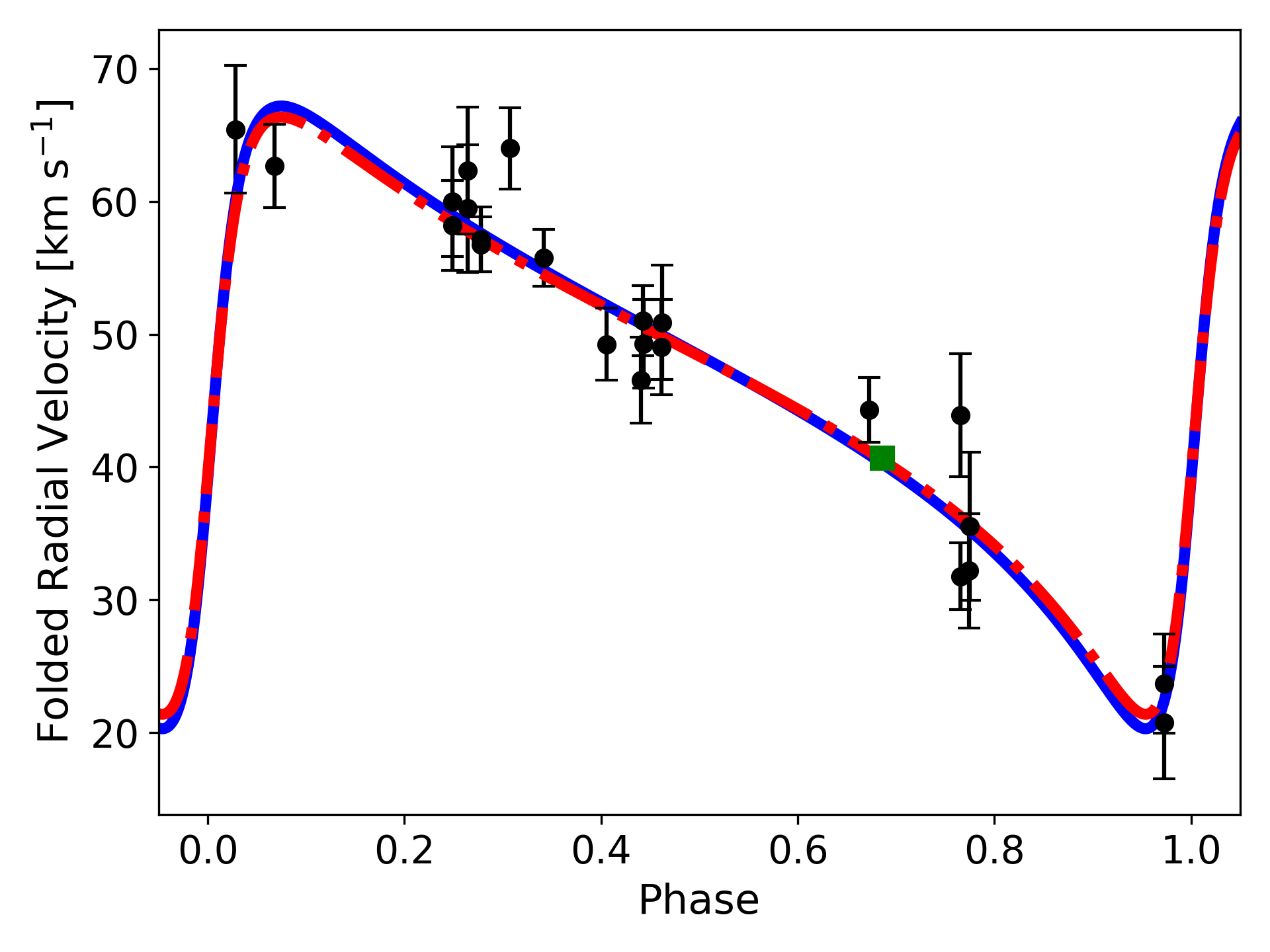}}
{\includegraphics[width=0.9\columnwidth]{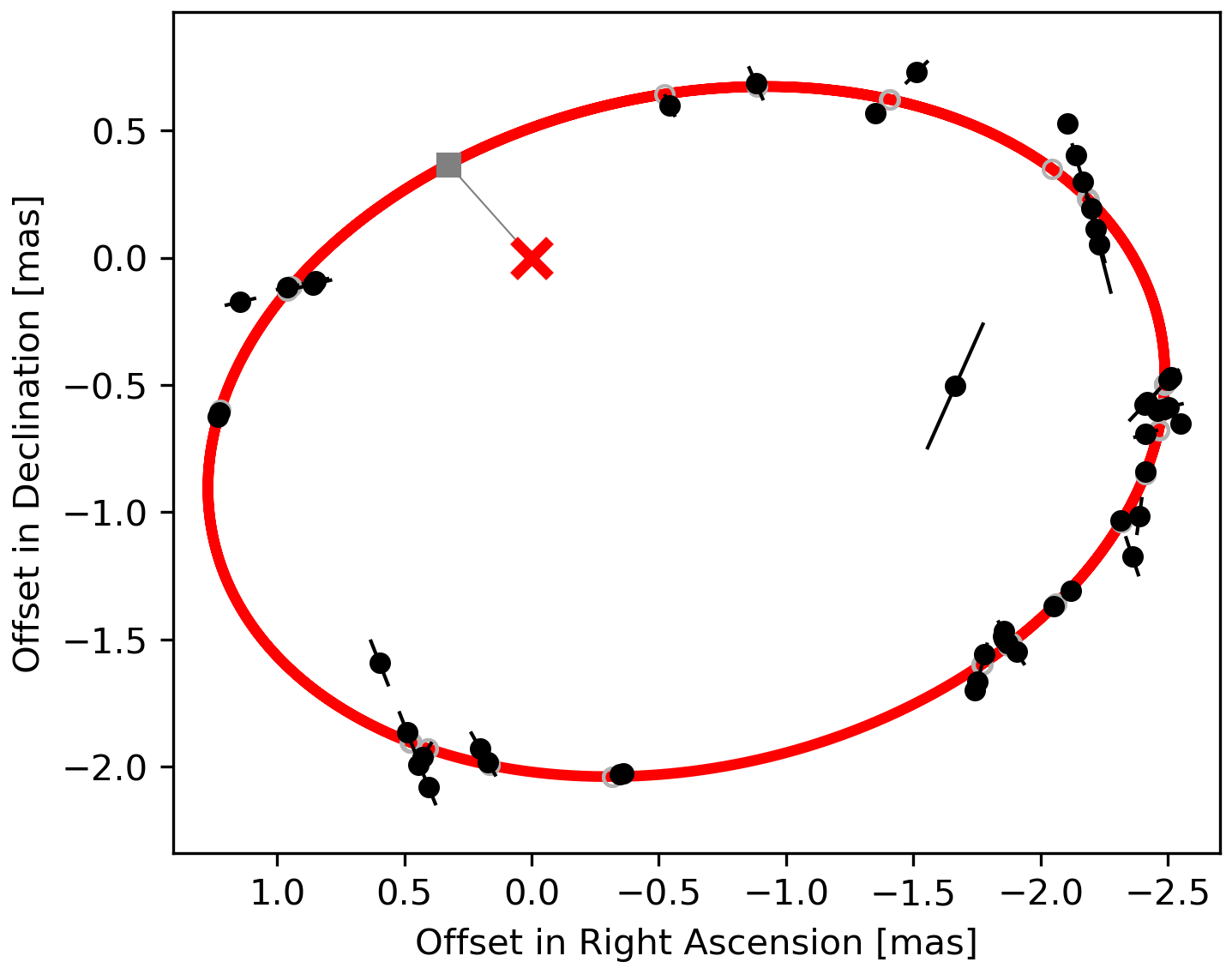}}
\caption{\label{fig:ns_cand_example}{\it Top panel}: Phase-folded radial velocity data of \sourceId{5136025521527939072}, together with the orbits using separately the astrometric (red dot-dashed line) and 
spectroscopic (blue solid line) orbital elements; the OHP/Sophie external measurements (green squared point) was not part of the fit. 
{\it Bottom panel}: \afterReferee{Along-scan residuals of the} mean epoch astrometric measurements (black symbols) \afterReferee{relative to the model positions (grey circles)} and \afterReferee{the} astrometric orbit (red solid line) of the same source. The red cross marks the focus of the orbit and the grey square \afterReferee{is} the periastron location. }
\end{figure}

The astrometric orbit and the phase-folded radial velocity of this source are shown in Figure \ref{fig:ns_cand_example}\footnote{The astrometric orbit figure was obtained on the basis of the \texttt{pystrometry} package \citep[\url{https://github.com/Johannes-Sahlmann/pystrometry},][]{johannes_sahlmann_2019_3515526}. Other examples and a description are given in \citet{DR3-DPACP-176}.}.
In order to validate the orbit, we have required observing time on the {\sc Sophie} spectrograph, 
mounted at the 1.93-meter telescope of the Observatoire de Haute-Provence (France). 
The {\sc Sophie} pipeline \citep{2009A&A...505..853B} together with a G2 mask
was used and we obtained a RV$=40.603\pm0.023$\kms with a FWHM = 9.274\kms on BJD 2459541.389492.
The consistency of this RV obtained about 4.5 years after the end of the Gaia data segment used for \gdrthree confirms the quality of the predicted orbit.

%
\subsection{Compact objects in \texorpdfstring{\fieldName{SB1}}{SB1} solutions}\label{compactSB1}
%

While the search for compact object companions above had been done using the astrometric orbits, we complete the search using the \fieldName{SB1} solution. 

\begin{figure}
\centering
{\includegraphics[width=\columnwidth]{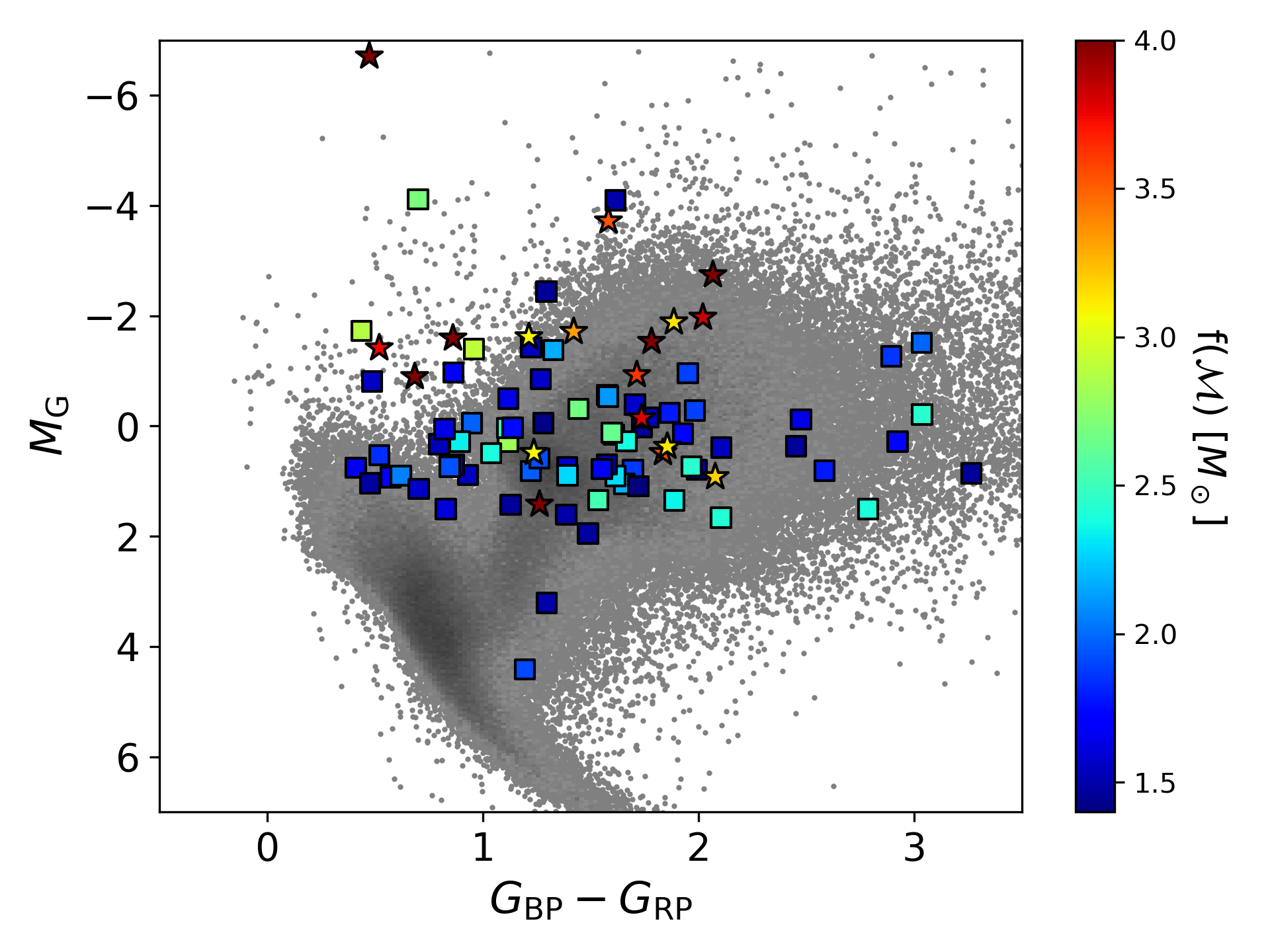}}
\caption{\label{fig:hr_sb1_fm} HRD of \fieldName{SB1} solutions with $f(\Mass) > 1.4\Msun$ and \fieldName{significance} $> 20$. Squared symbols are for sources with $1.4\Msun < f(\Mass) \leq 3\Msun$, star symbols for sources with $f(\Mass) > 3\Msun$. The background grey scale shows the
density distribution of all \fieldName{SB1} solutions.}
\end{figure}

The search of SB1 sources with large mass functions have been proposed since several decades as a way to identify candidates having a black hole or neutron star companion \citep{1969ApJ...156.1013T,1966SvA....10..251G}.
Among the SB1 sources 94 have \fieldName{significance} larger than 20 and $f(\Mass) > 1.4\Msun$, and 20 among them $f(\Mass) > 3\Msun$. The \fieldName{SB1} solution of \object{\sourceId{2006840790676091776}} shall be dismissed due to contamination by a nearby bright source. The inspection of RVS spectra of \object{\sourceId{5259215388421037696}} shows that
the source is probably an SB2 and the radial velocities, computed with an incorrect template, are most probably not correct, so the \fieldName{SB1}
solution of this star shall be discharged. 
\object{\sourceId{878555832642451968}} has a RV measurement in the LAMOST \citep{LamostDR1} survey (--100.86 \kms at BJD 2458068.5) which does not agree with the \fieldName{SB1} solution. 
A reanalysis of the RVS epoch RV shows that an alternative solution with a period of 8.438 days is also possible and in better agreement with LAMOST data. 
So the \fieldName{SB1} solution in the \gaia archive should be considered as dubious. 
In \figref{fig:hr_sb1_fm} we show the position of the sources with $f(\Mass) > 1.4\Msun$ in the color-magnitude diagram. The relevant data of the sources with $f(\Mass) > 3\Msun$ are reported in Table \ref{tab:sb1_giant_large_fm}.

\begin{table*}[htb]
    \caption{\fieldName{SB1} solutions with $f(\Mass) > 3\Msun$. The radii $R$ are from GSP-Phot or \fieldName{FLAME} when available. }
    \label{tab:sb1_giant_large_fm}
    \centering
\begin{tabular}{lccccccccc}
\hline
\hline
Gaia DR3  &  $f(\Mass)$ & Period &  $a_1\sin{i}$ & $R$ & $M_G$ & $G_{BP}-G_{RP}$ \\
          &  ($\Msun$) & (days) &  ($\Rsun$) & ($\Rsun$) &  &  \\
\hline
\object{4661290764764683776}\tablefootmark{a} & 13.67  & 204.930$\pm$0.862 & 349.49 & 60.52 & -6.707 & 0.474 &  \\
\object{5863544023161862144}\tablefootmark{b} &  7.80  &  10.605$\pm$0.001 &  40.25 &         & -1.531 & 1.783 & \\
\object{442992311418593664}\tablefootmark{c}  &  4.77  & 216.531$\pm$1.463 & 255.33 & 20.16 & -1.592 & 0.862 & \\
\object{206292746724589824}\tablefootmark{d}  &  4.53  & 347.002$\pm$0.364 & 343.50 & 42.25 & -2.741 & 2.067 \\
\object{5857059996952633984}\tablefootmark{e} &  4.25  & 155.085$\pm$0.228 & 196.63 &  7.42\tablefootmark{h} & -0.893 & 0.684 \\
\object{2174777963318889344} &  3.83  &  82.723$\pm$0.158 & 124.88 & 34.82 & -1.970 & 2.021 \\
\object{2031113506311851904}\tablefootmark{f} &  3.78  &  35.908$\pm$0.030 &  71.29 &          & -0.145 & 1.736 \\
\object{1996704839648530816} &  3.75  &   7.54574$\pm$0.00077 & 25.13 & 7.29 & -1.414 & 0.520 \\
\object{251157906379754496}  &  3.62  & 296.676$\pm$2.176 &  287.10 &    & -0.928 & 1.714 \\
\object{3331748140308820352} &  3.56  & 225.268$\pm$1.189 & 237.61 &     &  0.485 & 1.836 \\
\object{1828150428697001472} &  3.54  & 333.688$\pm$0.205 & 308.23 &           & -3.716 & 1.583 \\
\object{3112097229257687680}\tablefootmark{g} &  3.33  & 260.978$\pm$0.608 & 256.40 & 23.98\tablefootmark{h} & -1.711 & 1.422 \\
\object{5963629779180627968} &  3.19  &  11.369$\pm$0.002 &  31.30 & 11.86\tablefootmark{h} &  0.922 & 2.078 \\
\object{512307478642441984}\tablefootmark{f}  &  3.14  & 145.578$\pm$0.074 & 170.44 &         & -1.885 & 1.886 \\
\object{527155253604491392}  &  3.13  & 149.155$\pm$0.318 & 173.06 & 12.15\tablefootmark{h} &  0.371 & 1.856 \\
\object{5869320651099982464}\tablefootmark{f} &  3.09  &  63.924$\pm$0.011 &  97.95 &     & -1.615 & 1.214 \\
\object{2929565719083290240} &  3.09  &  32.473$\pm$0.014 &  62.31 & 7.84\tablefootmark{h}  &  0.484 & 1.237 \\
\hline
\end{tabular}
\tablefoot{ 
\tablefoottext{a}{\object{LHA 120-S 80}: B8Ie \citep{1978A&AS...31..243R} star in the LMC with emission lines. A distance of 49.59 kpc from \citet{2019Natur.567..200P} was used to compute absolute magnitude $M_G$.}
\tablefoottext{b}{\object{V878 Cen}: eclipsing binary star classified as Hot semi-detached system \citep{2013AN....334..860A} with the same period, confirmed by \gaia photometry.}
\tablefoottext{c}{\object{HIP 15429}: B5Ib star \citep{2012A&A...538A..76N}; belongs to PMa sample of \citet{2022A&A...657A...7K}.}
\tablefoottext{d}{\object{EM* GGA 311}: Emission-line Star \citep{1956BOTT....2n..19G}.}
\tablefoottext{e}{Classified as Algol type eclipsing binary from \gaia photometry with period of 1.17143 days.}
\tablefoottext{f}{\gaia light curve show ellipsoidal behavior.}
\tablefoottext{g}{\object{EM* RJHA 92}: Emission-line Star \citep{1989AJ.....98.1354R}.}
\tablefoottext{h}{Radius from \fieldName{FLAME}.}
}
\end{table*}

From \figref{fig:hr_sb1_fm}, we can see that, with few exceptions, all the selected sources are not main sequence stars.
It is thus challenging to determine the real nature of these systems because, if the primary is a giant star, it can easily outshine a companion on the main sequence. In some cases the secondary, even being on the main sequence, can be more massive than the primary, via mass transfer as in Algol-type systems \citep[see][for recent examples of black hole candidates dismissed as main sequence companions of stripped stars]{2022arXiv220306348E}. We recall that such systems are much more common objects than dormant black holes. 
Another possible explanation for such large mass function values is that the unseen companion is itself a binary composed by two main sequence stars. 
The understanding of the nature of the other systems would need a deeper analysis using external data and follow-up
observations, but such an analysis is beyond the scope of this paper.
We can, however, comment on some of these objects.
For \object{\sourceId{5863544023161862144}}, the presence of eclipses allows to classify it as an Algol system and to exclude the compact object companion hypothesis. The \gaia photometry of the source \object{\sourceId{5857059996952633984}} shows a light curve with eclipses of Algol type, however with a period of 1.17143 days, typical of Algol systems, in contrast with the period from the \fieldName{SB1} solution. By analysing the epoch radial velocities and the epoch photometry, we can exclude an aliasing, in both the \fieldName{SB1} solution and the photometry. A possible solution is that this source is a triple system where the RV data are from the outer component, and the eclipses involve the two inner components.

A particularly interesting source is \object{\sourceId{442992311418593664}} (\object{HIP 15429}), a B5Ib star \citep{2012A&A...538A..76N} with $f(\Mass)=4.77$~\Msun. In order to validate the orbit, this source was observed with the HERMES spectrograph mounted on the 1.2-meter Mercator telescope\footnote{\url{http://www.mercator.iac.es/}} \citep{Raskin2011} at the Roque de los Muchachos Observatory on La Palma Island. 
\afterReferee{A RV of $-47.2\pm3.5$~\kms\  at BJD 2459650.358 was obtained, compatible with the Gaia \fieldName{SB1} solution.}
\beforeReferee{A RV of --72.7 \kms\ at BJD 2459650.358 was obtained, in agreement with the \fieldName{SB1} solution.} The HERMES spectrum also confirms the spectral classification of \citet{2012A&A...538A..76N}.  Using 
the 3D extinction map of \cite{Lallement2019}, we estimate a dereddened absolute magnitude and color, $M_{\rm G,0}=-3.168$ and $G_{\rm BP,0}-G_{\rm RP,0}=0.073$, respectively. Comparing these values with the PARSEC evolutionary tracks \citep{Bressan12} for solar metallicity, we can interpret this source as being a $4.9\pm 0.3\Msun$ star, which just left the main sequence. The resulting minimum mass for the companion would be $10.4 \Msun$. 
Under the hypothesis that this is a \beforeReferee{ternary}\afterReferee{triple} system, the secondary would  then be a binary where at least one component  should have a mass equal to or larger than $5.2 \Msun$. But if all the stars in the system are coeval, this most massive star in the inner binary should be evolved too, and thus brighter than the primary, refuting the triple-star hypothesis.
Refuting the hypothesis that this is an Algol system is more difficult because even a $12 \Msun$ companion on the main sequence would be fainter than the B5Ib primary. 
However, we note that the absorption lines in the HERMES spectrum do not show any clear double profile despite the fact that the spectrum was obtained at the phase with maximum radial-velocity difference between the two components. Moreover there are only few known Algol systems with such a long period, \afterReferee{$\approx$ 216d,} and consisting of early-type stars ($\mu$ Sgr being such an example). Another possibility is that this source may have a dormant black hole companion. More observations and modeling will be needed to decide between the different hypotheses.
We finally note that the astrometric \fieldName{ruwe} parameter of this star is 2.10, which suggests that in the next release it will be possible to obtain an astrometric orbit. 

\begin{figure}
\centering
{\includegraphics[width=\columnwidth]{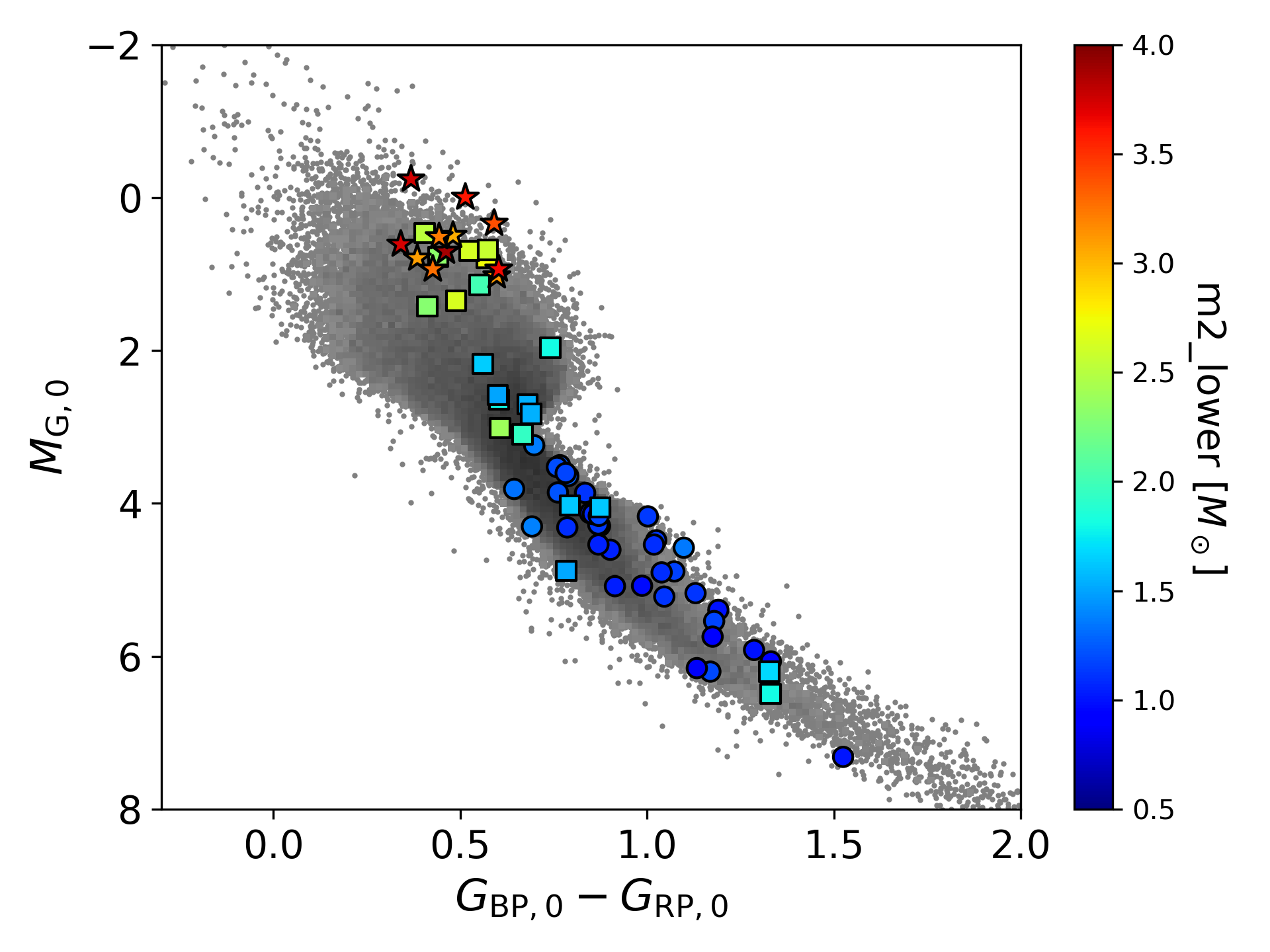}}
\caption{\label{fig:hr_m2_sb1} Dereddened HRD of compact object companion candidates from \fieldName{SB1} solutions on the main sequence. Circle symbols are for sources with \fieldName{m2\_lower}~$\leq 1.4\Msun$, square symbols are for sources with $1.4\Msun <$ \fieldName{m2\_lower} $\leq 3\Msun$, stars symbols for sources with \fieldName{m2\_lower} $ > 3\Msun$. The background grey scale shows the density distribution of all \fieldName{SB1} solutions classified as belonging to the main sequence by IsocLum.}
\end{figure}

We now move our attention to SB1 sources which can be classified as belonging to the main sequence, for which an estimate of the mass of the primary is provided by IsocLum. 
Restricting the search for compact object companions to main sequence primaries allows to reduce the number of false detections where the secondary is a normal main sequence star outshined by an evolved primary.
We then selected from the table presented in  \secref{Sect:masscat} the SB1 sources for which \fieldName{m2_lower} $>$ \fieldName{m1_upper} and \fieldName{significance} $>$ 20. 
This results in 68 sources; among them 15 have \fieldName{m2_lower} $> 3$\Msun and 25 $1.4\Msun <$ \fieldName{m2_lower} $\leq 3\Msun$.
Their position in the magnitude-colour diagram is shown in \figref{fig:hr_m2_sb1}.

It should be noted that spectroscopic radial velocity of stars cooler than 3875 K were not processed by the NSS pipeline. 
This introduces a selection on the mass of the primary, so that almost no SB1 is present in the catalogue with a primary mass below 0.6$\Msun$.
So, it is not possible with this selection to obtain candidates with companions belonging to the main population of white dwarfs.

Particularly interesting are sources with \fieldName{m2_lower} $> 3\Msun$, which could have a dormant black hole companion. 
We checked each of the candidates and we found that, among sources with \fieldName{m2_lower} $> 3\Msun$ the sources \object{\sourceId{548272473920331136}} and \object{\sourceId{6000420920026118656}} are known eclipsing binaries \citep{2008OEJV...83....1O,2013AN....334..860A}, classified as Algol type, while \object{\sourceId{1850548988047789696}} is an Algol type eclipsing binary detected by \gaia. 


We then checked the \gaia and TESS \citep{ricker14} photometry for the other 11 sources, and all show a modulation of the flux, in phase with the radial velocity, similar to the one expected from an ellipsoidal star\footnote{\object{\sourceId{2219809419798508544}} and \object{\sourceId{4514813786980451840}} are also classified as eclipsing, but the light curve is clearly ellipsoidal. See \citet{DR3-DPACP-170} for more details about this type of misclassification.}, confirming the binary nature. This ellipsoidal behaviour, however, suggests that these stars are rather evolved, but the small amplitude of the modulation (a few percents) tells us that the primaries do not fill their Roche lobe.
Using the relation between the mass ratio and the effective radius of the Roche lobe provided by \citet{1983ApJ...268..368E}, we get that the mass required for the primary to fill the Roche lobe, given the radii estimated with the IsocLum code, would be always below 0.1$\Msun$, which is too low for an evolved star and for the observed effective temperature. A more detailed modelling of the light curve of these objects is out of the scope of this article.

\afterReferee{The already known 17 X-ray binaries with BH companions \citep{corral16} have periods of ~0.3 - 5 days. We do expect the similar {\it dormant} binaries to have longer periods, and therefore the range we find here is consistent with the expected periods.} Our sources shall nevertheless be considered as candidates only, because they can have other explanations than having compact companions.
With the information that we have, we cannot rule out the possibility that these sources are Algol-type systems, however the bluest sources should be considered as more promising candidates of dormant black holes \afterReferee{because they are nearer the main sequence than redder sources, and therefore cannot easily outshine a MS companion}. Finally, we cannot exclude, as commented before, that the unseen companion is the inner binary composed by two MS stars. The periods of these 11 candidates range from 8.2 to 23.5 days, not short enough to exclude the \beforeReferee{ternary}\afterReferee{triple} system hypothesis.

\begin{figure}
\centering
{\includegraphics[width=\columnwidth]{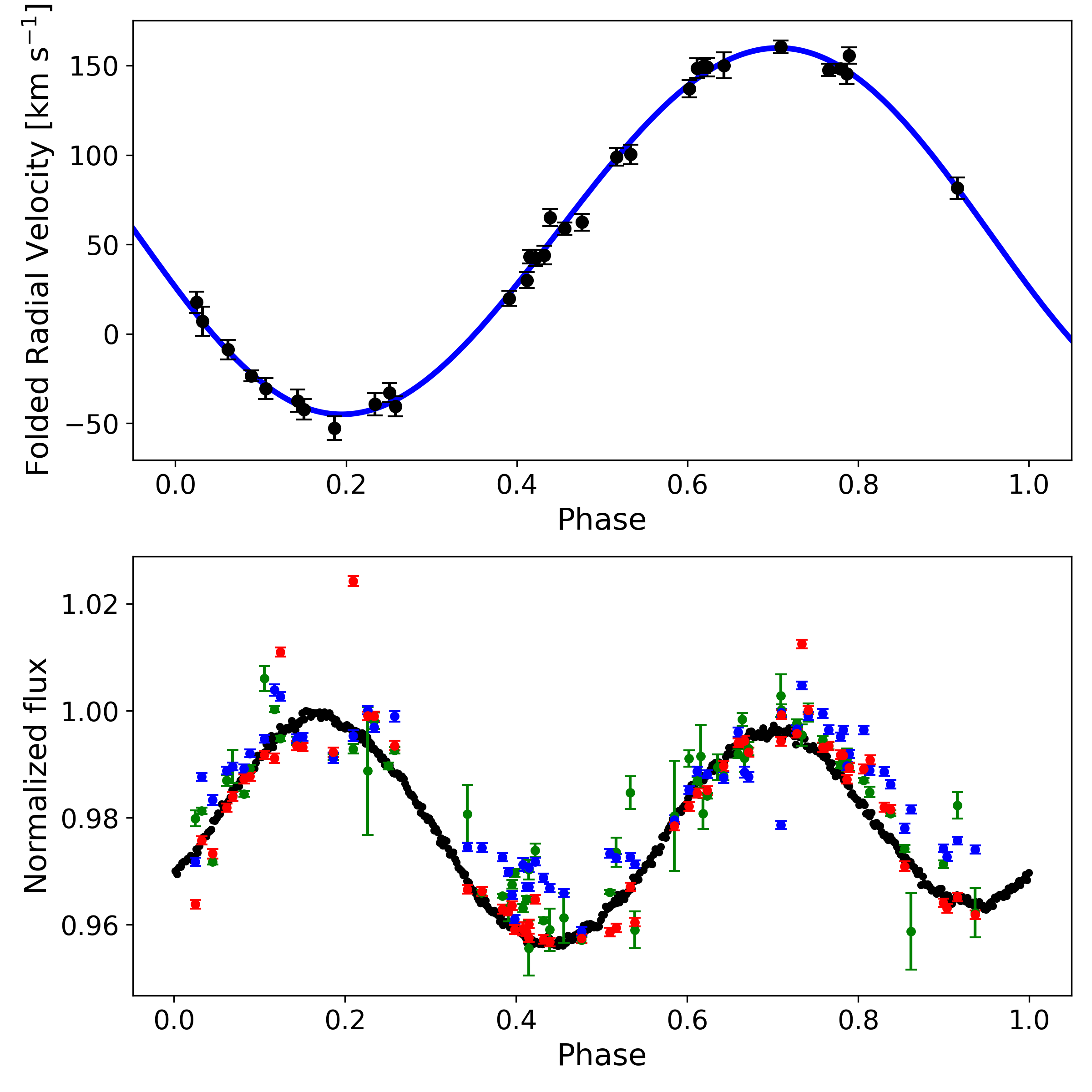}}
\caption{\label{fig:bh_cand_sb1_example}{\it Top panel}: Phase-folded radial velocity of \object{\sourceId{2966694650501747328}} \afterReferee{($P=10.398$d)}. {\it Bottom panel}: phase-folded TESS (black circles) and \gaia normalised flux (green, blue and red circles: flux in $G$, \gbp\ and \grp\ bands respectively).}
\end{figure}

The data of these 11 candidates are reported in Table \ref{tab:sb1_bh_cand}.
Figure \ref{fig:bh_cand_sb1_example} shows the phase-folded radial velocity of the source \object{\sourceId{2966694650501747328}}, together with its phase-folded TESS and \gaia normalised flux.

%

\begin{table*}[htb]
    \caption{Source candidates with compact object companions with \fieldName{m2_lower} $> 3\Msun$ and \fieldName{m2_lower} $>$ \fieldName{m1_upper} from \fieldName{SB1} solutions with \fieldName{significance}$>$20. The radius $R$ and dereddened $G_{BP,0}-G_{RP,0}$ colour were obtained with IsocLum.}
    \label{tab:sb1_bh_cand}
    \centering
\begin{tabular}{lccccccccc}
\hline
\hline
\gaia DR3  &  Period & $f(\Mass)$ & $\Mass_1$ & \fieldName{m2_lower} & $R$ & $a_1\sin{i}$ & $G_{BP,0}-G_{RP,0}$ \\
          &   (days) & ($\Msun$) & ($\Msun$) & ($\Msun$) & (\Rsun) &  (\Rsun) &\\
\hline
\object{2219809419798508544} & 10.8653$\pm$0.0025 & 1.287 & 1.949 & 3.129 & $3.95^{+0.23}_{-0.11}$ & 22.44 & 0.598 \\
\object{4514813786980451840} & 22.0204$\pm$0.0124 & 1.559 & 2.477 & 3.746 & $4.86^{+0.38}_{-0.36}$ & 38.30 & 0.368 \\
\object{5694373091078326784} & 12.8848$\pm$0.0040 & 1.660 & 2.161 & 3.708 & $3.82^{+0.19}_{-0.16}$ & 27.37 & 0.341 \\
\object{2966694650501747328} & 10.3980$\pm$0.0011 & 1.158 & 2.082 & 3.096 & $3.64^{+0.18}_{-0.05}$ & 21.04 & 0.385 \\
\object{948585824160038912}  &  8.2019$\pm$0.0015 & 1.472 & 2.016 & 3.283 & $3.67^{+0.09}_{-0.21}$ & 19.45 & 0.427 \\
\object{2197954362764248192} & 17.5097$\pm$0.0070 & 1.470 & 2.178 & 3.359 & $5.71^{+0.49}_{-1.37}$ & 32.24 & 0.591 \\
\object{2933630927108779776} & 14.7175$\pm$0.0007 & 1.225 & 2.178 & 3.221 & $4.39^{+0.19}_{-0.16}$ & 27.02 & 0.443 \\
\object{448452383082046208}  & 23.4939$\pm$0.0131 & 1.474 & 2.318 & 3.584 & $4.61^{+0.36}_{-0.43}$ & 39.25 & 0.514 \\
\object{5243109471519822720} & 14.9137$\pm$0.0005 & 1.565 & 2.014 & 3.649 & $4.23^{+0.09}_{-0.08}$ & 29.58 & 0.603 \\
\object{5536105058044762240} & 12.1766$\pm$0.0039 & 1.180 & 2.178 & 3.000 & $4.50^{+0.15}_{-0.17}$ & 23.52 & 0.481 \\
\object{6734611563148165632} & 14.3444$\pm$0.0016 & 1.695 & 2.103 & 3.883 & $4.22^{+0.12}_{-0.13}$ & 29.60 & 0.462 \\
\hline
\end{tabular}
\end{table*}

%


%

%
\section{Substellar companions}\label{sec:exoplanets} 
%
The two well-known categories of substellar companions, planets and brown dwarfs, have been for a few decades now the objective of long-term ground-based Doppler search programs in the Solar neighbourhood (e.g., \citealt{Cumming2008,Howard2010,Mayor2011,Bonfils2013,Butler2017,Rosenthal2021,Pinamonti2022}). The \gdrthree astrometric performance levels reach the sensitivity to detect substellar companions around a statistically significant number of stars, enabling first-time measurements of their three-dimensional orbital architectures and true masses.

\begin{figure}[htb]\begin{center}
\includegraphics[width=\columnwidth]{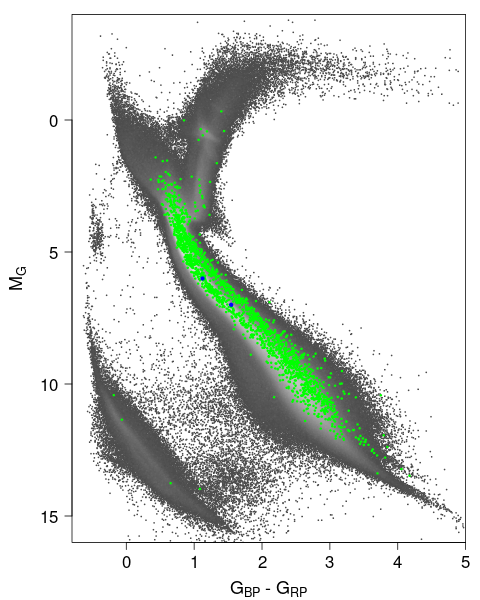}
\caption{H-R diagram of sources with low astrometric mass functions ($<0.001$\afterReferee{\Msun}; green dots);
the grey background is the DR3 low extinction HRD. A very large fraction
are not low-mass companions, rather binaries with a mass ratio similar 
to their flux ratio. The two blue sources are 
\object{HD\,12800} and \object{HD\,3221}, described in \secref{candidate_EP} while
the four WDs are discussed in  \secref{ssec:exoplanets_wd}.}\label{fig:low_astro_mass_ratio}
\end{center}\end{figure}

%
%
\subsection{Astrometry: substellar companions}\label{sec:f_m_select_substellar} 
%

A naive search for substellar companions detected by \gaia astrometry might simply select solutions with low values of the astrometric mass function, say $f(\Mass)<0.001$\afterReferee{\Msun}. However, inspection of \figref{fig:low_astro_mass_ratio} shows that a sizeable fraction of companions of sources with low $f(\Mass)$ do not have small secondary mass with a negligible flux ratio but rather have a flux ratio close to the mass ratio, leaving them to clearly stand above the main sequence.

Keeping this in mind, we browsed the catalogue of masses presented in \secref{Sect:masscat} to investigate the regime of astrometrically detected companions with lower mass bound $\Mass_c$ (assuming they contribute no flux) in the substellar regime, operationally defined as having $20\leq \Mass_c\leq 80$\Mjup and $\Mass_c\leq 20$\Mjup for brown dwarfs (BDs) and exoplanets (EPs), respectively. For a subset of sources with orbital solutions in the \gdrthree archive but without a companion mass estimate in our catalogue of masses, the information was derived based on primary mass estimates from the Starhorse catalogue \citep{Anders2022}.  

A total of 1843 BDs and 72 EPs were identified in the catalogue of companion masses. This includes 20 sources with \fieldName{AstroSpectroSB1} solution type that have upper bounds to the companion mass $<80$\Mjup, i.e. for which the assumption of negligible flux ratio is confirmed.
A small subset of 10 BDs were already known, identified in ground-based RV surveys for planets \cite{MaGe2014,Wilson2016,Kiefer2019,Dalal2021}. We report in Table \ref{tab:known_EP_BD_comp} the basic comparison between period and eccentricity from \gaia and the literature, minimum and true mass estimates. A total of 9 known EPs were also validated against literature sources, and we report the same information in Table \ref{tab:known_EP_BD_comp}. As an illustrative example, the astrometric orbital solution for \object{HD 81040 b} is shown in \url{https://www.cosmos.esa.int/web/gaia/iow_20220131}. Additional plots of \gdrthree orbits of substellar companions can also be found in \cite{DR3-DPACP-176}.



\def\Spm{\pm}
\begin{sidewaystable*}[!htbp]
 \caption{Known substellar companions with a confirmed mass in the planetary and brown dwarf regime, respectively. 
 \label{tab:known_EP_BD_comp}}
\centering
  \normalsize
  \begin{tabular}{llccccccccc}
  \hline
  \hline
\gaia DR3  &    Name & $\Mass_c\sin i$ & $\Mass_c$ & $i$ & $ P_\mathrm{lit}$ & $P_\mathrm{Gaia}$ & $e_\mathrm{lit}$ & $e_\mathrm{Gaia}$ & $a_0$ &Refs. \\
 & & $(\Mjup)$ & $(\Mjup)$ & (deg) &  (days) & (days) & & & (mas) & \\
  \hline  \noalign{\smallskip}
6421118739093252224 & \object{HD 175167 b}  & $7.8\Spm3.5$       & $9.5\pm0.9$  & $28\Spm19$  & $1290\Spm22$ & $898\Spm198$ & $0.54\Spm0.09$ & $0.19\Spm0.12$ & $0.22\Spm0.02$ &1 \\
4062446910648807168 & \object{HD 164604 b}  & $2.7\Spm1.3$       & $14.3\Spm5.5$ & $29\Spm19$  & $606\Spm9$  & $615\Spm12$ & $0.24\Spm0.14$ & $0.61\Spm0.34$ & $0.56\Spm0.22$ &1,2  \\
1594127865540229888 & \object{HD 132406 b}  & $5.6$        & $6.7\Spm2.1$  & $122\Spm14$ & $974\Spm39$ & $893\Spm251$ & $0.34\Spm0.09$ & $0.31\Spm0.29$ & $0.16\Spm0.04$ &3 \\
4745373133284418816 & \object{HR 810 b}     & $2.26\Spm0.18$  & $6.2\Spm0.5$  & $87\Spm6$   & $312\Spm5$ & $332\Spm6$ & $0.15\Spm0.05$ & $0.04\Spm0.2$ & $0.30\Spm0.02$  &4,5 \\
2367734656180397952 & \object{BD -17 0063 b} & $5.1\Spm0.12$   & $4.3\Spm0.5$  & $80\Spm6$   & $656\Spm0.6$ & $649\Spm36$ & $0.54\Spm0.005$ & $0.28\Spm0.22$ & $0.22\Spm0.02$ &6 \\
5855730584310531200 & \object{HD 111232 b}  & $6.8$        & $8.3\Spm0.6$  & $97\Spm4$   & $1143\Spm14$ & $882\Spm34$ & $0.20\Spm0.01$ & $0.50\Spm0.10$ & $0.51\Spm0.03$ &7 \\
637329067477530368 & \object{HD 81040 b}   & $6.8\Spm0.7$     & $7.9\Spm0.9$  & $108\Spm6$  & $1002\Spm7$ & $851\Spm113$ & $0.53\Spm0.04$ & $0.37\Spm0.15$ & $0.39\Spm0.03$ &8 \\
4976894960284258048 & \object{HD 142 b}     & $1.3\Spm0.2$       & $7.1\Spm1.0$  & $59\Spm7$   & $350\Spm4$ & $319\Spm7$ & $0.26\Spm0.18$ & $0.26\Spm0.23$ & $0.21\Spm0.03$ &5,9,10 \\
2603090003484152064 & \object{GJ 876 b}     & $2.1\Spm0.2$ & $3.6\Spm 0.4$  & $101\Spm8$  & $61.08\Spm0.01$ & $61.4\Spm0.2$ & $0.027\Spm0.002$ & $0.16\Spm0.15$ & $0.43\Spm0.05$  &11-18\\

  \hline  \noalign{\smallskip}
  \hline \noalign{\smallskip}
 
2651390587219807744 & \object{BD -00 4475 b} & $25\Spm2$   & $48.4\Spm7.6$   & $129\Spm7$ & $723.2\Spm0.7$      & $780\Spm84$     & $0.39\Spm0.01$ & $0.48\Spm0.11$ & $1.91\Spm0.28$ &19 \\
2778298280881817984 & \object{HD 5433 b}    & $49\Spm3$   & $53.8\Spm1.7$ & $12\Spm39$  & $576.6\Spm1.6$      & $576.7\Spm10.6$ & $0.81\Spm0.02$ & $0.46\Spm0.12$ & $1.04\Spm0.03$ &19 \\
3309006602007842048& \object{HD 30246 b}   & $55_{-8}^{+20}$ & $40.6\Spm8.3$ & $78\Spm2$  & $990\Spm6$          & $814\Spm141$    & $0.84\Spm0.08$ & $0.59\Spm0.10$ & $1.34\Spm0.24$ &19 \\
3750881083756656128 & \object{HD 91669 b}   & $30.6\Spm2.1$    & $43.2\Spm2.2$   & $58\Spm3$  & $497.5\Spm0.6$      & $500.4\Spm6.9$   & $0.448\Spm0.002$ & $0.32\Spm0.06$ & $0.73\Spm0.04$ &19 \\
3751763647996317056 & \object{HD 89707 b}   & $54_{-7}^{+8}$  & $82.5\Spm12.7$   & $54\Spm10$  & $298.5\Spm0.1$      & $297\Spm2$      & $0.90\Spm0.04$ & $0.68\Spm0.20$ & $1.82\Spm0.30$  &19\\
685029558383335168 & \object{HD 77065 b}   & $41\Spm2$        & $64.2\Spm5.1$   & $42\Spm3$  & $119.113\Spm0.003$  & $119.1\Spm0.2$  & $0.694\Spm0.0004$ & $0.70\Spm0.04$ & $1.04\Spm0.07$ &20\\
855523714036230016 & \object{HD 92320 b}   & $59.4\Spm4.0$    & $70\Spm3.1$   & $111\Spm2$ & $145.4\Spm0.01$     & $145.1\Spm0.3$  & $0.323\Spm0.001$ & $0.26\Spm0.05$ & $0.82\Spm0.01$  &20\\
824461960796102528 & \object{HD 82460 b}   & $73.2\Spm3.0$    & $62.5\Spm6.4$ & $66\Spm1$  & $590.9\Spm0.2$      & $579\Spm6$      & $0.84\Spm0.01$ & $0.73\Spm0.04$ & $1.63\Spm0.09$ &21\\
873616860770228352 & \object{BD +29 1539 b} & $59.7\Spm2.0$    & $60.7\Spm23.5$   & $120\Spm9$ & $175.87\Spm0.01$    & $173\Spm3$      & $0.275\Spm0.001$ & $0.43\Spm0.12$ & $0.61\Spm0.14$ &21 \\
5563001178343925376 & \object{HD 52756 b}   & $59.3\Spm2.0$    & $61.2\Spm8.6$   & $73\Spm4$  & $52.8657\Spm0.0001$ & $52.9\Spm0.1$   & $0.678\Spm0.0003$ & $0.54\Spm0.16$ & $0.54\Spm0.08$  &22\\
\hline
\end{tabular}
\tablebib{
(1)~\cite{Arriagada2010}; (2) \cite{Feng2019}; (3) \cite{daSilva2007}; (4) \cite{Kurster2000}; (5) \cite{Butler2006}; (6) \cite{Moutou2009}; (7) \cite{Mayor2004}; (8) \cite{Sozzetti2006};
(9) \cite{Tinney2002}; (10) \cite{Wittenmyer2012}; (11) \cite{2018A&A...609A.117T}; (12) \cite{2005ApJ...634..625R}; (13) \cite{2010ApJ...719..890R}; (14) \cite{2002ApJ...581L.115B}; (15) \cite{1998ApJ...505L.147M}; (16) \cite{Marcy2001}; (17) \cite{2010A&A...511A..21C}; (18) \cite{2016MNRAS.455.2484N}; (19) \cite{Dalal2021}; (20) \cite{Wilson2016}; (21) \cite{Kiefer2019}; (22) \citet{Sahlmann:2011fk}. 
}
\end{sidewaystable*}

%
\subsection{Astrometric masses: transition regimes}
%
The results of long-term Doppler surveys have allowed studying in some detail the shape of the mass distribution of relatively close-in ($a\lesssim 5$ au or so) companions to solar-type (F-G-K-type) stars, particularly in the two transition regimes between EPs and BDs and between BDs and stars. The most notable feature is the so-called \quoting{brown dwarf desert}: the (minimum) mass distribution has a clear decline moving from the planetary-mass to the BD-mass regime, reaches an apparent plateau with a minimum at $\sim40-50$\Mjup $(0.04-0.05 \Msun)$ and then rises again reaching the stellar-mass regime (e.g., \citealt{Grether2006,Sahlmann:2011fk,MaGe2014,Grieves2017}) 

\begin{figure}[!htb]
\begin{center}
\includegraphics[width=\columnwidth]{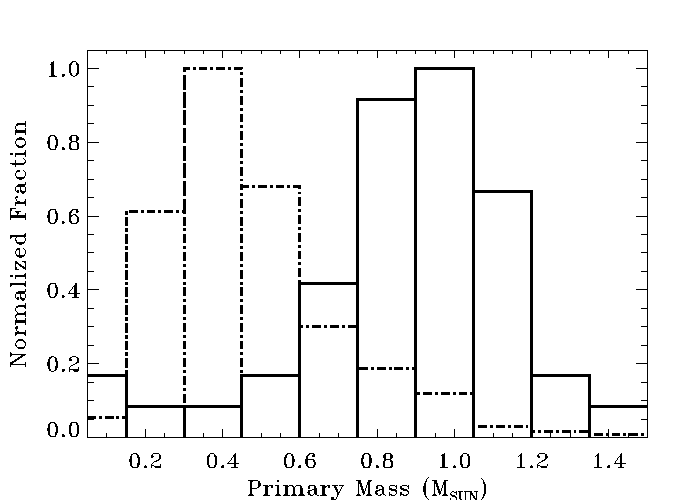}
\caption{Primary mass distributions for sources with astrometrically detected substellar companions with \fieldName{Orbital} (long-dashed histogram) and \fieldName{OrbitalTargetedSearch*} (solid-line histogram) solution types.}\label{fig:substellar_primaries}
\end{center}\end{figure}

Figure~\ref{fig:substellar_primaries} shows the distribution of primary masses for sources with astrometrically detected substellar-mass companions for the cases of NSS solution types (\fieldName{Orbital} and \fieldName{OrbitalTargetedSearch*}). The medians of the two distributions are 0.42\Msun and 0.91\Msun, respectively. The striking difference stems from the different ways the input lists for the two analysis channels were constructed (see \secref{sssec:cu4nss_intro_inputfiltering_astro} and \secref{sec:alternativeOrb}, and references therein). In particular, the bulk of sources with known solutions input to the alternative orbit determination algorithms is constituted by solar-type stars, and this is reflected in \figref{fig:substellar_primaries}. \beforeReferee{Given the still sub-optimal calibration levels in the bright-star regime for \gaia DR3, it is then}\afterReferee{The calibration levels in the bright-star regime is still sub-optimal for \gaia DR3; consequently it is} expected that nearby, relatively faint low-mass stars might be the sample of primaries around which the chances of detecting substellar companions are maximized, and this is also reflected in \figref{fig:substellar_primaries}. 

\begin{figure}[!htb]
\begin{center}
\includegraphics[width=0.95\columnwidth]{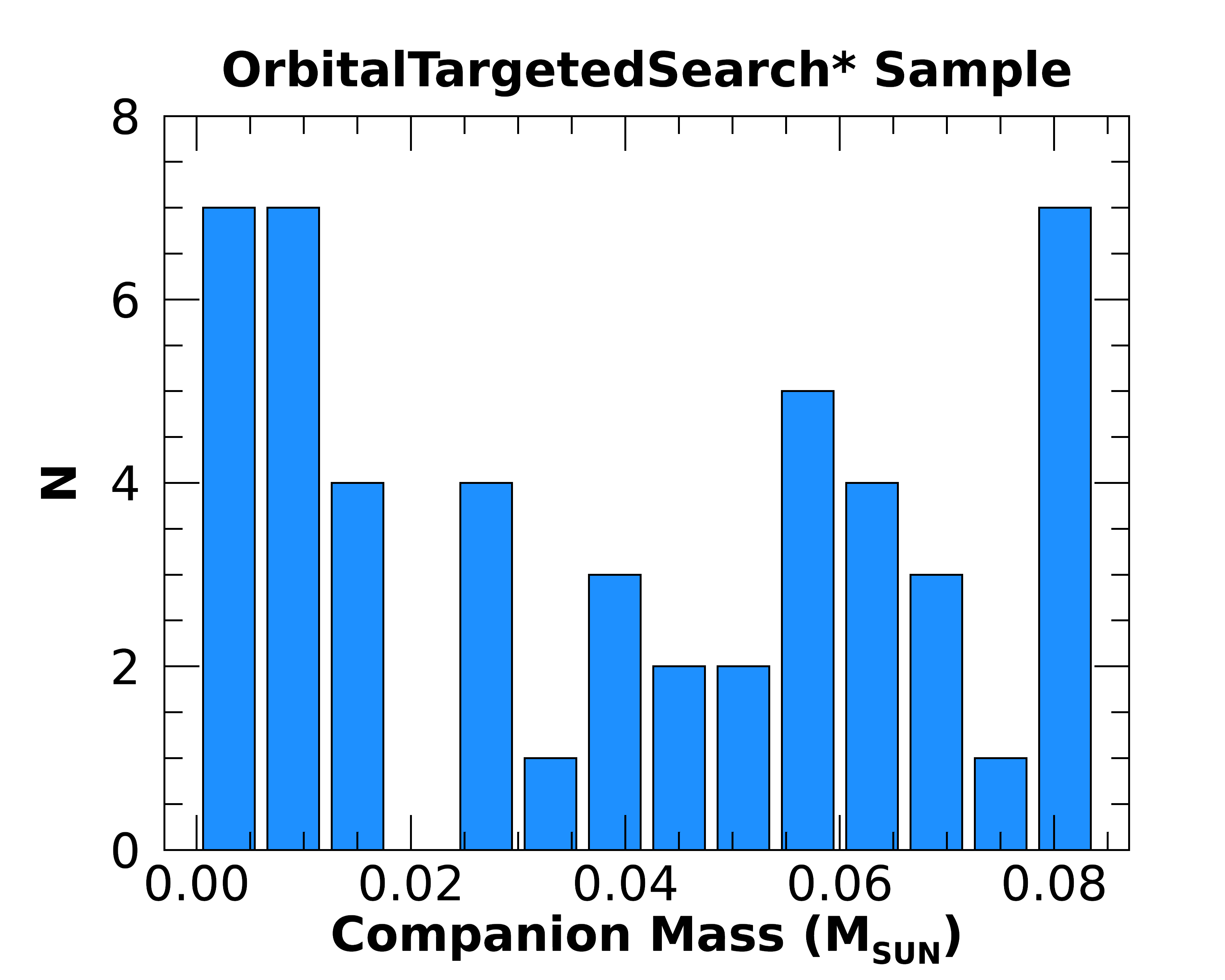}
\includegraphics[width=0.95\columnwidth]{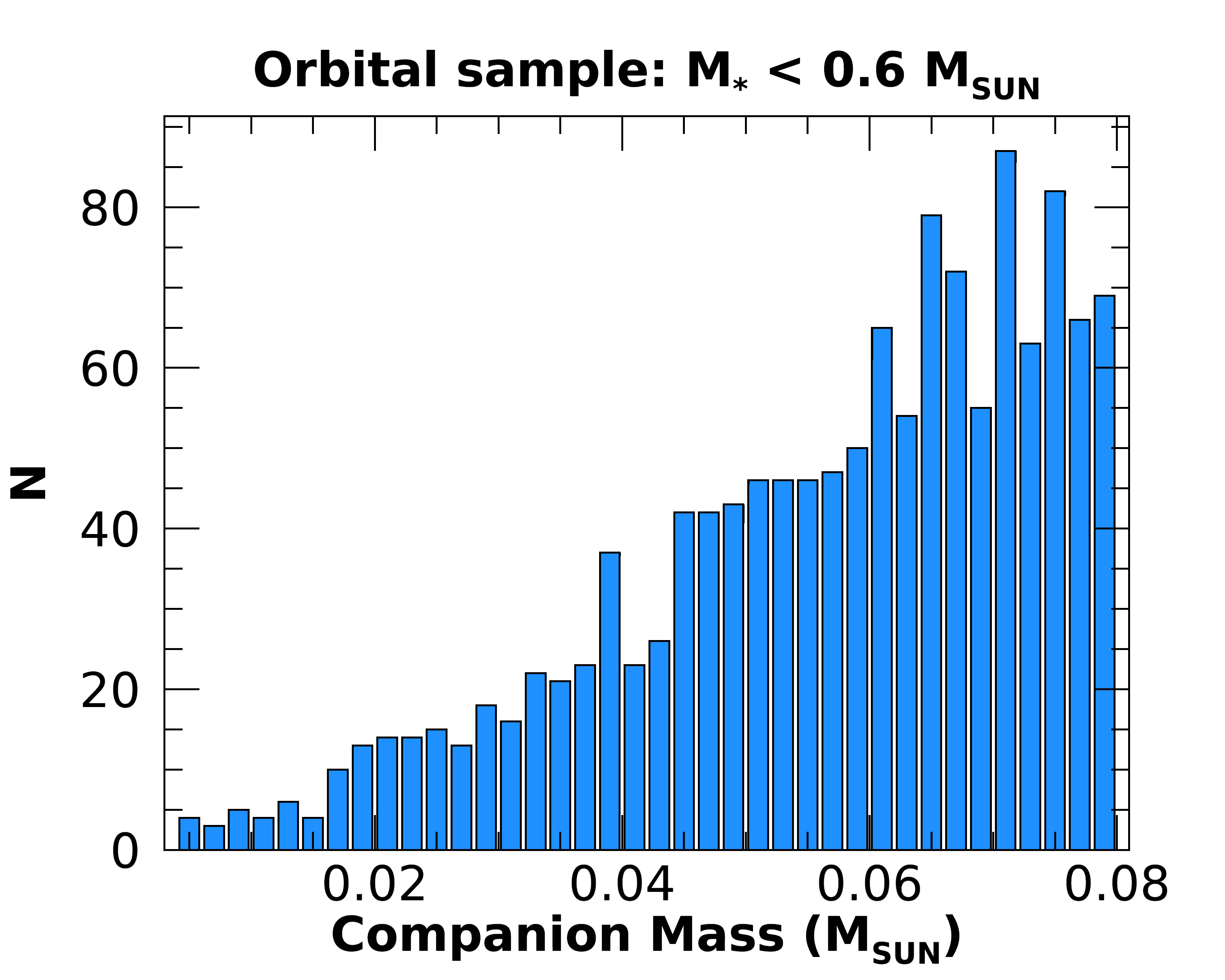}
\includegraphics[width=0.95\columnwidth]{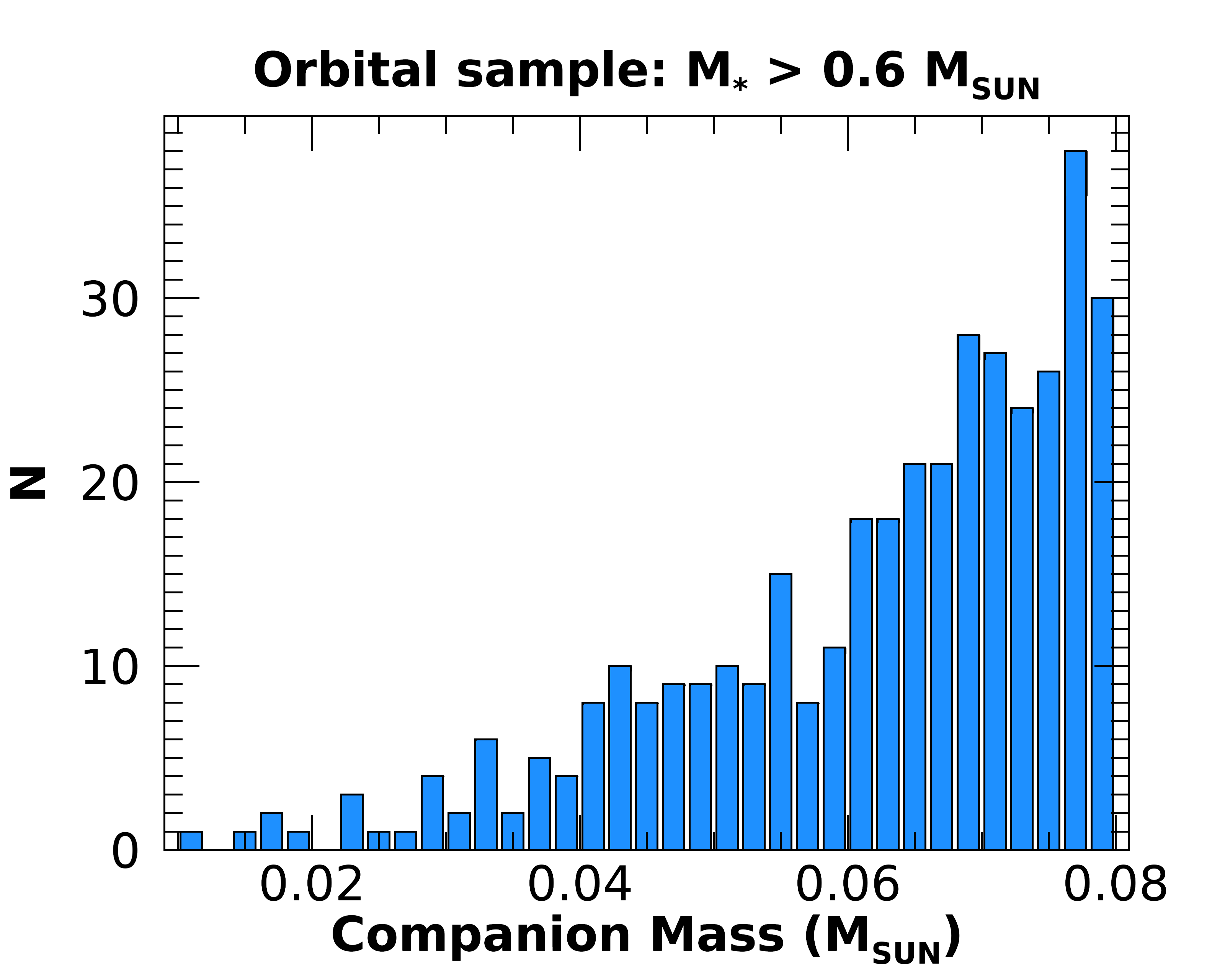}
\caption{{\it Top}: substellar companion mass distribution for the \fieldName{OrbitalTargetedSearch*} solution type. {\it Center}: The same, but for the \fieldName{Orbital} solution type, with a cut-off in the primary mass \Mass$_\star < 0.6$\Msun. {\it Bottom}: the same, but for the \fieldName{Orbital} solution type, with a cut-off in the primary mass \Mass$_\star > 0.6$\Msun.}\label{fig:substellar_secondaries}
\end{center}\end{figure}

The three panels of \figref{fig:substellar_secondaries} show the distribution of substellar companion masses for three samples. The distribution for the \fieldName{OrbitalTargetedSearch*} sample (top panel) corroborates the notion, already provided by Doppler surveys, of a minimum in occurrence at $\sim40$\Mjup close-in BDs around solar-type stars. For the first time, \gdrthree offers the opportunity to see the feature in the distribution based on true companion mass estimates, without the ambiguities inherent to studies of the population of substellar companions based on minimum mass values and/or simulation-driven upper mass limits (e.g., \citealt{Kiefer2021}, and references therein). 

The centre and bottom panels of \figref{fig:substellar_secondaries} show the equivalent distribution for the main NSS sample, split into two regimes of companions with substellar masses around M dwarfs (M$_\star < 0.6$\Msun) and higher-mass primaries, respectively. The first notable feature is the difference in slopes between the two samples rising toward the highest masses: a simple power-law fit returns $N\propto \Mass_c^{1.56}$ and $N\propto \Mass_c^{3.03}$ for companions around M dwarfs and higher-mass stars, respectively. A Kolmogorov-Smirnov (K-S) test indicates that the two distributions of substellar companion masses around M dwarfs and higher-mass stars in the center and bottom panels of \figref{fig:substellar_secondaries} have a $p$-value of $\sim1\times10^{-6}$, allowing to reject the hypothesis that they are drawn from the same distributions. Secondly, the occurrence of detected companions around higher-mass dwarfs appears approximately flat in the approximate range $0.04\lesssim \Mass_c\lesssim0.06$\Msun. For M dwarf primaries, the distribution continues to show a declining trend towards the lowest-mass end (corresponding to super-Jupiter-mass objects with $\Mass_c\lesssim 0.01$\Msun).

The above differences could in principle be mostly due to the not yet well-characterized sensitivity to substellar companions in different mass and orbital separation regimes of \gdrthree astrometry. 
However, a few considerations help reinforcing the idea that we are seeing, at least in part, intrinsic features in the distributions rather than effects due to the selection function. Most notably, 1) the flat shape of the distribution in the bottom panel of \figref{fig:substellar_secondaries} is consistent with that of the substellar companions distribution derived from RV surveys (e.g., \citealt{Kiefer2019}), and 2) the declining trend in the distribution towards the lowest-mass end in the central panel of \figref{fig:substellar_secondaries} is consistent with the well-established notion of a much lower frequency of giant planets around M dwarfs with respect to solar-type primaries (e.g., \citealt{Endl2006,Cumming2008,Bonfils2013,Pinamonti2022}). With \gdrthree we thus achieve the first-ever characterization of a conspicuous population of substellar companions with true mass estimates within typically 1-2 au of nearby M dwarfs.

%
\subsection{Substellar companion frequency in the 100 pc sample}
%
As low-mass stars provide the primary sample around which substellar companions have been detected with \gdrthree astrometry, we can attempt to derive a first-order estimate of their occurrence rate. Clearly, a detailed assessment of the \gaia sensitivity in terms of completeness (estimation of the number of missed companions) and reliability (estimate of the number of false detections) is warranted, but goes beyond the scope of this work, and will be presented elsewhere (Giacobbe et al. in prep.). 

\citet{EDR3-DPACP-81} have shown that \gdrthree is complete down to the M7 spectral sub-type within 100 pc from the Sun, with an M dwarf sample amounting to 218\,366 sources. The NSS sample encompasses 790 astrometrically detected companions with likely substellar masses around M dwarfs within 100 pc. Of these, the vast majority ($\sim94\%$) are BDs. Under the optimistic assumptions that 1) \gaia has homogeneous sensitivity to and is 100\% complete for BD companions across the M dwarf 100-pc sample, 2) none of the orbital solutions corresponding to BD companions around this sample are spurious, and 3) the companion does not contribute light, we can then make a first statement on the frequency of BDs around M dwarfs with $P\lesssim 1\,000$ d, which turns out to be $\sim0.3\%$. 

\citet{Dieterich2012} report a BD companion frequency around M dwarfs of $2.3^{+5.0}_{-0.7}\%$ for separation in the range 10-70 au. \citet{Bowler2015} find that such companions in the 10-100 au orbital radius range have an occurrence rate of $2.8^{+2.4}_{-1.5}\%$ around M dwarfs in young moving groups. \citet{Susemiehl2022} find a similar frequency ($2.7^{+1.0}_{-0.7}\%$) for field M dwarfs in the same separation interval. \citet{Winters2019} report a formally lower, but still compatible within the uncertainties, $1.3\pm 0.3\%$ frequency for BDs around M dwarfs within 25 pc in the separation range out to $\sim 300$\arcsec. 

Our occurrence rate estimate is likely underestimated (pending detailed assessment of the numbers of missed companions vs. those of spurious solutions and incorrectly classified objects), 
but nevertheless it is a clear example of the fact that \gdrthree provides critical constraints on the M dwarf binary fraction at close separations and very low mass ratios. Even with corrections for completeness and reliability still to be accounted for, the \gaia M dwarf sample in the Solar neighborhood with sensitivity to substellar companions within a few aus is orders of magnitude larger than those of all other spectroscopic surveys combined. 

%
\subsection{Astrometric masses: trends with stellar metallicity}
%
The number of new \gaia detections of substellar companions in the EP regime is still too small 
to provide an independent assessment of well-known trends of exoplanet frequency with stellar properties, such as the strong dependence of giant planet occurrence with metallicity (e.g., \citealt{FischerValenti2005,2009ApJ...697..544S,Santos2011,Mortier2012}; \citealt{Adibekyan2019}, and references therein). The population of likely BD companions is instead conspicuous, and amenable to verify the outcome of recent statistical investigations. 

As an example, \citet{MaGe2014} showed that, using the iron abundance relative to the Sun [Fe/H] as a proxy, the metallicity distribution of BD solar-type hosts has a median and standard deviation [Fe/H] = $-0.04\pm0.28$. Using the most recent compilation of BD companions based on \citet{Wilson2016,Kiefer2019,Dalal2021}, the corresponding values are [Fe/H] = $+0.01\pm0.25$. The stellar sample is thus not particularly metal-rich, as is the case of giant-planet hosts (median [Fe/H]$\sim+0.12$, e.g. \citealt{Adibekyan2019}), but is also not as metal-deficient as the typical field stars in the solar neighbourhood (depending on stellar sample, with median [Fe/H] in the range [--0.10,--0.15], see e.g. \citealt{Nordstrom2004,Raghavan2010,Sousa2011,Adibekyan2019}). 

In \cite{DR3-DPACP-104} recipes are outlined for selection of sources with global metallicity ([M/H]) of good and intermediate quality determined based on \gaia data. A total of 17\,129 sources with an astrometric orbital solution were selected to have intermediate-quality [M/H] from the archive. Unsurprisingly, the overwhelming majority of [M/H] determinations is for the brighter solar-type stars, therefore the typical M-dwarf primary with a substellar companion does not have a metallicity determination. However, we find [M/H]$=-0.02\pm0.29$ for a sample of 143 F-G-K-type BD hosts. For reference, applying a more strict recipe for good-quality [M/H] returns a sample of 74 sources with [M/H]$=+0.01\pm0.27$, both estimates are in excellent agreement, and indeed indistinguishable, with literature results. If no quality constraints are used, 327 sources with a substellar companion and a \gaia-derived metallicity have [M/H]$=-0.25\pm0.49$, which is indicative of the need to restrict ourselves to the regime of primaries with better-calibrated metallicities. 

%
\subsection{Known substellar objects: statistics and notable examples}
%
The comparison between \gaia orbital solutions and companion mass estimates for known substellar companions and literature results, small as the sample might be, is interesting for a number of reasons.  

First of all, a quick look at Table \ref{tab:known_EP_BD_comp} allows us to underline how the angular orbit size for known BDs is always $\gtrsim 0.5$ mas, while the opposite, with a few exceptions, holds for the known EPs, \object{HD 132406 b} being the record holder with the smallest measured angular semi-major axis: $a_1 = 136\pm40$ $\mu$as. Overall, for both EPs and BDs there is a tendency to underestimate orbital periods longer than the time-span of \gaia observations. Not unexpectedly, orbital eccentricities are typically more loosely constrained by \gaia astrometry with respect to those from Doppler spectroscopy, and very high-eccentricity orbits ($e\gtrsim0.8$) are typically underestimated. The loss of accuracy in the long-period and high-eccentricity regimes are known effects, already quantified via detailed simulations by \cite{2008A&A...482..699C}, and further discussed in e.g. \cite{DR3-DPACP-176}.

Inspection of Table \ref{tab:known_EP_BD_comp} also shows that there is no simple mapping of $\Mass_c\sin i$ into $\Mass_c$ given the derived $i$ value. \afterReferee{For some objects (e.g. \object{HD 132406 b}, \object{HD 81040 b}, and \object{HD 52756 b}), the minimum mass estimate translates into a larger true mass in agreement with the determined value of inclination, but in other cases (e.g., \object{HD 164604 b}, \object{HR 810 b}, and \object{HD 142 b}), $\Mass_c$ is estimated to be much larger than the $\sin i$ value would infer it to be, or even lower (e.g., \object{HD 30246 b} and \object{HD 82460 b}) than $\Mass_c\sin i$. When the parameter uncertainties are taken into account, the discrepancies are typically not very statistically significant, but the effect will nevertheless require to be understood}. As both minimum and true mass estimates depend on the assumptions made for the mass of the primary, part of the reason for the discrepancy might be due to the heterogeneity of methods used to derive the latter. However, the more fundamental explanation is likely to be found in the 
overall limitations of \gdrthree detection sensitivity, including the selection effects and biases introduced by astrometric NSS processing, as discussed in  \secref{ssec:astrometric_orbit_distributions} and Appendix \ref{sec:appendix_analytical_signal_dispersion}, particularly in the limit of relatively low astrometric signal-to-noise ratios and sub-optimal redundancy in the number of visibility periods with respect to the number of model parameters\footnote{The typical number of visibility periods is only twice the number of fitted parameters in an orbital model  \citep[see Appendix \ref{sec:appendix_analytical_signal_dispersion}, and also][]{DR3-DPACP-176}}.

Among the companions with a derived mass in the planetary regime, the case of \object{GJ 876 b} stands out. The planet constitutes one of the earliest radial-velocity discoveries in the field, first announced by \cite{1998ApJ...505L.147M} as a gas giant with ${\cal M}_c\sin i \sim 2$\Mjup orbiting a mid-M dwarf in the backyard of the Sun ($d=4.67$ pc). \citet{2002ApJ...581L.115B}, using HST/FGS data, published the astrometric orbit of GJ 876 b (constrained by the RV solution), determining an orbit size of $0.25\pm 0.06$ mas, an inclination of $84^\circ$ and a true mass very close to the minimum mass limit. The GJ 876 planetary system was subsequently found to host 4 planets, the hot super Earth GJ 876 d with period of $\sim2$ d, the two Jupiter-type planets GJ 876 c,b with periods of $\sim30$ and $\sim 61$ d, respectively, and the Neptune-mass companion GJ 876 e with a period of $\sim125$ d. The three outermost companions are dynamically interacting, locked in a 1:2:4 Laplace mean-motion resonance, which has been the subject of many studies (e.g., \citealt{2005ApJ...634..625R,2010ApJ...719..890R,2010A&A...511A..21C,2016MNRAS.455.2484N,2018A&A...609A.117T}). The more recent investigations, based on dynamical considerations, infer a close-to-coplanar configuration for the three interacting planets, and a likely inclination of GJ 876 b $\sim50-60^\circ$. This implies a true mass ${\cal M}_c\sim2.3-2.7$\Mjup. The amplitude of the astrometric perturbation determined with \gaia, $0.43\pm0.05$ mas, is larger than that measured by HST/FGS, and discrepant at the $2.3\sigma$ level. The inferred mass, 3.6\Mjup, is correspondingly larger, and also in this case the derived inclination $i=101^\circ$ does not allow for a simple mapping from the $\Mass_c\sin i$ value. 

A number of known companions, which in the literature have minimum mass estimates in the planetary regime, appear in Table \ref{tab:nssmass} with much higher true mass estimates from Gaia. These appear in the \gdrthree archive as validated orbital solutions of type \fieldName{OrbitalTargetedSearchValidated}. A few cases of  particular interest are discussed below: 

\begin{enumerate}
    \item \object{HD 114762} (\sourceId{3937211745905473024}): The first substellar companion candidate around a solar-type star with minimum mass $\Mass_c \sin{i}=0.011\pm0.001\Msun$ was inferred from radial velocity variations by \citet{Latham:1989rt}. \gdrone\ noise modelling resulted in a considerably higher companion mass estimate of $0.103^{+0.030}_{-0.025}\Msun$ \citep{Kiefer:2019us}. 
    The \gdrthree\ orbital solution has a period of $83.73\pm0.12$ d in agreement with the radial-velocity orbit and an orbit size of $a_0={1.80}\pm0.07$\,mas. Using the primary mass estimate from Table \ref{tab:nssmass}, the inferred companion mass is $\Mass_c={0.21}\pm0.01\Msun$, using standard linear propagation of the uncertainties. \gdrthree\ therefore  establishes that the companion is a low-mass M dwarf and not a substellar object;
    
    \item \object{HD 164604} (\sourceId{4062446910648807168}): \citet{Arriagada2010} announced a low-confidence detection of a $\Mass_c\sin i\sim2.7$\Mjup companion on a $606\pm9$ d orbit, whose parameters were then refined by \citet{Feng2019} who reported $\Mass_c\sin i = 1.99\pm 0.26$\Mjup and $P=641\pm9$ d. The \gdrthree\ orbital solution has $P=615\pm12$ d (in agreement at the $1.7\sigma$ and $0.6\sigma$ level with \citet{Feng2019} and \citet{Arriagada2010}, respectively) and $a_0=0.56\pm 0.22$\,mas. The inferred companion mass is $\Mass_c=14.3\pm5.5$\Mjup;
    
    \item \object{HD 162020} (\sourceId{5957920668132624256}): \citet{Udry2002} published the discovery of a $P=8.428$ d, slightly eccentric ($e=0.28$) $\Mass_c\sin i\sim14$\Mjup companion, whose minimum mass was recently updated to $9.8\pm2.7$\Mjup by \citet{Stassun:2017ud}. The \gdrthree\ orbital solution has $P=8.429\pm0.001$ d, $e=0.23\pm0.05$, and $a_0=0.91\pm 0.03$\,mas. The detected companion is a low-mass star with $\Mass_c=0.39\pm0.02$\Msun; 
    
    \item \object{KIC 7917485} (\sourceId{2075978592919858432}): \citet{Murphy2016} published the detection of a $\Mass_c\sin i = 11.8^{+0.8}_{-0.6}$\Mjup companion on a $P=840^{+22}_{-20}$ d orbit around the Delta Scuti, A-type star Kepler-1648, based on a pulsation timing variations technique. The \gdrthree\ orbital solution has $P=810\pm28$ d and $a_0=0.42\pm 0.02$\,mas. At a distance of 1.38 kpc, the companion turns out to be an M dwarf with $\Mass_c=0.55\pm0.03$\Msun. 

\end{enumerate}

%
\beforeReferee{\subsection{Validated planetary-mass companions}}
\afterReferee{\subsection{Validated orbital solutions that can imply new exoplanet discoveries}}

%
Two sources have validated astrometric orbital solutions \citep[see][for details]{DR3-DPACP-176} that imply the presence of previously-unpublished planetary-mass companions \afterReferee{if a `binary scenario' can be excluded. In such scenario the small apparent orbit size would be caused by a binary star with components of similar mass and brightness ratios. These sources are:} 
\beforeReferee{, which we discuss below: } 

\begin{enumerate}
    \item \object{HIP 66074} (\gdrthree\ 1712614124767394816): The \gdrthree orbital solution has $P=297\pm2.8$ d, $e=0.46\pm0.17$, $a_0=0.21\pm0.03$ mas. Given the primary mass estimate corresponding to an M0 dwarf, the inferred companion mass \afterReferee{in the exoplanet scenario} is $\Mass_c=7.3\pm1.1$\Mjup. 
%
    \item \object{HIP 28193} (\gdrthree\ 2884087104955208064):
    The orbital period, semi-major axis and eccentricity of this new exoplanet are $827\pm50$ d, $e=0.07\pm0.10$, and $a_0=0.25\pm0.02$ mas, respectively. Using the K-dwarf primary mass from Table \ref{tab:nssmass}, the inferred companion \afterReferee{in the exoplanet scenario} is a super-Jupiter with a mass of $5.3\pm0.6\Mjup$. 
    We note that, had we used Monte Carlo resampling, the semi-major axis distribution would have been asymmetric with larger uncertainties, and this would have been the case for the companion-mass distribution as well. This is an example of a solution with a poorly-constrained eccentricity ($e=0.07\pm0.10$) for which the uncertainties in the Thiele-Innes coefficients are likely overestimated and Monte-Carlo resampling is not advisable \citep[cf.][]{DR3-DPACP-127, DR3-DPACP-176}.
\end{enumerate}

\afterReferee{A third source (\gdrthree\ 1035000055055287680, \object{HIP 40497}) that initially also fell into this category, has been identified as SB2 in the literature \citep{2007A&A...466.1089B}, see also the discussion in \citet{DR3-DPACP-176}. This illustrates that the risk of confusing the binary and exoplanet scenarios is real when only considering the \gaia astrometric orbit.

For \object{HIP 66074} and \object{HIP 28193} we can, however, make use of auxiliary radial-velocity information. To evaluate the binary scenario where the companion is a main-sequence star, we used the mass-luminosity relationships of \citet{Henry:1993aa} to estimate that the two components must have masses that agree within a few percent to be compatible with both the orbital parameters and the photocentre orbit size. Consequently, the primary component's RV semi-amplitudes of \object{HIP 66074} and \object{HIP 28193} would be $K_1\sim\!20$ km\,s$^{-1}$ and $K_1\sim9$ km s$^{-1}$, respectively, clearly incompatible with the high-precision RVs used for validating the orbital solution that have dispersions that are three orders of magnitude smaller \citep{Butler2017, DR3-DPACP-176}. Similarly, the \gdrthree \fieldName{radial_velocity_error}, computed from the dispersion of individual \gaia RV measurements, is $0.15$ km s$^{-1}$ and $0.16$ km s$^{-1}$, respectively, which lies in the first percentile for sources with $G<12$, and therefore also appears incompatible with the binary scenario. The blending of the SB2 spectra could possibly lead to such suppressed RV variability. In the case of \object{HIP 28193}, the small uncertainties in the ground-based RVs coupled with the estimated FWHM of the underlying cross-correlation functions of $\lesssim9$ km s$^{-1}$ speak against this possibility\footnote{At the times of minimum/maximum RV, the separation between the primary and companion RV should be $\sim 18$ km s$^{-1}$, i.e.\ wider than the spectroscopic cross-correlation function.}.

In terms of absolute magnitudes of these systems, the difference between the exoplanet and binary scenario amounts to $\sim$ 0.8 mag, which because of the width of the observed HRD can also not be used to definitely rule out the binary scenario.

In the Hipparcos-\gaia catalogues of accelerations produced by \citet{Brandt2021} and \citet{2022A&A...657A...7K} no statistically significant PMa is reported for \object{HIP 66074} (S/N$\, \sim1$), while a moderately high PMa (S/N$\sim10$) at the \gaia mean epoch is found for \object{HIP 28193}. For the relatively short-period orbit of the companion around \object{HIP 66074}, the inferred companion mass is approximately in line with a S/N$\sim1$ in the proper motion difference. In the case of the longer-period orbit of the companion around \object{HIP 28193}, the PMa value this might point to a companion with a larger mass than the one inferred nominally. An alternative possibility would be that the \gaia orbit has significantly underestimated the true period. 

To summarize, the most likely scenario for both \object{HIP 66074} and \object{HIP 28193} is therefore the presence of a newly-discovered giant exoplanet. A more detailed analysis and probably more auxiliary data are needed to definitely rule out the binary scenario. When that is achieved,} these are to be considered as the first \gaia astrometric planet detections and the first examples of confirmed exoplanet discoveries with the astrometry technique.

%
\subsection{Candidates with substellar masses: statistics and notable examples}\label{candidate_EP}
%

Among the substellar mass candidates with solution type  \fieldName{OrbitalTargetedSearch}, a few are worth particular mention: 

\begin{enumerate}

\item \object{HD 12800} (\gdrthree\ 522135261462534528): the \gdrthree orbital solution for this bright source (54 Cas) has $P=401\pm12$ d, $a_0=0.25\pm0.05$ mas, and $\Mass_c=5.6\pm1.4$\Mjup. This is the only candidate companion around a main-sequence solar-type star with a mass well in the planetary regime. 

\item \object{HD 3221} (\gdrthree\ 4901802507993393664): the \gdrthree orbital solution for this bright solar-type star has $P=476\pm5$ d, $a_0=0.36\pm0.01$ mas, and $\Mass_c=14.2\pm0.6$\Mjup. The primary is a fast-rotating ($v\sin i\sim70$ \kms), very young star with an estimated age of $\sim10-30$ Myr in the Tucana/Horologium association. The candidate companion, at the planet-brown dwarf mass boundary according to a classical definition \citep{Burrows2001}, if confirmed, would be the first of this type in an orbital separation regime virtually inaccessible to Doppler and direct imaging surveys.

\end{enumerate}

As mentioned above, the population of substellar mass candidates with solution type \fieldName{OrbitalTargetedSearch*} is typically found around solar-type primaries. It is interesting to note how the eccentricity distributions for EPs and BDs in this sample are marginally different based on K-S test (p-value of 0.04),
with the more massive BDs ($\Mass_c\gtrsim50$ \Mjup) having a median $e\sim0.5$, while EPs and lower-mass BDs ($\Mass_c\lesssim40$ \Mjup) have typically $e\sim0.3$. This is in agreement with \citet{MaGe2014}, and with the notion that the former might correspond to the low-mass tail of objects formed like stars while the latter sample would map the high-mass tail of objects formed like planets.

The population of substellar mass candidates with solution type \fieldName{Orbital} is instead predominantly found around low-mass M primaries. In this case, the eccentricity distributions of candidate EPs and BDs detected around this sample are indistinguishable, with typically $e\sim0.4$ in both cases. This might indicate that most of the detected companions might have formed in the same way, and that some of them are actually of intrinsically larger mass  \afterReferee{compared to the lower mass bounds adopted in this work as discussed in Sect. \ref{sec:f_m_select_substellar}}. 

\subsection{Substellar companion candidates to white dwarfs}\label{ssec:exoplanets_wd}
%
The \TBOTable table contains 38 orbital solutions for sources on the white dwarf (WD) sequence, all of them correspond to astrometric orbits. Assuming a fixed mass of $0.6\,M_\sun$ for the WD host and that the companion is dark, there are four sources with substellar companion candidates. These are  \object{\sourceId{2813020961166816512}} (\object{LP 522-46}) with $\Mass_c\!\sim\!34\Mjup$,  \object{\sourceId{2098419251579450880}} with $\Mass_c\!\sim\!34\Mjup$, and \sourceId{6471102606408911360} (\object{L 279-25}) with $\Mass_c\!\sim\!22\Mjup$. The last one \sourceId{4698424845771339520} (\object{WD 0141-675}) stands out as being part of the 10 pc sample and with a companion candidate in the planetary-mass range. 

In \figref{fig:double_degenerates} these four sources are marked with red symbols. Two sources (\object{LP 522-46} and \object{L 279-25}) are located above the hydrogen sequence, which suggests that the companion could itself be degenerate and luminous enough to cause an excess in luminosity and at the same time dilute the astrometric orbit signal. The companions of these sources are therefore likely neither dark nor substellar. On the other hand, \object{\sourceId{2098419251579450880}} and \object{WD 0141-675} lie within the hydrogen sequence and therefore \figref{fig:double_degenerates} appears consistent with the interpretation of these sources having substellar companions. 

\object{WD 0141-675} in particular was included in the OrbitalTargetedSearch sample \citep{DR3-DPACP-176} because it is nearby and metal-polluted \citep{Debes:2010aa}, hence represents a promising target to search for the presence of an orbiting giant planet that could act as the much sought-after perturber of the circumstellar material \citep{Debes:2012aa}. 
The \gdrthree\ orbital solution has a period of $33.65\pm0.05$ d and with a WD mass of $0.57\pm0.03\,M_\sun$ \citep{Subasavage:2017wc}, the estimated planet mass is $\Mass_c={9.26}^{+2.64}_{-1.15}\Mjup$, where we accounted for all parameter covariances using Monte-Carlo resampling. The resulting $\Mass_c$ distribution is asymmetric with its mode at $\sim\!8.3\Mjup$.

With only a handful of known giant planets orbiting white dwarfs \citep{Veras:2021vy, Blackman:2021wi}, the \gaia discovery of a super-Jupiter candidate planet orbiting \object{WD 0141-675} is remarkable. If the \gdrthree\ orbit is confirmed and other possible scenarios, e.g.\ some kind of WD binary system with nearly equal-brightness components (\object{WD 0141-675} is marked with an orange square in \figref{fig:double_degenerates}), can be excluded this would represent the discovery of the most nearby planet-hosting WD and the first giant planet around a metal-enriched WD, which will make it an important test for our understanding of the fate of stars and their planetary systems. 

%
\subsection{Radial Velocity: Substellar companions}
%

\begin{figure}[!htb]
\begin{center}
\includegraphics[width=0.95\columnwidth]{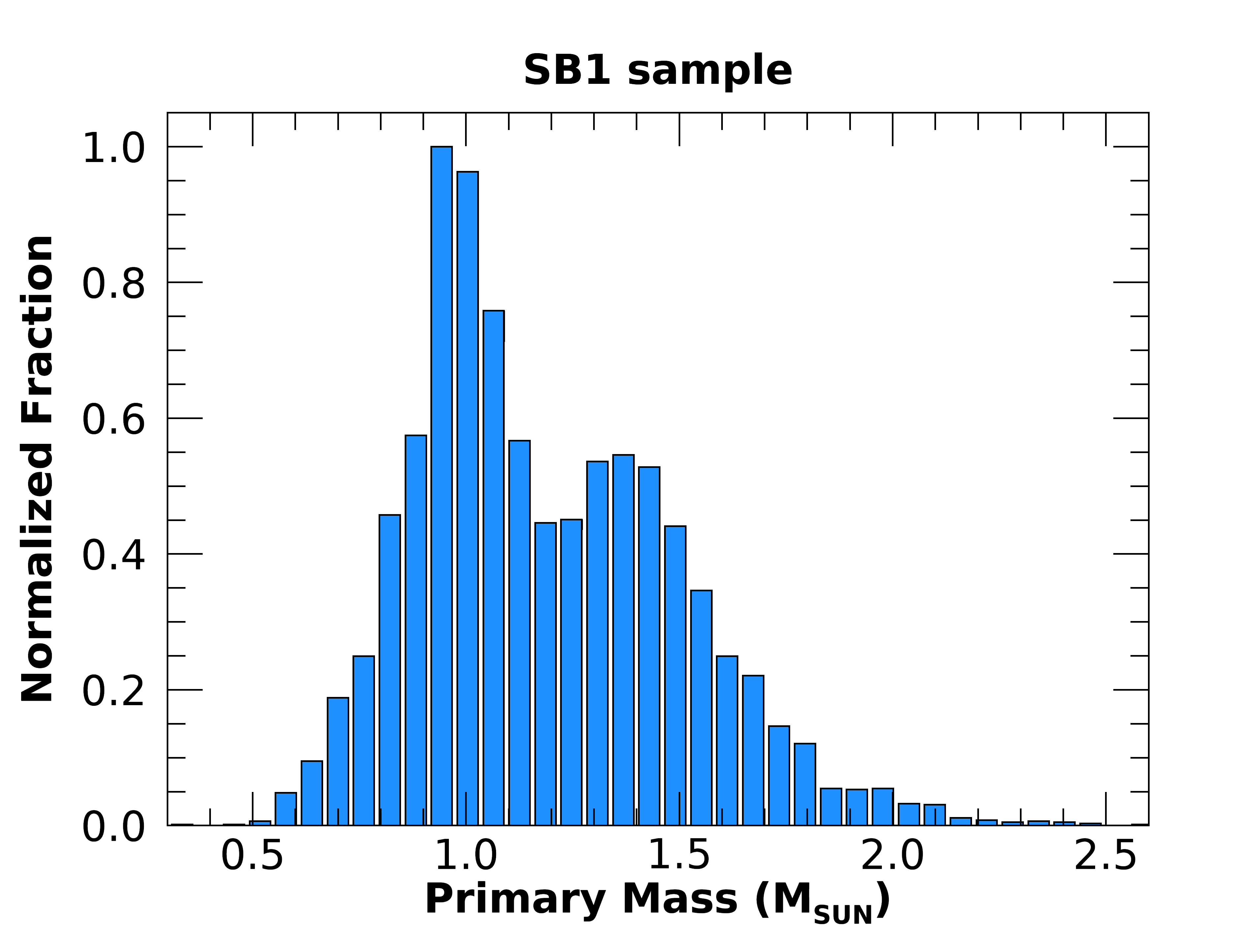}
\includegraphics[width=0.95\columnwidth]{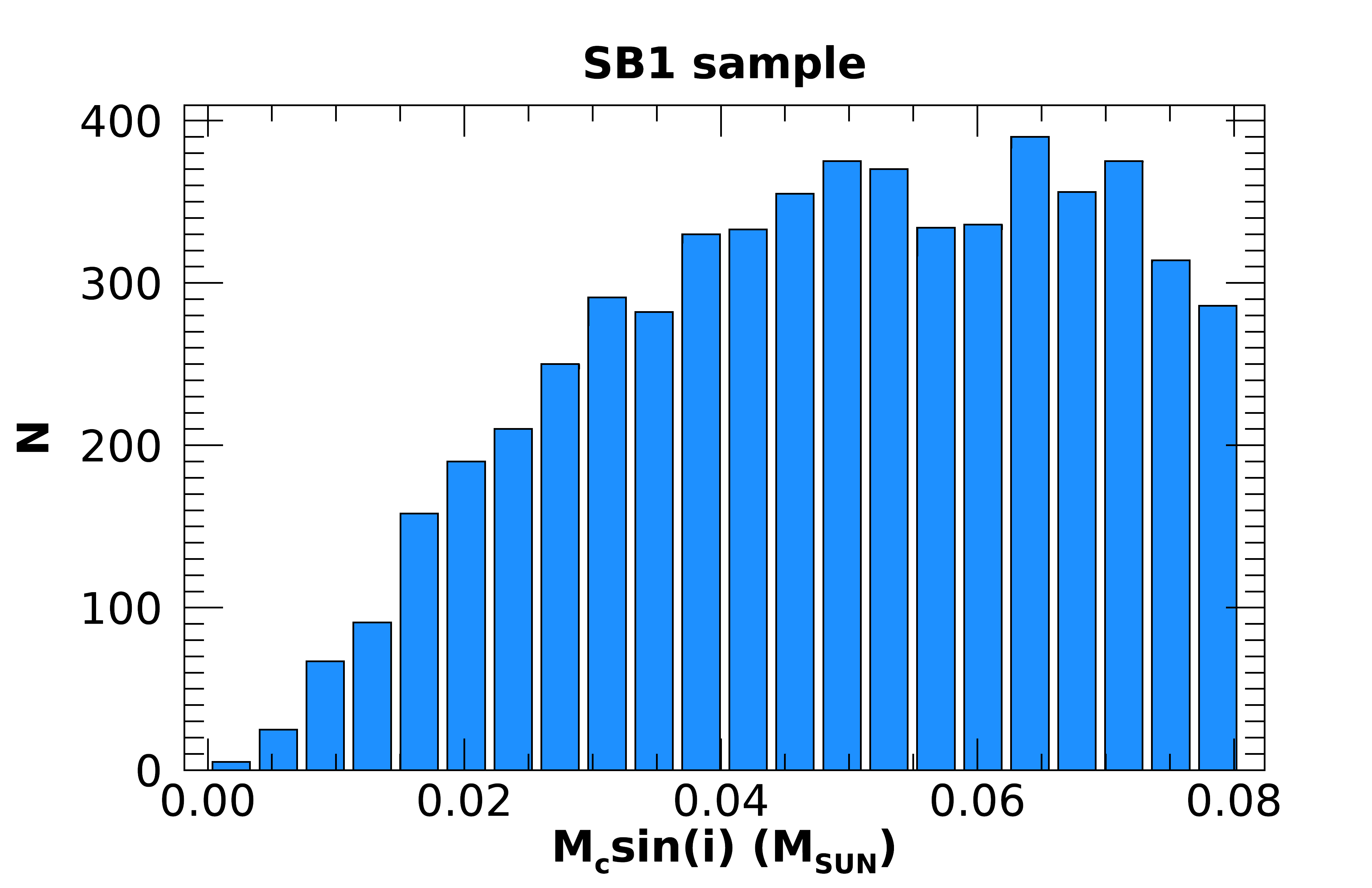}
\caption{{\it Top}: Primary mass distribution for sources with \fieldName{SB1} solution type and inferred minimum companion  masses in the substellar regime. {\it Bottom}: substellar companion mass distribution for the \fieldName{SB1} solution type. }\label{fig:substellar_SB1}
\end{center}\end{figure}

Out of $6\times10^4$ minimum mass estimates for \fieldName{SB1} solutions (see Table \ref{tab:nssmass}), about 10\% (5723) have $\Mass_c\sin i < 0.08$\Msun, and about 10\% of these (437) have $\Mass_c\sin i < 0.02$\Msun. Not unexpectedly, the mass distribution for the primaries (see \figref{fig:substellar_SB1}, top panel) has a median of $\sim1.0$\Msun, similar to that of the primary mass distribution of \fieldName{OrbitalTargetedSearch*} sources rather, but with a significantly larger contribution from bright, earlier-type stars. The distribution of $\Mass_c\sin i$, shown in the lower panel of \figref{fig:substellar_SB1}, contains an expected feature, i.e. the decline in numbers of companions at the low-mass end due to the intrinsic lack of sensitivity of \gaia RVS to RV amplitudes of signals typically induced by planetary-mass companions (significantly $< 1$ km s$^{-1}$). On the other hand, the number of higher-mass substellar companions appears constant, all the way into the low-mass star regime, and independently on primary mass. This is unexpected, as an intrinsically lower frequency of intermediate-mass BDs is observed in RV surveys \citep{MaGe2014}, particularly in the short-period regime to which \gaia radial-velocities are sensitive. The eccentricity distributions of \fieldName{SB1} companions with mass estimates in the EP and BD regimes appear entirely indistinguishable, independently of primary mass, and this also is not in agreement with the \citet{MaGe2014} analysis. 

For $>80\%$ of the \fieldName{SB1} sample of close-in companions with $\Mass_c\sin i < 0.08$ \Msun the orbital solutions have $P<10$ days, and this fraction grows to $>90\%$ for companions with $\Mass_c\sin i < 0.02$ \Msun. Such companions, which are detected with typical significance of their $K_1$-values below 10, typically correspond to primaries with \fieldName{ruwe} $> 1.4$, but they are not expected to be those responsible for high \fieldName{ruwe} values, as the typical sizes of the induced astrometric perturbations would escape detection by \gaia \citep[e.g.][]{2020MNRAS.496.1922B}. As discussed in \secref{sssec:cu4nss_intro_outputfiltering} a sizeable fraction of these short-period orbits are actually kind of aliases of longer-period ones. For example, the companions reported around HIP 24329, HD 35956, and HD 8691 have $\Mass_c\sin i \sim 3$ \Mjup, $\sim11$ \Mjup and $\sim14$ \Mjup, respectively, and $P=0.63$ d, 3.02, and 3.77 d, respectively. However, these sources also have \fieldName{Orbital} solutions with P$=1499$ d, 1203 d, and 581 d. The latter $P$ values very closely match the published periods of the spectroscopic orbits for the three stars reported by \citet{Wilson2016}, \citet{Katoh2013}, and \citet{Sperauskas2019}, respectively. 

The sample of short-period \fieldName{SB1} orbits with minimum masses corresponding to substellar companions should therefore be considered with caution, although not all orbital solutions can be wrong or spurious. For example, a cross-check with the NASA Exoplanet Archive shows that the spectroscopic orbit of WASP-18b \afterReferee{(\object{\sourceId{4955371367334610048}})}, a transiting super-Jupiter with $\Mass_c\sim10$ \Mjup and $P=0.94$ d discovered by \citet{Hellier2009} around the Hyades-age F6 dwarf HD 10069, is recovered as an \fieldName{SB1} with the correct period and RV semi-amplitude ($K_1\sim1.8$ km s$^{-1}$). This source has also been detected in the \gaia photometry \citep{DR3-DPACP-162} and is present in the \fieldName{vari_planetary_transit} table.

The list of most exoplanet candidates detected by \gaia using either astrometry, transit or radial velocities is published in \url{https://www.cosmos.esa.int/web/gaia/exoplanets}.


%
\section{Multiple stars }
%

Although the NSS pipeline in \gdrthree produce solutions of binary stars, the results can also be used to uncover higher multiplicity stars. \beforeReferee{ternary}\afterReferee{Triple} (and higher multiplicity) stars are of particular interest. The study of the architecture and dynamics of
hierarchical stellar systems provides precious information on the mechanisms at work during star formation. For example, the fact that the orbits of a hierarchical system are coplanar would indicate that the
stars are formed in a viscous accretion disk, while their mass ratios shed light on the disk fragmentation mechanism \citep[see e.g.][]{2017ApJ...844..103T}.

\subsection{Multiplicity from spectroscopic and astrometric solutions}\label{sec:multi_sb_orbital}

A first way to find multiple stars is to look at the sources for which the pipeline produced an astrometric \fieldName{Orbital} solution and an \afterReferee{\fieldName{SB1/SB2}} solution which were not combined. In fact,
given that the astrometry is sensitive to long periods/larger orbits, while the spectroscopy is sensitive to shorter periods, in the case of a \beforeReferee{ternary}\afterReferee{triple} star the astrometry would detect the outer period, while the spectroscopy will unveil the inner period. 

\begin{figure}[htb]
\begin{center}
\includegraphics[width=\columnwidth]{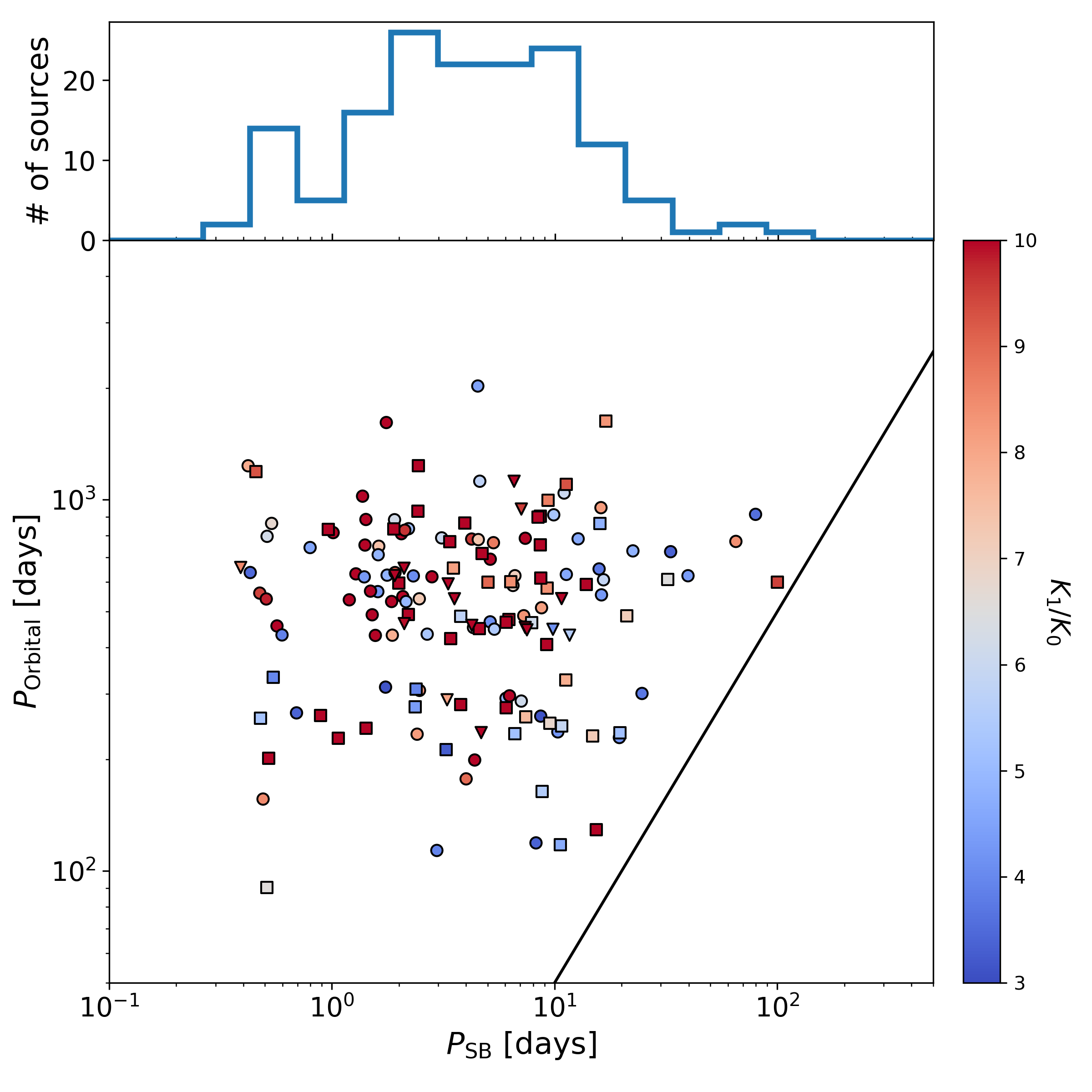}
\caption{Distribution of outer vs inner periods for \beforeReferee{ternary}\afterReferee{triple} systems found matching \fieldName{Orbital} with \fieldName{SB1} (circles), \fieldName{SB2} (squares) and 
\fieldName{SB2C} solutions (triangles), coloured by the ratio of spectroscopic over astrometric
semi-amplitudes. The solid line shows the limit $P_\mathrm{Orbital} = 5\; P_\mathrm{SB}$. Top panel: Integrated distribution of inner periods.}\label{fig:period_period_triples_orb_sb}
\end{center}\end{figure}

Many of the sources for which the \fieldName{Orbital} and \fieldName{SB1/SB1C} solution was found, but not combined by the pipeline are, however, not \beforeReferee{ternary}\afterReferee{triple} stars. In many cases they have similar periods but were not combined because of the inconsistency of the other orbital parameters. There are also many cases where the SB solution is actually some kind of alias of the astrometric period; these cases can be spotted noting that the semi-amplitude $K_0$ of the astrometric motion (defined by \equref{eq:semi-amplK12}, substituting $a_0/\varpi$ for $a_1$) is similar to the semi-amplitude $K_1$ of the radial velocity curve. These confusing cases have typically a significance lower than 10.
We then identify genuine \beforeReferee{ternary}\afterReferee{triple} stars by selecting those for which the significance of both astrometric and spectroscopic solution is larger than 10, $K_1 > 3\cdot K_0$ and $P_\mathrm{Orbital} > 5\cdot P_\mathrm{SB}$.
With this selection we obtain 81 \beforeReferee{ternary}\afterReferee{triple} systems from matching \fieldName{Orbital} with \fieldName{SB1} solutions, 55 from matching with \fieldName{SB2}, 16 from matching with \fieldName{SB2C} and none from matching with \fieldName{SB1C}.
The distribution of the outer vs inner periods of these sources is reported in \figref{fig:period_period_triples_orb_sb}.

One can also find multiple stars by looking at the sources for which the pipeline produced an astrometric acceleration solution and an \fieldName{SB1/SB2} solution. 
In this case, the astrometric acceleration would detect outer periods which are around or longer than the length of \gaia observations ($\sim 1\,000$ days), while the SB solutions detect the inner periods. 

In order to avoid that the astrometric acceleration and the SB solutions are in reality of the same orbit, we selected only SB solutions with a period $<$ 300 days. We also restrain our search to SB solution with \fieldName{significance} $>20$ to avoid to be polluted by some kind of aliasing.

\begin{figure}[htb]
\begin{center}
\includegraphics[width=\columnwidth]{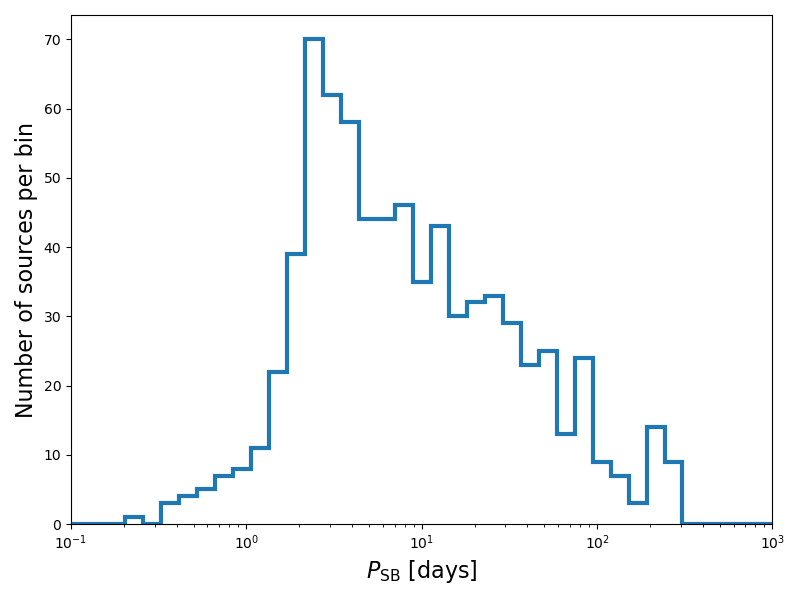}
\caption{Distribution of inner periods for \beforeReferee{ternary}\afterReferee{triple} systems found comparing astrometric acceleration and SB solutions.}\label{fig:period_hist_triples_acc_sb}
\end{center}\end{figure}

Distribution of inner periods for \beforeReferee{ternary}\afterReferee{triple} systems found matching astrometric acceleration and SB solutions is shown in \figref{fig:period_hist_triples_acc_sb}. We can note that the mode of the distribution is at around 3 days.

\subsection{Multiplicity in wide visual binaries}

Another method to discover \beforeReferee{ternary}\afterReferee{triple} or higher multiplicity stars consists in using catalogues of wide visual binaries and see if one of the two components is detected as binary.
We started from the \citet[][]{2021MNRAS.506.2269E} catalogue, and selected sources with $\mathcal{R}< 0.01$ (0.08\% contamination from chance-alignment), crossmatched with  \TBOTable solutions and spectroscopic trends, and we found 10\,063 systems for which one of the two components is a non-single star and 52 systems where both components are non-single stars.
For 10 of the first group, the non single component is actually a \beforeReferee{ternary}\afterReferee{triple} star (\fieldName{Orbital}+\fieldName{SB1/SB2}, selected as in \secref{sec:multi_sb_orbital}), making them a \beforeReferee{quaternary}\afterReferee{quadruple} system. 

\begin{figure}[htb]
\begin{center}
\includegraphics[width=\columnwidth]{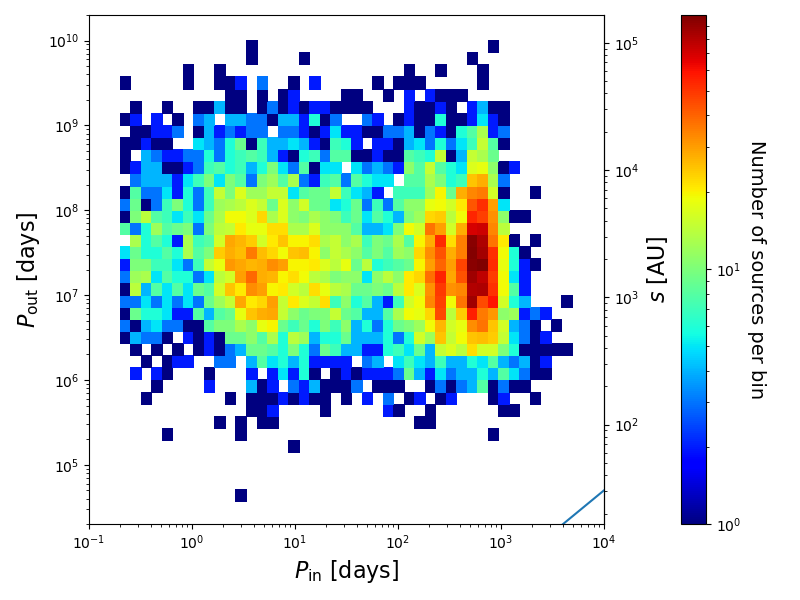}
\caption{Distribution of outer vs inner periods for \beforeReferee{ternary}\afterReferee{triple} systems found in El-Badry catalogue. \afterReferee{Left scale: period, right scale: separation. The line, bottom right, shows the limit $P_\mathrm{out} = 5\; P_\mathrm{in}$.} }\label{fig:period_period_elbadry}
\end{center}\end{figure}

\begin{figure}[htb]
\begin{center}
\includegraphics[width=\columnwidth]{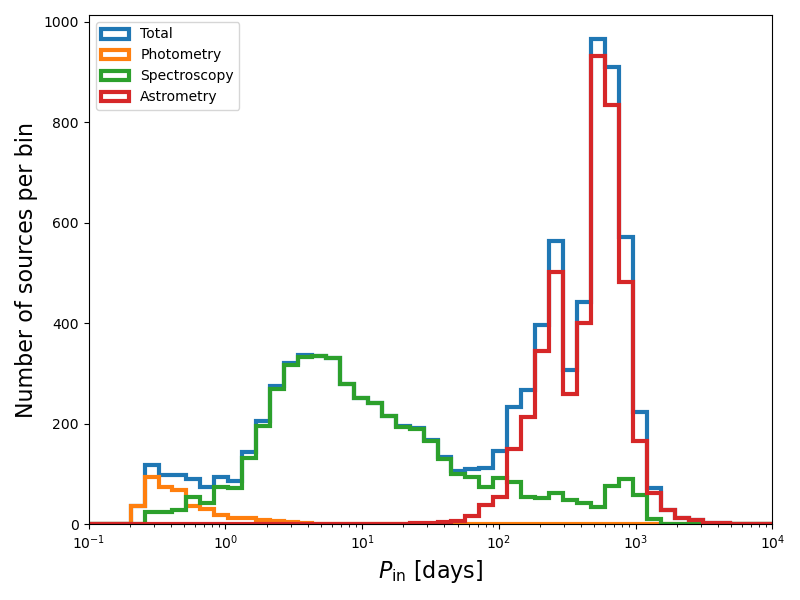}
\caption{Distribution of inner periods for \beforeReferee{ternary}\afterReferee{triple} systems found in El-Badry catalogue.}\label{fig:period_inner_elbadry}
\end{center}\end{figure}

Figure~\ref{fig:period_period_elbadry} shows the distribution of the outer vs inner periods for the \beforeReferee{ternary}\afterReferee{triple} systems found in El-Badry catalogue,
where the outer period $P_\mathrm{out}$ is computed from the separation $s$ (provided in El-Badry catalogue) as
$P_\mathrm{out}=\sqrt{s^3/7.496\times 10^{-6}(\Mass_1+\Mass_2)}$ and assuming $\Mass_1+\Mass_2=1.5\Msun$, while the inner period $P_\mathrm{in}$ is the period in the NSS solution.

The drop in the distribution for separations below 200 au is certainly due to selection effects introduced by the processing due to the blending of the two components in a large range of scanning angles. The cut at longer inner periods is dictated by the limited baseline of \gaia observations. The shortening of the maximum inner period with the increase of separation, is due to the fact that sources at large separations are also at larger distances, where the astrometric signal of the internal orbit drops.

We note from \figref{fig:period_period_elbadry} that the distribution of the inner period is multimodal. As shown in \figref{fig:period_inner_elbadry}, this is in part due to the different methods of orbital detection (highlighted with different colours).  However, if we compare the distribution of spectroscopic solutions in this sample with respect to the whole NSS solutions (\figref{fig:dist-TBO-period}), we can see that the overabundance of solutions with 2 days $< P_\mathrm{in} <30$ days is real. This overabundance is also visible in the Multiple Star catalogue by \citet{2018ApJS..235....6T}, figure 7. \citet{2004RMxAC..21....7T} attributed the overdensity he observed at $P_\mathrm{in}< 7$ days to dynamical interaction between the orbits and consequent tidal interaction within the inner couple, as suggested by \citet{1998MNRAS.300..292K}.

This short analysis shows that the study of multiple systems will greatly benefit from the results of the NSS catalogue.

%
\section{Conclusions}
%
On 24 April 1610, Galileo brought his telescope to demonstrate its performances to his 
opponents and other scholars in Bologna. Martin Hork\'y, Kepler's student writes: 
``{\it I tested the instrument of Galileo's in a thousands ways}''..., ``{\it below it works 
wonderfully; in the heavens, it deceives one as some fixed stars are seen double.}'' 
\citep{Feyerabend}. This is how the discovery of first telescopic double stars failed 
and how easy it is to consider that facts can be contradicted by one's prejudices. 

Four hundreds and twelve years later, the validations and the results obtained so far suggest that NSS
entries in \gdrthree are binaries, not \gaia telescope artefacts, although to be fair, 
it is also expected that this catalogue contains spurious solutions.
We thus need to start this conclusion by stressing that the counterpart of so large a 
material is some unavoidable contamination that should be kept in mind when analysing the results. 

In particular the selection of distribution tails (e.g. small periods, small or large mass functions) 
is likely to preferentially select the wrong solutions, as happened to us in the course
of this verification paper.
The abundant partially resolved sources and their impact on the astrometry and epoch radial velocities 
conspire with the scanning law periodic motions to produce solutions that look like bona fide unresolved binaries. 
To cope with this, increasing the threshold 
on the significance of the solutions appears satisfactory but may well represent a Pyrrhic victory, 
as this decreases drastically the sample sizes.

The NSS catalogue is already the result of a drastic selection of sources.
The impact of the total selection effects due successively to the input list, 
to the data processing, and posterior source filtering need to be taken into
account, and the statistical studies of the NSS sample will require dedicated studies. 
This has not been attempted here, although the main points have probably been mentioned.
There are subjects which are already known to represent pitfalls, for example
the acceleration solutions which should not be used for physical interpretations.

Although the main impact of the NSS catalogue originates from the simultaneous presence
of the main kind of binaries, and consequently an impressive coverage of binary periods,
the novelty is mainly brought by the exquisite astrometric precision, allowing the detection of many astrometric binaries.
There is however one strength and one weakness concerning astrometric orbits: 
if the astrometric and the photometric effects are not decoupled, 
the actual size of the orbit of the primary cannot be found. 
This is difficult, but when this can be done, then both masses and luminosities
of the components become available.
Masses that have been estimated here try to take this into account, but the combination
with external data, as some examples have shown, will prove very useful if not 
sometimes mandatory.

Using either astrometric orbits, spectroscopic orbits, or both, the whole H-R diagram can be studied with an abundant material. Obviously, some analyses may be more interesting than others. Here, new ultracool dwarf binaries are found, and the small mass ratios can also be studied down to the substellar domain. True masses are found for substellar companions and two new super-Jupiter candidates may have been found. Twin degenerates and one WD hosting a super-Jupiter companion are also proposed. 
Concerning compact companions with larger masses, 
potential companions are present in the mass range of neutron stars or black holes, but which may also be Algol or \beforeReferee{ternary}\afterReferee{triple} systems: indeed, at this step, we stress that most findings here should be considered as tentative; although verifications have been done, further analyses are warranted.
The eccentricity-period relation will also undoubtedly be under scrutiny although interesting substructures for giants are already shown here. The detection of ellipsoidal variables mistaken as long-period variables, or the detection of sources found in a rare evolutionary stage like EL CVn, underlines the potential of acquiring both photometric and orbital data.
This is another testimony that \gaia is an impressively complete observatory in orbit.

Without any detailed content study nor modelling, that will be the work of the scientific 
exploitation to come, the multiple topics tackled in the sections above, while only
a very preliminary, tentative exploration of the \gdrthree binaries,
clearly demonstrate its scientific potential.
Paraphrasing Aaron Levenstein about statistics, what these analyses reveal is suggestive, 
what they conceal may be essential.

%
\begin{acknowledgements}
This work is dedicated to the memory of our late colleague Dimitri Pourbaix 
who suggested the title of this article.
He had been for many years leader of the Gaia CU4 Coordination Unit 
and passed away before seeing the Gaia DR3 NSS outcome. This catalogue will
remain in a large part the testimony of his involvement.\\


We would like to thank the referee Andrei Tokovinin, and the DPAC reporter, Lennart Lindegren, for their useful suggestions.\\

This work presents results from the European Space Agency (ESA) space mission \gaia. \gaia\ data are being processed by the \gaia\ Data Processing and Analysis Consortium (DPAC). Funding for the DPAC is provided by national institutions, in particular the institutions participating in the \gaia\ MultiLateral Agreement (MLA). The \gaia\ mission website is \url{https://www.cosmos.esa.int/gaia}. The \gaia\ archive website is \url{https://archives.esac.esa.int/gaia}.
Further acknowledgements are given in Appendix~\ref{acknowgaia}.\\

This publication has also made use of observations collected with the
SOPHIE spectrograph on the 1.93-m telescope at
Observatoire de Haute-Provence (CNRS), France (program 21B.PNPS.AREN) using
support by the French Programme National de Physique Stellaire (PNPS).
We thank all the staff of Haute-Provence Observatory for
their support at the 1.93-m telescope and on SOPHIE. 
Observations have also been obtained with the Mercator Telescope and 
the HERMES spectrograph, which is supported by the Research Foundation – Flanders 
(FWO), Belgium, the Research Council of KULeuven, Belgium, the Fonds National de la Recherche
Scientifique (F.R.S.-FNRS), Belgium, the Royal Observatory of Belgium, the Observatoire de 
Gen\`eve, Switzerland and the Th\"uringer Landessternwarte Tautenburg, Germany.\\

This project has received funding from the European Research Council (ERC) under the European Union’s Horizon 2020 research and innovation programme (grant agreement No. [951549]).

Ts.M. research was supported by Grant No. 2016069 of the United States-Israel Binational Science Foundation (BSF) and by Grant No. I-1498-303.7/2019 of the German-Israeli Foundation for Scientific Research and Development (GIF).


\end{acknowledgements}
%

\bibliographystyle{aa} 
\bibliography{dpac,biblio} 

\begin{appendix}
%
\section{\label{acknowgaia}}
%


The \gaia\ mission and data processing have financially been supported by, in alphabetical order by country:
\begin{itemize}
\item the Algerian Centre de Recherche en Astronomie, Astrophysique et G\'{e}ophysique of Bouzareah Observatory;
\item the Austrian Fonds zur F\"{o}rderung der wissenschaftlichen Forschung (FWF) Hertha Firnberg Programme through grants T359, P20046, and P23737;
\item the BELgian federal Science Policy Office (BELSPO) through various PROgramme de D\'{e}veloppement d'Exp\'{e}riences scientifiques (PRODEX) grants, the Fonds Wetenschappelijk Onderzoek through grant VS.091.16N, the Fonds de la Recherche Scientifique (FNRS), and the Research Council of Katholieke Universiteit (KU) Leuven through grant C16/18/005 (Pushing AsteRoseismology to the next level with TESS, GaiA, and the Sloan DIgital Sky SurvEy -- PARADISE);  
\item the Brazil-France exchange programmes Funda\c{c}\~{a}o de Amparo \`{a} Pesquisa do Estado de S\~{a}o Paulo (FAPESP) and Coordena\c{c}\~{a}o de Aperfeicoamento de Pessoal de N\'{\i}vel Superior (CAPES) - Comit\'{e} Fran\c{c}ais d'Evaluation de la Coop\'{e}ration Universitaire et Scientifique avec le Br\'{e}sil (COFECUB);
\item the Chilean Agencia Nacional de Investigaci\'{o}n y Desarrollo (ANID) through Fondo Nacional de Desarrollo Cient\'{\i}fico y Tecnol\'{o}gico (FONDECYT) Regular Project 1210992 (L.~Chemin);
\item the National Natural Science Foundation of China (NSFC) through grants 11573054, 11703065, and 12173069, the China Scholarship Council through grant 201806040200, and the Natural Science Foundation of Shanghai through grant 21ZR1474100;  
\item the Tenure Track Pilot Programme of the Croatian Science Foundation and the \'{E}cole Polytechnique F\'{e}d\'{e}rale de Lausanne and the project TTP-2018-07-1171 `Mining the Variable Sky', with the funds of the Croatian-Swiss Research Programme;
\item the Czech-Republic Ministry of Education, Youth, and Sports through grant LG 15010 and INTER-EXCELLENCE grant LTAUSA18093, and the Czech Space Office through ESA PECS contract 98058;
\item the Danish Ministry of Science;
\item the Estonian Ministry of Education and Research through grant IUT40-1;
\item the European Commission’s Sixth Framework Programme through the European Leadership in Space Astrometry (\href{https://www.cosmos.esa.int/web/gaia/elsa-rtn-programme}{ELSA}) Marie Curie Research Training Network (MRTN-CT-2006-033481), through Marie Curie project PIOF-GA-2009-255267 (Space AsteroSeismology \& RR Lyrae stars, SAS-RRL), and through a Marie Curie Transfer-of-Knowledge (ToK) fellowship (MTKD-CT-2004-014188); the European Commission's Seventh Framework Programme through grant FP7-606740 (FP7-SPACE-2013-1) for the \gaia\ European Network for Improved data User Services (\href{https://gaia.ub.edu/twiki/do/view/GENIUS/}{GENIUS}) and through grant 264895 for the \gaia\ Research for European Astronomy Training (\href{https://www.cosmos.esa.int/web/gaia/great-programme}{GREAT-ITN}) network;
\item the European Cooperation in Science and Technology (COST) through COST Action CA18104 `Revealing the Milky Way with \gaia (MW-Gaia)';
\item the European Research Council (ERC) through grants 320360, 647208, and 834148 and through the European Union’s Horizon 2020 research and innovation and excellent science programmes through Marie Sk{\l}odowska-Curie grant 745617 (Our Galaxy at full HD -- Gal-HD) and 895174 (The build-up and fate of self-gravitating systems in the Universe) as well as grants 687378 (Small Bodies: Near and Far), 682115 (Using the Magellanic Clouds to Understand the Interaction of Galaxies), 695099 (A sub-percent distance scale from binaries and Cepheids -- CepBin), 716155 (Structured ACCREtion Disks -- SACCRED), 951549 (Sub-percent calibration of the extragalactic distance scale in the era of big surveys -- UniverScale), and 101004214 (Innovative Scientific Data Exploration and Exploitation Applications for Space Sciences -- EXPLORE);
\item the European Science Foundation (ESF), in the framework of the \gaia\ Research for European Astronomy Training Research Network Programme (\href{https://www.cosmos.esa.int/web/gaia/great-programme}{GREAT-ESF});
\item the European Space Agency (ESA) in the framework of the \gaia\ project, through the Plan for European Cooperating States (PECS) programme through contracts C98090 and 4000106398/12/NL/KML for Hungary, through contract 4000115263/15/NL/IB for Germany, and through PROgramme de D\'{e}veloppement d'Exp\'{e}riences scientifiques (PRODEX) grant 4000127986 for Slovenia;  
\item the Academy of Finland through grants 299543, 307157, 325805, 328654, 336546, and 345115 and the Magnus Ehrnrooth Foundation;
\item the French Centre National d’\'{E}tudes Spatiales (CNES), the Agence Nationale de la Recherche (ANR) through grant ANR-10-IDEX-0001-02 for the `Investissements d'avenir' programme, through grant ANR-15-CE31-0007 for project `Modelling the Milky Way in the \gaia era’ (MOD4Gaia), through grant ANR-14-CE33-0014-01 for project `The Milky Way disc formation in the \gaia era’ (ARCHEOGAL), through grant ANR-15-CE31-0012-01 for project `Unlocking the potential of Cepheids as primary distance calibrators’ (UnlockCepheids), through grant ANR-19-CE31-0017 for project `Secular evolution of galaxies' (SEGAL), and through grant ANR-18-CE31-0006 for project `Galactic Dark Matter' (GaDaMa), the Centre National de la Recherche Scientifique (CNRS) and its SNO \gaia of the Institut des Sciences de l’Univers (INSU), its Programmes Nationaux: Cosmologie et Galaxies (PNCG), Gravitation R\'{e}f\'{e}rences Astronomie M\'{e}trologie (PNGRAM), Plan\'{e}tologie (PNP), Physique et Chimie du Milieu Interstellaire (PCMI), and Physique Stellaire (PNPS), the `Action F\'{e}d\'{e}ratrice \gaia' of the Observatoire de Paris, the R\'{e}gion de Franche-Comt\'{e}, the Institut National Polytechnique (INP) and the Institut National de Physique nucl\'{e}aire et de Physique des Particules (IN2P3) co-funded by CNES;
\item the German Aerospace Agency (Deutsches Zentrum f\"{u}r Luft- und Raumfahrt e.V., DLR) through grants 50QG0501, 50QG0601, 50QG0602, 50QG0701, 50QG0901, 50QG1001, 50QG1101, 50\-QG1401, 50QG1402, 50QG1403, 50QG1404, 50QG1904, 50QG2101, 50QG2102, and 50QG2202, and the Centre for Information Services and High Performance Computing (ZIH) at the Technische Universit\"{a}t Dresden for generous allocations of computer time;
\item the Hungarian Academy of Sciences through the Lend\"{u}let Programme grants LP2014-17 and LP2018-7 and the Hungarian National Research, Development, and Innovation Office (NKFIH) through grant KKP-137523 (`SeismoLab');
\item the Science Foundation Ireland (SFI) through a Royal Society - SFI University Research Fellowship (M.~Fraser);
\item the Israel Ministry of Science and Technology through grant 3-18143 and the Tel Aviv University Center for Artificial Intelligence and Data Science (TAD) through a grant;
\item the Agenzia Spaziale Italiana (ASI) through contracts I/037/08/0, I/058/10/0, 2014-025-R.0, 2014-025-R.1.2015, and 2018-24-HH.0 to the Italian Istituto Nazionale di Astrofisica (INAF), contract 2014-049-R.0/1/2 to INAF for the Space Science Data Centre (SSDC, formerly known as the ASI Science Data Center, ASDC), contracts I/008/10/0, 2013/030/I.0, 2013-030-I.0.1-2015, and 2016-17-I.0 to the Aerospace Logistics Technology Engineering Company (ALTEC S.p.A.), INAF, and the Italian Ministry of Education, University, and Research (Ministero dell'Istruzione, dell'Universit\`{a} e della Ricerca) through the Premiale project `MIning The Cosmos Big Data and Innovative Italian Technology for Frontier Astrophysics and Cosmology' (MITiC);
\item the Netherlands Organisation for Scientific Research (NWO) through grant NWO-M-614.061.414, through a VICI grant (A.~Helmi), and through a Spinoza prize (A.~Helmi), and the Netherlands Research School for Astronomy (NOVA);
\item the Polish National Science Centre through HARMONIA grant 2018/30/M/ST9/00311 and DAINA grant 2017/27/L/ST9/03221 and the Ministry of Science and Higher Education (MNiSW) through grant DIR/WK/2018/12;
\item the Portuguese Funda\c{c}\~{a}o para a Ci\^{e}ncia e a Tecnologia (FCT) through national funds, grants SFRH/\-BD/128840/2017 and PTDC/FIS-AST/30389/2017, and work contract DL 57/2016/CP1364/CT0006, the Fundo Europeu de Desenvolvimento Regional (FEDER) through grant POCI-01-0145-FEDER-030389 and its Programa Operacional Competitividade e Internacionaliza\c{c}\~{a}o (COMPETE2020) through grants UIDB/04434/2020 and UIDP/04434/2020, and the Strategic Programme UIDB/\-00099/2020 for the Centro de Astrof\'{\i}sica e Gravita\c{c}\~{a}o (CENTRA);  
\item the Slovenian Research Agency through grant P1-0188;
\item the Spanish Ministry of Economy (MINECO/FEDER, UE), the Spanish Ministry of Science and Innovation (MICIN), the Spanish Ministry of Education, Culture, and Sports, and the Spanish Government through grants BES-2016-078499, BES-2017-083126, BES-C-2017-0085, ESP2016-80079-C2-1-R, ESP2016-80079-C2-2-R, FPU16/03827, PDC2021-121059-C22, RTI2018-095076-B-C22, and TIN2015-65316-P (`Computaci\'{o}n de Altas Prestaciones VII'), the Juan de la Cierva Incorporaci\'{o}n Programme (FJCI-2015-2671 and IJC2019-04862-I for F.~Anders), the Severo Ochoa Centre of Excellence Programme (SEV2015-0493), and MICIN/AEI/10.13039/501100011033 (and the European Union through European Regional Development Fund `A way of making Europe') through grant RTI2018-095076-B-C21, the Institute of Cosmos Sciences University of Barcelona (ICCUB, Unidad de Excelencia `Mar\'{\i}a de Maeztu’) through grant CEX2019-000918-M, the University of Barcelona's official doctoral programme for the development of an R+D+i project through an Ajuts de Personal Investigador en Formaci\'{o} (APIF) grant, the Spanish Virtual Observatory through project AyA2017-84089, the Galician Regional Government, Xunta de Galicia, through grants ED431B-2021/36, ED481A-2019/155, and ED481A-2021/296, the Centro de Investigaci\'{o}n en Tecnolog\'{\i}as de la Informaci\'{o}n y las Comunicaciones (CITIC), funded by the Xunta de Galicia and the European Union (European Regional Development Fund -- Galicia 2014-2020 Programme), through grant ED431G-2019/01, the Red Espa\~{n}ola de Supercomputaci\'{o}n (RES) computer resources at MareNostrum, the Barcelona Supercomputing Centre - Centro Nacional de Supercomputaci\'{o}n (BSC-CNS) through activities AECT-2017-2-0002, AECT-2017-3-0006, AECT-2018-1-0017, AECT-2018-2-0013, AECT-2018-3-0011, AECT-2019-1-0010, AECT-2019-2-0014, AECT-2019-3-0003, AECT-2020-1-0004, and DATA-2020-1-0010, the Departament d'Innovaci\'{o}, Universitats i Empresa de la Generalitat de Catalunya through grant 2014-SGR-1051 for project `Models de Programaci\'{o} i Entorns d'Execuci\'{o} Parallels' (MPEXPAR), and Ramon y Cajal Fellowship RYC2018-025968-I funded by MICIN/AEI/10.13039/501100011033 and the European Science Foundation (`Investing in your future');
\item the Swedish National Space Agency (SNSA/Rymdstyrelsen);
\item the Swiss State Secretariat for Education, Research, and Innovation through the Swiss Activit\'{e}s Nationales Compl\'{e}mentaires and the Swiss National Science Foundation through an Eccellenza Professorial Fellowship (award PCEFP2\_194638 for R.~Anderson);
\item the United Kingdom Particle Physics and Astronomy Research Council (PPARC), the United Kingdom Science and Technology Facilities Council (STFC), and the United Kingdom Space Agency (UKSA) through the following grants to the University of Bristol, the University of Cambridge, the University of Edinburgh, the University of Leicester, the Mullard Space Sciences Laboratory of University College London, and the United Kingdom Rutherford Appleton Laboratory (RAL): PP/D006511/1, PP/D006546/1, PP/D006570/1, ST/I000852/1, ST/J005045/1, ST/K00056X/1, ST/\-K000209/1, ST/K000756/1, ST/L006561/1, ST/N000595/1, ST/N000641/1, ST/N000978/1, ST/\-N001117/1, ST/S000089/1, ST/S000976/1, ST/S000984/1, ST/S001123/1, ST/S001948/1, ST/\-S001980/1, ST/S002103/1, ST/V000969/1, ST/W002469/1, ST/W002493/1, ST/W002671/1, ST/W002809/1, and EP/V520342/1.
\end{itemize}

The \gaia\ project and data processing have made use of:
\begin{itemize}
\item the Set of Identifications, Measurements, and Bibliography for Astronomical Data \citep[SIMBAD,][]{2000AAS..143....9W}, the `Aladin sky atlas' \citep{2000A&AS..143...33B,2014ASPC..485..277B}, and the VizieR catalogue access tool \citep{2000A&AS..143...23O}, all operated at the Centre de Donn\'{e}es astronomiques de Strasbourg (\href{http://cds.u-strasbg.fr/}{CDS});
\item the National Aeronautics and Space Administration (NASA) Astrophysics Data System (\href{http://adsabs.harvard.edu/abstract_service.html}{ADS});
\item the SPace ENVironment Information System (SPENVIS), initiated by the Space Environment and Effects Section (TEC-EES) of ESA and developed by the Belgian Institute for Space Aeronomy (BIRA-IASB) under ESA contract through ESA’s General Support Technologies Programme (GSTP), administered by the BELgian federal Science Policy Office (BELSPO);
\item the software products \href{http://www.starlink.ac.uk/topcat/}{TOPCAT}, \href{http://www.starlink.ac.uk/stil}{STIL}, and \href{http://www.starlink.ac.uk/stilts}{STILTS} \citep{2005ASPC..347...29T,2006ASPC..351..666T};
\item Matplotlib \citep{Hunter:2007};
\item IPython \citep{PER-GRA:2007};  
\item Astropy, a community-developed core Python package for Astronomy \citep{2018AJ....156..123A};
\item \textsc{SCIPY} \citep{Jones:2001aa},
\item \textsc{NUMPY} \citep{Oliphant2007}, 
\item \textsc{PANDAS} \citep{jeff_reback_2022_6408044},
\item R \citep{RManual};
\item Vaex \citep{2018A&A...618A..13B};
\item the \hip-2\ catalogue \citep{2007A&A...474..653V}. The \hip\ and \tyc\ catalogues were constructed under the responsibility of large scientific teams collaborating with ESA. The Consortia Leaders were Lennart Lindegren (Lund, Sweden: NDAC) and Jean Kovalevsky (Grasse, France: FAST), together responsible for the \hip\ Catalogue; Erik H{\o}g (Copenhagen, Denmark: TDAC) responsible for the \tyc\ Catalogue; and Catherine Turon (Meudon, France: INCA) responsible for the \hip\ Input Catalogue (HIC);  
\item the \tyctwo\ catalogue \citep{2000A&A...355L..27H}, the construction of which was supported by the Velux Foundation of 1981 and the Danish Space Board;
\item The \tyc\ double star catalogue \citep[TDSC,][]{2002A&A...384..180F}, based on observations made with the ESA \hip\ astrometry satellite, as supported by the Danish Space Board and the United States Naval Observatory through their double-star programme;
\item data products from the Two Micron All Sky Survey \citep[2MASS,][]{2006AJ....131.1163S}, which is a joint project of the University of Massachusetts and the Infrared Processing and Analysis Center (IPAC) / California Institute of Technology, funded by the National Aeronautics and Space Administration (NASA) and the National Science Foundation (NSF) of the USA;
\item the ninth data release of the AAVSO Photometric All-Sky Survey (\href{https://www.aavso.org/apass}{APASS}, \citealt{apass9}), funded by the Robert Martin Ayers Sciences Fund;
\item the first data release of the Pan-STARRS survey \citep{panstarrs1,panstarrs1b,panstarrs1c,panstarrs1d,panstarrs1e,panstarrs1f}. The Pan-STARRS1 Surveys (PS1) and the PS1 public science archive have been made possible through contributions by the Institute for Astronomy, the University of Hawaii, the Pan-STARRS Project Office, the Max-Planck Society and its participating institutes, the Max Planck Institute for Astronomy, Heidelberg and the Max Planck Institute for Extraterrestrial Physics, Garching, The Johns Hopkins University, Durham University, the University of Edinburgh, the Queen's University Belfast, the Harvard-Smithsonian Center for Astrophysics, the Las Cumbres Observatory Global Telescope Network Incorporated, the National Central University of Taiwan, the Space Telescope Science Institute, the National Aeronautics and Space Administration (NASA) through grant NNX08AR22G issued through the Planetary Science Division of the NASA Science Mission Directorate, the National Science Foundation through grant AST-1238877, the University of Maryland, Eotvos Lorand University (ELTE), the Los Alamos National Laboratory, and the Gordon and Betty Moore Foundation;
\item the second release of the Guide Star Catalogue \citep[GSC2.3,][]{2008AJ....136..735L}. The Guide Star Catalogue II is a joint project of the Space Telescope Science Institute (STScI) and the Osservatorio Astrofisico di Torino (OATo). STScI is operated by the Association of Universities for Research in Astronomy (AURA), for the National Aeronautics and Space Administration (NASA) under contract NAS5-26555. OATo is operated by the Italian National Institute for Astrophysics (INAF). Additional support was provided by the European Southern Observatory (ESO), the Space Telescope European Coordinating Facility (STECF), the International GEMINI project, and the European Space Agency (ESA) Astrophysics Division (nowadays SCI-S);
\item the eXtended, Large (XL) version of the catalogue of Positions and Proper Motions \citep[PPM-XL,][]{2010AJ....139.2440R};
\item data products from the Wide-field Infrared Survey Explorer (WISE), which is a joint project of the University of California, Los Angeles, and the Jet Propulsion Laboratory/California Institute of Technology, and NEOWISE, which is a project of the Jet Propulsion Laboratory/California Institute of Technology. WISE and NEOWISE are funded by the National Aeronautics and Space Administration (NASA);
\item the first data release of the United States Naval Observatory (USNO) Robotic Astrometric Telescope \citep[URAT-1,][]{urat1};
\item the fourth data release of the United States Naval Observatory (USNO) CCD Astrograph Catalogue \citep[UCAC-4,][]{2013AJ....145...44Z};
\item the sixth and final data release of the Radial Velocity Experiment \citep[RAVE DR6,][]{2020AJ....160...83S,rave6a}. Funding for RAVE has been provided by the Leibniz Institute for Astrophysics Potsdam (AIP), the Australian Astronomical Observatory, the Australian National University, the Australian Research Council, the French National Research Agency, the German Research Foundation (SPP 1177 and SFB 881), the European Research Council (ERC-StG 240271 Galactica), the Istituto Nazionale di Astrofisica at Padova, the Johns Hopkins University, the National Science Foundation of the USA (AST-0908326), the W.M.\ Keck foundation, the Macquarie University, the Netherlands Research School for Astronomy, the Natural Sciences and Engineering Research Council of Canada, the Slovenian Research Agency, the Swiss National Science Foundation, the Science \& Technology Facilities Council of the UK, Opticon, Strasbourg Observatory, and the Universities of Basel, Groningen, Heidelberg, and Sydney. The RAVE website is at \url{https://www.rave-survey.org/};
\item the first data release of the Large sky Area Multi-Object Fibre Spectroscopic Telescope \citep[LAMOST DR1,][]{LamostDR1};
\item the K2 Ecliptic Plane Input Catalogue \citep[EPIC,][]{epic-2016ApJS..224....2H};
\item the ninth data release of the Sloan Digitial Sky Survey \citep[SDSS DR9,][]{SDSS9}. Funding for SDSS-III has been provided by the Alfred P. Sloan Foundation, the Participating Institutions, the National Science Foundation, and the United States Department of Energy Office of Science. The SDSS-III website is \url{http://www.sdss3.org/}. SDSS-III is managed by the Astrophysical Research Consortium for the Participating Institutions of the SDSS-III Collaboration including the University of Arizona, the Brazilian Participation Group, Brookhaven National Laboratory, Carnegie Mellon University, University of Florida, the French Participation Group, the German Participation Group, Harvard University, the Instituto de Astrof\'{\i}sica de Canarias, the Michigan State/Notre Dame/JINA Participation Group, Johns Hopkins University, Lawrence Berkeley National Laboratory, Max Planck Institute for Astrophysics, Max Planck Institute for Extraterrestrial Physics, New Mexico State University, New York University, Ohio State University, Pennsylvania State University, University of Portsmouth, Princeton University, the Spanish Participation Group, University of Tokyo, University of Utah, Vanderbilt University, University of Virginia, University of Washington, and Yale University;
\item the thirteenth release of the Sloan Digital Sky Survey \citep[SDSS DR13,][]{2017ApJS..233...25A}. Funding for SDSS-IV has been provided by the Alfred P. Sloan Foundation, the United States Department of Energy Office of Science, and the Participating Institutions. SDSS-IV acknowledges support and resources from the Center for High-Performance Computing at the University of Utah. The SDSS web site is \url{https://www.sdss.org/}. SDSS-IV is managed by the Astrophysical Research Consortium for the Participating Institutions of the SDSS Collaboration including the Brazilian Participation Group, the Carnegie Institution for Science, Carnegie Mellon University, the Chilean Participation Group, the French Participation Group, Harvard-Smithsonian Center for Astrophysics, Instituto de Astrof\'isica de Canarias, The Johns Hopkins University, Kavli Institute for the Physics and Mathematics of the Universe (IPMU) / University of Tokyo, the Korean Participation Group, Lawrence Berkeley National Laboratory, Leibniz Institut f\"ur Astrophysik Potsdam (AIP),  Max-Planck-Institut f\"ur Astronomie (MPIA Heidelberg), Max-Planck-Institut f\"ur Astrophysik (MPA Garching), Max-Planck-Institut f\"ur Extraterrestrische Physik (MPE), National Astronomical Observatories of China, New Mexico State University, New York University, University of Notre Dame, Observat\'ario Nacional / MCTI, The Ohio State University, Pennsylvania State University, Shanghai Astronomical Observatory, United Kingdom Participation Group, Universidad Nacional Aut\'onoma de M\'{e}xico, University of Arizona, University of Colorado Boulder, University of Oxford, University of Portsmouth, University of Utah, University of Virginia, University of Washington, University of Wisconsin, Vanderbilt University, and Yale University;
\item the second release of the SkyMapper catalogue \citep[SkyMapper DR2,][Digital Object Identifier 10.25914/5ce60d31ce759]{2019PASA...36...33O}. The national facility capability for SkyMapper has been funded through grant LE130100104 from the Australian Research Council (ARC) Linkage Infrastructure, Equipment, and Facilities (LIEF) programme, awarded to the University of Sydney, the Australian National University, Swinburne University of Technology, the University of Queensland, the University of Western Australia, the University of Melbourne, Curtin University of Technology, Monash University, and the Australian Astronomical Observatory. SkyMapper is owned and operated by The Australian National University's Research School of Astronomy and Astrophysics. The survey data were processed and provided by the SkyMapper Team at the  Australian National University. The SkyMapper node of the All-Sky Virtual Observatory (ASVO) is hosted at the National Computational Infrastructure (NCI). Development and support the SkyMapper node of the ASVO has been funded in part by Astronomy Australia Limited (AAL) and the Australian Government through the Commonwealth's Education Investment Fund (EIF) and National Collaborative Research Infrastructure Strategy (NCRIS), particularly the National eResearch Collaboration Tools and Resources (NeCTAR) and the Australian National Data Service Projects (ANDS);
\item the \gaia-ESO Public Spectroscopic Survey \citep[GES,][]{GES_final_release_paper_1,GES_final_release_paper_2}. The \gaia-ESO Survey is based on data products from observations made with ESO Telescopes at the La Silla Paranal Observatory under programme ID 188.B-3002. Public data releases are available through the \href{https://www.gaia-eso.eu/data-products/public-data-releases}{ESO Science Portal}. The project has received funding from the Leverhulme Trust (project RPG-2012-541), the European Research Council (project ERC-2012-AdG 320360-Gaia-ESO-MW), and the Istituto Nazionale di Astrofisica, INAF (2012: CRA 1.05.01.09.16; 2013: CRA 1.05.06.02.07).
\end{itemize}

The GBOT programme  uses observations collected at (i) the European Organisation for Astronomical Research in the Southern Hemisphere (ESO) with the VLT Survey Telescope (VST), under ESO programmes
092.B-0165,
093.B-0236,
094.B-0181,
095.B-0046,
096.B-0162,
097.B-0304,
098.B-0030,
099.B-0034,
0100.B-0131,
0101.B-0156,
0102.B-0174, and
0103.B-0165;
%
%
and (ii) the Liverpool Telescope, which is operated on the island of La Palma by Liverpool John Moores University in the Spanish Observatorio del Roque de los Muchachos of the Instituto de Astrof\'{\i}sica de Canarias with financial support from the United Kingdom Science and Technology Facilities Council, and (iii) telescopes of the Las Cumbres Observatory Global Telescope Network.


%
\section{Parameters describing the orbital motion}\label{sec:model_parameters}
%
The parameters relative to the orbital motion as presented in the \TBOTable
are introduced in what follows. More details can be found
in the online documentation and articles accompanying the data release
\cite{DR3-DPACP-163, DR3-DPACP-176, DR3-DPACP-178, DR3-DPACP-179} for astrometric,  spectroscopic, and eclipsing binaries respectively.

%
\subsection{Astrometry}\label{ssec:model_parameters_astrometry}
%
The \gaia along-scan astrometric measurement $w$ (abscissa) for a binary system can be modelled by the combination of a single-source model $w_\mathrm{ss}$, describing the standard astrometric motion of the system's barycentre, and a Keplerian model $w_\mathrm{k1}$.\\
The single-source model can be written as 
\begin{equation}\label{eq:single_source_model}
w_\mathrm{ss} = (  \Delta\alpha^{\star} + \mu_{\alpha^\star} \, t ) \, \sin \psi + (  \Delta\delta + \mu_\delta \, t ) \, \cos \psi + \varpi \, f_\varpi, 
\end{equation}
where $ \Delta\alpha^{\star}= \Delta\alpha \cos{\delta}$ and $\Delta\delta$ are small offsets in equatorial coordinates from some fixed reference point, $\mu_{\alpha^\star}$, $\mu_\delta$ are proper motions in these coordinates, $t$ is time, $\varpi$ is the parallax, $f_\varpi$ is the parallax factor, and $\psi$ is the scan angle. The astrometric motion corresponding to a Keplerian orbit of a binary system has generally seven independent parameters, the Campbell elements. These
are the period $P$, the epoch of periastron passage $T_0$, the eccentricity $e$, the inclination $i$, the ascending node $\Omega$, the argument of periastron $\omega$, and the semi-major axis of the photocentre $a_0$.
The Thiele-Innes coefficients $A, B, F, G$ are defined as
\begin{equation}
\begin{array}{ll}
\! \! A \, =& \! \! \! a_0 \; (\cos \omega \cos \Omega - \sin \omega \sin \Omega \cos i) \\
\! \! B \, =& \! \! \! a_0 \; (\cos \omega \sin \Omega + \sin \omega \cos \Omega \cos i) \\
\! \! F \, =& \! \! -a_0 \; (\sin \omega \cos \Omega + \cos \omega \sin \Omega \cos i) \\
\! \! G \, =& \! \! -a_0 \; (\sin \omega \sin \Omega - \cos \omega \cos \Omega \cos i). 
\end{array}
\label{eq:cu4nss_astrobin_orbital_ABFG}
\end{equation} 
The elliptical rectangular coordinates $X$ and $Y$ are functions of eccentric anomaly $E$ and eccentricity: 
\begin{eqnarray}
E - e \sin E &=& \frac{2\pi}{P} (t-T_0)\\
X &=& \cos E - e\\
Y &=& \sqrt{1-e^2} \sin E \, \, .
\end{eqnarray}
The single Keplerian model can then be written as 
\begin{equation}\label{eq:k1_model}
w_\mathrm{k1} = (B \, X + G \, Y) \sin \psi + (A \, X + F \, Y) \cos \psi.
\end{equation}
The combined model $w^\mathrm{(model)}$ for the \gaia along-scan astrometry is
\begin{equation}\label{eq:abscissa1}
\begin{split}
w^\mathrm{(model)} =&\, w_\mathrm{ss} + w_\mathrm{k1} \\
 =&\, ( \Delta\alpha^{\star} + \mu_{\alpha^\star} \, t ) \, \sin \psi + ( \Delta\delta + \mu_\delta \, t ) \, \cos \psi + \varpi \, f_\varpi \\
 &+\, (B \, X + G \, Y) \sin \psi + (A \, X + F \, Y) \cos \psi.
\end{split}\end{equation}
This model has been extensively used for modelling the \hip epoch data of non-single stars \citep[e.g.][]{Sahlmann:2011fk}.

%
\subsection{Spectroscopy}\label{ssec:model_parameters_spectroscopy}
%
The radial motion of the primary is given by
\begin{equation}
\mathrm{RV}_1(t) \, = \, \gamma \, + \, K_1 \left[ \cos(v(t)+\omega) \, + \, e \, \cos(\omega) \, \right],
\end{equation}
with $v(t)$ the true anomaly deriving from the eccentric anomaly $E$ by
\begin{equation}
\tan{\frac{v}{2}} \, = \, \sqrt{\frac{1+e}{1-e}} \, \, \tan{\frac{E}{2}},
\end{equation}
and with the semi-amplitude of the primary (resp. secondary)
\begin{eqnarray}
\label{eq:semi-amplK12}
K_{1,2} = {\frac{\kappa}{P}} {\frac{a_{1,2}\sin i}{\sqrt{1-e^2}}},
\end{eqnarray}
where $\kappa \sim {10879}$ when $K_i$ is expressed in {\kms}, $P$ in days
and $a_i$ the semi-major axis of the primary (resp. secondary) in au.

This model has been used to produce the NSS solutions for \fieldName{SB1} and \fieldName{SB2}.
When astrometry and spectroscopy have been combined 
(\fieldName{AstroSpectroSB1} solutions),
in order to avoid correlations with the Campbell elements,
it was easier to complete the Thiele-Innes elements with 
\begin{equation}
\begin{array}{ll}
\! \! C \, =& \! \! +a_1\sin\omega\sin i \\
\! \! H \, =& \! \! +a_1\cos\omega\sin i ,
\end{array}
\label{equ:cu4nss_combined_processing_ti}
\end{equation} 
and these parameters have been published instead.

%
\section{Analytic orbit detection sensitivity as a function of orbit inclination \label{sec:appendix_analytical_signal_dispersion}}
%
\begin{figure*}[htb]\begin{center}
  \includegraphics[width=0.7\textwidth]{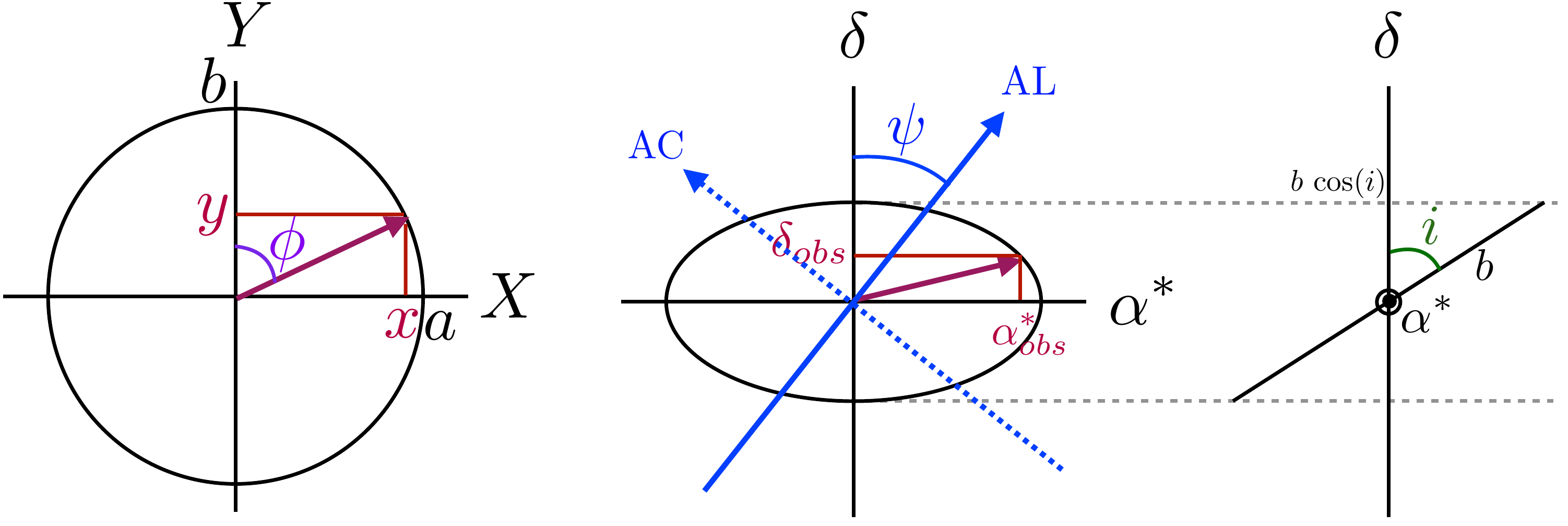} 
 \caption{\emph{Left:} Orbital plane ellipse with width $a$ and height $b$ (the circular case $a=b=1$ is shown). \emph{Middle:} Sky-projected plane with scan angle $\psi$ and projected orbit due to inclination $i$ shown in the right panel.}
\label{fig:inclModel}
\end{center}\end{figure*} 

Our toy orbit model is shown in \figref{fig:inclModel}. 
To keep the expressions simple we aligned the ellipse major-axis $a$ with $\alpha^*$ and introduced the inclination as rotation around the $\alpha^*$-axis. 
The orbital-phase ($\phi$)-dependent position $\left(x(t),y(t)\right)$ is projected to the on-sky position $\alpha^*,\delta$ via:
\begin{eqnarray}
\alpha^* (\phi) &=& x(\phi) \quad \quad \quad = a\, \sin{\phi} \\
\delta (\phi) &=& y(\phi) \ \cos{i} \  \ \ = b\, \cos{\phi} \ \cos{i}
\end{eqnarray} 
The orbit signal in the along-scan (AL, i.e.\ along $\psi$) and across-scan (AC, rotated 90 degrees counter-clockwise) directions is    
\begin{eqnarray}
\vec{d_\text{AL}} &=&
\begin{pmatrix}\alpha^* (\phi) \\  \delta \ \  (\phi)\end{pmatrix}  \begin{pmatrix}\sin{\psi} \\ \cos{\psi}\end{pmatrix} ^{T} \\
\vec{d_\text{AC}} &=&
\begin{pmatrix}\alpha^* (\phi) \\  \delta \ \  (\phi)\end{pmatrix}  \begin{pmatrix}-\cos{\psi} \\ \ \  \ \sin{\psi}\end{pmatrix} ^{T}
\end{eqnarray} 
To derive the RMS of $\vec{d_\text{AL}}$ and $\vec{d_\text{AC}}$, we make the following simplifying assumptions: (a) The orbit is circular, i.e.\ $a=b$: The angular velocity $\dot \phi$ is constant and integration over $\phi$ is straightforward.
(b) At each observation the orbital phase $\phi$ is random: The orbital-average value can be obtained  by simply integrating over $\phi$.
(c) At each observation the scan-angle $\psi$ is random. The mean square amplitude of $\vec{d_\text{AL}}$ for a given scan angle $\psi$ under assumptions (a) and (b) reduces to 
\begin{eqnarray}
\hat d^2_{\text{AL},\psi} &=& \frac{1}{2\pi}\int^{2\pi}_0 \ \| \vec{d_\text{AL}} \|^2 \ d\phi \nonumber \\ &=&\frac{1}{2\pi}   \int^{2\pi}_0 \ \left(  a \, \sin{\phi} \, \sin{\psi} +  a \, \cos{\phi} \, \cos{i} \, \cos{\psi} \right)^2 \ d\phi \nonumber \\
&=&  \frac{a^2}{2}  \left( \sin^2{\psi} +  \cos^2{\psi} \cos^2{i}  \right), 
\end{eqnarray} 
where we used the identity
\begin{equation}\label{ }
 \int^{2\pi}_0 \ \left(  u \, \sin{x} +  v \, \cos{x}  \right)^2 \ dx= \pi \left( u^2 + v^2 \right).
\end{equation}
Similarly we obtain in the AC direction
\begin{equation}
\hat d^2_{\text{AC},\psi} =  \frac{a^2}{2}  \left( \cos^2{\psi} +  \sin^2{\psi} \cos^2{i}  \right).
\end{equation}
Applying assumption (c) we get
\begin{eqnarray}
\hat d^2_{\text{AL}} &=& \frac{1}{2\pi}\int^{2\pi}_0 \ \hat d^2_{\text{AL},\psi} \ d\psi \nonumber \\
&=&  \frac{a^2}{4\pi} \, \int^{2\pi}_0 \ \sin^2{\psi} +  \cos^2{\psi} \cos^2{i} \ d\psi \nonumber \\
&=& \  \frac{a^2}{4}  \, \left( 1 +  \cos^2{i}  \right) 
\end{eqnarray} 
and in AC
\begin{equation}
\hat d^2_{\text{AC}} = \  \frac{a^2}{4}  \left( 1 +  \cos^2{i}  \right). 
\end{equation}
Assuming constant weights $W_\text{AL}$ and $W_\text{AC}$ for the AL and AC observations, respectively, the resulting RMS level is
\begin{equation}
\text{RMS}(i)  = \sqrt{ \hat d^2_{\text{AL}} + \hat d^2_{\text{AC}} } = \frac{a}{2} \, \sqrt{  \left( W_\text{AL} + W_\text{AC} \right) \, \left( 1 +  \cos^2{i}  \right) }. \label{eq:rmsInclDep}
\end{equation}
It is instructive to examine the two extreme scenarios:
\begin{enumerate}
  \item One-dimensional astrometry, i.e.\ $W_\text{AL}=1$ and $W_\text{AC}=0$, which approximates the \gdrthree\ astrometric solution: we obtain  
\begin{equation}\label{eq:signal_rms_analytic_1d}
\text{RMS}_\mathrm{1D}(i)  = \frac{a}{2} \, \sqrt{ 1 +  \cos^2{i} },
\end{equation}
which results in a $\sqrt{2}$ reduction for edge-on systems compared to the face-on configuration. 
  \item Two-dimensional astrometry with equal weights, i.e. $W_\text{AL}=W_\text{AC}=1$, which approximates conventional CCD imaging:
  \begin{equation}\label{eq:signal_rms_analytic_2d}
\text{RMS}_\mathrm{2D}(i)  = \frac{a}{2}  \, \sqrt{ 2 \left( 1 +  \cos^2{i}  \right)},
\end{equation}
i.e. same inclination-dependency as one-dimensional observations but with a $\sqrt{2}$ boost in RMS because the number of observations is doubled\footnote{This two-dimensional case was formally derived with a random rotation of the frame per observation, which does not correspond to fixed-axis orientation CCD observation. In our toy model, the scan-angle integral takes care of considering the random ($\Omega$-)orientation of the orbit since we assumed the rotation axis is aligned with $\alpha^\star$. Alternatively the sky rotation angle reminiscent of $\Omega$ could be introduced and integrated over to achieve this averaging, and then there is no need to integrate over scan angle for a fixed orientation telescope. As long as $\psi$ and/or $\Omega$ are assumed to be randomly oriented without range restriction, the simpler model used here is accurate.}.
  \end{enumerate}

%
\section{Biases and uncertainties of Thiele-Innes parameters}
%
%
\subsection{Biases introduced by fitting Thiele-Innes parameters to noisy data}\label{ssec:thiele_innes_biases}
%
All \gaia processing pipelines for fitting astrometric orbits\footnote{There are three flavours: (1) the \quoting{binary pipeline}  \citep{DR3-DPACP-163} (2) the Genetic Algorithm channel of the \quoting{exoplanet pipeline} \citep{DR3-DPACP-176} (3) the Markov Chain Monte Carlo channel of the \quoting{exoplanet pipeline}.} follow a similar parametrisation scheme \citep[\secref{ssec:model_parameters_astrometry}; DR3 documentation;][]{DR3-DPACP-163,DR3-DPACP-176}: the three non-linear parameters $P$, $e$, and $T_0$ are fitted using different algorithms but the four remaining parameters are represented by the Thiele-Innes coefficients $A, B, F, G$, which linearise part of the equations.

Since the features in the distributions of DR3 orbit parameters hinted at signal-dependent effects, we simulated the $A, B, F, G$ recovery performance as a function of noise level. We applied the same simulation approach as in \secref{ssec:astrometric_orbit_sensitivity} to compute the along-scan signal $w_\mathrm{k1}$ of the Keplerian orbit (\equref{eq:k1_model}) for an ensemble of binary systems with the same distance, period, and semi-major axis. We computed the rectangular coordinates $X$ and $Y$ according to the simulated parameters and then solved the linear equation
\begin{equation}\label{eq:abscissa2}
w_\mathrm{k1} + \epsilon_w= (B \, X + G \, Y) \sin \psi + (A \, X + F \, Y) \cos \psi 
\end{equation}
for the unknown $A, B, F, G$, where $\epsilon_w$ is a random noise term that we added to the noise-less simulated abscissa. For simplicity, we chose uniform abscissa uncertainties $\sigma_w$ and zero covariances. We used a standard matrix-inversion solver\footnote{\url{https://github.com/Johannes-Sahlmann/linearfit}} to determine the solution.

For the noise term $\epsilon_w$, we chose its amplitude relative to the simulated semi-major axis, which was the same for all simulated systems ($a_0=0.059$ mas, see \secref{ssec:astrometric_orbit_sensitivity}) but independent of the number of simulated observation epochs for a given source. For a signal-to-noise of S/N$ = 5$, for example, a random noise term was added to $w$ according to a normal distribution with dispersion $\epsilon_{w,5} = a_0/5$. This setup does intentionally not account for the inclination-dependent sensitivity (\secref{ssec:astrometric_orbit_sensitivity}).

\begin{figure}[htb]\begin{center}
\includegraphics[width = 0.8\columnwidth]{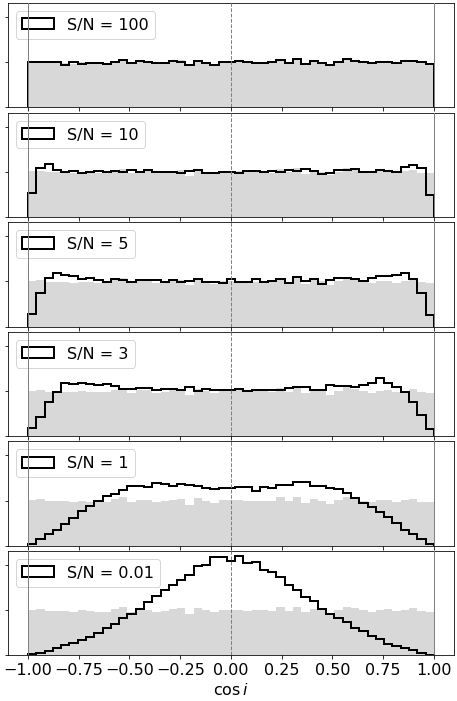}
\caption{Simulated (filled grey) and recovered (solid line) distributions of $\cos{i}$ for six levels of signal-to-noise as explained in the text. The vertical dashed line at $i=90\degr$ indicates the edge-on configurations.}
\label{fig:simulation_cosi_biases}\end{center}\end{figure}

Figure~\ref{fig:simulation_cosi_biases} show the simulated and recovered distributions of $\cos{i}$ for various levels of noise. At high S/N=100 the simulated distribution is recovered but for progressively smaller S/N a lack of face-on configurations starts to appear and the inclinations extracted from the fitted $A, B, F, G$ are biased towards edge-on configurations. At very small S/N=0.01, where the $A, B, F, G$  are essentially unconstrained except for their amplitude term dependent on $a_0$, the $\cos{i}$ distribution is peaked at $\cos{i}=0$, i.e.\ edge-on configurations.

The apparent dearth of face-on configurations for \fieldName{Orbital} solutions (\figref{fig:orbit_param_distributions_wds_comparison}) is therefore the consequence of extracting the Thiele-Innes parameters $A, B, F, G$ from noisy data. The dependence on S/N expected from the simulations is observed in the actual data as well (\figref{fig:orbit_param_distributions_200pc}). Since the DR3 solutions contain a continuum of S/N levels, the corresponding $\cos{i}$ distribution is a superposition of the distributions for individual S/N levels.

\begin{figure}[htb]\begin{center}
\includegraphics[width = 0.8\columnwidth]{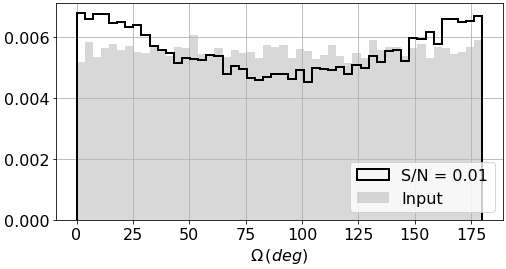}
\caption{Simulated (filled grey) and recovered (solid line) distributions of $\Omega$ at S/N=0.01.}
\label{fig:simulation_ascOMEGA_bias}\end{center}\end{figure}

Figure \ref{fig:simulation_ascOMEGA_bias} shows the recovered $\Omega$ distribution in the lowest S/N case we simulated. The suppression of orbits with $\Omega=90\degr$ observed for \fieldName{Orbital} solutions is reproduced by the simulation. This modulation becomes very weak for higher levels of S/N. We conclude that this feature can therefore also be attributed to a bias introduced by fitting the $A, B, F, G$ coefficients.

\begin{figure}[htb]\begin{center}
\includegraphics[width = 0.8\columnwidth]{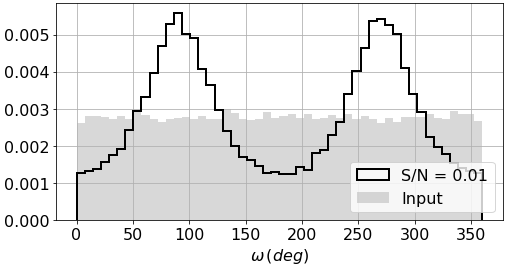}
\caption{Simulated (filled grey) and recovered (solid line) distributions of $\omega$ for non-circular orbits with $e=0.5$ at S/N=0.01.}
\label{fig:simulation_omega_bias}\end{center}\end{figure}

To simulate the effect on $\omega$ we simulated non-circular orbits with eccentricity $e=0.5$, all other parameters remained the same. Figure \ref{fig:simulation_omega_bias} shows a bimodal modulation of the recovered $\omega$ distribution.
However, the minima at $\omega=0\degr$ or $180\degr$ do not match the ones observed for \fieldName{Orbital} solutions at $\omega=90\degr$ or $270\degr$. Instead, they reproduce the minima seen in the orb6 solutions shown in \figref{fig:orbit_param_distributions_wds_comparison}.

%
\subsection{Selection effect causing the modulation in the \texorpdfstring{$\omega$}{omega} distribution}\label{ssec:omega_selection_effect}
\begin{figure}[htb]\begin{center}
\includegraphics[width = 0.9\columnwidth]{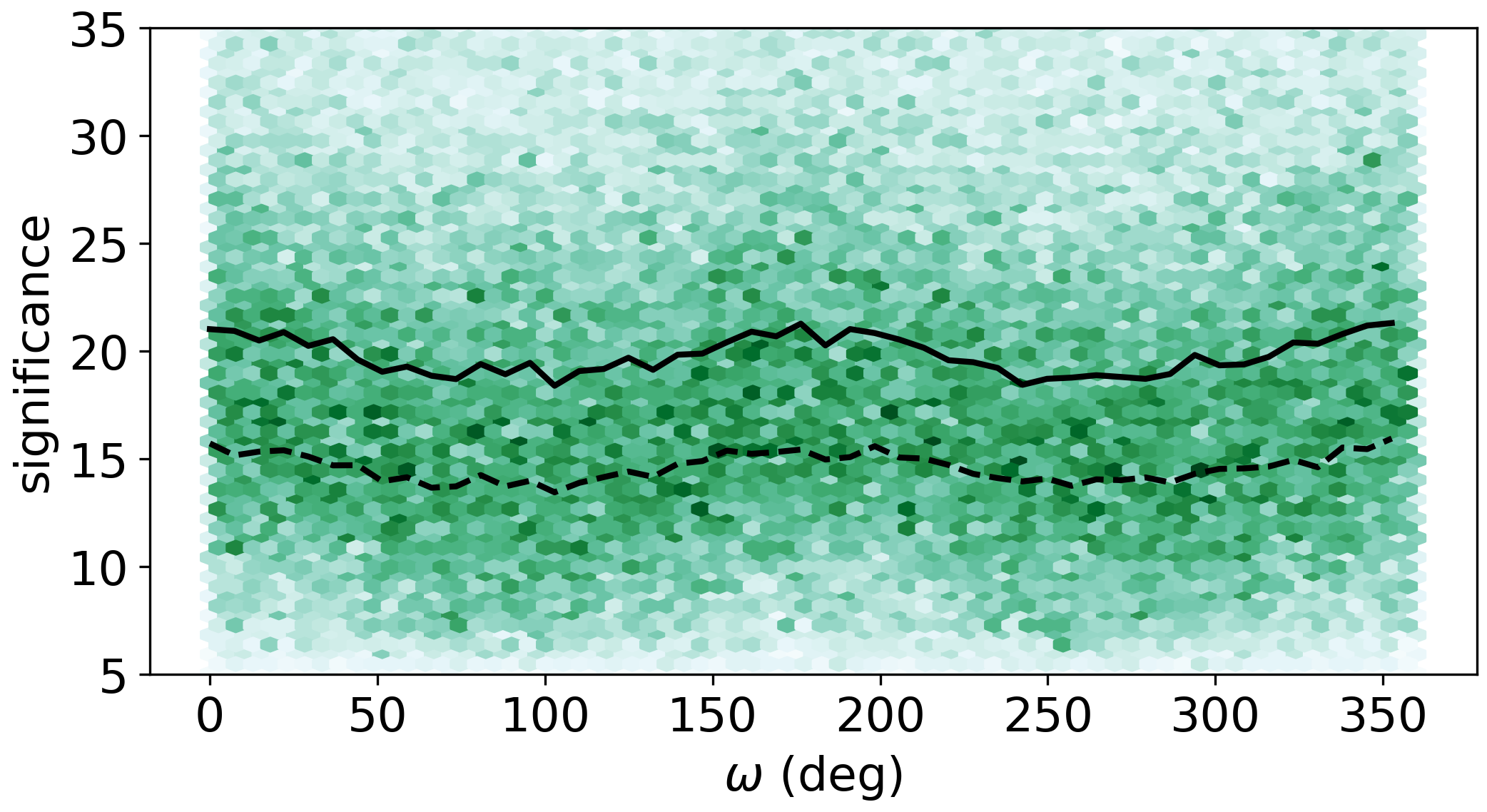}
\caption{Density histogram of published significance as a function of $\omega$ for DR3 \fieldName{Orbital} solutions. The solid curve shows the running median value. The dashed curve shows the median of the Monte-Carlo-resampled significance.}
\label{fig:significance_versus_omega}\end{center}\end{figure}

%
The accepted \fieldName{Orbital} solutions have to pass a period-dependent threshold on their significance, defined as the ratio between the semi-major axis and its uncertainty $a_0 / \sigma_{a_0}$ \citep{DR3-DPACP-163}. Figure \ref{fig:significance_versus_omega} shows that there is a bimodal correlation between the published significance and $\omega$ with minima at $\omega=90\degr$ or $270\degr$. The application of an $\omega$-independent significance threshold therefore leads to a suppression of solutions around these values. We argue that this is the explanation for the observed modulation in the $\omega$ distribution of \gaia \fieldName{Orbital} solutions.

%
\subsection{Geometric element conversion from Thiele-Innes parameters in the presence of uncertainties.}\label{ssec:thiele_innes_uncertainties}
%
The archive tables list the fitted astrometric-orbit parameters $P$, $e$, $T_0$, $A, B, F, G$ with their formal uncertainties and the \fieldName{corr_vec} field contains the correlation strengths between those parameters. When converting $A, B, F, G$ to geometric orbit elements and determining their uncertainties we therefore have to account for their covariances. This can be done with classical error propagation using some form of linearisation of the parameter dependencies or by Monte Carlo simulations. As also pointed out in \citet{DR3-DPACP-127}, the distributions of the geometric parameters are often non-Gaussian, which favours the latter approach.

We used \fieldName{Python} implementations of both methods\footnote{\url{https://github.com/Johannes-Sahlmann/pystrometry} \citep{johannes_sahlmann_2019_3515526}} to investigate the value and uncertainty estimates of $i$, $\Omega$, and $\omega$ for the non-circular \fieldName{Orbital} solution in DR3. 
One finding is that the inclination uncertainty generally increases towards face-on configurations and that the linearised uncertainty estimate is usually comparable to the Monte-Carlo estimate, except for large uncertainties $\gtrsim10\degr$ where the former is larger. There are only few cases where the discrepancy between the two different inclination estimates is larger than the linearised uncertainty.

The significance estimates using linearised error propagation can be different from alternative estimates that use Monte-Carlo resampling for obtaining $\sigma_{a_0}'$. 
These differences typically become important for solutions with poorly-constrained eccentricities (i.e.\ $e/\sigma_e < 1$) for which Monte Carlo resampling is not recommended because of overestimated variances of the Thiele-Innes
coefficients \citep{DR3-DPACP-127, DR3-DPACP-176}. 
In \figref{fig:significance_versus_omega} we see that the correlation between the Monte-Carlo-resampled semi-major axis significance and $\omega$ is slightly weaker, i.e.\ has a smaller-amplitude variation, than for the published significance.

%
\section{Primary mass computation\label{sec:homemademass}}
%

We use the PARSEC isochrones\footnote{http://stev.oapd.inaf.it/cmd V3.6} \citep{Bressan12} with ages $\tau$ by steps of 0.01 in $\log(\tau)$ and metallicity [M/H] from --2 to 0.4~dex by steps of 0.05. Considering that the mass distribution of the isochrones is not uniform and that the age sampling is in log, weights need to be applied to each isochrone point $i$:
$P(i) = P(\Mass,\tau,[M/H]) = P(\Mass) P(\tau) P([M/H])$. For $P(\Mass)$ we use the \citet{Chabrier01} IMF, for $P(\tau)$ a flat star formation rate and for $P([M/H])$ a Gaussian centered on zero with a dispersion of 0.05 so that by default the solar isochrones will drive the mass determination. 
The default mass sampling being too sparse at the bottom of the main sequence, we interpolated the isochrones with finer mass steps. We removed the lowest mass stars which present a step feature in the isochrone H-R diagram which is not observed in the \gdrthree\ HRD, so that the masses can only be derived for $M_G<14.4$ which corresponds to $\Mass>0.1$\Msun. We also removed points outside a very crude age-metallicity relation that is $[M/H]<-0.4-0.05\tau$ for $\tau>10$ or $[M/H]>0.5-0.05\tau$.  

Our observables are the absolute magnitude $M_G$ and the colour $(BP-RP)_0$, noted below $\tilde{O}$.
Through Bayes's theorem and considering that the isochrone point $i$ contains the $\Mass$ information, we have:
\begin{equation}
 P(\Mass|\tilde{O}) \propto \sum_i P(\tilde{O}|i) P(i)
 \label{eq:isocmass}
\end{equation}
To allow for small systematics in the \gaia photometry, we quadratically add a 0.01~mag error to the $G$,\, \gbp, \grp\ formal magnitude errors.  

To avoid the presence of outliers in \fieldName{azero_gspphot}, especially at the bottom of the main-sequence \citep{DR3-DPACP-127}, we use the 3D extinction map of \cite{Lallement2019} to provide an estimate of the extinction $A_0$. Most of the sources in the main-sequence are within the completeness limit of this map. A 10\% relative error on the extinction with a minimum error of 0.01 is assumed. We derive the extinction coefficients $k_G$, $k_{BP}$, $k_{RP}$ from the EDR3 extinction law\footnote{\url{https://www.cosmos.esa.int/web/gaia/edr3-extinction-law}}.

We know that the position on the H-R diagram provides a direct estimate on the mass only for main-sequence stars. We therefore remove all isochrone points with PARSEC label $=0$ (pre-main sequence stage) before applying Eq.~\ref{eq:isocmass} and only provide a mass estimate for stars with more than half of the isochrone points at 3$\sigma$ from the observables with PARSEC label=1 (main sequence stage).  When several flux ratios are tested, the label of the smallest valid one is used for the main-sequence star selection. Therefore no mass estimate is provided for giant stars and mass estimates for pre-main-sequence stars will not be valid.

For each flux ratio $F_2/F_1$ tested, we obtain the absolute magnitude to be compared with the isochrones with:
\begin{equation}
 M_G = G - k_G\ A_0+5+5 \log(\varpi/1000)+2.5 \log(1+\frac{F_2}{F_1})
\end{equation}
and the colour simply with $G_{\rm BP}-G_{\rm RP}-(k_{BP}-k_{RP}) A_0$. Note that we do not consider the change in colour induced by the presence of the secondary. The mass is then fully driven by $M_G$. We consider that the resulting mass distribution obtained through Eq.~\ref{eq:isocmass} is valid if we have more than 5 isochrone points within 3$\sigma$ in both magnitude and colour to be compared with our star and if the closest isochrone point has a $\chi^2$ p-value larger than 0.01. 

\begin{figure}[htb]\begin{center}
\includegraphics[width=0.9\columnwidth]{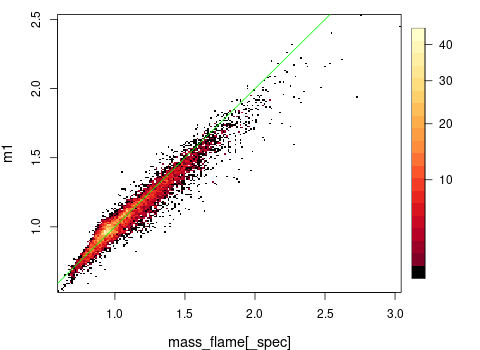}
\caption{Density plot of the comparison between the primary masses derived here with the \fieldName{FLAME} ones for systems with \dt{fluxratio} $<0.01$.}
\label{fig:massesVflame}
\end{center}\end{figure}

Our derived masses are compared with the \fieldName{FLAME} ones for systems with a small flux ratio in \figref{fig:massesVflame}. As expected those are fully consistent within the errors quoted, with less than 1\% of 5$\sigma$ outliers. For the small masses, we see the over-estimation trend of the \fieldName{FLAME} masses due to the over-estimation of the GSP-Phot extinction for stars with $M_G>7$  \citep{DR3-DPACP-127}, corresponding to $\Mass<0.7$\Msun. 

\begin{figure}[htb]\begin{center}
\includegraphics[width=0.9\columnwidth]{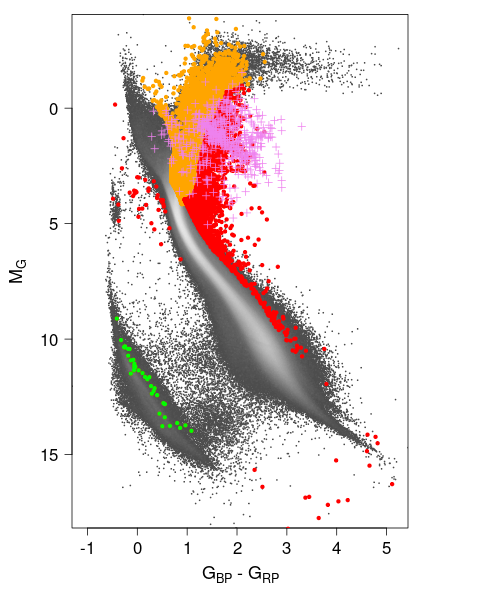}
\caption{H-R diagram of the NSS astrometric solutions without a primary mass estimate. The grey background is the HRD of the full DR3 low extinction stars \citep[$A_0<0.05$ according to][]{Lallement2019}, equivalent of Fig.~5 of \cite{DR2-DPACP-31}). The orange points are identified as giants. The red points do not have isochrone match. Violet + do not have an extinction estimate and un-corrected magnitude and colour are therefore used for those. Green points are the white dwarfs with a mass of 0.65$\pm0.16\Msun$ assumed for \tabref{tab:nssmass}.}
\label{fig:m1missed}
\end{center}\end{figure}

Figure~\ref{fig:m1missed} shows the location in the HRD of the sources with an astrometric solution but no primary mass estimate obtained from our procedure. They have all been corrected by the extinction except those without an extinction estimate. Apart from giants, young stars above the main sequence are missed by construction as well as a few sub-dwarfs and sources too faint to have a reliable \gbp.

%

\section{Acronyms}\label{sec:acronyms}
{\small
\begin{tabular}{ll}
\hline\hline
\textbf{Acronym}  &  \textbf{Description}  \\ 
2MASS & Two-Micron All Sky Survey \\
ADQL&Astronomical Data Query Language \\
AGB&Asymptotic Giant Branch (star) \\
AL & ALong scan (direction) \\
AP & Astrophysical Parameters \\
BH & Black hole \\
BP & Gaia Blue Photometer \\
CMD & Colour Magnitude Diagram \\
DoF&Degree(s) of Freedom \\
DPAC & Data Processing and Analysis Consortium \\
DR1 & Gaia Data Release 1 \\
DR2 & Gaia Data Release 2 \\
DR3 & Gaia Data Release 3 \\
EB&Eclipsing Binary \\
EDR3 & Gaia Early Data Release 3 \\
FLAME & Final Luminosity Age Mass Estimator \\
FWHM&Full Width at Half-Maximum \\
GALEX&GALaxy Evolution eXplorer \\
GoF & Goodness of Fit \\
GSPPhot&Generalised Stellar Parametriser PHOTometry \\
GSPSpec&Generalised Stellar Parametriser SPECtroscopy \\
GUCD & Gaia Ultra-cool Dwarf\\
HealPix & Hierarchical Equal Area isoLatitude Pixelisation \\
HPM & High Proper Motion \\
HRD & Hertzsprung-Russell diagram\\
IMF&Initial Mass Function \\
LMC&Large Magellanic Cloud \\
LPV & Long Period Variables \\
MAD & Median Absolute Deviation \\
MS&Main Sequence (star) \\
NLS & Non linear spectro \\
NS & Neutron star \\
NSS&Non-Single Star \\
PMa&proper motion anomaly \\
RGB&Red Giant Branch (star) \\
RMS&Root-Mean-Square \\
RP & Gaia Red Photometer \\
RUWE & Re-normalised unit-weight error \\
RV & Radial Velocity \\
RVS & Radial Velocity Spectrometer \\
SB & Spectroscopic Binary  \\
SB1 & Single-line Spectroscopic Binary  \\
SB2 & Double-line Spectroscopic Binary \\
SB*C & Circular orbit \\
SNR & Signal-to-Noise ratio (also denoted SN and S/N)\\
TBO & Two body orbits \\
UCD&Ultra cool dwarf (star) \\
UV&UltraViolet \\
VIMF & variable-induced movers fixed \\
WD & White dwarf\\
\hline
\end{tabular} 
}


%
\end{appendix}
\end{document}